\documentclass[final]{cdr}
\pdfoutput=1            
\graphicspath{ {figures/}{volume-physics/figures/}{volume-physics/generated/} }

\def\expshort{DUNE\xspace}
\def\explong{The Deep Underground Neutrino Experiment\xspace}

\def\thedocsubtitle{Long-Baseline Neutrino Facility (LBNF) and 
Deep Underground Neutrino Experiment (DUNE)}%
\def\cdrtitle{Conceptual Design Report}

\def\volphystitle{Volume 2: The Physics Program for DUNE at LBNF\xspace}

\def\volintro{Volume 1: \textit{The LBNF and DUNE Projects\xspace}}
\def\volphys{Volume 2: \textit{The Physics Program for DUNE at LBNF\xspace}}
\def\vollbnf{Volume 3: \textit{The Long-Baseline Neutrino Facility for DUNE\xspace}}
\def\voldune{Volume 4: \textit{The DUNE Detectors at LBNF\xspace}}

\newcommand{\dm}[1]{$\Delta m^2_{#1}$\xspace}
\newcommand{\sinstt}[1]{$\sin^22\theta_{#1}$\xspace}

\newcommand{\deltacp}{$\delta_{\rm CP}$\xspace}

\newcommand{\numu}{$\nu_\mu$\xspace}
\newcommand{\nue}{$\nu_e$\xspace}
\newcommand{\nutau}{$\nu_\tau$\xspace}
\newcommand{\anumu}{$\bar\nu_\mu$\xspace}
\newcommand{\anue}{$\bar\nu_e$\xspace}

\newcommand{\mdeltacp}{\delta_{\rm CP}}

\newcommand{\superk}{Super--Kamiokande\xspace}

\newcommand{\minerva}{MINER$\nu$A\xspace}

\def\Ar39{$^{39}$Ar}

\DeclareSIUnit \kton {\kilo\tonne}
\DeclareSIUnit \kt {\kilo\tonne}
\DeclareSIUnit \Mt {\mega\tonne}
\DeclareSIUnit \eV {\electronvolt}
\DeclareSIUnit \keV {\kilo\electronvolt}
\DeclareSIUnit \MeV {\mega\electronvolt}
\DeclareSIUnit \GeV {\giga\electronvolt}
\DeclareSIUnit \km {\kilo\meter}
\DeclareSIUnit \kW {\kilo\watt}
\DeclareSIUnit \MW {\mega\watt}
\DeclareSIUnit \MHz {\mega\hertz}
\DeclareSIUnit \mrad {\milli\radian}
\DeclareSIUnit \year {year}
\DeclareSIUnit \POT {POT}
\DeclareSIUnit \sig {$\sigma$}
\DeclareSIUnit\parsec{pc}
\DeclareSIUnit\lightyear{ly}
\DeclareSIUnit\foot{ft}
\DeclareSIUnit\ft{ft}
\def\ktyr{\si[inter-unit-product=\ensuremath{{}\cdot{}}]{\kt\year}\xspace}
\def\Mtyr{\si[inter-unit-product=\ensuremath{{}\cdot{}}]{\Mt\year}\xspace}

\def\ktMWyr{\si[inter-unit-product=\ensuremath{{}\cdot{}}]{\kt\MW\year}\xspace}

\newcommand{\SIadj}[2]{\SI[number-unit-product = -]{#1}{#2}}
\newcommand{\ktadj}[1]{\SIadj{#1}{\kt}}
\newcommand{\kmadj}[1]{\SIadj{#1}{\km}}
\newcommand{\keVadj}[1]{\SIadj{#1}{\keV}}
\newcommand{\MeVadj}[1]{\SIadj{#1}{\MeV}}
\newcommand{\GeVadj}[1]{\SIadj{#1}{\GeV}}
\newcommand{\MWadj}[1]{\SIadj{#1}{\MW}}

\hypersetup{
    pdftitle={\expshort CDR \thedocsubtitle},
    pdfauthor={\expshort Collaboration},
    final=true,
    colorlinks=false,
    linktocpage=true,
    linkbordercolor=blue,
    citebordercolor=green,
    urlbordercolor=magenta,
    filecolor=black,
    pdfpagemode=UseOutlines,
    pdfborderstyle={/S/U},  
}

\renewcommand\thedoctitle{\volphystitle}
\def\titleextra{\includegraphics[width=0.55\textwidth]{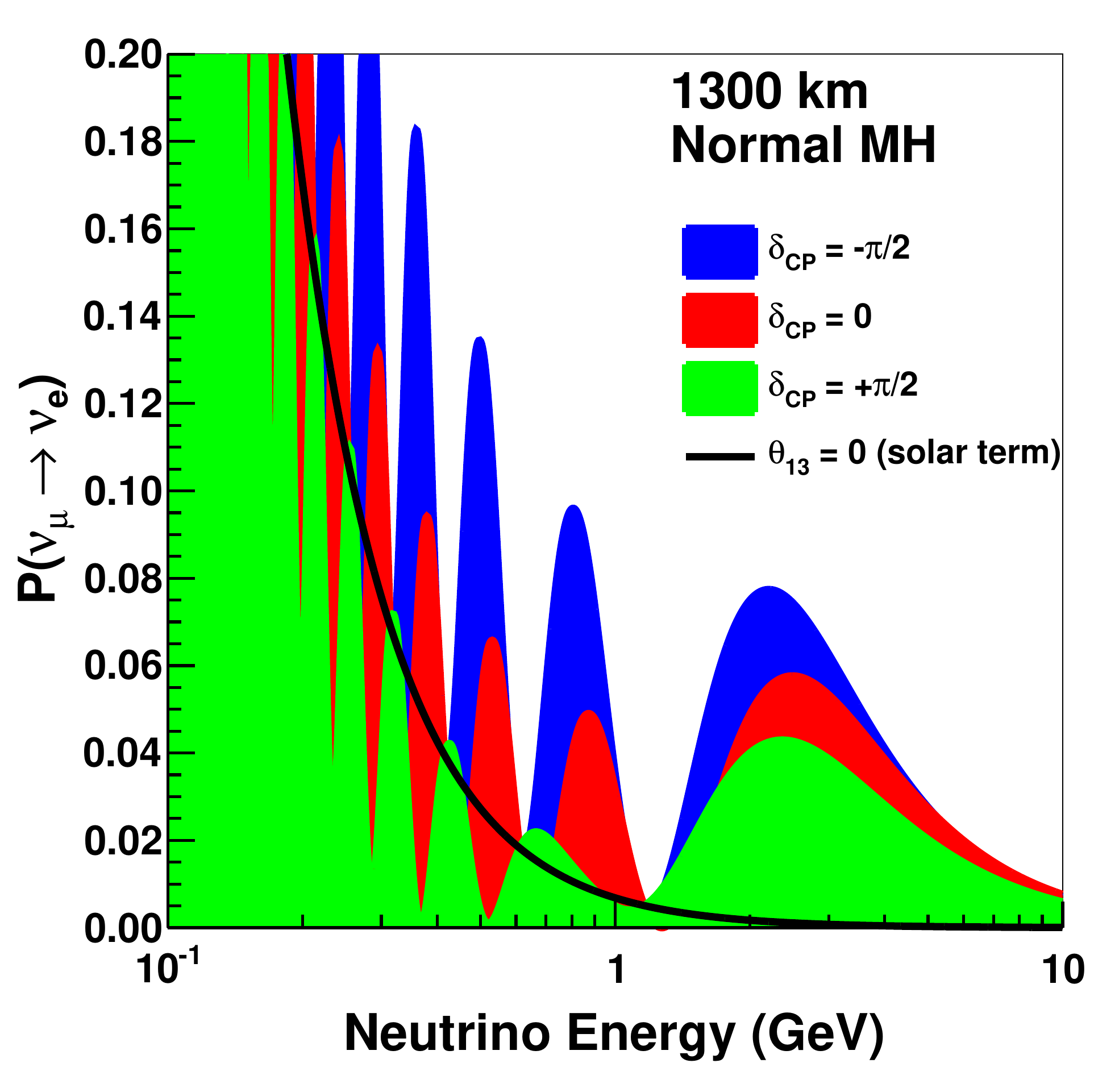}}
\begin{document}


\pagestyle{titlepage}

\begin{center}
   {\Huge  \thedocsubtitle}  

  \vspace{5mm}

  {\Huge  \cdrtitle}  

  \vspace{10mm}

 {\LARGE \thedoctitle}

  \vspace{15mm}

\titleextra

  \vspace{10mm}
  \today
    \vspace{15mm}
    
\end{center}

\cleardoublepage



# The DUNE Collaboration

R. Acciarri[43], M. A. Acero[10], M. Adamowski[43], C. Adams[144], P. Adamson[43], S. Adhikari[77], Z. Ahmad[135],
C. H. Albright[43], T. Alion[120], E. Amador[131], J. Anderson[8], K. Anderson[43], C. Andreopoulos[76], M. Andrews[43],
R. Andrews[43], I. Anghel[62], J. d. Anjos[99], A. Ankowski[137], M. Antonello[72], A. Aranda Fernandez[31], A. Ariga[13],
T. Ariga[13], D. Aristizabal[75], E. Arrieta-Diaz[87], K. Aryal[131], J. Asaadi[129], D. Asner[103], M. S. Athar[6], M. Auger[13],
A. Aurisano[29], V. Aushev[68], D. Autiero[63], M. Avila[131], J. J. Back[139], X. Bai[122], B. Baibussinov[106], M. Baird[61],
A. B. Balantekin[142], B. Baller[43], P. Ballett[41], B. Bambah[52], M. Bansal[107], S. Bansal[107], G. J. Barker[139],
W. A. Barletta[83], G. Barr[102], N. Barros[109], L. Bartoszek[78], A. Bashyal[131], M. Bass[102], F. Bay[133], J. Beacom[100],
B. R. Behera[107], G. Bellettini[110], V. Bellini[26], O. Beltramello[16], P. A. Benetti[108], A. Bercellie[114], M. Bergevin[19],
E. Berman[43], H. Berns[20], R. Bernstein[43], S. Bertolucci[16], B. Bhandari[49], V. Bhatnagar[107], B. Bhuyan[60], J. Bian[90],
K. Biery[43], M. Bishai[15], T. Blackburn[128], A. Blake[73], F. d. M. Blaszczyk[14], E. Blaufuss[82], B. Bleakley[124],
E. Blucher[28], V. Bocean[43], F. Boffelli[108], J. Boissevain[78], S. Bolognesi[117], T. Bolton[65], M. Bonesini[89], T. Boone[33],
C. Booth[118], S. Bordoni[53], M. Borysova[68], B. Bourguille[53], S. B. Boyd[139], D. Brailsford[73], A. Brandt[131],
J. Bremer[16], S. Brice[43], C. Bromberg[87], G. Brooijmans[34], G. Brown[131], R. Brown[15], G. Brunetti[43], X. Bu[43],
N. Buchanan[33], H. Budd[114], B. Bugg[130], P. Calafiura[74], E. Calligarich[108], E. Calvo[17], L. Camilleri[34],
M. Campanelli[77], C. Cantini[147], B. Carls[43], R. Carr[34], M. Cascella[77], C. Castromonte[44], E. Catano Mur[62],
F. Cavanna[43], S. Centro[106], A. Cervera Villanueva[136], V. B. Chandratre[12], A. Chatterjee[131], S. Chattopadhyay[43],
S. Chattopadhyay[135], L. Chaussard[63], S. Chembra[52], H. Chen[15], K. Chen[15], M. Chen[21], D. Cherdack[33], C. Chi[34],
S. Childress[43], S. Choubey[47], B. C. Choudhary[38], G. Christodoulou[76], C. Christofferson[122], E. Church[103],
D. Cianci[81], D. Cline[21], T. Coan[125], A. Cocco[93], J. Coelho[134], P. Cole[58], G. Collin[83], J. M. Conrad[83],
M. Convery[115], R. Corey[122], L. Corwin[122], J. Cranshaw[8], P. Crivelli[147], D. Cronin-Hennessy[90], A. Curioni[89],
J. Cushing[28], D. L. Adams[15], D. Dale[58], S. R. Das[60], T. Davenne[116], G. S. Davies[61], J. Davies[128], J. Dawson[2],
K. De[131], A. de Gouvea[97], J. K. de Jong[102], P. de Jong[92], P. De Lurgio[8], M. Decowski[92], A. Delbart[117],
C. Densham[116], R. Dharmapalan[8], N. Dhingra[107], S. Di Luise[147], M. Diamantopoulou[9], J. S. Diaz[61],
G. Diaz Bautista[114], M. Diwan[15], Z. Djurcic[8], J. Dolph[15], G. Drake[8], D. Duchesneau[71], M. Duvernois[142],
H. Duyang[120], D. A. Dwyer[74], S. Dye[48], S. Dytman[111], B. Eberly[115], R. Edgecock[50], D. Edmunds[87], S. Elliott[78],
M. Elnimr[21], S. Emery[117], E. Endress[105], S. Eno[82], A. Ereditato[13], C. O. Escobar[43], J. Evans[81], A. Falcone[108],
L. Falk[128], A. Farbin[131], C. Farnese[106], Y. Farzan[57], A. Fava[106], L. Favilli[18], J. Felde[82], J. Felix[46], S. Fernandes[4],
L. Fields[43], A. Finch[73], M. Fitton[116], B. Fleming[144], T. Forest[58], J. Fowler[40], W. Fox[61], J. Fried[15], A. Friedland[78],
S. Fuess[43], B. Fujikawa[74], A. Gago[105], H. Gallagher[134], S. Galymov[63], T. Gamble[118], R. Gandhi[47],
D. Garcia-Gamez[81], S. Gardiner[20], G. Garvey[78], V. M. Gehman[74], A. Gendotti[147], G. d. Geronimo[15], C. Ghag[77],
P. Ghoshal[47], D. Gibin[106], I. Gil-Botella[17], R. Gill[15], D. Girardelli[25], A. Giri[55], S. Glavin[109], D. Goeldi[13],
S. Golapinni[65], M. Gold[96], R. A. Gomes[44], J. J. Gomez Cadenas[136], M. C. Goodman[8], D. Gorbunov[56],
S. Goswami[47], N. Graf[111], N. Graf[115], M. Graham[115], E. Gramelini[144], R. Gran[91], C. Grant[20], N. Grant[139],
V. Greco[26], H. Greenlee[43], L. Greenler[142], C. Greenley[79], M. Groh[29], S. Grullon[109], T. Grundy[73], K. Grzelak[138],
E. Guardincerri[78], V. Guarino[8], E. Guarnaccia[137], G. P. Guedes[42], R. Guenette[102], A. Guglielmi[106], A. T. Habig[91],
R. W. Hackenburg[15], A. Hackenburg[144], H. Hadavand[131], R. Haenni[13], A. Hahn[43], M. D. Haigh[139], T. Haines[78],
T. Hamernik[43], T. Handler[130], S. Hans[15], D. Harris[43], J. Hartnell[128], T. Hasegawa[67], R. Hatcher[43],
A. Hatzikoutelis[130], S. Hays[43], E. Hazen[14], M. Headley[123], A. Heavey[43], K. Heeger[144], J. Heise[123], K. Hennessy[76],
J. Hewes[81], A. Higuera[49], T. Hill[58], A. Himmel[43], M. Hogan[33], P. Holanda[25], A. Holin[77], W. Honey[73],
S. Horikawa[147], G. Horton-Smith[65], B. Howard[61], J. Howell[43], P. Hurh[43], J. Huston[87], J. Hylen[43], R. Imlay[79],
J. Insler[79], G. Introzzi[108], D. Ioanisyan[145], A. Ioannisian[145], K. Iwamoto[114], A. Izmaylov[136], C. Jackson[131],
D. E. Jaffe[15], C. James[43], E. James[43], F. Jediny[36], C. Jen[137], A. Jhingan[107], S. Jiménez[17], J. H. Jo[127],
M. Johnson[43], R. Johnson[29], J. Johnstone[43], B. J. Jones[83], J. Joshi[15], H. Jostlein[43], C. K. Jung[127], T. Junk[43],
A. Kaboth[59], R. Kadel[74], T. Kafka[134], L. Kalousis[137], Y. Kamyshkov[130], G. Karagiorgi[81], D. Karasavvas[9],
Y. Karyotakis[71], A. Kaur[107], P. Kaur[107], B. Kayser[43], N. Kazaryan[145], E. Kearns[14], P. Keener[109], S. Kemboi[131],
E. Kemp[25], S. H. Kettell[15], M. Khabibullin[56], M. Khandaker[58], A. Khotjantsev[56], B. Kirby[15], M. Kirby[43],
J. Klein[109], T. Kobilarcik[43], S. Kohn[19], G. Koizumi[43], A. Kopylov[56], M. Kordosky[141], L. Kormos[73], U. Kose[16],
V. A. Kostelecký[61], M. Kramer[19], I. Kreslo[13], R. Kriske[90], W. Kropp[21], Y. Kudenko[56], V. A. Kudryavtsev[118],
S. Kulagin[56], A. Kumar[107], G.K. Kumar[69], J. Kumar[48], L. Kumar[107], T. Kutter[79], A. Laminack[4], K. Lande[109],
C. Lane[39], K. Lang[132], F. Lanni[15], J. Learned[48], P. Lebrun[43], D. Lee[78], H. Lee[114], K. Lee[21], W. M. Lee[43],
M. A. Leigui de Oliveira[1], Q. Li[43], S. Li[15], S. Li[100], X. Li[127], Y. Li[15], Z. Li[40], J. Libo[120], C. S. Lin[74], S. Lin[33],
J. Ling[15], J. Link[137], Z. Liptak[32], D. Lissauer[15], L. Littenberg[15], B. Littlejohn[54], Q. Liu[78], T. Liu[125], S. Lockwitz[43],
N. Lockyer[43], T. Loew[74], M. Lokajicek[3], K. Long[59], M. D. L. Lopes[43], J. P. Lopez[32], J. Losecco[98], W. Louis[78],

J. Lowery[61], M. Luethi[13], K. B. Luk[19], B. Lundberg[43], T. Lundin[43], X. Luo[144], T. Lux[53], J. Lykken[43],
A. A. Machado[72], J. R. Macier[43], S. Magill[8], G. Mahler[15], K. Mahn[87], M. Malek[59], M. Malhotra[12], D. Malon[8],
F. Mammoliti[26], S. Mancina[8], S. K. Mandal[38], S. Mandodi[52], S. L. Manly[114], A. Mann[134], A. Marchionni[43],
W. Marciano[15], C. Mariani[137], J. Maricic[48], A. Marino[32], M. Marshak[90], C. Marshall[114], J. Marshall[24],
J. Marteau[63], J. Martin-Albo[102], S. Martinez[54], S. Matsuno[48], J. Matthews[79], C. Mauger[78], K. Mavrokoridis[76],
D. Mayilyan[145], E. Mazzucato[117], N. McCauley[76], E. McCluskey[43], N. McConkey[118], K. McDonald[112],
K. S. McFarland[114], A. M. McGowan[114], C. McGrew[127], R. McKeown[141], D. McNulty[58], R. McTaggart[124],
A. Mefodiev[56], M. Mehrian[33], P. Mehta[95], D. Mei[121], O. Mena[136], S. Menary[146], H. Mendez[85], A. Menegolli[108],
G. Meng[106], Y. Meng[21], H. Merritt[59], D. Mertins[4], M. Messier[61], W. Metcalf[79], M. Mewes[61], H. Meyer[140],
T. Miao[43], R. Milincic[48], W. Miller[90], G. Mills[78], O. Mineev[56], O. Miranda[30], C. S. Mishra[43], S. R. Mishra[120],
B. Mitrica[51], D. Mladenov[16], I. Mocioiu[104], R. Mohanta[52], N. Mokhov[43], C. Montanari[16], D. Montanari[43],
J. Moon[83], M. Mooney[15], C. Moore[43], J. Morfin[43], B. Morgan[139], C. Morris[49], W. Morse[15], Z. Moss[83], C. Mossey[43],
C. A. Moura[1], J. Mousseau[86], L. Mualem[23], M. Muether[140], S. Mufson[61], S. Murphy[147], J. Musser[61], R. Musser[131],
Y. Nakajima[74], D. Naples[111], J. Navarro[10], D. Navas[17], J. Nelson[141], M. Nessi[16], M. Newcomer[109], Y. Ng[131],
R. Nichol[77], T. C. Nicholls[116], K. Nikolics[147], E. Niner[61], B. Norris[43], F. Noto[16], P. Novakova[15], P. Novella[136],
J. Nowak[73], M. S. Nunes[25], H. O'Keeffe[73], R. Oldeman[24], R. Oliveira[25], T. Olson[134], Y. Onishchuk[68], J. Osta[43],
T. Ovsjannikova[56], B. Page[87], S. Pakvasa[48], S. Pal[118], O. Palamara[43], A. Palazzo[84], J. Paley[43], C. Palomares[17],
E. Pantic[20], V. Paolone[111], V. Papadimitriou[43], J. Park[137], S. Parke[43], Z. Parsa[15], S. Pascoli[41], R. Patterson[23],
S. Patton[74], T. Patzak[2], B. Paulos[142], L. Paulucci[1], Z. Pavlovic[43], G. Pawloski[90], S. Peeters[128], E. Pennacchio[63],
A. Perch[77], G. N. Perdue[43], L. Periale[147], J. D. Perkin[118], H. Pessard[71], G. Petrillo[43], R. Petti[120], A. Petukhov[122],
F. Pietropaolo[106], R. Plunkett[43], S. Pordes[43], M. Potekhin[15], R. Potenza[26], B. Potukuchi[64], N. Poudyal[121],
O. Prokofiev[43], N. Pruthi[107], P. Przewlocki[94], D. Pushka[43], X. Qian[15], J. L. Raaf[43], R. Raboanary[7], V. Radeka[15],
A. Radovic[141], G. Raffelt[84], I. Rakhno[43], H. T. Rakotondramanana[7], L. Rakotondravohitra[7], Y. A. Ramachers[139],
R. Rameika[43], J. Ramsey[78], A. Rappoldi[108], G. Raselli[108], P. Ratoff[73], B. Rebel[43], C. Regenfus[147],
J. Reichenbacher[122], D. Reitzner[43], A. Remoto[71], A. Renshaw[49], S. Rescia[15], M. Richardson[118], K. Rielage[78],
K. Riesselmann[43], M. Robinson[118], L. Rochester[115], O. B. Rodrigues[25], P. Rodrigues[114], B. Roe[86], M. Rosen[48],
R. M. Roser[43], M. Ross-Lonergan[41], M. Rossella[108], A. Rubbia[147], C. Rubbia[45], R. Rucinski[43],
C. Rudolph von Rohr[13], B. Russell[144], D. Ruterbories[114], R. Saakyan[77], N. Sahu[55], P. Sala[88], N. Samios[15],
F. Sanchez[53], M. Sanchez[62], B. Sands[112], S. Santana[85], R. Santorelli[17], G. Santucci[127], N. Saoulidou[9],
A. Scaramelli[88], H. Schellman[101], P. Schlabach[43], R. Schmitt[43], D. Schmitz[28], J. Schneps[134], K. Scholberg[40],
A. Schukraft[43], J. Schwehr[33], E. Segreto[72], S. Seibert[109], J. A. Sepulveda-Quiroz[62], F. Sergiampietri[147],
L. Sexton-Kennedy[43], D. Sgalaberna[147], M. Shaevitz[34], J. Shahi[107], S. Shahsavarani[131], P. Shanahan[43],
S. U. Shankar[47], R. Sharma[15], R. K. Sharma[113], T. Shaw[43], R. Shrock[127], I. Shyrma[68], N. Simos[15], G. Sinev[40],
I. Singh[107], J. Singh[107], J. Singh[80], V. Singh[11], G. Sinnis[78], W. Sippach[34], D. Smargianaki[16], M. Smy[21],
E. Snider[43], P. Snopok[54], J. Sobczyk[143], H. Sobel[21], M. Soderberg[129], N. Solomey[140], W. Sondheim[78], M. Sorel[136],
A. Sousa[29], K. Soustruznik[27], J. Spitz[86], N. J. Spooner[118], M. Stancari[43], I. Stancu[4], D. Stefan[16], H. M. Steiner[74],
J. Stewart[15], J. Stock[122], S. Stoica[51], J. Stone[14], J. Strait[43], M. Strait[28], T. Strauss[43], S. Striganov[43], R. Sulej[94],
G. Sullivan[82], Y. Sun[48], L. Suter[8], C. M. Sutera[26], R. Svoboda[20], B. Szczerbinska[37], A. Szelc[81],
S. Söldner-Rembold[81], R. Talaga[8], M. Tamsett[128], S. Tariq[43], E. Tatar[58], R. Tayloe[61], C. Taylor[78], D. Taylor[123],
K. Terao[34], M. Thiesse[118], J. Thomas[77], L. F. Thompson[118], M. Thomson[24], C. Thorn[15], M. Thorpe[116], X. Tian[120],
D. Tiedt[122], S. C. Timm[43], A. Tonazzo[2], T. Tope[43], A. Topkar[12], F. R. Torres[25], M. Torti[108], M. Tortola[136],
F. Tortorici[26], M. Toups[83], C. Touramanis[76], M. Tripathi[20], I. Tropin[43], Y. Tsai[115], K. V. Tsang[74], R. Tsenov[119],
S. Tufanli[144], C. Tull[74], J. Turner[41], M. Tzanov[79], E. Tziaferi[9], Y. Uchida[59], J. Urheim[61], T. Usher[115], M. Vagins[66],
P. Vahle[141], G. A. Valdiviesso[5], L. Valerio[43], Z. Vallari[127], J. Valle[136], R. Van Berg[109], R. Van de Water[78],
P. Van Gemmeren[8], F. Varanini[106], G. Varner[48], G. Vasseur[117], K. Vaziri[43], G. Velev[43], S. Ventura[106],
A. Verdugo[17], T. Viant[147], T. V. Vieira[25], C. Vignoli[72], C. Vilela[127], B. Viren[15], T. Vrba[36], T. Wachala[70],
D. Wahl[142], M. Wallbank[118], N. Walsh[20], B. Wang[125], H. Wang[22], L. Wang[121], T. Wang[41], T.K. Warburton[118],
D. Warner[33], M. Wascko[59], D. Waters[77], T. B. Watson[131], A. Weber[116], M. Weber[13], W. Wei[121], A. Weinstein[62],
D. Wells[122], D. Wenman[142], M. Wetstein[62], A. White[131], L. Whitehead[49], D. Whittington[61], M. Wilking[127],
J. Willhite[43], P. Wilson[43], R. J. Wilson[33], L. Winslow[83], P. Wittich[35], S. Wojcicki[126], H. H. Wong[19], K. Wood[120],
E. Worcester[15], M. Worcester[15], S. Wu[147], T. Xin[62], C. Yanagisawa[127], S. Yang[29], T. Yang[43], K. Yarritu[78], J. Ye[125],
M. Yeh[15], N. Yershov[56], K. Yonehara[43], B. Yu[15], J. Yu[131], J. Zalesak[3], A. Zalewska[70], B. Zamorano[128], L. Zang[118],
A. Zani[16], G. Zavala[46], G. Zeller[43], C. Zhang[15], C. Zhang[121], E. D. Zimmerman[32], M. Zito[117], and R. Zwaska[43]

[1]*ABC Federal University, Santo André –SP, 09210-580, Brazil*
[2]*APC-Paris, Batiment Condorcet; 10, rue Alice Domon et L'eonie Duquet; F-75205 Paris CEDEX 13, France*
[3]*Institute of Physics ASCR, v. v. i. , Na Slovance 2; 182 21 Praha 8, Czech Republic*
[4]*University of Alabama (Tuscaloosa), Tuscaloosa, AL 35487-0324, USA*
[5]*University Federal de Alfenas em Poços de Caldas, 11999, CEP 37715-900 Poços de Caldas-MG, Brazil*

[6] Aligarh Muslim University, Department of Physics Aligarh-202002, India

[7] Antananarivo, Présidence de l'Université d'Antananarivo: BP 566, Antananarivo 101, Madagascar

[8] Argonne National Lab., Argonne, IL 60439, USA

[9] University of Athens, University Campus, Zografou GR 157 84 Greece, Greece

[10] Universidad del Atlantico, Km 7 antigua vi a Puerto Colombia, Barranquilla, Colombia

[11] Banaras Hindu University, Dept. of Physics, Varanasi UP 221005, India

[12] Bhabha Atmoic Research Center, Trombay, Mumbai, India

[13] University of Bern, Lab for High Energy Physics; Sidlerstrasse 5; CH-3012 Bern, Switzerland

[14] Boston University, Boston, MA 02215, USA

[15] Brookhaven National Lab., Upton, NY 11973-5000, USA

[16] CERN, European Organization for Nuclear Research European Laboratory for Particle Physics; 1211 Gen'eve 23, Switzerland

[17] CIEMAT, División de Física de Partículas; Avenida Complutense 40; E-28040 Madrid, Spain

[18] CNR Pisa, 1, 56124 Pisa PI, Italy

[19] University of California (Berkeley), Berkeley, CA 94720-7300, USA

[20] University of California (Davis), Davis, CA 95616, USA

[21] University of California (Irvine), Irvine, CA 92697-4575, USA

[22] University of California (Los Angeles), Los Angeles, CA 90095-1547, USA

[23] California Inst. of Tech., MC 356-48; Pasadena, CA 91125, USA

[24] University of Cambridge, JJ Thomson Avenue, Cambridge CB3 0HE, UK

[25] University de Campinas, Av. Sérgio Buarque de Holanda, 777 CEP 13083-859 Campinas-SP, Brazil

[26] University di Catania, Via Santa Sofia, I-95123 Catania, Italy

[27] Institute of Particle and Nuclear Physics of the Faculty of Mathematics and Physics of the Charles University in Prague, V Holešovičkách 747/2; 180 00 Praha 8-Libeň, Czech Republic

[28] University of Chicago, Chicago, IL 60637-1434, USA

[29] University of Cincinnati, Cincinnati, OH 45221-0011, USA

[30] Cinvestav, 07360 Ciudad de México, D.F., Mexico

[31] Universidad de Colima, Facultad de Ciencias Bernal Diaz del Castillo 340 Colonia Villa San Sebastian Colima, Colima, Mexico

[32] University of Colorado, Boulder, CO 80309, USA

[33] Colorado State University, Fort Collins, CO 80523, USA

[34] Columbia University, New York, NY 10027, USA

[35] Cornell University, Laboratory for Elementary-Particle Physics; Newman Lab; Ithaca, NY 14853–5001, USA

[36] Czech Technical University in Prague, Brehova 7; 115 19 Praha 1, Czech Republic

[37] Dakota State University, Madison, SD 57042, USA

[38] University of Delhi, Department of Physics and Astrophysics, Delhi 110007, India

[39] Drexel University, Philadelphia, PA 19104, USA

[40] Duke University, Durham, NC 27706, USA

[41] University of Durham, Institute for Particle Physics Phenomenology; Dept. of Physics, Ogden Centre for Fund. Physics; South Road; Durham DH1 3LE, UK

[42] University Estadual de Feira de Santana, S/N;44036-900, Feira de Santana-BA, Brazil

[43] Fermi National Accelerator Lab, Batavia, IL 60510-0500, USA

[44] University Federal de Goias, Goiania, GO, Brazil

[45] Gran Sasso Science Institute, 7, L'Aquila, Italy

[46] Universidad de Guanajuato, Gto., C.P.37000, Mexico

[47] Harish-Chandra Research Institute, Jhunsi, Allahabad 211 019, India

[48] University of Hawaii, Honolulu, HI 96822-2219, USA

[49] University of Houston, Houston, TX 77204, USA

[50] Huddersfield, Huddersfield, West Yorkshire HD1 3DH, UK

[51] Horia Hulubei National Institute of Physiscs and Nuclear Engineering, Horia Hulubei National Institute of Physiscs and Nuclear Engineering, Bucharest-Magurele, Romania

[52] University of Hyderabad, Gachibowli, Hyderabad - 500 046, India

[53] Institut de Fìsica d'Altes Energies (IFAE); Campus Universitat Autonoma de Barcelona , E-08193 Cerdanyola del Valles (Barcelona), Spain

[54] Illinois Institute of Technology , Room 182LS; Chicago, IL 60616, USA

[55] IIT Hyderabad, Kandi, Hyderabad - 502205, India

[56] Institute for Nuclear Research of the Russian Academy of Sciences (INR RAS), 60th October Anniversary Prosp. 7a Moscow, Russia 117312, Russia

[57] Institute for Research in Fundamental Sciences (IPM), Tehran, Iran. Postal code: 19538-33511 PO Box:

19395-5531, Iran

[58] Idaho State University, Department of Physics 921 South 8th Ave. Stop 8106 Pocatello, ID 83209-8106, USA

[59] Imperial College of Science Tech. & Medicine, Blackett Lab.; Prince Consort Road; London SW7 2BZ, UK

[60] Indian Institute of Technology Guwahati, Guwahati, 781 039, India

[61] Indiana University, Bloomington, IN 47405-7105, USA

[62] Iowa State University, Ames, IA 50011, USA

[63] Institut de Physique Nucleaire de Lyon (IPNL), Rue E. Fermi 4 69622 Villeurbanne, France

[64] University of Jammu, Physics Department, JAMMU-180006, India

[65] Kansas State University, Manhattan, KS 66506, USA

[66] Kavli IPMU, University of Tokyo, Kashiwa Shi, Chiba 277-8568, Japan

[67] KEK, High Energy Accelerator Research Organization 1–1 Oho, Tsukuba-shi; Ibaraki-ken 305-0801, Japan

[68] KYIV National University, Department of Nuclear Physics, 64, 01601 Kyiv, Ukraine

[69] Koneru Lakshmaiah, India

[70] Krakow, Golebia 24, 31-007 Kraków, Poland

[71] Lab. d'Annecy-le-Vieux de Phys. des Particules, BP 110; F-74941 Annecy-le-Vieux CEDEX, France

[72] Laboratori Nazionali del Gran Sasso, I-67010 Assergi, AQ, Italy

[73] Lancaster University, Bailrigg, Lancaster LA1 4YB, UK

[74] Lawrence Berkeley National Lab., Berkeley, CA 94720-8153, USA

[75] Université de Liège, Bat B5, Sart Tilman B-4000, Belgium

[76] University of Liverpool, L69 7ZE, Liverpool, UK

[77] University College London, London, WC1E 6BT, UK

[78] Los Alamos National Laboratory, Los Alamos, NM 87545, USA

[79] Louisiana State University, Baton Rouge, LA 70803-4001, USA

[80] University of Lucknow, Department of Physics Lucknow 226007 Uttar Pradesh, India

[81] University of Manchester, Oxford Road, Manchester M13 9PL, UK

[82] University of Maryland, College Park, MD 20742-4111, USA

[83] Massachusetts Institute of Technology, Cambridge, MA 02139-4307, USA

[84] Max Planck MPP, Max-Planck-Institut fuer Physik (Werner-Heisenberg-Institut) Foehringer Ring 6 80805 Muenchen, Germany

[85] University of Puerto Rico, Box 9016; Mayaguez, PR 00681-9000, USA

[86] University of Michigan, 450 Church Street Ann Arbor, Michigan 48109-1040, USA

[87] Michigan State University, East Lansing, MI 48824, USA

[88] University di Milano, INFN Sezione di Milano, I-20133 Milano, Italy

[89] INFN Sezione di Milano Bicocca, Piazza della Scienza 3, 20126 Milano, Italy

[90] University of Minnesota (Twin Cities), Minneapolis, MN 55455, USA

[91] University of Minnesota (Duluth), Duluth, MN 55812, USA

[92] NIKHEF, Science Park, Amsterdam, Netherlands

[93] Istituto Nazionale di Fisica Nucleare - Sezione di Napoli, Complesso Universitario di Monte S. Angelo, I-80126 Napoli, Italy

[94] National Centre for Nuclear Research, A. Soltana 7, 05 400 Otwock, Poland

[95] Jawaharlal Nehru University, New Delhi 110067, INDIA, India

[96] University of New Mexico, MSC07 4220; Albuquerque, NM 87131, USA

[97] Northwestern University, Evanston, Il 60208, USA

[98] University of Notre Dame, Notre Dame, IN 46556-5670, USA

[99] Observatorio Nacional, R. Lourival Bpo., 89 - Madruga, Vassouras - RJ, 27700-000, Brazil

[100] Ohio State University, Dept. of Physics; 191 W. Woodruff Ave.; Columbus, OH 43210, USA

[101] Oregon State University, Dept. of Physics; 301 Weniger Hall; Corvallis, OR 97331–6507, USA

[102] University of Oxford, Oxford, OX1 3RH, UK

[103] Pacific Northwest National Lab, , USA

[104] Pennsylvania State University, PMB 264; University Park, PA 16802-6300, USA

[105] PUCP, Av. Universitaria 1801, Lima, Peru

[106] University of Padova, Dip. Fisica e Astronomia G. Galilei and INFN Sezione di Padova, I-35131 Padova, Italy

[107] Panjab University, Chandigarh, 160014 U.T., India

[108] University of Pavia, INFN Sezione di Pavia, I-27100 Pavia, Italy

[109] University of Pennsylvania, Philadelphia, PA 19104-6396, USA

[110] University di Pisa, Theor. Division; Largo B. Pontecorvo 3, Ed. B-C; I-56127 Pisa, Italy

[111] University of Pittsburgh, Pittsburgh, PA 15260, USA

[112] Princeton University, Princeton, New Jersey 08544-0708, USA

[113] Punjab Agri. University, Centre for High Energy Physics; Lahore - 54590, India

[114] University of Rochester, Rochester, NY 14627-0171, USA
[115] SLAC National Accelerator Laboratory, Menlo Park, CA 94025, USA
[116] STFC Rutherford Appleton Laboratory, Harwell Oxford, Didcot OX11 0QX, UK
[117] CEA/Saclay, IPhT; Inst. de Physique Theorique; Orme des Merisiers, Point Courrier 136; F-91191 Gif-sur-Yvette CEDE, France
[118] University of Sheffield, Sheffield, S3 7RH, UK
[119] University of Sofia, Atomic Physics Dept., Faculty of Physics; 5 James Bourchier Blvd.; BG-1164 Sofia, Bulgaria
[120] University of South Carolina, Columbia, SC 29208, USA
[121] University of South Dakota, Vermillion, SD 57069, USA
[122] South Dakota School of Mines and Technology, Rapid City, SD 57701, USA
[123] South Dakota Science And Technology Authority, Lead, SD 57754, USA
[124] South Dakota State University, Brookings, SD 57007, USA
[125] Southern Methodist University, Dallas, TX 75275, USA
[126] Stanford University, Varian Physics Bldg.; 382 Via Pueblo Mall; Stanford, CA 94305–4060, USA
[127] Stony Brook University, Nucleon Decay and Neutrino Physics Group Department of Physics and Astronomy Stony Brook University Stony Brook, NY 11794-3800, USA
[128] University of Sussex, Brighton, BN1 9RH, UK
[129] Syracuse University, Syracuse, NY 13244-1130, USA
[130] University of Tennessee at Knoxville, Knoxville, TN, USA
[131] University of Texas (Arlington), Arlington, TX 76019, USA
[132] University of Texas (Austin), Austin, TX 78712-0264, USA
[133] TUBITAK Space Technologies Research Institute , TR-06800, Ankara, Turkey
[134] Tufts University, Medford, MA 02155, USA
[135] Variable Energy Cyclotron Centr, 1/AF, Bidhannagar Kolkata - 700 064 West Bengal, India
[136] Instituto de Fisica Corpuscular, C/Catedratico Jose Beltran, 2 E-46980 Paterna (Valencia), Spain
[137] Virginia Tech., Blacksburg, VA 24061-0435, USA
[138] University of Warsaw, Faculty of Physics ul. Pasteura 5 02-093 Warsaw, Poland
[139] University of Warwick, Coventry CV4 7AL, UK
[140] Wichita State University, Physics Division 1845 Fairmount St. Wichita, KS 67260-0032, USA
[141] College of William and Mary, Williamsburg, VA 23187-8795, USA
[142] University of Wisconsin, Madison, WI 53706, USA
[143] Wroclaw University, Plac Maxa Borna 9, 50-204 Wroclaw, Poland
[144] Yale University, New Haven, CT 06520, USA
[145] Yerevan Institute for Theoretical Physics and Modeling, Halabian Str. 34; Yerevan 0036, Armenia
[146] York, Physics and Astronomy Dept.; 4700 Keele St.; Toronto M3J 1P3, Canada
[147] ETH Zurich, HPK F 23 Schafmattstr. 20 8093 Zürich, Switzerland

2015-11-25



\renewcommand{\familydefault}{\sfdefault}
\renewcommand{\thepage}{\roman{page}}
\setcounter{page}{0}

\pagestyle{plain} 


\setcounter{tocdepth}{2}
\textsf{\tableofcontents}

\textsf{\listoffigures}

\textsf{\listoftables}

\printnomenclature

\iffinal\else
\textsf{\listoftodos}
\clearpage
\fi

\renewcommand{\thepage}{\arabic{page}}
\setcounter{page}{1}

\pagestyle{fancy}

\renewcommand{\chaptermark}[1]{%
\markboth{Chapter \thechapter:\ #1}{}}
\fancyhead{}
\fancyhead[RO,LE]{\textsf{\footnotesize \thechapter--\thepage}}
\fancyhead[LO,RE]{\textsf{\footnotesize \leftmark}}

\fancyfoot{}
\fancyfoot[RO]{\textsf{\footnotesize LBNF/DUNE Conceptual Design Report}}
\fancyfoot[LO]{\textsf{\footnotesize \thedoctitle}}
\fancypagestyle{plain}{}

\renewcommand{\headrule}{\vspace{-4mm}\color[gray]{0.5}{\rule{\headwidth}{0.5pt}}}

\nomenclature{$\mathcal{O}(n)$}{of order $n$}
\nomenclature{3D}{3 dimensional (also 1D, 2D, etc.)} 
\nomenclature{CDR}{Conceptual Design Report}
\nomenclature{CF}{Conventional Facilities}
\nomenclature{CP}{product of charge and parity transformations}
\nomenclature{CPT}{product of charge, parity and time-reversal transformations}
\nomenclature{CPV}{violation of charge and parity symmetry}
\nomenclature{DAQ}{data acquisition}
\nomenclature{DOE}{U.S. Department of Energy}
\nomenclature{DUNE}{Deep Underground Neutrino Experiment}
\nomenclature{ESH}{Environment, Safety and Health}
\nomenclature{eV}{electron volt, unit of energy (also keV, MeV, GeV, etc.)}
\nomenclature{FD}{far detector}
\nomenclature{FGT}{Fine-Grained Tracker}
\nomenclature{FSCF}{far site conventional facilities}
\nomenclature{NSCF}{near site conventional facilities}
\nomenclature{GUT}{grand unified theory}
\nomenclature{\ktyr}{exposure (without beam), expressed in kilotonnes times years}
\nomenclature{\ktMWyr}{exposure, expressed in kilotonnes $\times$ megawatts $\times$ years, based on 56\% beam uptime and efficiency} 
\nomenclature{L}{level, indicates depth in feet underground at the far site, e.g., 4850L}
\nomenclature{LAr}{liquid argon}
\nomenclature{LArTPC}{liquid argon time-projection chamber}
\nomenclature{LBL}{long-baseline (physics)}
\nomenclature{LBNF}{Long-Baseline Neutrino Facility}
\nomenclature{MH}{mass hierarchy}
\nomenclature{MI}{Main Injector (at Fermilab)}
\nomenclature{ND}{near neutrino detector}
\nomenclature{NDS}{Near Detector Systems; refers to the collection of detector systems at the near site }
\nomenclature{near detector}{except in Volume 4 Chapter 7, \textit{near detector} refers to the \textit{neutrino} detector system in the NDS}
\nomenclature{POT}{protons on target}
\nomenclature{QA}{quality assurance}
\nomenclature{SM}{Standard Model of particle physics}
\nomenclature{t}{metric ton, written \textit{tonne} (also kt)}
\nomenclature{tonne}{metric ton}
\nomenclature{TPC}{time-projection chamber (not used as `total project cost' in the CDR)}

\nomenclature{CKM}{(CKM matrix) Cabibbo-Kobayashi-Maskawa matrix, also known as
quark mixing matrix} 
\nomenclature{C.L.}{confidence level}
\nomenclature{octant}{any of the eight parts into which 4$\pi$ is divided by three mutually perpendicular axes; the range of the PMNS angles is $0$ to $\pi/2$, which spans only two of the eight octants}
\nomenclature{PMNS}{(PMNS matrix) Pontecorvo-Maki-Nakagawa-Sakata matrix, also known as
the lepton or neutrino mixing matrix} 

\nomenclature{CC}{charged current (interaction)}
\nomenclature{DIS}{deep inelastic scattering}
\nomenclature{FSI}{final-state interactions}
\nomenclature{GEANT4}{GEometry ANd Tracking, a platform for the simulation of the passage of particles through matter using Monte Carlo methods} 
\nomenclature{GENIE}{Generates Events for Neutrino Interaction Experiments (an object-oriented neutrino Monte Carlo generator)} 
\nomenclature{MC}{Monte Carlo (detector simulation methods)}
\nomenclature{QE}{quasi-elastic (interaction)}
\nomenclature{BR}{branching ratio}
\nomenclature{DM}{dark matter}
\nomenclature{DSNB}{Diffuse Supernova Neutrino Background}
\nomenclature{GLoBES}{General Long-Baseline Experiment Simulator (software package)}
\nomenclature{L/E}{length-to-energy ratio}
\nomenclature{LRI}{long-range interactions}
\nomenclature{$M_{\odot}$}{solar mass}
\nomenclature{NC}{neutral current (interaction)}
\nomenclature{NH}{normal (mass) hierarchy}
\nomenclature{NSI}{nonstandard interactions}
\nomenclature{MSW}{Mikheyev-Smirnov-Wolfenstein (effect)}
\nomenclature{SME}{Standard-Model Extension}
\nomenclature{SUSY}{supersymmetry}
\nomenclature{WIMP}{weakly-interacting massive particle}


\chapter{Introduction to LBNF and DUNE}
\label{ch:physics-overview}


\section{An International Physics Program}

The global neutrino physics community is developing a multi-decade
physics program to measure unknown parameters of the Standard Model of
particle physics and search for new phenomena.  The program will be carried out as an international,
leading-edge, dual-site experiment for neutrino science and proton decay studies, which 
is known as the Deep Underground Neutrino Experiment (DUNE).
The detectors for this experiment will be designed, built, commissioned and operated by the international DUNE Collaboration. The facility required to support this experiment, the Long-Baseline Neutrino Facility (LBNF), is hosted by Fermilab and its design and construction is organized as a DOE/Fermilab project incorporating international partners. Together LBNF and DUNE will comprise the world's highest-intensity neutrino beam at Fermilab, in Batavia, IL, a high-precision near detector on the Fermilab site, a massive liquid argon time-projection chamber (LArTPC) far detector installed deep underground at the Sanford Underground Research Facility (SURF) \SI{1300}{\km} away in Lead, SD, and all of the conventional and technical facilities necessary to support the beamline and detector systems.

The strategy for executing the experimental program presented in this Conceptual 
Design Report (CDR) has been developed to meet the requirements 
set out in the P5 report~\cite{p5report} and takes into account the recommendations of the European Strategy for Particle Physics~\cite{ESPP-2012}. It adopts a model where U.S. and international funding agencies 
share costs on the DUNE detectors, and CERN and other participants provide in-kind contributions 
to the supporting infrastructure of LBNF. LBNF and DUNE will be tightly coordinated as DUNE collaborators 
design the detectors and infrastructure that will carry out the scientific program.
  
The scope of LBNF is
\begin{itemize}
\item an intense neutrino beam aimed at the far site
\item conventional facilities at both the near and far sites
\item cryogenics infrastructure to support the DUNE detector at the far site
\end{itemize}

The DUNE detectors include
\begin{itemize}
\item a high-performance neutrino detector and beamline 
measurement system
located a few hundred meters downstream of the neutrino source
\item a massive liquid argon time-projection chamber (LArTPC) neutrino detector located deep underground at the far site
\end{itemize}

With the facilities provided by LBNF and the detectors provided by
DUNE, the DUNE Collaboration proposes to mount a focused attack on the
puzzle of neutrinos with broad sensitivity to neutrino oscillation
parameters in a single experiment.  The focus of the scientific
program is the determination of the neutrino mass hierarchy and the
explicit demonstration of leptonic CP violation, if it exists, by
precisely measuring differences between the oscillations of muon-type
neutrinos and antineutrinos into electron-type neutrinos and
antineutrinos, respectively. Siting the far detector deep underground
will provide exciting additional research opportunities in nucleon
decay, studies utilizing atmospheric neutrinos, and neutrino
astrophysics, including measurements of neutrinos from a core-collapse
supernova should such an event occur in our galaxy during the
experiment's lifetime.

\section{The LBNF/DUNE Conceptual Design Report Volumes}

\subsection{A Roadmap of the CDR}

The LBNF/DUNE CDR describes the proposed physics program and 
technical designs at the conceptual design stage.  At this stage, the design is
still undergoing development and the CDR therefore presents a \textit{reference design} 
for each element as well as \textit{alternative designs} that are under consideration.

The CDR is composed of four volumes and is supplemented by several annexes that 
provide details on the physics program and technical designs. The volumes are as follows

\begin{itemize}
\item \volintro{} provides an executive summary of and strategy for the experimental 
program and introduces the CDR.
\item \volphys{} outlines the scientific objectives and describes the physics studies that 
the DUNE Collaboration will undertake to address them.
\item \vollbnf{} describes the LBNF Project, which includes design and construction of the 
beamline at Fermilab, the conventional facilities at both Fermilab and SURF, and the cryostat
 and cryogenics infrastructure required for the DUNE far detector.
\item \voldune{} describes the DUNE Project, which includes the design, construction and 
commissioning of the near and far detectors. 
\end{itemize}

More detailed information for each of these volumes is provided in a set of annexes listed on the \href{https://web.fnal.gov/project/LBNF/SitePages/CD-1-R%20Reports%20and%20Documents.aspx}{\textit{CD-1-R Reports and Documents} page}. 


\subsection{About this Volume}

\volphys\, outlines the science objectives in
Chapter~\ref{ch:physics-goals}, describes each of the areas of study in the following
chapters, and concludes with a summary.

The LBNF/DUNE science objectives are categorized as primary, ancillary and additional, 
with the primary objectives driving the experiment and facility designs that together will 
also enable pursuit of the ancillary objectives. 
Pursuit of the additional goals may require technological developments beyond the current designs.

Chapters~\ref{ch:physics-lbnosc},~\ref{ch:physics-atmpdk} and~\ref{ch:physics-snblowe}
describe the physics program for the DUNE far detector in the areas of long-baseline neutrino
oscillations, nucleon decay, atmospheric neutrinos and detection of supernova neutrino
bursts and low-energy neutrinos; they also discuss the requirements that these studies 
impose on the detector design.

Chapter~\ref{ch:physics-nd} discusses the role that the near detector plays in the overall DUNE
physics program and the requirements that it must satisfy, and describes the measurements and 
new physics searches that the near detector will enable on its own.


\cleardoublepage

\chapter{LBNF/DUNE Scientific Goals}
\label{ch:physics-goals}

\section{Overview of Goals}

LBNF/DUNE will address fundamental questions key to our understanding of the Universe. These include
\begin{itemize}
   \item {\bf What is the origin of the matter-antimatter asymmetry in the Universe?} Immediately after
                    the Big Bang, matter and antimatter were created equally, but now matter dominates.
                    By studying the properties of neutrino and antineutrino oscillations to determine if charge-parity (CP) symmetry is violated in the lepton sector, 
LBNF/DUNE 
                    will pursue the current most promising avenue for understanding this asymmetry.
   \item {\bf What are the fundamental underlying symmetries of the Universe?} The patterns of mixings and masses between the particles of the Standard Model are not understood. By making precise measurements of the mixing between the neutrinos and the ordering of neutrino masses and comparing these with the quark sector, LBNF/DUNE could reveal new underlying symmetries of the Universe.
  \item{\bf  Is there a Grand Unified Theory of the Universe?} Results from a range of experiments suggest that the
                 physical forces observed today were unified into one force at the birth of the Universe.
                Grand Unified Theories (GUTs), which attempt to describe the unification of forces,
                predict that protons should decay, a process that has never been observed. DUNE will 
                search for proton decay in the range of proton lifetimes predicted by a wide range of GUT models.
   \item{\bf How do supernovae explode and what new physics will we learn from a neutrino burst?}
   Many of the heavy elements that are the key components of life were created in the super-hot cores of collapsing stars. DUNE would be able to detect the neutrino bursts from core-collapse supernova within our galaxy (should any occur). Measurements of the time, flavor and energy structure of the neutrino burst will be critical for understanding the dynamics of this important astrophysical phenomenon, as well as providing information on neutrino properties and other particle physics.
\end{itemize}


The LBNF/DUNE scientific objectives are categorized into: the \textit{primary science program}, addressing the key science questions highlighted by the particle physics project prioritization panel (P5); 
a high-priority {\textit{ancillary science program} that is 
enabled by the construction of LBNF and DUNE; and \textit{additional scientific objectives}, that may require developments 
of the LArTPC technology. The goals of the primary science program define the high-level requirements for LBNF and the 
DUNE detectors. The ancillary science program provides further requirements, specifically on the design of the near 
detector, required for the full scientific exploitation of this world leading facility.

\section{The Primary Science Program}

The primary science program of the LBNF/DUNE experiment focuses on fundamental open questions in neutrino and astroparticle physics: 
\begin{itemize}
  \item precision measurements of the parameters that govern $\nu_{\mu} \rightarrow \nu_\text{e}$ and
           $\overline{\nu}_{\mu} \rightarrow \overline{\nu}_\text{e}$ oscillations with the goal of
  \subitem -- measuring the charge-parity (CP) violating phase $\delta_\text{CP}$ --- where a value differing from zero or $\pi$ would represent the discovery of CP-violation in the leptonic sector, providing a possible explanation for the matter-antimatter asymmetry in the universe;
  \subitem -- determining the neutrino mass ordering (the sign of $\Delta m^2_{31} \equiv m_3^2-m_1^2$), often referred to as the neutrino \textit{mass hierarchy};  
  \subitem -- precision tests of the three-flavor neutrino oscillation paradigm through studies of muon neutrino disappearance 
    and electron neutrino appearance in both $\nu_\mu$ and $\overline{\nu}_{\mu}$ beams, including the 
    measurement of the mixing angle $\theta_{23}$ and the determination of the octant in which this angle lies;
    \item search for proton decay in several important decay modes, for example $\text{p}\rightarrow\text{K}^+\overline{\nu}$, where the observation of proton decay would represent a ground-breaking discovery in physics, providing a portal to Grand Unification of the forces;
    \item detection and measurement of the $\nu_\text{e}$ flux from a core-collapse supernova within our galaxy, should any occur during the lifetime of the DUNE experiment.
\end{itemize}

\section{The Ancillary Science Program}

The intense neutrino beam from LBNF, the massive DUNE LArTPC far detector and the highly capable DUNE near detector provide a rich ancillary science program, beyond the primary mission of the experiment. The ancillary science program includes:
\begin{itemize}
     \item other accelerator-based neutrino flavor transition measurements with sensitivity to Beyond Standard Model (BSM) physics, such as:    
            \subitem -- non-standard interactions (NSIs);
             \subitem -- the search for sterile neutrinos at both the near and far sites;
             \subitem -- measurements of tau neutrino appearance;
     \item measurements of neutrino oscillation phenomena using atmospheric neutrinos;
     \item a rich neutrino interaction physics program utilizing the DUNE near detector, including:
         \subitem -- a wide-range of measurements of neutrino cross sections;
         \subitem -- studies of nuclear effects, including neutrino final-state interactions;
         \subitem -- measurements of the structure of nucleons;      
         \subitem -- measurement of $\sin^2\theta_\text{W}$;  
     \item  and the search for signatures of dark matter.
\end{itemize} 
Furthermore, a number of previous breakthroughs in particle physics have been serendipitous, in the sense that they were beyond the
original scientific objectives of their experiments. The intense LBNF neutrino beam and novel capabilities for both 
the DUNE near and far detectors will probe new regions of parameter space for both the accelerator-based and astrophysical frontiers, 
providing the opportunity for discoveries that are not currently anticipated.

\section{Additional Scientific Objectives}
There are a number of opportunities that could be enabled by developments/improvements to the LArTPC detector technology over the course of the DUNE installation. These include:
\begin{itemize}
      \item measurements of neutrino oscillation phenomena and of solar physics using solar neutrinos;
      \item detection and measurement of the diffuse supernova neutrino flux;
      \item measurement of neutrinos from astrophysical sources at energies from 
      gamma-ray bursts, active galactic nuclei, black-hole and neutron-star mergers, or other transient sources.
\end{itemize}

\cleardoublepage

\chapter{Long-Baseline Neutrino Oscillation Physics}
\label{ch:physics-lbnosc}

\section{Overview and Theoretical Context} 
\label{sec:physics-lbnosc-context}

The Standard Model of particle physics presents a remarkably accurate
description of the elementary particles and their
interactions. However, its limitations pose deeper questions about
Nature. With the discovery of the Higgs boson at CERN, the Standard
Model would be ``complete'' except for the discovery of neutrino
mixing, which indicated neutrinos had a very small but nonzero
mass. In the Standard Model, the simple Higgs mechanism is responsible
for both quark and charged lepton masses, quark mixing and
charge-parity (CP) violation. However, the small size of neutrino
masses and their relatively large mixing bears little resemblance to
quark masses and mixing, suggesting that different physics -- and
possibly different mass scales -- in the two sectors may be present,
thus motivating precision study of mixing and CP violation in the
lepton sector of the Standard Model.

DUNE plans to pursue a detailed study of neutrino mixing, resolve the
neutrino mass ordering, and search for CP violation in the lepton
sector by studying the oscillation patterns of
high-intensity \numu and \anumu 
beams measured over a long baseline.  Neutrino oscillation arises from
mixing between the flavor 
\nue, \numu, \nutau and mass $(\nu_1,\, \nu_2,\, \nu_3)$ eigenstates
of neutrinos.
In direct correspondence with mixing in the quark sector, the transformations
between basis states is expressed in the form of a complex unitary
matrix, known as the \textit{PMNS mixing matrix}: 

\begin{equation}
\left(\begin{array}{ccc} \nu_e \\ \nu_\mu \\ \nu_\tau \end{array} \right)= 
\underbrace{
  \left(\begin{array}{ccc}
      U_{e 1} &  U_{e 2} & U_{e 3} \\ 
      U_{\mu1} &  U_{\mu2} & U_{\mu 3} \\ 
      U_{\tau 1} &  U_{\tau 2} & U_{\tau 3} 
    \end{array} \right)
}_{U_{\rm PMNS}} \left(\begin{array}{ccc} \nu_1 \\ \nu_2 \\ \nu_3 \end{array} \right).
\label{eqn:pmns0}
\end{equation}
The PMNS matrix in full generality depends on just three mixing angles
and a CP-violating phase\footnote{There are two additional CP phases (Majorana phases), but they are unobservable in the oscillation processes.}.  The mixing angles and phase are designated
as $(\theta_{12},\, \theta_{23},\, \theta_{13})$ and
\deltacp.
This matrix can be parameterized as the product of three
two-flavor mixing matrices as follows~\cite{Schechter:1980gr}, where $c_{\alpha \beta}=\cos \theta_{\alpha \beta}$ and $s_{\alpha
  \beta}=\sin \theta_{\alpha \beta}$:

\begin{equation}
U_{\rm PMNS} = 
  \underbrace{
    \left( \begin{array}{ccc}
        1 & 0 & 0 \\ 
        0 & c_{23} & s_{23} \\ 
        0 & -s_{23} & c_{23}
    \end{array} \right)
  }_{\rm I}
\underbrace{
  \left( \begin{array}{ccc}
      c_{13} & 0 & e^{-i\mdeltacp}s_{13} \\ 
      0 & 1 & 0 \\ 
      -e^{i\mdeltacp}s_{13} & 0 & c_{13}
  \end{array} \right) 
}_{\rm II}
\underbrace{
 \left( \begin{array}{ccc}
      c_{12} & s_{12} & 0 \\ 
      -s_{12} & c_{12} & 0 \\ 
      0 & 0 & 1
  \end{array} \right) 
}_{\rm III}
\label{eqn:pmns}
\end{equation}
The parameters of the PMNS
matrix determine the probability amplitudes of the neutrino
oscillation phenomena that arise from mixing.  The frequency of neutrino oscillation 
depends on the difference in the squares of the neutrino
masses, $\Delta m^{2}_{ij} \equiv m^{2}_{i} - m^{2}_{j}$; a set of three
neutrino mass states implies two independent mass-squared differences
(the ``solar'' mass splitting, $\Delta m^{2}_{21}$, and the ``atmospheric'' mass splitting, 
$\Delta m^{2}_{31}$), where $\Delta m^{2}_{31} = \Delta m^{2}_{32} + \Delta m^{2}_{21}$. The ordering of the
mass states is known as the \emph{neutrino mass hierarchy}. An ordering of
$m_1 < m_2 < m_3$ is known as the \emph{normal hierarchy} since it matches
the mass ordering of the charged leptons in the Standard Model, whereas an ordering of $m_3 < m_1 < m_2$
is referred to as the \emph{inverted hierarchy}.

The entire complement of neutrino experiments to date has measured
five of the mixing parameters~\cite{Gonzalez-Garcia:2014bfa,Capozzi:2013csa,Forero:2014bxa}: the three angles $\theta_{12}$,
$\theta_{23}$ and (recently) $\theta_{13}$, and the two mass differences
$\Delta m^{2}_{21}$ and $\Delta m^{2}_{31}$. The sign of $\Delta
m^{2}_{21}$ is known, but not that of $\Delta m^{2}_{31}$, which 
is the crux of the 
mass hierarchy ambiguity.
The values of $\theta_{12}$ and $\theta_{23}$ are large, while 
$\theta_{13}$ is smaller. The value of \deltacp is unknown.
The absolute values of the entries of the PMNS matrix, which
contains information on the strength of flavor-changing weak decays in
the lepton sector, can be expressed in approximate form as
\begin{equation}
|U_{\rm PMNS}|\sim \left(\begin{array}{ccc} 0.8 & 0.5 & 0.1 \\ 0.5 & 0.6 & 0.7 \\ 0.3 & 0.6 & 0.7\end{array} \right).
\label{eq:pmnsmatrix}
\end{equation}
using values for the mixing angles given in Table~\ref{tab:oscpar_nufit}.
The three-flavor-mixing scenario for neutrinos is now well
established. However, the mixing parameters are not known to the same precision 
as are those in the
corresponding quark sector, and several important quantities, including
the value of \deltacp and the sign of the large mass splitting, are
still undetermined. 

The relationships between the values of the parameters in the neutrino
and quark sectors suggest that mixing in the two sectors is
qualitatively different. Illustrating this difference, the value of
the entries of the CKM quark-mixing matrix (analogous to the PMNS matrix for
neutrinos, and thus indicative of the strength of flavor-changing weak
decays in the quark sector) can be expressed in approximate form as
\begin{equation}
|V_{\rm CKM}|\sim \left(\begin{array}{ccc} 1 & 0.2 & 0.004\\ 0.2 & 1 & 0.04 \\ 0.008 & 0.04 & 1\end{array} \right).
\label{eq:ckmmatrix}
\end{equation}
for comparison to the entries of the PMNS matrix given in Equation~\ref{eq:pmnsmatrix}.
As discussed in \cite{King:2014nza}, the question of why the quark mixing angles are
smaller than the lepton mixing angles is an important part of the 
flavor pattern question. 

To quote the discussion in~\cite{deGouvea:2013onf}, ``while the CKM
matrix is almost proportional to the identity matrix plus
hierarchically ordered off-diagonal elements, the PMNS matrix is far
from diagonal and, with the possible exception of the $U_{e3}$
element, all elements are ${\cal O}(1)$.''
It is important here to note that the smaller of the lepton
mixing angles is of similar magnitude to the larger of the quark mixing parameters, namely the Cabibbo angle~\cite{Boucenna:2012xb}.
One theoretical method often used to address this question involves the use of non-Abelian discrete
subgroups of $SU(3)$ as flavor symmetries; the popularity of this method 
is due in part from
the fact that these symmetries can give rise to the nearly \emph{tri-bi-maximal}\footnote{Tri-bi-maximal mixing refers to a form of the neutrino mixing matrix with effective bimaximal mixing of $\nu_\mu$ and $\nu_\tau$
at the atmospheric scale ($L/E \sim$ \SI{500}{\km / \GeV}) and effective trimaximal
mixing for $\nu_e$ with $\nu_\mu$ and $\nu_\tau$ 
at the solar scale ($L/E \sim$ \SI{15000}{\km / \GeV})~\cite{Harrison:2002er}.} 
structure of the PMNS matrix.
Whether employing these flavor symmetries or other methods,
any theoretical principle that attempts to describe the fundamental
symmetries implied by the observed organization of quark and neutrino
mixing --- such as those proposed in unification models --- leads to
testable predictions such as sum rules between CKM and PMNS
parameters~\cite{King:2014nza,deGouvea:2013onf,Mohapatra:2005wg,Albright:2006cw}.
Data on the patterns of neutrino mixing 
are already proving crucial in the quest for a 
relationship between quarks and leptons and their seemingly arbitrary generation
structure.  

Clearly much work remains in order to complete the standard three-flavor 
mixing picture, particularly 
with regard to $\theta_{23}$ (is it less than, greater than, or equal
to $45^\circ$?), mass hierarchy (normal or inverted?) 
and \deltacp.
Additionally, there is 
great value in obtaining a set of measurements for multiple parameters 
\emph{from a single experiment}, so that correlations and systematic 
uncertainties can be handled properly.  Such an experiment would also be 
well positioned to extensively test the standard picture of three-flavor mixing.  
DUNE is designed to be this experiment.

\section{Expected Event Rate and Sensitivity Calculations}
\label{sec:physics-lbnosc-senscalc}

The oscillation probability of \numu $\rightarrow$ \nue through matter in a constant density
approximation is,  
to first order~\cite{Nunokawa:2007qh}:
\begin{eqnarray}
P(\nu_\mu \rightarrow \nu_e) & \simeq & \sin^2 \theta_{23} \sin^2 2 \theta_{13} 
\frac{ \sin^2(\Delta_{31} - aL)}{(\Delta_{31}-aL)^2} \Delta_{31}^2\\ \nonumber
& & + \sin 2 \theta_{23} \sin 2 \theta_{13} \sin 2 \theta_{12} \frac{ \sin(\Delta_{31} - aL)}{(\Delta_{31}-aL)} \Delta_{31} \frac{\sin(aL)}{(aL)} \Delta_{21} \cos (\Delta_{31} + \mdeltacp)\\ \nonumber
& & + \cos^2 \theta_{23} \sin^2 2 \theta_{12} \frac {\sin^2(aL)}{(aL)^2} \Delta_{21}^2, \\ \nonumber
\label{eqn:appprob}
\end{eqnarray}
where $\Delta_{ij} = \Delta m^2_{ij} L/4E_\nu$, $a = G_FN_e/\sqrt{2}$, $G_F$ is the Fermi constant, $N_e$ is the number density of electrons in the Earth, $L$ is the baseline in km, and $E_\nu$ is the neutrino energy in GeV. 
In the equation above, both \deltacp and $a$ 
switch signs in going from the
$\nu_\mu \to \nu_e$ to the $\bar{\nu}_\mu \to \bar{\nu}_e$ channel; i.e.,
a neutrino-antineutrino asymmetry is introduced both by CP violation (\deltacp)
and the matter effect ($a$). The origin of the matter effect asymmetry 
is simply the presence of electrons and absence of positrons in the Earth.  
In the few-GeV energy range, the asymmetry from the matter effect increases with baseline as the neutrinos
pass through more matter, therefore an experiment with a longer baseline will be
more sensitive to the neutrino mass hierarchy. For baselines longer than 
$\sim$\SI{1200}\km, the degeneracy between the asymmetries from matter
and CP-violation effects can be resolved~\cite{Bass:2013vcg}; hence DUNE, with a baseline of $\sim$\SI{1300}\km, 
will be able to unambiguously
determine the neutrino mass hierarchy \textit{and} measure the value of \deltacp~\cite{Diwan:2004bt}. 

The electron neutrino appearance probability, $P(\nu_\mu \rightarrow \nu_e)$, 
is shown in 
Figure~\ref{fig:oscprob} 
at a baseline of \SI{1300}\km{} as a function of neutrino 
energy for several values of \deltacp. As this figure illustrates, the value 
of \deltacp affects both the amplitude and frequency of
the oscillation. The difference in probability amplitude
for different values of \deltacp is larger at higher oscillation nodes, which 
correspond to energies less than 1.5~GeV. Therefore, a broadband experiment, 
capable of measuring not only the rate of \nue appearance but of mapping out the 
spectrum of observed oscillations down to energies of at least 500~MeV, 
is desirable~\cite{Diwan:2003bp}. Since there are terms proportional to $\sin\mdeltacp$ in Equation~\ref{eqn:appprob},
changes to the value of \deltacp induce opposite changes to \nue and
\anue appearance probabilities, so a beam that is capable of operating in
neutrino mode (forward horn current) and antineutrino mode (reverse horn current)
is also a critical component of the experiment.

\begin{cdrfigure}[Appearance probability for neutrinos and antineutrinos]{oscprob}{The appearance probability at a baseline of 1300~km,
  as a function of neutrino energy, for \deltacp = $-\pi/2$ (blue), 
  0 (red), and $\pi/2$ (green), for neutrinos (left) and antineutrinos
  (right), for normal hierarchy. The black line indicates the oscillation
  probability if $\theta_{13}$ were equal to zero.}
\includegraphics[width=0.45\linewidth]{energy_nu_no.pdf}
\includegraphics[width=0.45\linewidth]{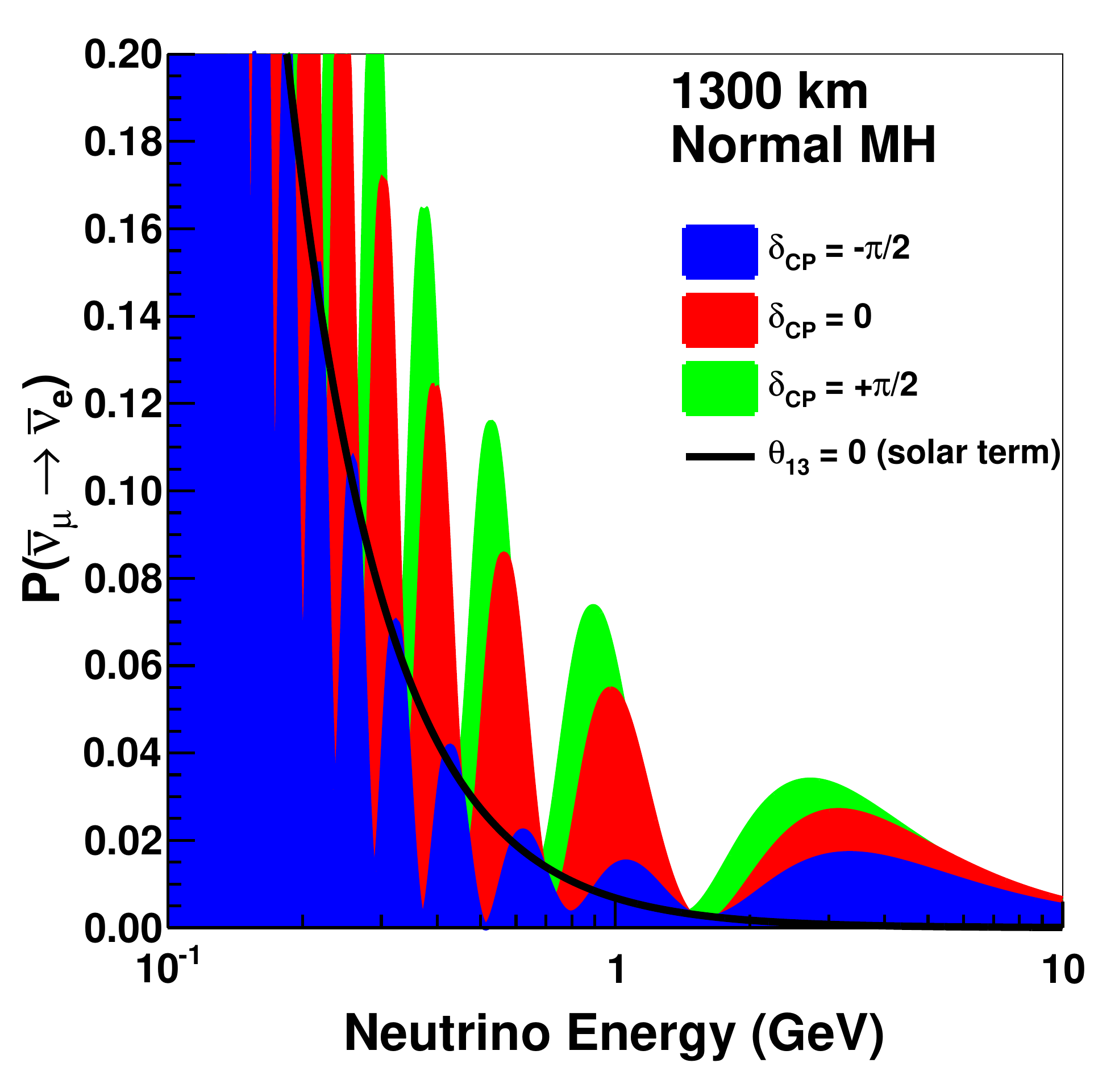}
\end{cdrfigure}

The experimental sensitivities presented here are estimated using
GLoBES\cite{Huber:2004ka,Huber:2007ji}. GLoBES takes neutrino beam
fluxes, cross sections, and detector-response parameterization as
inputs. This document presents a range of possible physics
sensitivities depending on the design of the neutrino beam, 
including the proton beam energy and power used.  
The beam power as a function of proton beam energy from the
PIP-II upgrades and the number of protons-on-target per year assumed
in the sensitivities are shown in Table~\ref{tab:beam_req_pot}. These
numbers assume a combined uptime and efficiency of the FNAL
accelerator complex and the LBNF beamline of 56\%. 
\begin{cdrtable}[Expected POT per year at various primary proton beam
  momenta.]{ccc}{beam_req_pot}{Expected POT per year at various
    primary proton beam momenta.}
Proton Momentum (GeV/c) & Expected Beam Power (MW) & Expected POT/year
\\
\toprowrule 
120 & 1.2 & \num{1.1e21} \\ \colhline
80 & 1.07 & \num{1.47e21} \\ \colhline
60 & 1.03 & \num{1.89e21} \\
\end{cdrtable}
A conservative estimate of sensitivity is calculated using neutrino
fluxes produced from a detailed GEANT4 beamline simulation that is
based on the reference design of the beamline as presented in
\vollbnf.  Neutrino fluxes from a simulation based on an optimized
beam design are used to show the goal sensitivity.  There is a range
of design options that produce sensitivities in between the
sensitivity of the reference beam design and the optimized beam design,
and further optimization is possible. The actual flux will depend upon
details of the hadron production and focusing design; optimization of
the beam design to maximize experimental sensitivity is a critical
aspect of the experiment design.  Table~\ref{tab:beamparam} summarizes
the key properties of the GEANT4 beamline simulations used to
produce fluxes for the sensitivity studies.  The main differences
between the two beam designs are the target geometry, horn current,
horn design and decay pipe length; the choice of horn design has the
biggest effect on the
sensitivity. Section~\ref{sec:physics-lbnosc-beam} describes the
beamline simulations in more detail and explores the potential
improvements that could be achieved by variations in the reference
beam design.

\begin{cdrtable}[Beamline parameters for CDR reference and optimized designs]{lcc}{beamparam}{A comparison of the beamline parameters assumed for the CDR Reference Design flux and the Optimized Design flux used in the sensitivity calculations presented in this chapter.  Section~\ref{sec:physics-lbnosc-beam-req} provides the details.}
Parameter & CDR Reference Design & Optimized Design\\
\toprowrule 
 Proton Beam Energy & 80~GeV & 80~GeV \\  \colhline
 Proton Beam Power & 1.07~MW & 1.07~MW\\ \colhline
 Target & Graphite & Graphite \\ \colhline
 Horn Current & 230~kA & 297~kA \\ \colhline
 Horn Design & NuMI-style & Genetic Optimization \\  \colhline
 Decay Pipe Length & 204~m & 241~m \\ \colhline
 Decay Pipe Diameter & 4~m & 4~m\\
\end{cdrtable}

The signal for \nue appearance is an excess of charged-current (CC)
\nue and \anue interactions over the expected background in the far
detector.  The background to \nue appearance is composed of: (1) CC
interactions of \nue and \anue intrinsic to the beam; (2)
misidentified \numu and \anumu CC events; (3) neutral current (NC)
backgrounds and (4) $\nu_\tau$ and $\bar{\nu}_\tau$ CC events in which
the $\tau$'s decay leptonically into electrons/positrons. NC and
$\nu_\tau$ backgrounds are due to interactions of higher-energy
neutrinos but they contribute to backgrounds mainly at low energy,
which is important for the sensitivity to CP violation.

The LArTPC performance parameters that go into the GLoBES calculation
are generated using the DUNE Fast Monte Carlo (MC) simulation, which
is described in detail in~\cite{Adams:2013qkq}.  The Fast MC combines
the simulated flux, the GENIE neutrino interaction
generator~\cite{Andreopoulos:2009rq}, and a parameterized detector
response that is used to simulate the reconstructed energy and
momentum of each final-state particle. The detector response
parameters used to determine the reconstructed quantities\footnote{The assumptions
on detector response used in the Fast MC are preliminary, and are
expected to improve as the full detector simulation advances and
more information on the performance of LArTPC detectors becomes
available.} are
summarized in Table~\ref{tab:fastmc_detectorinputs}.  The simulated energy
deposition of the particles in each interaction is then used to
calculate reconstructed kinematic quantities (e.g., the neutrino
energy). Event sample classifications ($\nu_e$ CC-like, $\nu_{\mu}$
CC-like, or NC-like), including mis-ID rates, are determined by the
identification of lepton candidates. Lepton candidates are selected
based on a variety of criteria including particle kinematics,
detector thresholds, and probabilistic estimates of particle fates. To
reduce the NC and $\nu_{\tau}$ CC backgrounds in the $\nu_e$ CC-like
sample, an additional discriminant is formed using reconstructed
transverse momentum along with reconstructed neutrino and hadronic
energy as inputs to a k-Nearest-Neighbor (kNN) machine-learning
algorithm.  Figures~\ref{fig:smearing_nue} and \ref{fig:smearing_nc} show
the true-to-reconstructed energy smearing matrices extracted from the
Fast MC and used as inputs to GLoBES.  Figure~\ref{fig:eff} shows the
analysis sample detection $\times$ selection efficiencies for the
various signal and background modes used by GLoBES, also extracted
from the Fast MC.

\begin{cdrtable}[Fast MC Detector Response Summary]{lclc}{fastmc_detectorinputs}{Summary of the single-particle
    far detector response used in the Fast MC. For some particles, the response depends upon behavior or
    momentum, as noted in the table. If a muon or a pion that is mis-identified as a muon is
    contained within the detector, the momentum is smeared based on track length.
    Exiting particles are smeared based on the contained energy.
    For neutrons with momentum <~1~GeV/c,
    there is a 10\% probability that the particle will escape detection, so the reconstructed energy is
    set to zero. For neutrons that are detected, the reconstructed energy is taken to be 60\%
    of the deposited energy after smearing.}
  Particle type & Detection & Energy/Momentum & Angular \\ \rowtitlestyle
  & Threshold (KE) & Resolution & Resolution \\ \toprowrule
  $\mu^{\pm}$ & 30 MeV & Contained track: track length  & 1$^\circ$\\
              &        & Exiting track: 30\%            & \\ \colhline
  $\pi^{\pm}$ & 100 MeV & $\mu$-like contained track:  track length & 1$^\circ$\\
    &         & $\pi$-like contained track: 5\% & \\
  &           & Showering or exiting: 30\% & \\ \colhline
  e$^{\pm}$/$\gamma$ & 30 MeV & 2\% $\oplus$ 15\%/$\sqrt{E}$[GeV] & 1$^\circ$ \\ \colhline
  p & 50 MeV & p<400 MeV/c: 10\% & 5$^\circ$ \\
  &        & p>400 MeV/c: 5\% $\oplus$ 30\%/$\sqrt{E}$[GeV] & \\ \colhline
  n & 50 MeV & 40\%/$\sqrt{E}$[GeV] & 5$^\circ$ \\ \colhline
  other & 50 MeV & 5\% $\oplus$ 30\%/$\sqrt{E}$[GeV] & 5$^\circ$ \\ \colhline
  \end{cdrtable}

\begin{cdrfigure}[\nue and \anue True-to-reconstructed energy smearing matrices]{smearing_nue}{True-to-reconstructed energy smearing matrices for \nue (top), \numu (center) and \nutau (bottom) CC interactions extracted from the Fast MC and used as inputs to GLoBES.  Left: Used in the \nue appearance sample. Right: Used in the \numu disappearance samples}
 \includegraphics[width=0.4\textwidth]{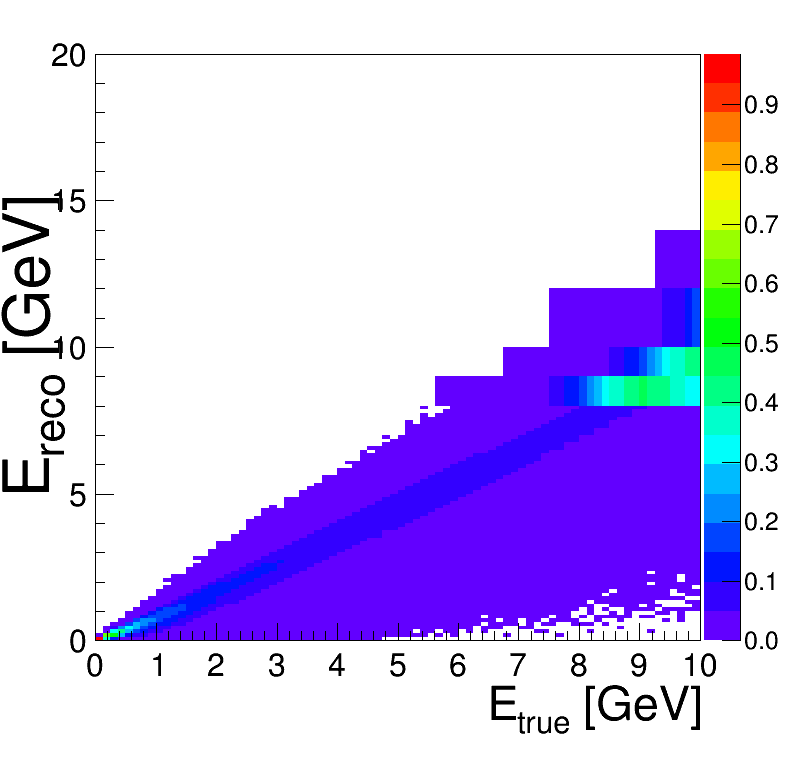}
\includegraphics[width=0.4\textwidth]{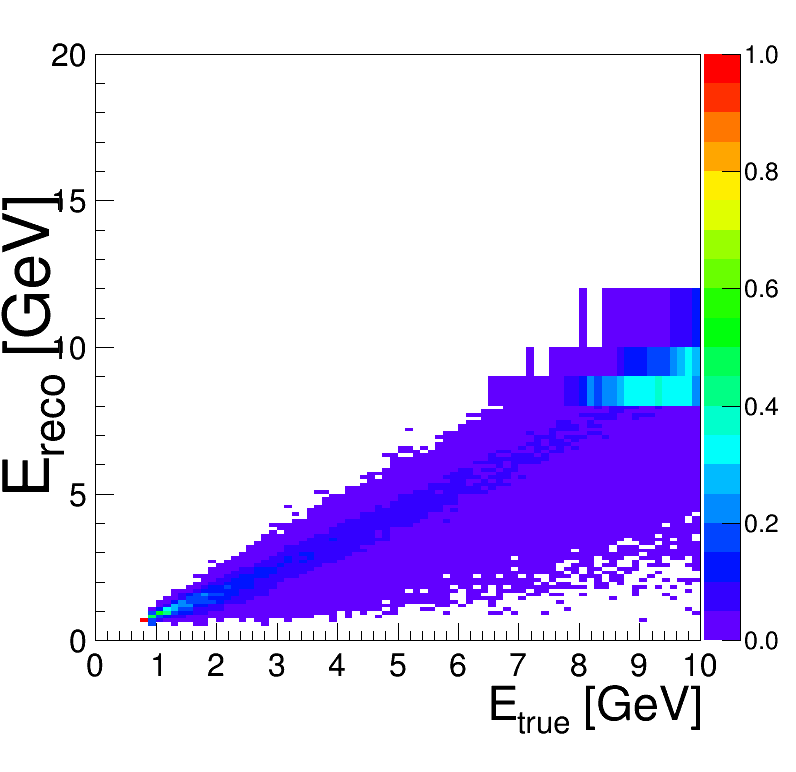}
 \includegraphics[width=0.4\textwidth]{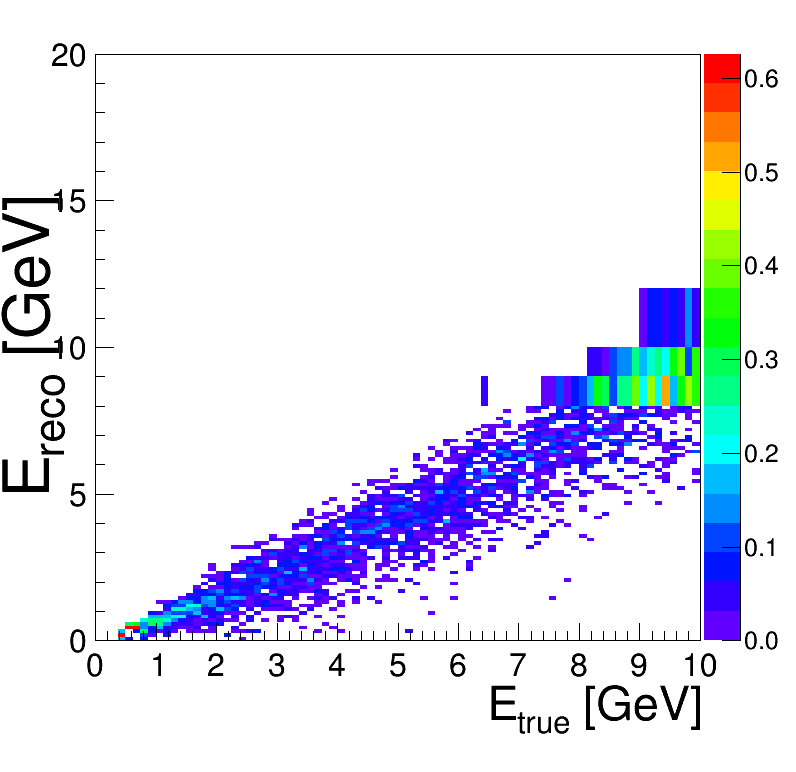}
\includegraphics[width=0.4\textwidth]{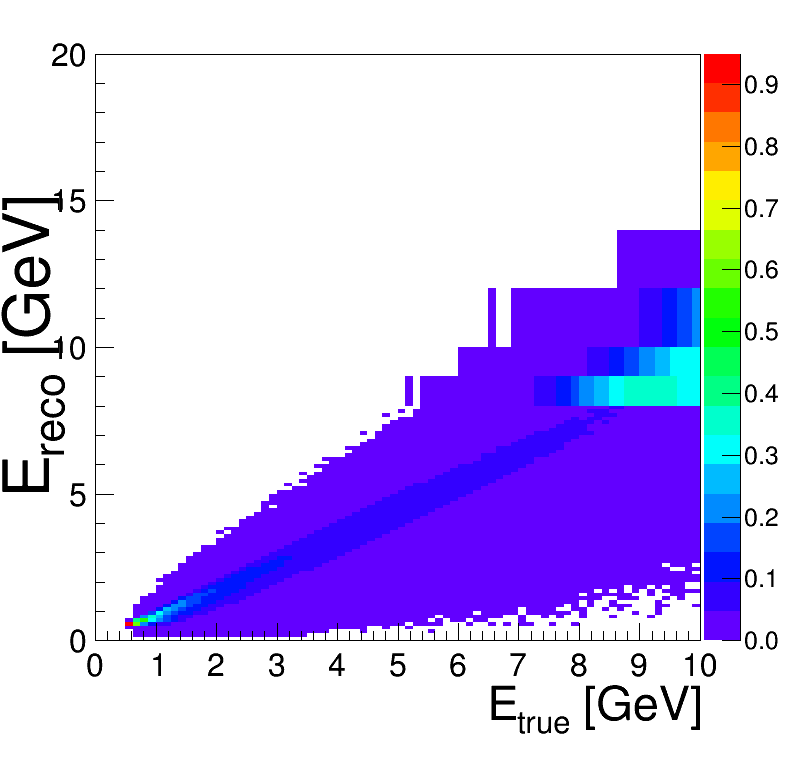}
 \includegraphics[width=0.4\textwidth]{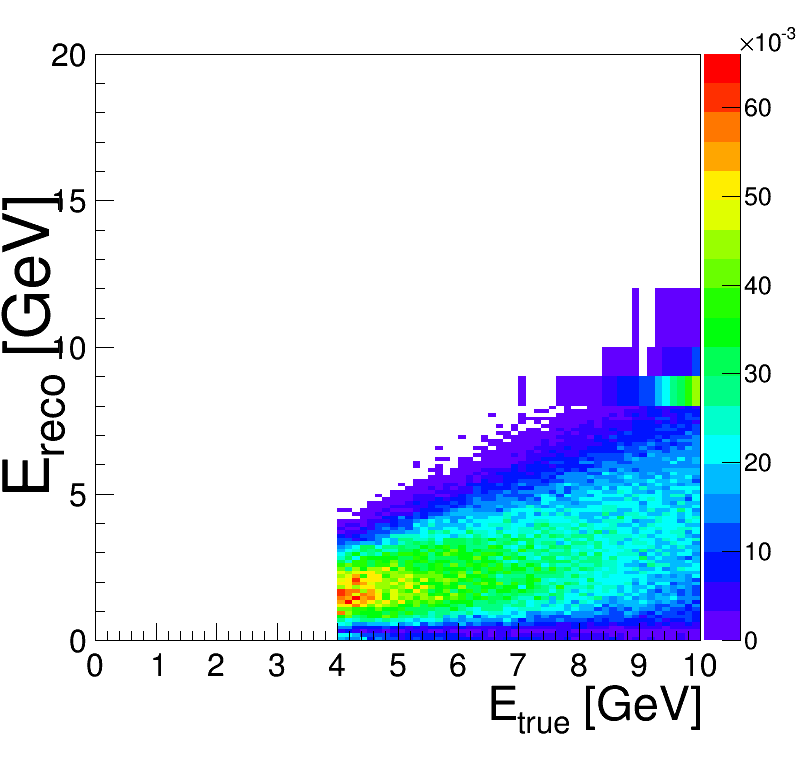}
 \includegraphics[width=0.4\textwidth]{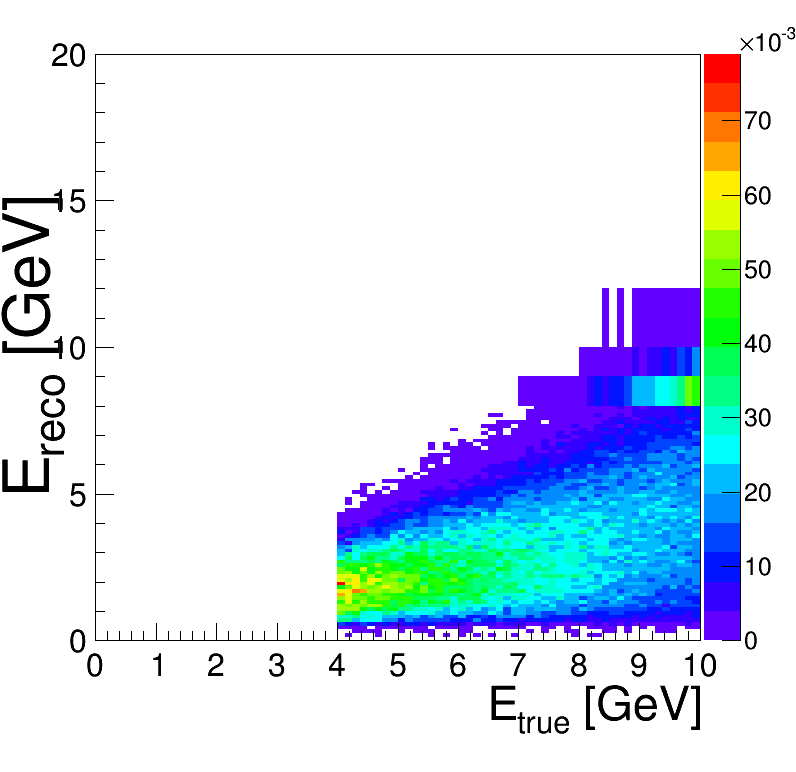}
\end{cdrfigure}



\begin{cdrfigure}[NC True-to-reconstructed energy smearing matrices]{smearing_nc}{True-to-reconstructed energy smearing matrices for NC interactions extracted from the Fast MC and used as inputs to GLoBES.  Left: Used in the \nue appearance sample.  Right: Used in the \numu disappearance sample.}
 \includegraphics[width=0.49\textwidth]{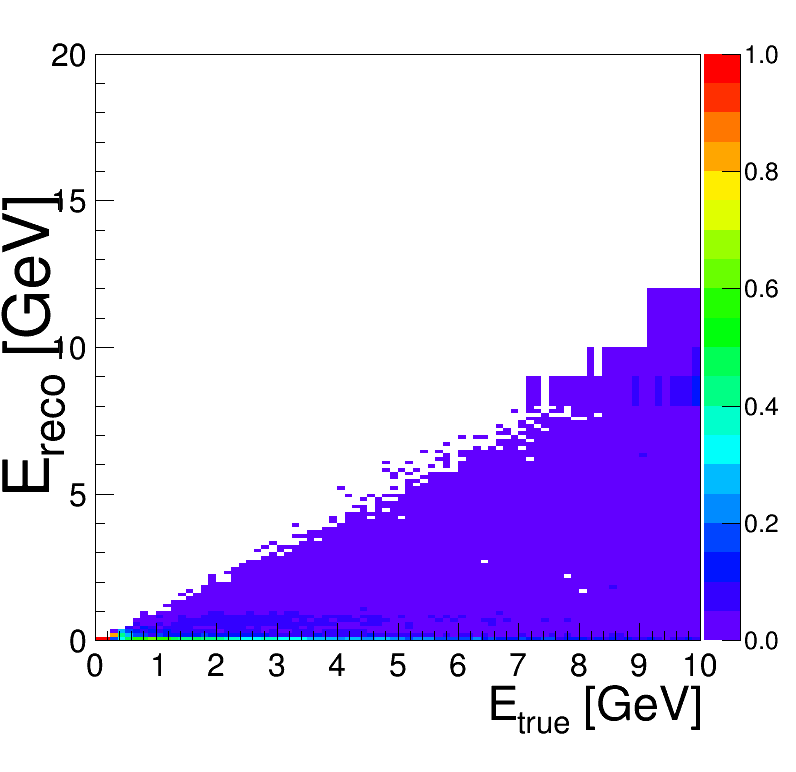}
 \includegraphics[width=0.49\textwidth]{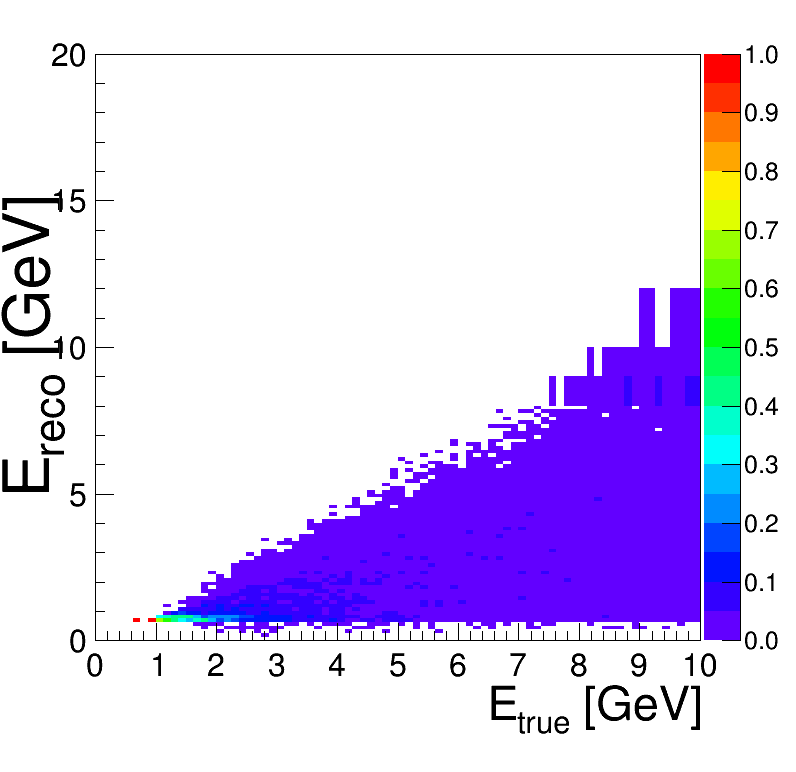}
\end{cdrfigure}

\begin{cdrfigure}[Analysis sample detection $\times$ selection efficiencies]{eff}{Analysis sample detection $\times$ selection efficiencies for the various signal and background modes extracted from the Fast MC and used as inputs to GLoBES.  Top: Used in the \nue appearance sample. Bottom: Used in the \numu disappearance sample.  Left: Neutrino beam mode.  Right: Antineutrino beam mode.  The NC backgrounds (and \numu CC backgrounds for the appearance mode) have been increased by a factor of 10 for visibility.}
 \includegraphics[width=0.49\textwidth]{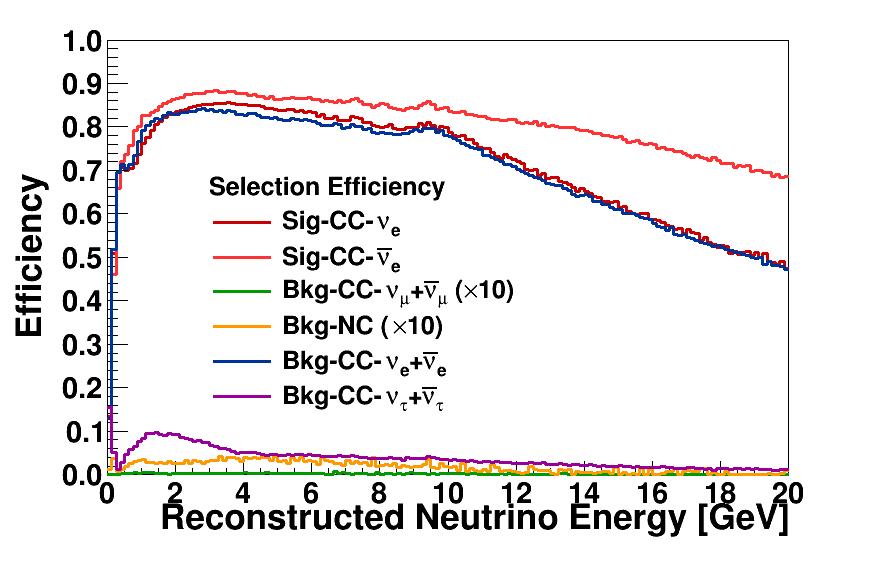}
 \includegraphics[width=0.49\textwidth]{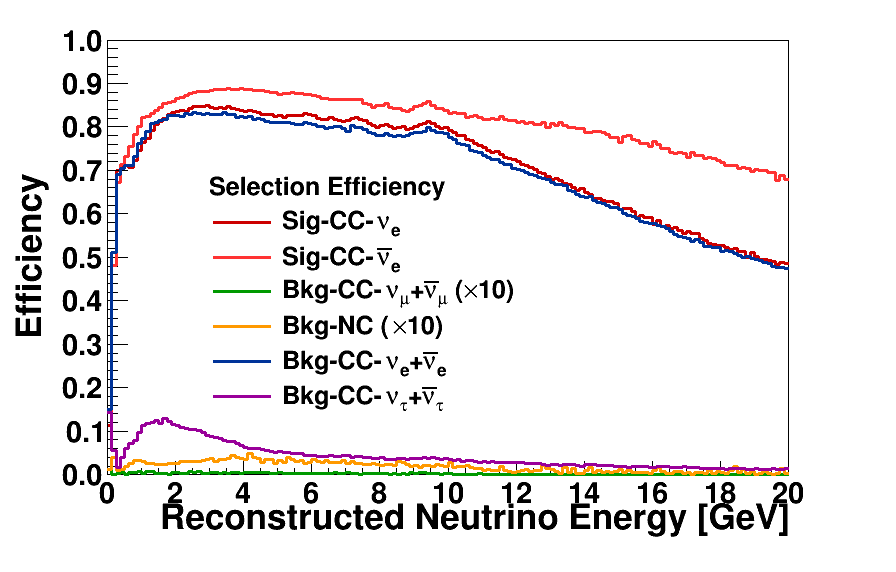}
 \includegraphics[width=0.49\textwidth]{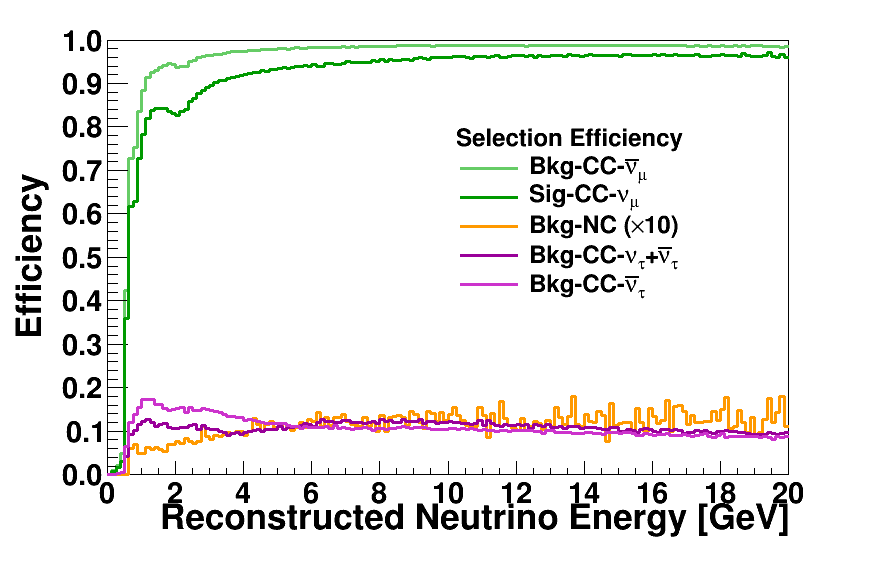}
 \includegraphics[width=0.49\textwidth]{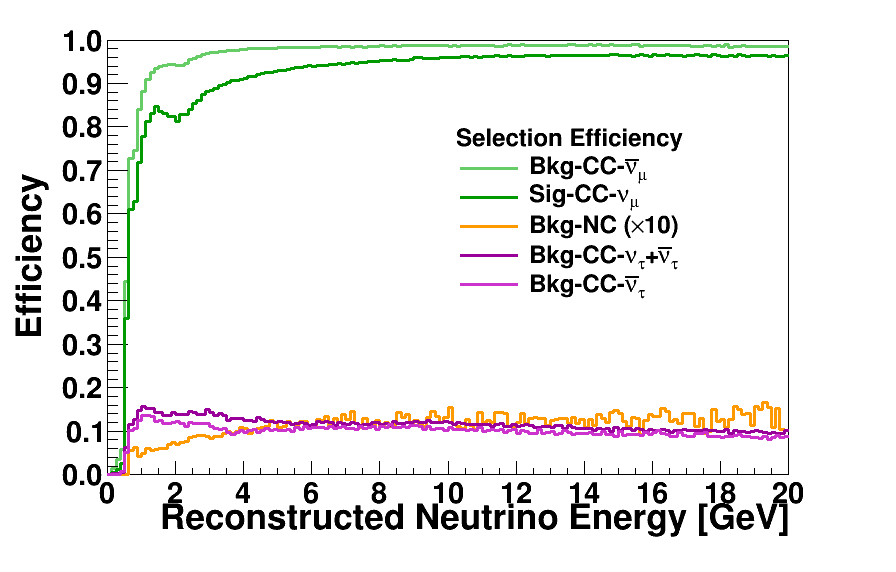}
\end{cdrfigure}

The cross section inputs to GLoBES have been generated using GENIE
2.8.4~\cite{Andreopoulos:2009rq}.  The neutrino oscillation parameters
and the uncertainty on those parameters are taken from the
Nu-Fit~\cite{Gonzalez-Garcia:2014bfa} global fit to neutrino data; the
values are given in Table~\ref{tab:oscpar_nufit}.  (See also
\cite{Capozzi:2013csa} and \cite{Forero:2014bxa} for other recent
global fits.) Most of the sensitivities in this chapter are shown
assuming normal hierarchy; this is an arbitrary choice for simplicity
of presentation.

\begin{cdrtable}[Oscillation parameter values and relative uncertainties]{lcc}{oscpar_nufit}{Central value and relative uncertainty of neutrino oscillation parameters from a global fit~\cite{Gonzalez-Garcia:2014bfa} to neutrino oscillation data. Because the probability distributions are somewhat non-Gaussian (particularly for $\theta_{23}$), the relative uncertainty is computed using 1/6 of the 3$\sigma$ allowed range from the fit, rather than the 1$\sigma$ range.   For $\theta_{23}$ and $\Delta m^2_{31}$, the best-fit values and uncertainties depend on whether normal mass hierarchy (NH) or inverted mass hierarchy (IH) is assumed.}
Parameter &    Central Value & Relative Uncertainty \\
\toprowrule
$\theta_{12}$ & 0.5843 & 2.3\% \\ \colhline
$\theta_{23}$ (NH) & 0.738  & 5.9\% \\ \colhline
$\theta_{23}$ (IH) & 0.864  & 4.9\% \\ \colhline
$\theta_{13}$ & 0.148  & 2.5\% \\ \colhline
$\Delta m^2_{21}$ & 7.5$\times10^{-5}$~eV$^2$ & 2.4\% \\ \colhline
$\Delta m^2_{31}$ (NH) & 2.457$\times10^{-3}$~eV$^2$ &  2.0\% \\ \colhline
$\Delta m^2_{31}$ (IH) & -2.449$\times10^{-3}$~eV$^2$ &  1.9\% \\
\end{cdrtable}

Figures~\ref{fig:appspectra} and~\ref{fig:disspectra} show the
expected event rate for \nue appearance and \numu disappearance,
respectively, including expected flux, cross section, and oscillation
probabilities as a function of neutrino energy at a baseline of
\num{1300}~km. The spectra are shown for a \SI{150}~\ktMWyr{} exposure each for
neutrino and antineutrino beam mode, for a total \SI{300}~\ktMWyr{}
exposure.  The optimized beam design results in an increased signal
rate in the lower-energy region. Tables~\ref{tab:apprates}
and~\ref{tab:disrates} give the integrated rate for the
$\nu_e$ appearance and $\nu_\mu$ disappearance spectra,
respectively.  The spectra and rates are shown for both the reference
beam design and the optimized beam design.

\begin{cdrfigure}[\nue and \anue appearance spectra]{appspectra}{\nue and \anue appearance spectra: Reconstructed energy distribution of selected $\nu_e$ CC-like events assuming a \SI{150}~\ktMWyr{} exposure in the neutrino-beam mode (left) and antineutrino-beam mode (right), for a total \SI{300}~\ktMWyr{} exposure.  The plots assume normal mass hierarchy and $\mdeltacp = 0$.  The spectra are shown for both the CDR reference beam design and the optimized beam design as described in Section~\ref{sec:physics-lbnosc-beam-req}.}
 \includegraphics[width=0.49\textwidth]{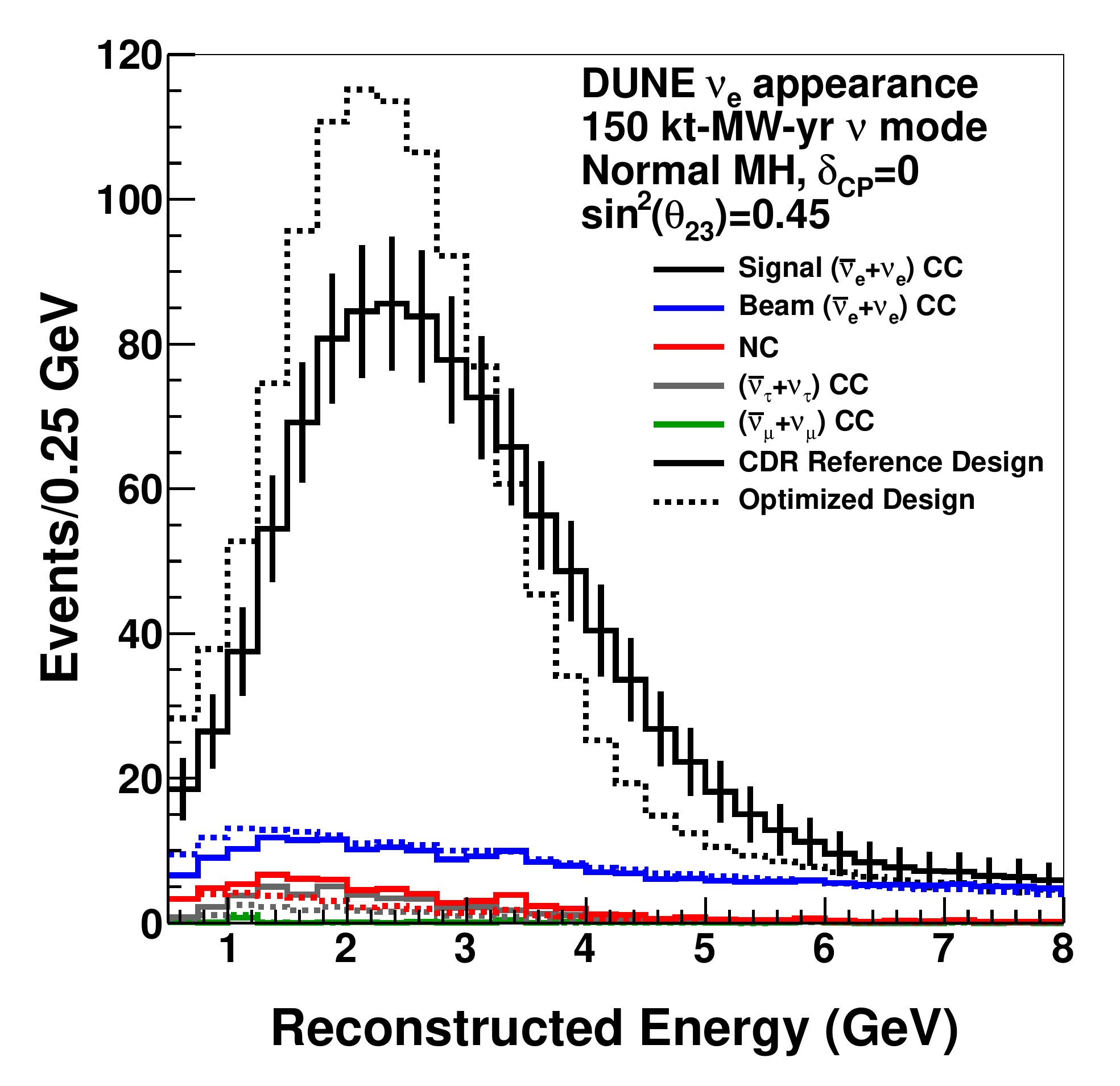}
 \includegraphics[width=0.49\textwidth]{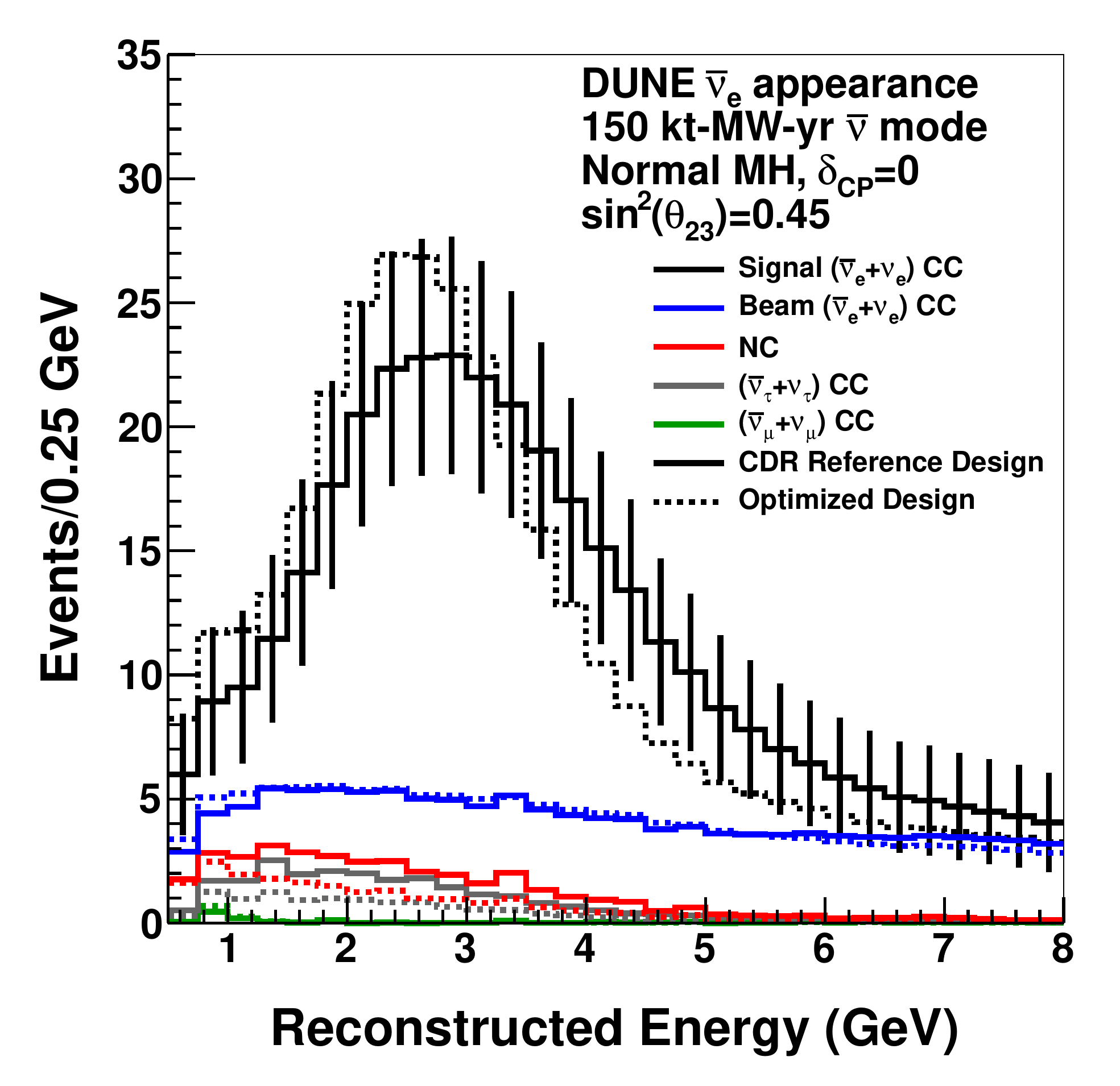}
\end{cdrfigure}

\begin{cdrfigure}[\numu and \anumu disappearance spectra]{disspectra}{\numu and \anumu disappearance spectra: Reconstructed energy distribution of selected $\nu_{\mu}$ CC-like events assuming a \SI{150}~\ktMWyr{} exposure in the neutrino-beam mode (left) and antineutrino-beam mode (right), for a total \SI{300}~\ktMWyr{} exposure.  The plots assume normal mass hierarchy and $\mdeltacp = 0$.  The spectra are shown for both the CDR reference beam design and the optimized beam design as described in Section~\ref{sec:physics-lbnosc-beam-req}.}
 \includegraphics[width=0.49\textwidth]{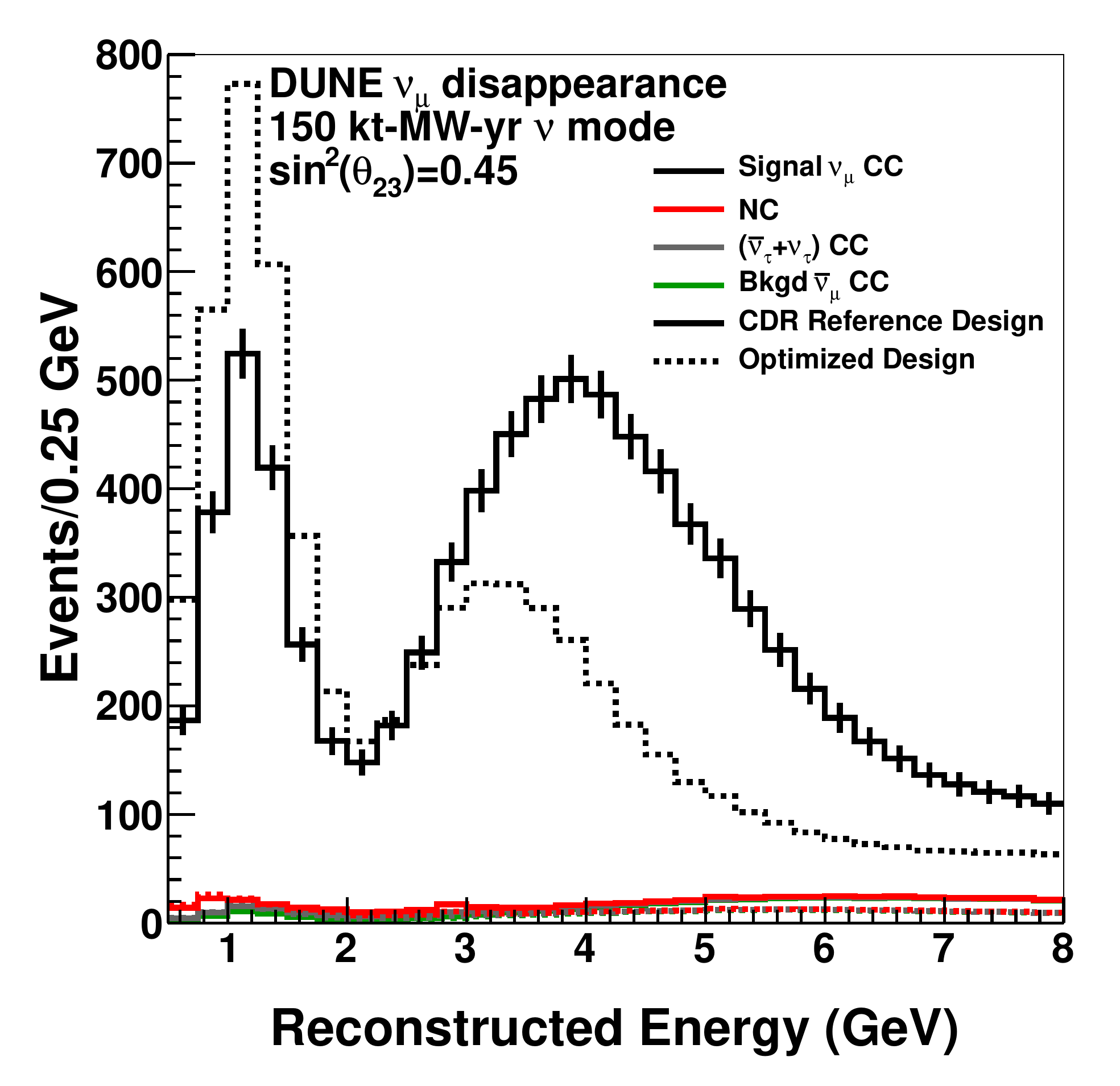}
 \includegraphics[width=0.49\textwidth]{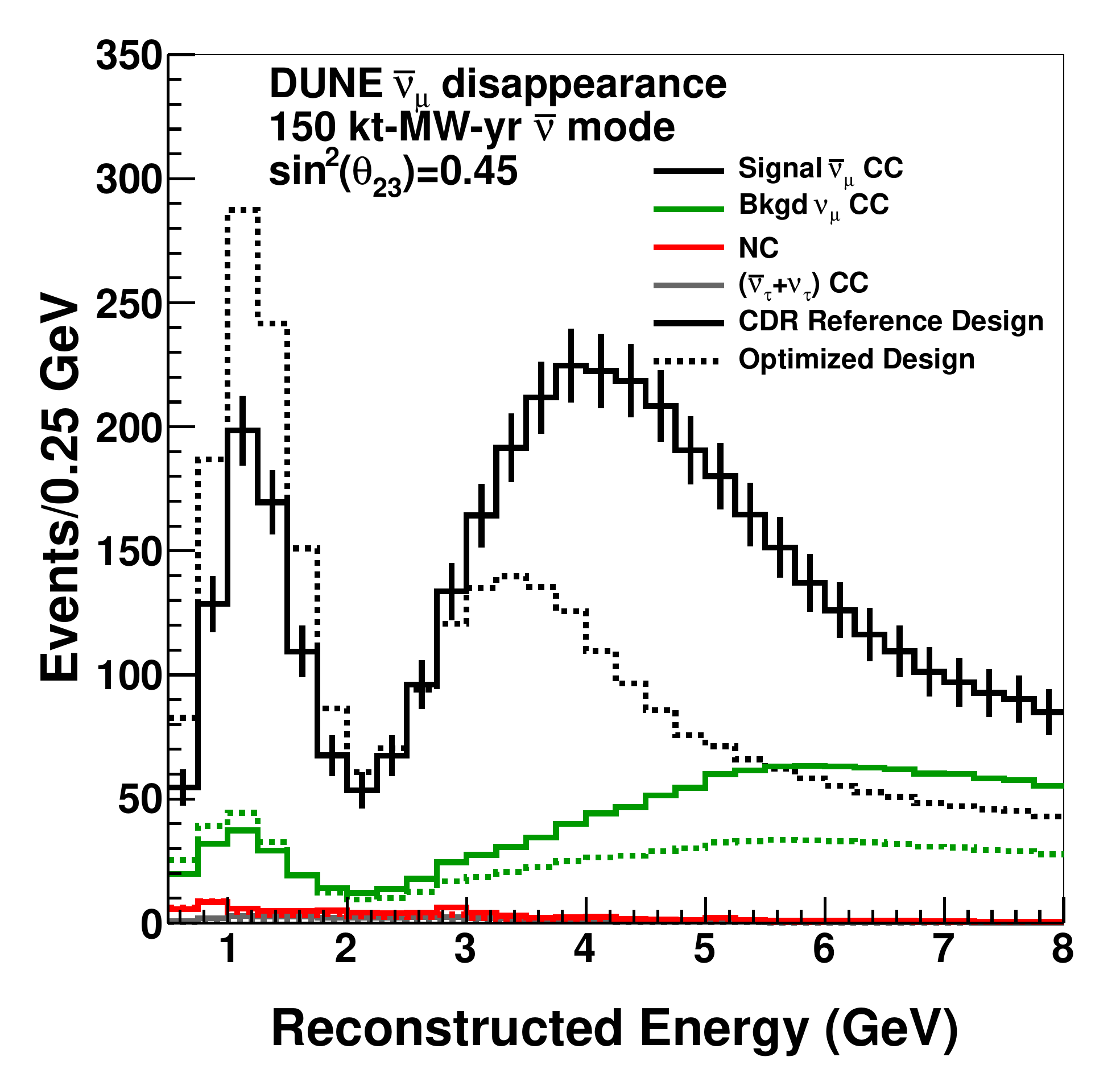}
\end{cdrfigure}

\begin{cdrtable}[\nue and \anue appearance rates]{lcc}{apprates}{\nue and \anue appearance rates: Integrated rate of selected $\nu_e$ CC-like events between 0.5 and 8.0~GeV assuming a \SI{150}~\ktMWyr{} exposure in the neutrino-beam mode and antineutrino-beam mode.  The signal rates are shown for both normal mass hierarchy (NH) and inverted mass hierarchy (IH), and all the background rates assume normal mass hierarchy.  All the rates assume $\mdeltacp = 0$, and the rates are shown for both the CDR reference beam design and the optimized beam as described in Section~\ref{sec:physics-lbnosc-beam-req}.}
  & CDR Reference Design & Optimized Design\\
  \toprowrule
 $\nu$ mode (\SI{150}~\ktMWyr{}) & & \\
 \colhline 
 \nue Signal NH (IH) & 861 (495) & 945 (521)\\
 \anue Signal NH (IH) & 13 (26) & 10 (22)\\
  \colhline
 Total Signal NH (IH) & 874 (521) & 955 (543) \\
  \colhline 
 Beam $\nu_{e}+\bar{\nu}_{e}$ CC Bkgd & 159 & 204 \\
 NC Bkgd & 22 & 17 \\
 $\nu_{\tau}+\bar{\nu}_{\tau}$ CC Bkgd & 42 & 19 \\
 $\nu_{\mu}+\bar{\nu}_{\mu}$ CC Bkgd & 3 & 3 \\
  \colhline
 Total Bkgd & 226 & 243 \\
 \toprowrule
 $\bar{\nu}$ mode (\SI{150}~\ktMWyr{}) & & \\
 \colhline 
 \nue Signal NH (IH) & 61 (37) & 47 (28)\\
 \anue Signal NH (IH) & 167 (378) & 168 (436)\\
  \colhline
 Total Signal NH (IH) & 228 (415) & 215 (464) \\
  \colhline 
 Beam $\nu_{e}+\bar{\nu}_{e}$ CC Bkgd & 89 & 105 \\
 NC Bkgd & 12 & 9 \\
 $\nu_{\tau}+\bar{\nu}_{\tau}$ CC Bkgd & 23 & 11 \\
 $\nu_{\mu}+\bar{\nu}_{\mu}$ CC Bkgd & 2 & 2 \\
  \colhline 
 Total Bkgd & 126 & 127 \\
\end{cdrtable}

\begin{cdrtable}[\numu and \anumu disappearance rates]{lcc}{disrates}{\numu and \anumu disappearance rates: Integrated rate of selected $\nu_{\mu}$ CC-like events between 0.5 and 20.0~GeV assuming a \SI{150}~\ktMWyr{} exposure in the neutrino-beam mode and antineutrino-beam mode.  The rates are shown for normal mass hierarchy and $\mdeltacp = 0$, and the rates are shown for both the CDR reference beam design and the optimized beam as described in Section~\ref{sec:physics-lbnosc-beam-req}.}
  & CDR Reference Design & Optimized Design\\
  \toprowrule
 $\nu$ mode (\SI{150}~\ktMWyr{}) & & \\
 \colhline 
 \numu Signal & 10842 & 7929 \\
 \colhline 
  \anumu CC Bkgd & 958 & 511 \\
 NC Bkgd & 88 & 76 \\
 $\nu_{\tau}+\bar{\nu}_{\tau}$ CC Bkgd & 63 & 29 \\
 \toprowrule
 $\bar{\nu}$ mode (\SI{150}~\ktMWyr{}) & & \\
\colhline 
 \anumu Signal & 3754 & 2639 \\
\colhline 
  \numu CC Bkgd & 2598 & 1525 \\
 NC Bkgd & 50 & 41 \\
 $\nu_{\tau}+\bar{\nu}_{\tau}$ CC Bkgd & 39 & 18 \\
\end{cdrtable}

Sensitivities to the neutrino mass hierarchy and the degree of CP
violation are obtained by simultaneously fitting the $\nu_\mu
\rightarrow \nu_\mu$, $\bar{\nu}_\mu \rightarrow \bar{\nu}_\mu$,
$\nu_\mu \rightarrow \nu_e$, and $\bar{\nu}_\mu \rightarrow
\bar{\nu}_e$ oscillated spectra.  It is assumed that 50\% of the total
exposure comes in neutrino beam mode and 50\% in antineutrino beam
mode.  A 50\%/50\% ratio of neutrino to antineutrino data has been
shown to produce a nearly optimal sensitivity, and small deviations
from this (e.g., 40\%/60\%, 60\%/40\%) produce negligible changes in
the sensitivity.

The neutrino oscillation parameters are all allowed to vary,
constrained by a Gaussian prior with 1$\sigma$ width 
as given by the relative uncertainties shown in
Table~\ref{tab:oscpar_nufit}.  The effect of systematic uncertainty is
approximated using signal and background normalization uncertainties,
which are treated as 100\% uncorrelated among the four samples.  The
baseline systematic uncertainty estimates and the effect of
considering larger signal and background normalization uncertainties,
as well as some energy-scale uncertainties are discussed in
Section~\ref{sec:physics-lbnosc-beamnd-req}.

In these fits, experimental sensitivity is quantified using a test
statistic, $\Delta\chi^2$, which is calculated by comparing the
predicted spectra for alternate hypotheses.  These quantities are
defined, differently for neutrino mass hierarchy and CP-violation
sensitivity, to be:
\begin{eqnarray}
\Delta\chi^2_{MH} & = & \chi^2_{IH} - \chi^2_{NH}\textrm{ (true normal hierarchy),}\label{eq:dx2_MH}\\ 
\Delta\chi^2_{MH} & = & \chi^2_{NH} - \chi^2_{IH}\textrm{ (true inverted hierarchy),}\\
\Delta\chi^2_{CPV} & = & Min[\Delta\chi^2_{CP}(\mdeltacp^{test}=0),\Delta\chi^2_{CP}(\mdeltacp^{test}=\pi)]\textrm{, where} \\
\Delta\chi^2_{CP} & = & \chi^2_{\mdeltacp^{test}} - \chi^2_{\mdeltacp^{true}}.\label{eq:dx2_CP} \\ \nonumber
\end{eqnarray}
Since the true value of $\mdeltacp$ is unknown, a scan is performed
over all possible values of $\mdeltacp^{true}$. Both the neutrino mass
hierarchy and the $\theta_{23}$ octant are also assumed to be unknown and
are varied in the fits, with the lowest value of $\Delta\chi^2$ thus
obtained used to estimate the sensitivities.

A ``typical
experiment'' is defined as one with the most probable data given a set of input
parameters, i.e., in which no statistical fluctuations have been
applied. 
 In this case, the predicted spectra and the true spectra are
identical; for the example of CP violation,
$\chi^2_{\mdeltacp^{true}}$ is identically zero and the
$\Delta\chi^2_{CP}$ value for a typical experiment is given by
$\chi^2_{\mdeltacp^{test}}$.

\section{Mass Hierarchy}
\label{sec:physics-lbnosc-mh}

The \kmadj{1300} baseline establishes one of DUNE's key strengths:
sensitivity to the matter effect. This effect leads to a large
asymmetry in the $\nu_\mu\to \nu_e$ versus $\bar{\nu}_\mu \to
\bar{\nu}_e$ oscillation probabilities, the sign of which depends
on the mass hierarchy (MH).  At 1300~km this asymmetry is
approximately $\pm 40\%$ in the region of the peak flux; this is
larger than the maximal possible CP-violating asymmetry associated
with \deltacp, meaning that both the MH and \deltacp can be determined
unambiguously with high confidence within the same experiment using
the beam neutrinos.  DUNE's goal is to determine the MH with a
significance of at least $\sqrt{\overline{\Delta\chi^{2}}} = 5$ for
all \deltacp values using beam neutrinos.  Concurrent analysis of the
corresponding atmospheric-neutrino samples will improve the precision
with which the MH is resolved.

Figure~\ref{fig:mh_nominal} shows the significance with which the MH
can be determined as a function of the value of \deltacp, for an
exposure of \SI{300}~\ktMWyr, which corresponds to seven years of data
(3.5 years in neutrino mode plus 3.5 years in antineutrino mode) with
a 40-kt detector and a 1.07-MW 80-GeV beam.  For this exposure, the MH
is determined with a minimum significance of
$\sqrt{\overline{\Delta\chi^{2}}} = 5$ for 100\% of the \deltacp
values for the optimized beam design and nearly 100\% of \deltacp
values for the CDR reference beam design.
Figure~\ref{fig:mh_exposure} shows the significance with which the MH
can be determined for 0\% (most optimistic), 50\% and 100\% of
\deltacp values as a function of exposure.  Minimum exposures of
approximately \SI{400}~\ktMWyr{} and \SI{230}~\ktMWyr{} are required
to determine the MH with a significance of
$\sqrt{\overline{\Delta\chi^2}} = 5$ for 100\% of \deltacp values for
the CDR reference beam design and the optimized beam design,
respectively.

\begin{cdrfigure}[Mass hierarchy sensitivity for a \SI{300}~\ktMWyr{} exposure]{mh_nominal}{The significance with which the mass hierarchy can be determined as a function of the value of \deltacp for an exposure of \SI{300}~\ktMWyr{} assuming normal MH (left) or inverted MH (right).  The shaded region represents the range in sensitivity due to potential variations in the beam design.}
 \includegraphics[width=0.49\textwidth]{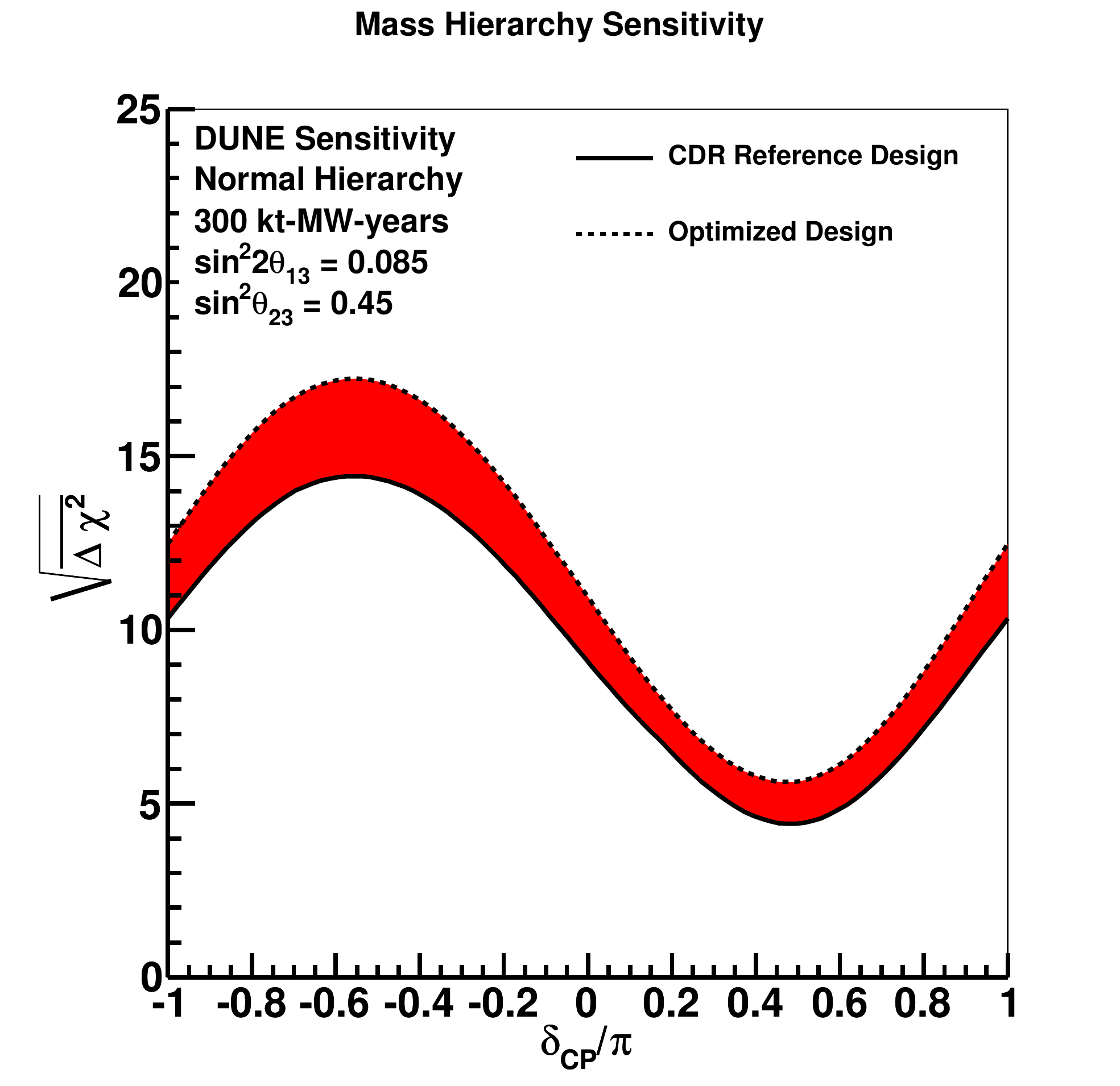}
 \includegraphics[width=0.49\textwidth]{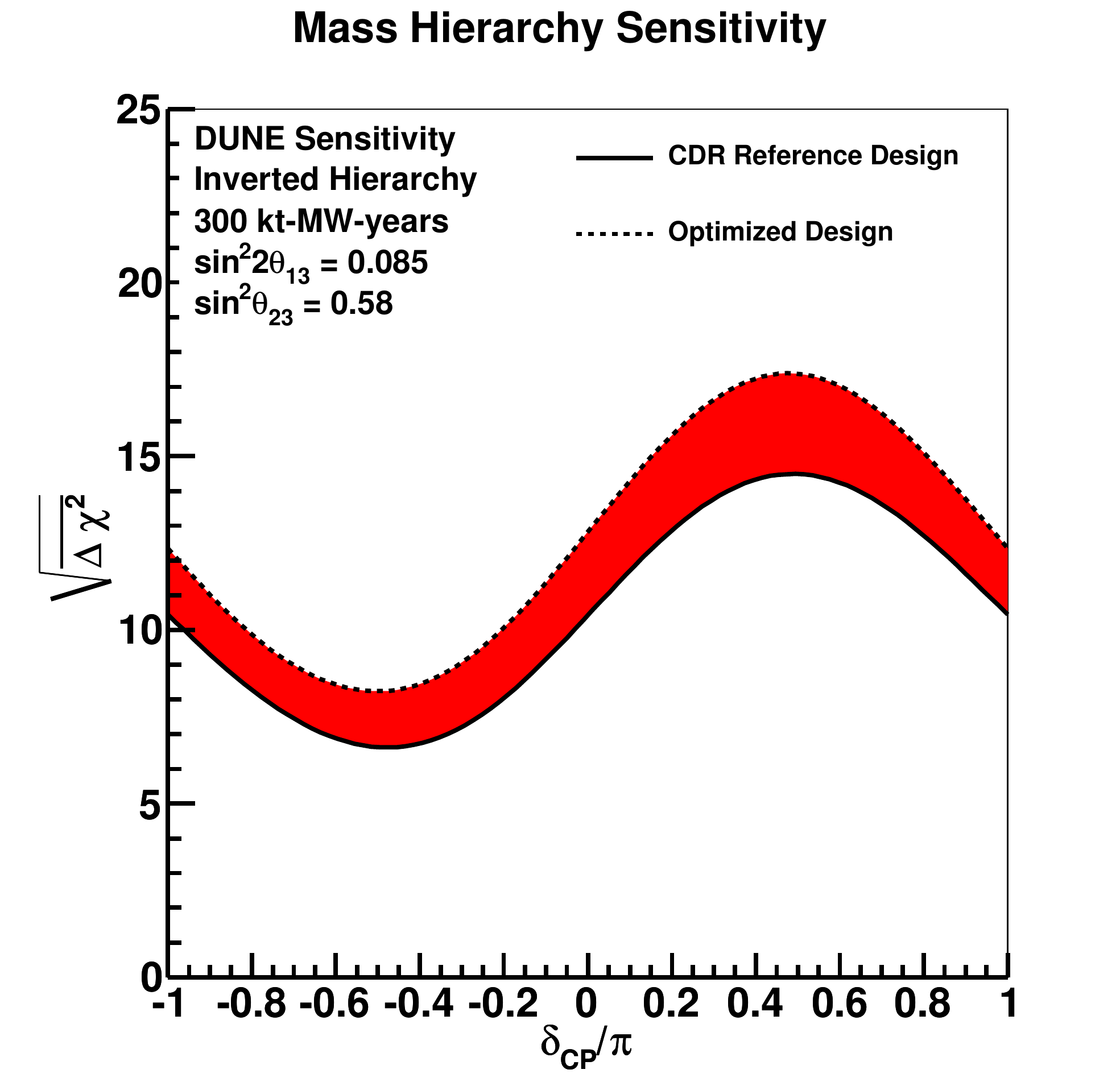}
\end{cdrfigure}

\begin{cdrfigure}[Mass hierarchy sensitivity as a function of exposure]{mh_exposure}{The minimum significance with which the mass hierarchy can be determined for all values of \deltacp (100\%), 50\% and in the most optimistic scenario (0\%) as a function of exposure.  The two different shaded bands represent the different sensitivities due to variations in the beam design. This plot assumes normal mass hierarchy. (The inverted hierarchy case is very similar.) }
 \includegraphics[width=\textwidth]{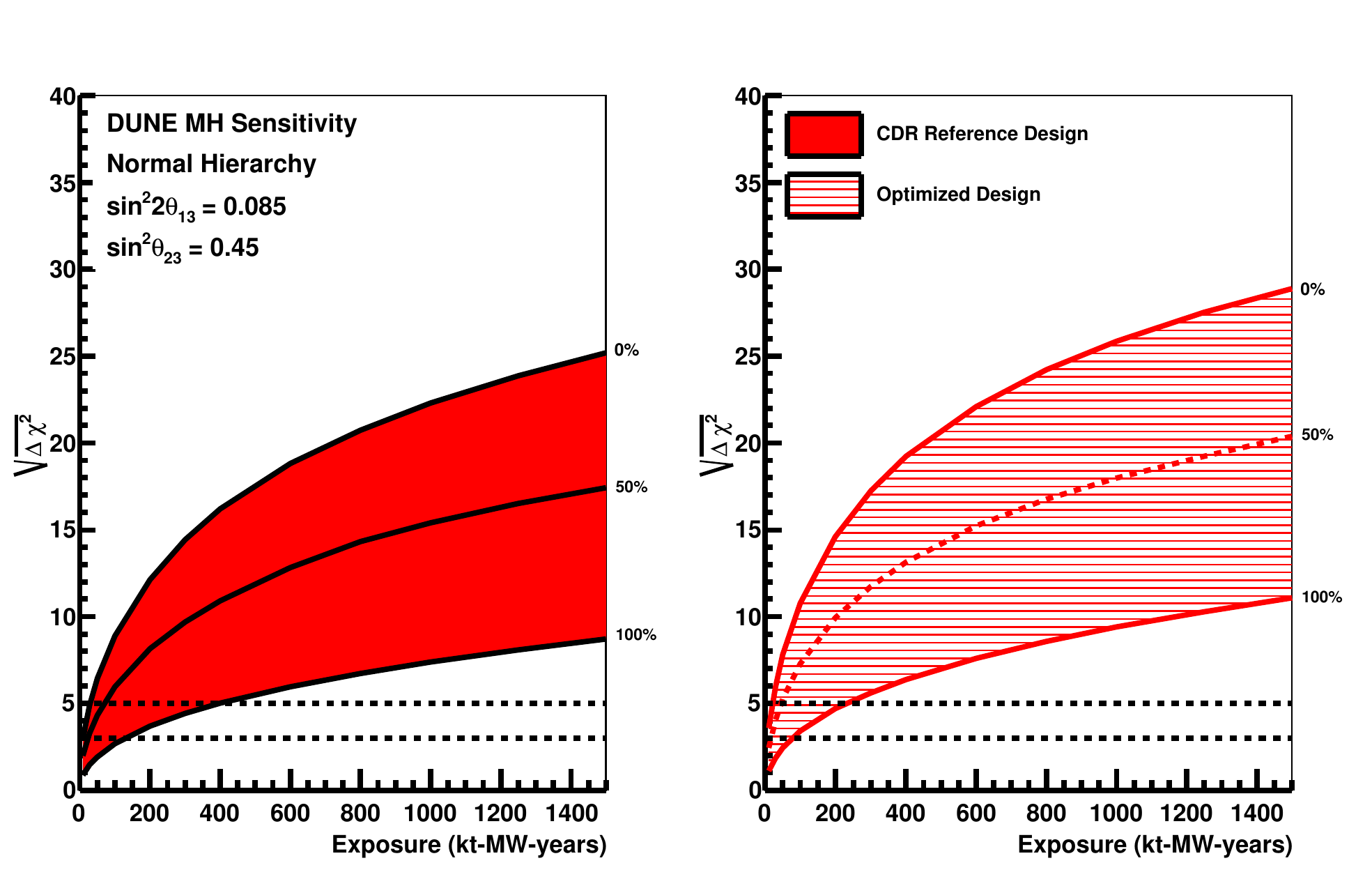}
\end{cdrfigure}

Figures~\ref{fig:mh_theta23}, \ref{fig:mh_theta13}, and
\ref{fig:mh_deltamsq} show the variation in the MH sensitivity due to
different values of $\theta_{23}$, $\theta_{13}$, and \dm{31} within
the allowed ranges.  The value of $\theta_{23}$ has the biggest impact
on the sensitivity, and the least favorable scenario corresponds to a
true value of \deltacp in which the MH asymmetry is maximally offset
by the leptonic CP asymmetry, and where, independently,
$\sin^2{\theta_{23}}$ takes on a value at the low end of its
experimentally allowed range.

\begin{cdrfigure}[Variation in MH sensitivity due to $\theta_{23}$]{mh_theta23}{The variation in the MH sensitivity due to different values of $\theta_{23}$ within the allowed range.  In this figure, the nominal value of $\sin^2\theta_{23} = 0.45$ provides a significance of at least $\sqrt{\overline{\Delta\chi^{2}}} = 5$ for all values of \deltacp. (See Figure~\ref{fig:mh_exposure} for the possible range of exposures to achieve this level of significance.) The significance decreases for all values of \deltacp as $\sin^2\theta_{23}$ gets smaller.}
 \includegraphics[width=0.7\textwidth]{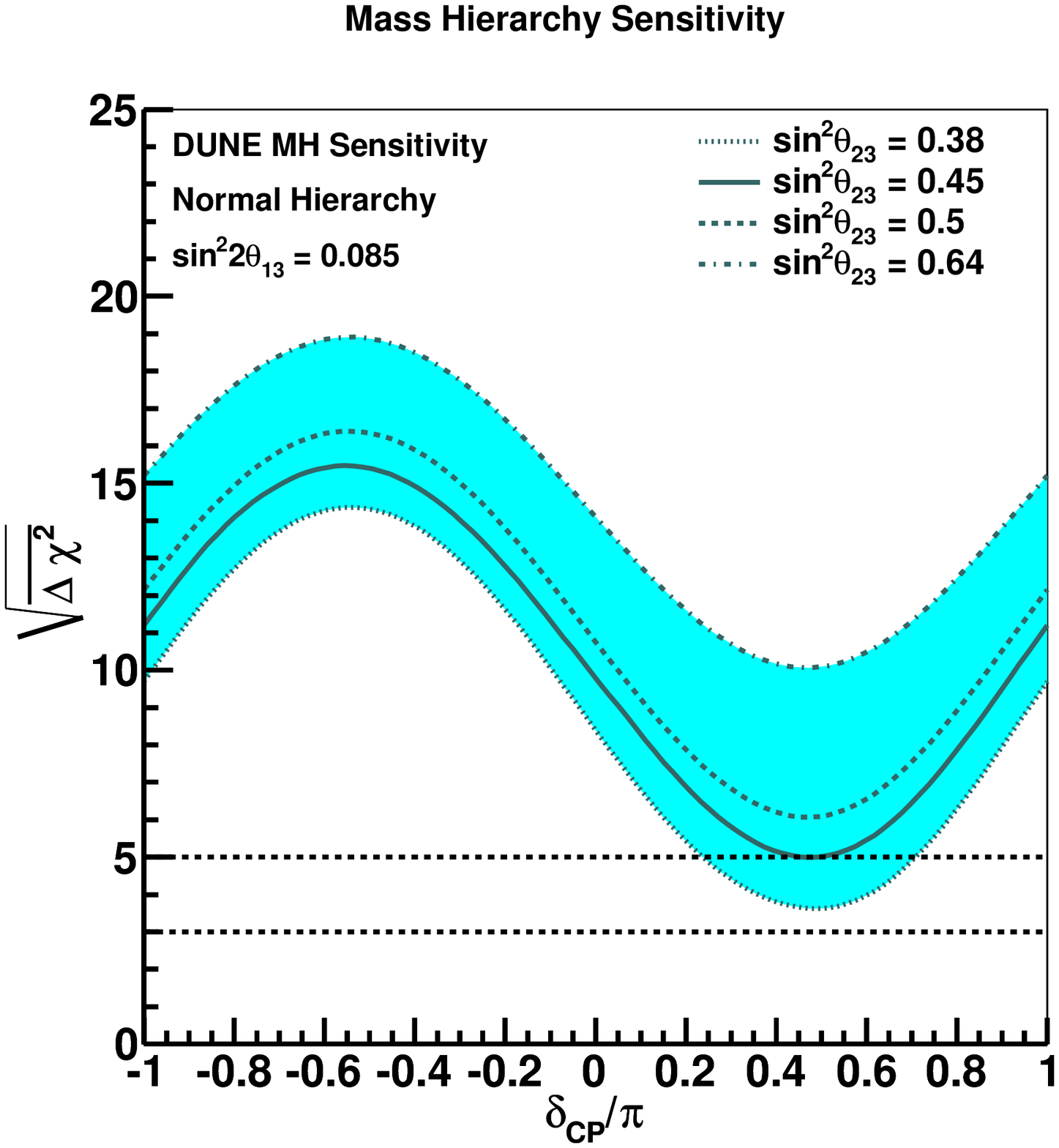}
\end{cdrfigure}

\begin{cdrfigure}[Variation in MH sensitivity due to $\theta_{13}$]{mh_theta13}{The variation in the MH sensitivity due to different values of $\theta_{13}$ within the allowed range.  In this figure, he nominal value of $\sin^22\theta_{13} = 0.085$ provides a significance of at least $\sqrt{\overline{\Delta\chi^{2}}} = 5$ for all values of \deltacp.  (See Figure~\ref{fig:mh_exposure} for the possible range of exposures to achieve this level of significance.)}
 \includegraphics[width=0.7\textwidth]{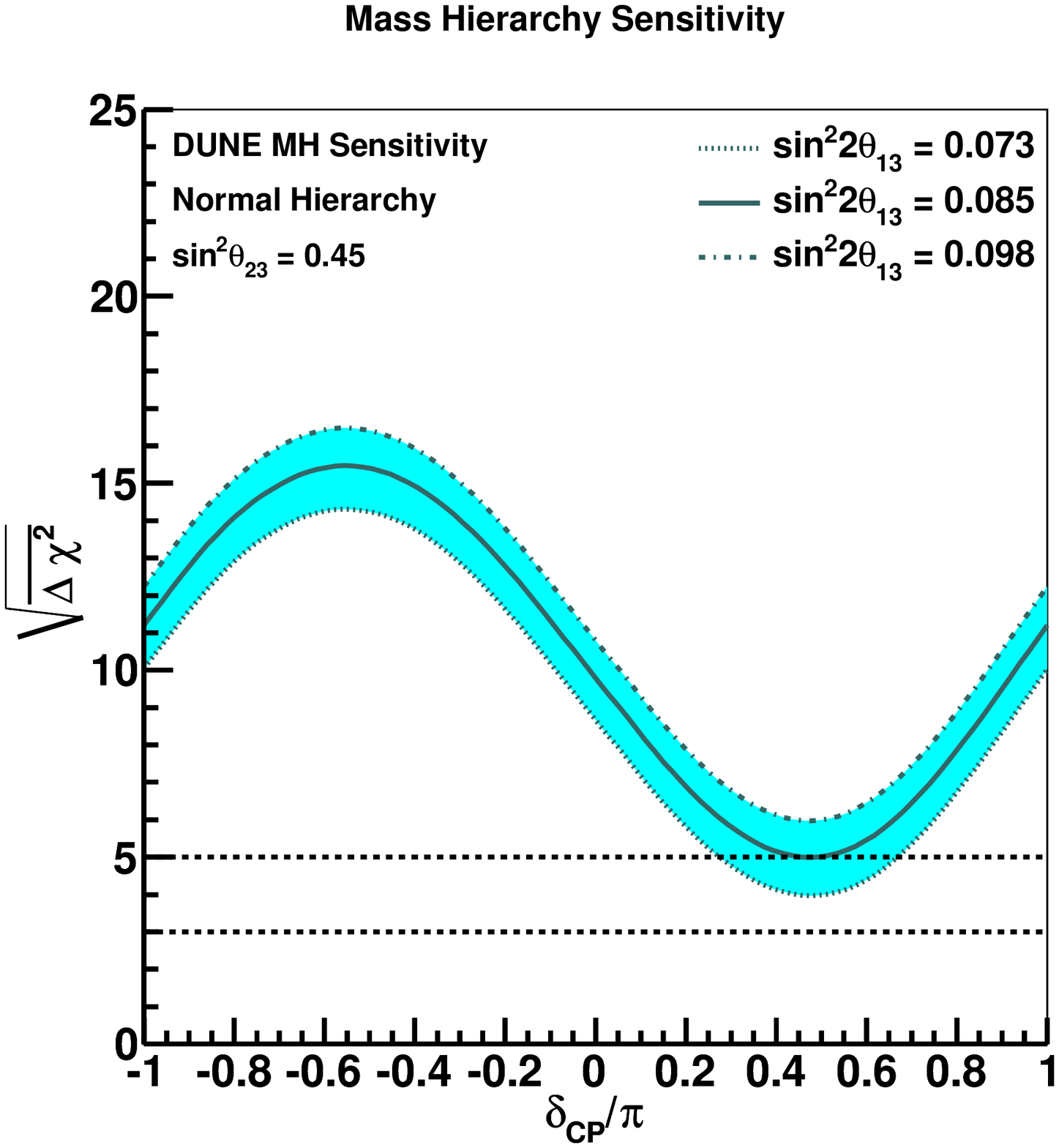}
\end{cdrfigure}

\begin{cdrfigure}[Variation in MH sensitivity due to $\Delta m^{2}_{31}$]{mh_deltamsq}{The variation in the MH sensitivity due to different values of \dm{31} within the allowed range.  In this figure, the nominal value of \dm{31} = $2.46\times 10^{-3}$~eV$^2$ provides a significance of at least $\sqrt{\overline{\Delta\chi^{2}}} = 5$ for all values of \deltacp.  (See Figure~\ref{fig:mh_exposure} for the possible range of exposures to achieve this level of significance.)}
 \includegraphics[width=0.7\textwidth]{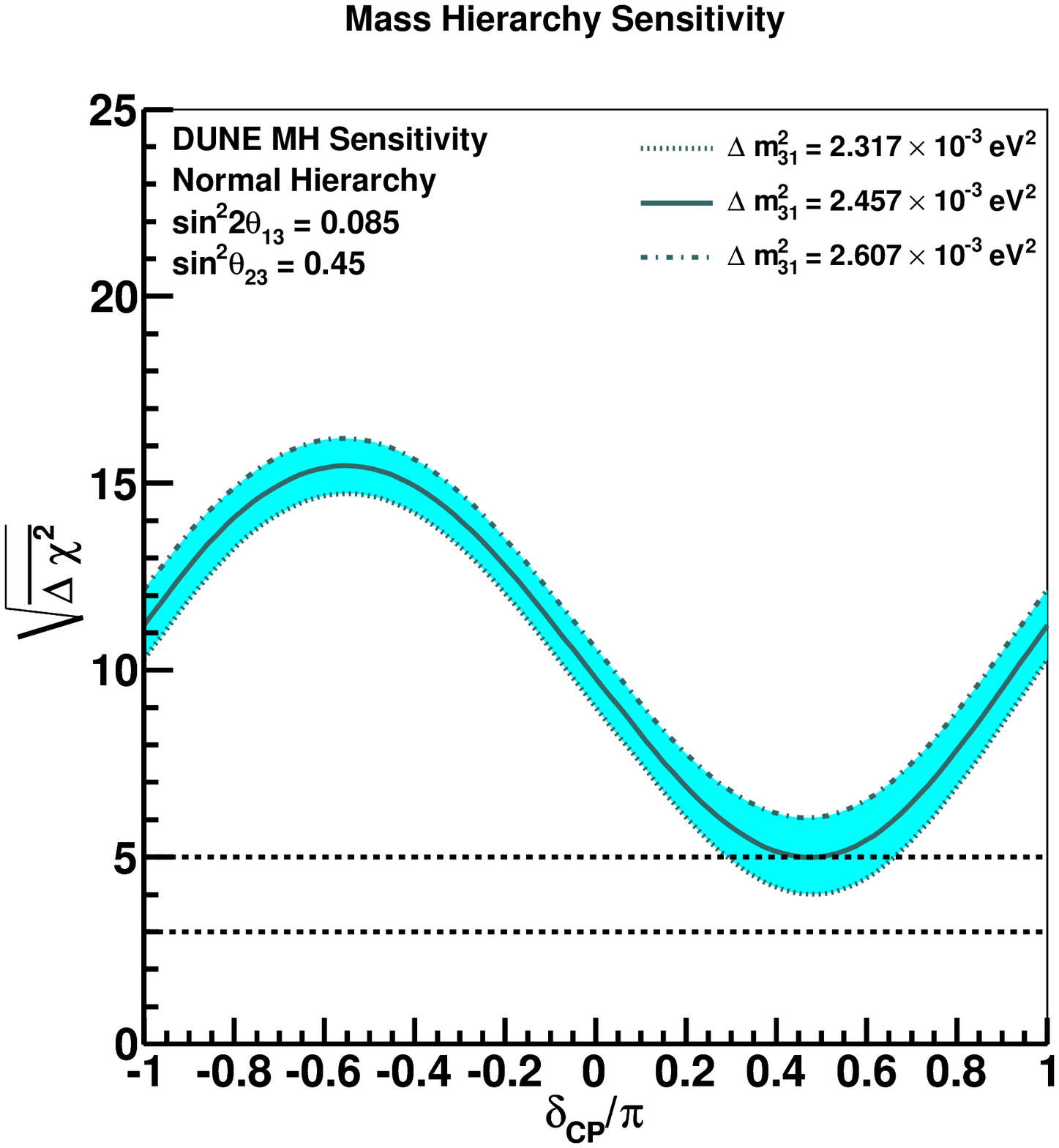}
\end{cdrfigure}
 
Studies have indicated that special attention must be paid to the statistical interpretation of MH sensitivities~\cite{Qian:2012zn,Blennow:2013oma}.
In general, if an experiment is repeated many times, a distribution of $\Delta\chi^2$
values will appear due to statistical fluctuations.
It is usually assumed that the $\Delta \chi^2$ metric follows the expected chi-squared
function for one degree of freedom, which has a mean of
$\overline{\Delta\chi^2}$ and can be interpreted using a Gaussian
distribution with a standard deviation of
$\sqrt{|\overline{\Delta\chi^2}|}$.
In assessing the MH sensitivity of future experiments, it is common practice to generate
a simulated data set (for an assumed true MH) that does not include statistical fluctuations. 
In this typical case, $\overline{\Delta\chi^2}$ is reported as the expected sensitivity, 
where $\overline{\Delta\chi^2}$ is representative of the mean value of $\Delta\chi^2$ that 
would be obtained in an ensemble of experiments for a particular true MH.  
With the exception of Figure~\ref{fig:mhstats}, the sensitivity plots
in this document have been generated using this method.
However, studies in~\cite{Qian:2012zn,Blennow:2013oma}
show that, in the case of the mass hierarchy
determination, the $\Delta \chi^2$ metric {\em does not} follow the expected chi-squared
function for one degree of freedom.  Rather, these studies show that
when the observed counts in the experiment are large enough,
the distribution of $\Delta\chi^2$ used here approximately follows
a Gaussian distribution with a
mean and standard deviation of $\overline{\Delta\chi^2}$ and
$2\sqrt{|\overline{\Delta\chi^2}|}$, respectively. Because the distribution is atypical, the interpretation of 
test statistic values in terms of confidence intervals is different than in the standard case.

The effect of statistical fluctuations in the MH measurement is shown
in Figure~\ref{fig:mhstats}.  The colored bands show the possible
range in the significance of a MH determination when statistical
fluctuations are included for a measurement that would yield a
significance of $\sqrt{\overline{\Delta\chi^{2}}} = 5$ for 100\% of
\deltacp values in our standard treatment (the solid blue line).  Also
shown in Figure~\ref{fig:mhstats} are horizontal lines that specify
the confidence level of an experiment that measures a particular value
of $\sqrt{\Delta \chi^2}$, following the convention
in~\cite{Qian:2012zn}. An experiment that measures $\sqrt{\Delta
  \chi^2} = 5$ (black dashed line) has a 1-\num{3.7e-6} probability of
determining the correct MH, while an experiment that
measures$\sqrt{\Delta \chi^2} = 3$ (blue dashed line) has a 98.9\%
probability of determining the correct MH. An experiment that measures
$\sqrt{\Delta \chi^2} = 0$ (cyan dashed line) has a 50\% probability
of determining the correct MH.  In this case, both hypotheses (normal
or inverted hierarchy) fit the data equally well, and the probability
of guessing correctly is 50\%.

\begin{cdrfigure}[MH sensitivity including statistical fluctuations]{mhstats}{
  The  sensitivity, given by  $\sqrt{\Delta T}=\sqrt{\Delta\chi^2}$ for a typical experiment 
  (solid blue line), is compared to the bands within which
  68\% (green) and 95\% (yellow) of experiments are expected to fall due to statistical fluctuations.
  The solid blue line (representing a minimum significance of $\sqrt{\Delta T} = 5$ for 100\% of \deltacp
   values) is the expected sensitivity in our standard treatment.
   (See Figure~\ref{fig:mh_exposure} for the possible range of exposures to achieve this level of significance.) 
  The dashed lines show the values of the $\sqrt{\Delta T}$ metric an experiment must measure for the probability of determining
  the correct neutrino MH to be 50\%~(cyan), 98.9\%~(blue), or 
  1 to \num{3.7e-6}~(black), following the convention in~\cite{Qian:2012zn}.  
  In the legend, the numbers corresponding to the dashed lines indicate 
  [probability of determining MH {\em incorrectly}] vs. [probability of determining the MH {\em correctly}].}
 \includegraphics[width=0.7\linewidth]{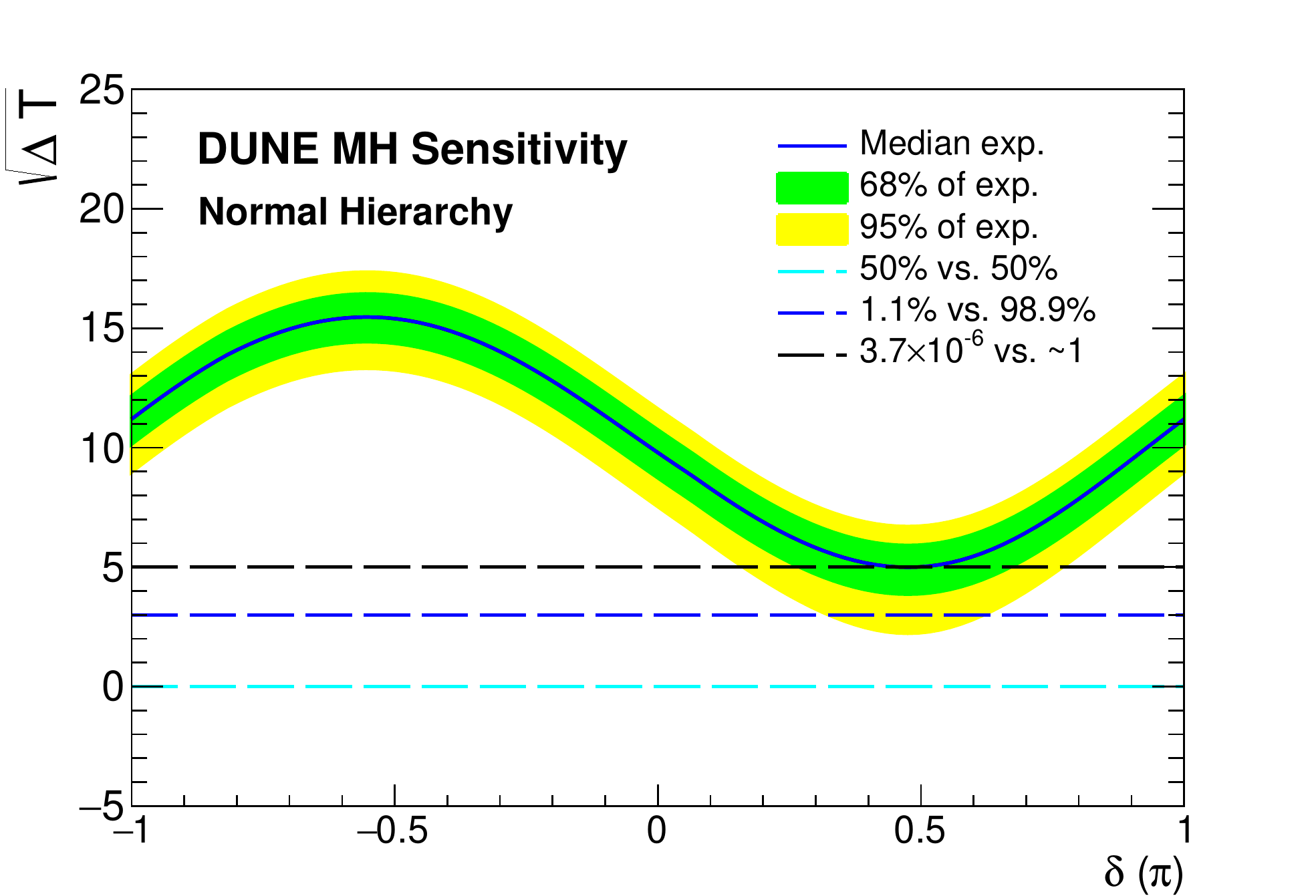}
\end{cdrfigure}

\section{CP-Symmetry Violation}
\label{sec:physics-lbnosc-cpv}

In the particular parameterization of the PMNS matrix shown in
Equation~\ref{eqn:pmns}, the middle factor labeled ``II'' describes
the mixing between the $\nu_1$ and $\nu_3$ mass states, and depends on
the CP-violating phase \deltacp.  With the recent measurement of
$\theta_{13}$, it is now known that the minimal conditions required
for measuring \deltacp in the three-flavor framework have been met;
all three mixing angles are nonzero, and there are two distinct mass
splittings.  In the approximation for the electron neutrino appearance
probability given in Equation~\ref{eqn:appprob}, expanding the middle
term results in the presence of CP-odd terms (dependent on $\sin
\mdeltacp$) that have opposite signs in $\nu_{\mu} \rightarrow \nu_e$
and $\bar{\nu}_{\mu} \rightarrow \bar{\nu}_e$ oscillations.
For $\mdeltacp \neq 0$ or $\pi$, these terms introduce an asymmetry in
neutrino versus antineutrino oscillations. The magnitude of the
CP-violating terms in the oscillation depends most directly on the
size of the Jarlskog Invariant~\cite{Jarlskog:1985cw}, a function that
was introduced to provide a measure of CP violation independent of the
mixing-matrix parameterization. In terms of the parameterization
presented in Equation~\ref{eqn:pmns}, the Jarlskog Invariant is:
\begin{equation}
J_{CP}^{\rm PMNS} \equiv \frac{1}{8} \sin 2 \theta_{12} \sin 2 \theta_{13}
\sin 2 \theta_{23} \cos \theta_{13} \sin \mdeltacp.
\end{equation}
The relatively large values of the mixing angles in the lepton sector imply that
leptonic CP-violation effects may be quite large ---  
depending on the value of the phase \deltacp, which is currently unknown. 
Experimentally, it is unconstrained at the 3$\sigma$ level by the global fit~\cite{Gonzalez-Garcia:2014bfa}.
Given the current best-fit values of the mixing angles~\cite{Gonzalez-Garcia:2014bfa} and assuming normal hierarchy,
\begin{equation}
J_{CP}^{\rm PMNS} \approx 0.03 \sin \mdeltacp.
\end{equation}
This is in sharp contrast to the very small mixing in the quark sector,  
which leads to a very small value of the corresponding quark-sector
Jarlskog Invariant~\cite{Beringer:1900zz},
\begin{equation}
J_{CP}^{\rm CKM} \approx 3 \times 10^{-5},
\end{equation}
despite the large value of $\delta^{\rm CKM}_{CP}\approx70^{\circ}$.

The variation in the $\nu_\mu \rightarrow
\nu_e$ oscillation probability (Equation~\ref{eqn:appprob}) with the value of \deltacp
indicates that it is experimentally possible to measure the value of
\deltacp at a fixed baseline using only the observed shape of the
$\nu_\mu \rightarrow \nu_e$ {\em or} the 
$\bar{\nu}_\mu \rightarrow \bar{\nu}_e$
appearance signal measured over an energy range that encompasses at
least one full oscillation interval. A measurement of the value of
$\mdeltacp \neq 0 \ {\rm or} \ \pi$, assuming that neutrino mixing follows the three-flavor model, would imply CP violation.  

The CP asymmetry,
$\mathcal{A}_{CP}$, is defined as 
\begin{equation}
\label{eqn:cp-asymm}
 \mathcal{A}_{CP} = \frac{P(\nu_\mu \rightarrow \nu_e) -
  P(\bar{\nu}_\mu \rightarrow \bar{\nu}_e)}{P(\nu_\mu \rightarrow
  \nu_e) + P(\bar{\nu}_\mu \rightarrow \bar{\nu}_e)}.
\end{equation}
In the three-flavor model the asymmetry can be approximated to leading
order in $\Delta m_{21}^2$ as~\cite{Marciano:2006uc}:
\begin{equation}
\mathcal{A}_{CP} \sim \frac{\cos \theta_{23} \sin 2 \theta_{12}
  {\sin \mdeltacp}}{\sin \theta_{23} \sin \theta_{13}}
\left(\frac{\Delta m^2_{21} L}{ 4 E_{\nu}}\right) + {\rm matter
  \ effects}
\label{eqn:cpasym}
\end{equation}
Regardless of the measured value obtained for \deltacp, the explicit
observation of the asymmetry $\mathcal{A}_{CP}$ in $\nu_{\mu}
\rightarrow \nu_e$ and $\bar{\nu}_{\mu} \rightarrow
\bar{\nu}_e$ oscillations is sought to directly demonstrate the
leptonic CP-violation effect.  A measurement of \deltacp that is
inconsistent with the measurement of $\mathcal{A}_{CP}$ according to
Equation~\ref{eqn:cpasym} could be evidence of physics beyond the
standard three-flavor model.  Furthermore, for long-baseline
experiments such as DUNE where the neutrino beam propagates through
the Earth's mantle, the leptonic CP-violation effects must be
disentangled from the matter effects, discussed in
Section~\ref{sec:physics-lbnosc-mh}.

Figure~\ref{fig:cpv_nominal} shows the significance with which the CP
violation ($\mdeltacp \neq 0 \ {\rm or} \ \pi$) can be determined as a
function of the value of \deltacp for an exposure of \SI{300}~\ktMWyr,
which corresponds to seven years of data (3.5 years in neutrino mode
plus 3.5 years in antineutrino mode) with a 40-kt detector and a
1.07-MW 80-GeV beam.  Figure~\ref{fig:cpv_exposure} shows the significance
with which CP violation can be determined for 25\%, 50\% or 75\% of \deltacp
values as a function of exposure.
Table~\ref{tab:cpv_requiredexposure} lists the minimum exposure
required to determine CP violation with a significance of 5$\sigma$
for 50\% of \deltacp values or 3$\sigma$ for 75\% of \deltacp values
for both the CDR reference beam design and the optimized beam design.
The CP-violation sensitivity as a function of \deltacp as shown in
Figure~\ref{fig:cpv_nominal} has a characteristic double peak
structure because the significance of a CP-violation measurement
necessarily drops to zero where there is no CP violation: at the
CP-conserving values of $-\pi,~0,~{\rm and}~\pi$.  Therefore, unlike
the MH determination, it's not possible for any experiment to provide
100\% coverage in \deltacp for a CP-violation measurement because CP-violation effects vanish at certain values of \deltacp.

\begin{cdrfigure}[CP-violation sensitivity for a \SI{300}~\ktMWyr{} exposure]{cpv_nominal}{The significance with which the CP violation can be determined as a function of the value of \deltacp for an exposure of \SI{300}~\ktMWyr{} assuming normal MH (left) or inverted MH (right).  The shaded region represents the range in sensitivity due to potential variations in the beam design.}
 \includegraphics[width=0.49\textwidth]{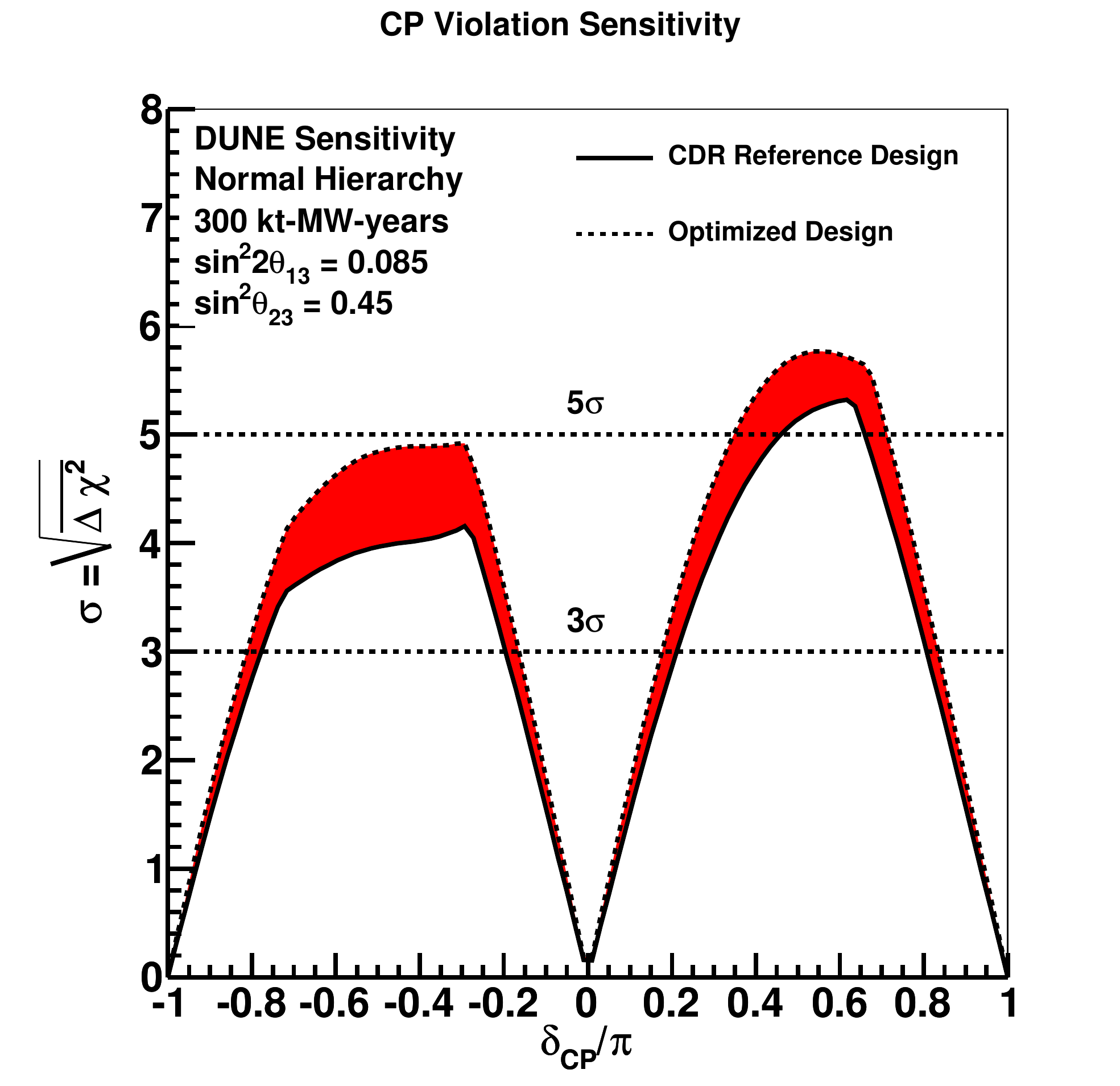}
 \includegraphics[width=0.49\textwidth]{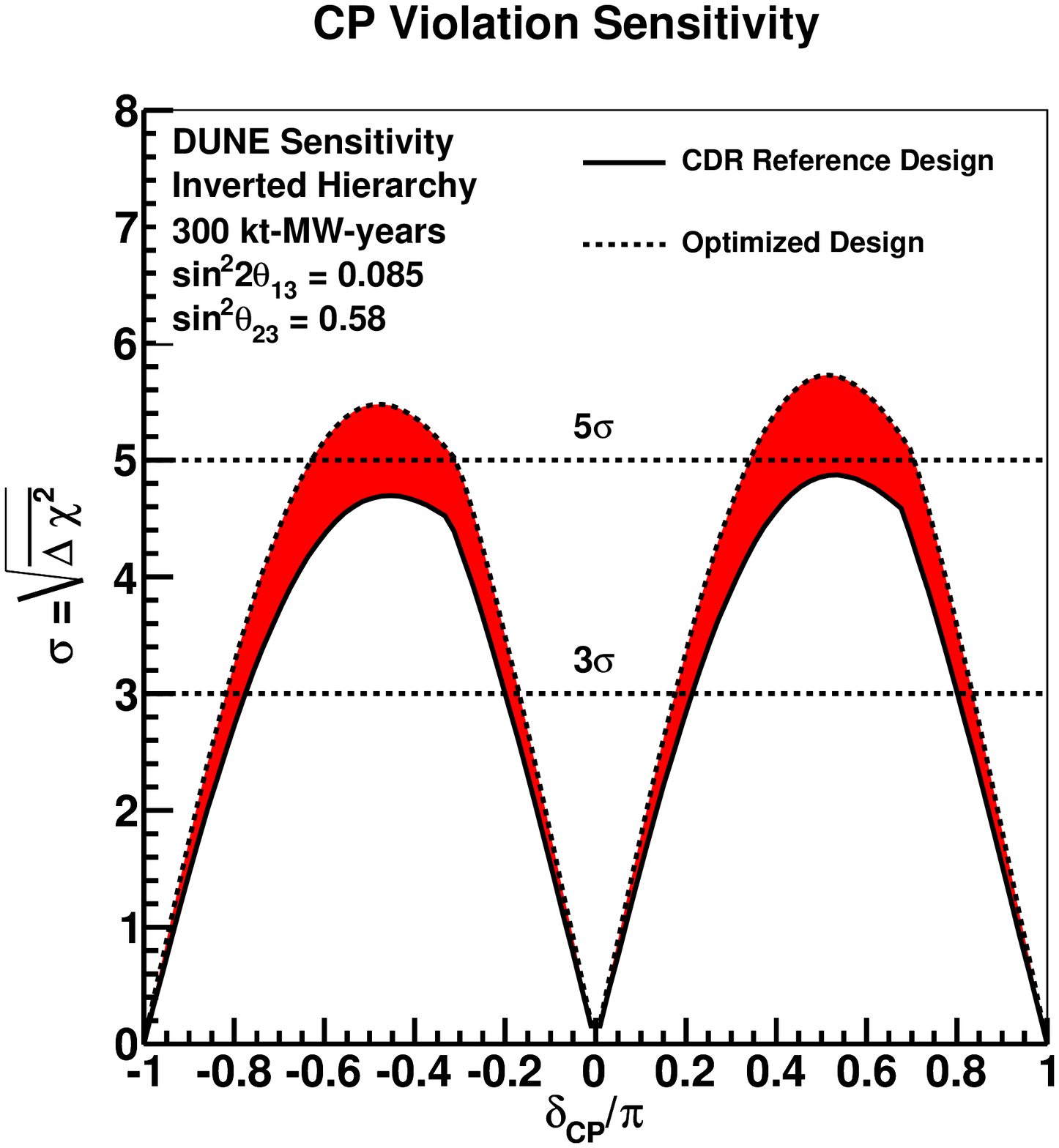}
\end{cdrfigure}

\begin{cdrfigure}[CP-violation sensitivity as a function of exposure]{cpv_exposure}{The minimum significance with which CP violation can be determined for 25\%, 50\% and 75\%  of \deltacp values as a function of exposure.  The two different shaded bands represents the different sensitivities due to potential variations in the beam design. This plot assumes normal mass hierarchy. }
\includegraphics[width=\textwidth]{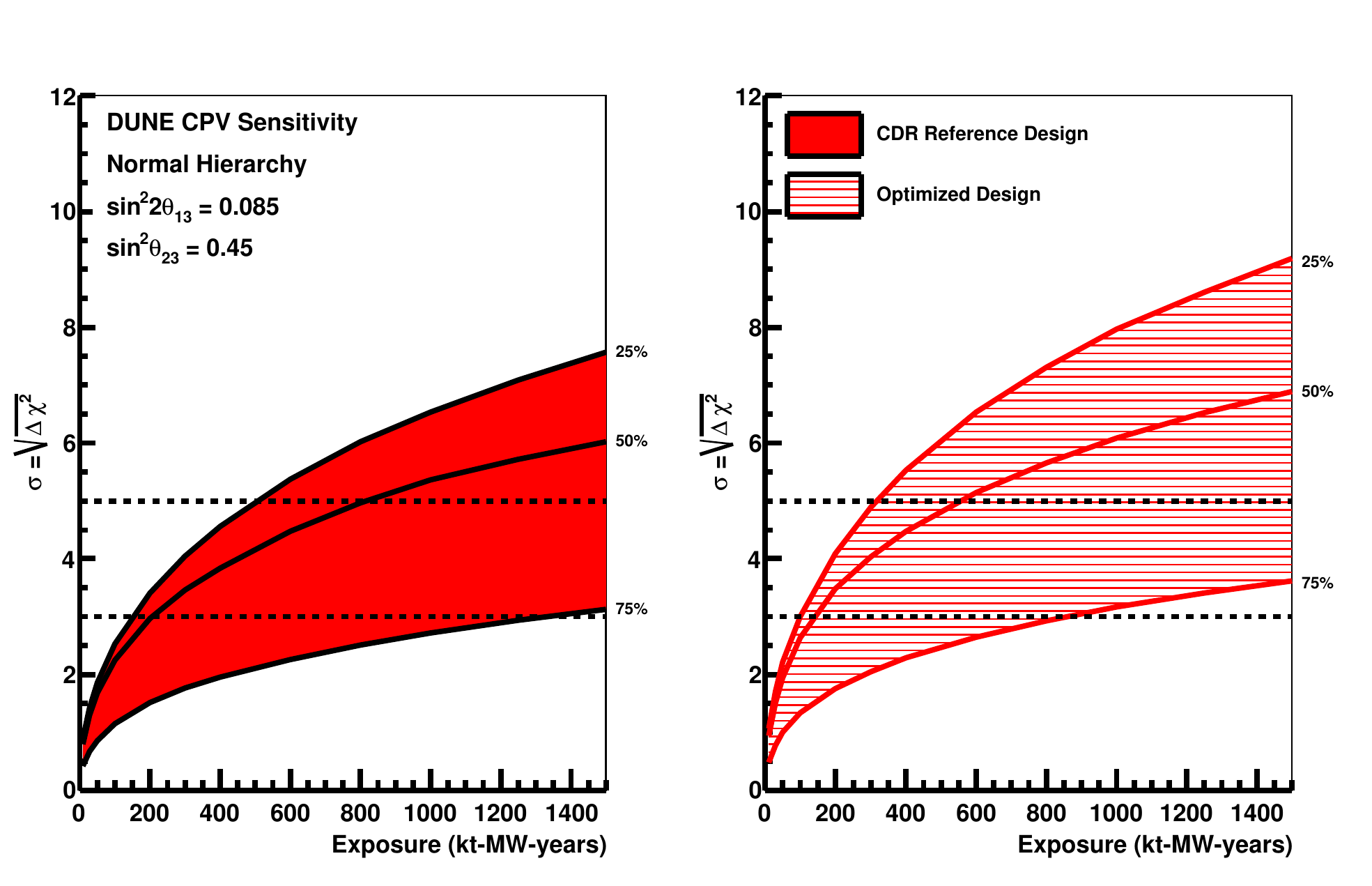}
\end{cdrfigure}

\begin{cdrtable}[Required exposure for a CP-violation measurement]{lcc}{cpv_requiredexposure}{The minimum exposure required to determine CP violation with a significance of 3$\sigma$ for 75\% of \deltacp values or 5$\sigma$ for 50\% of \deltacp values for the CDR reference beam design and the optimized beam design.}
 Significance & CDR Reference Design & Optimized Design\\
 \toprowrule
 3$\sigma$ for 75\% of \deltacp values & \SI{1320}~\ktMWyr{} & \SI{850}~\ktMWyr{} \\ \colhline 
 5$\sigma$ for 50\% of \deltacp values & \SI{810}~\ktMWyr{} & \SI{550}~\ktMWyr{}\\
\end{cdrtable}

Figures~\ref{fig:cpv_theta23}, \ref{fig:cpv_theta13}, and \ref{fig:cpv_deltamsq} show the variation in the CP sensitivity due to different values of $\theta_{23}$, $\theta_{13}$, and \dm{31} within the allowed ranges.  The value of $\theta_{23}$ has the biggest impact on the sensitivity, and the least favorable scenario corresponds to a value of $\sin^2{\theta_{23}}$ at the high end of its 
experimentally allowed range.

\begin{cdrfigure}[Variation in CP sensitivity due to $\theta_{23}$]{cpv_theta23}{The variation in the CP sensitivity due to different values of $\theta_{23}$ within the allowed range.  In this figure, the nominal value of $\sin^2\theta_{23} = 0.45$ provides a significance of at least 3$\sigma$ for 75\% of \deltacp values. (See Figure~\ref{fig:cpv_exposure} for the possible range of exposures to achieve this level of significance.) The significance decreases for all values of \deltacp as $\sin^2\theta_{23}$ gets larger.}
 \includegraphics[width=0.7\textwidth]{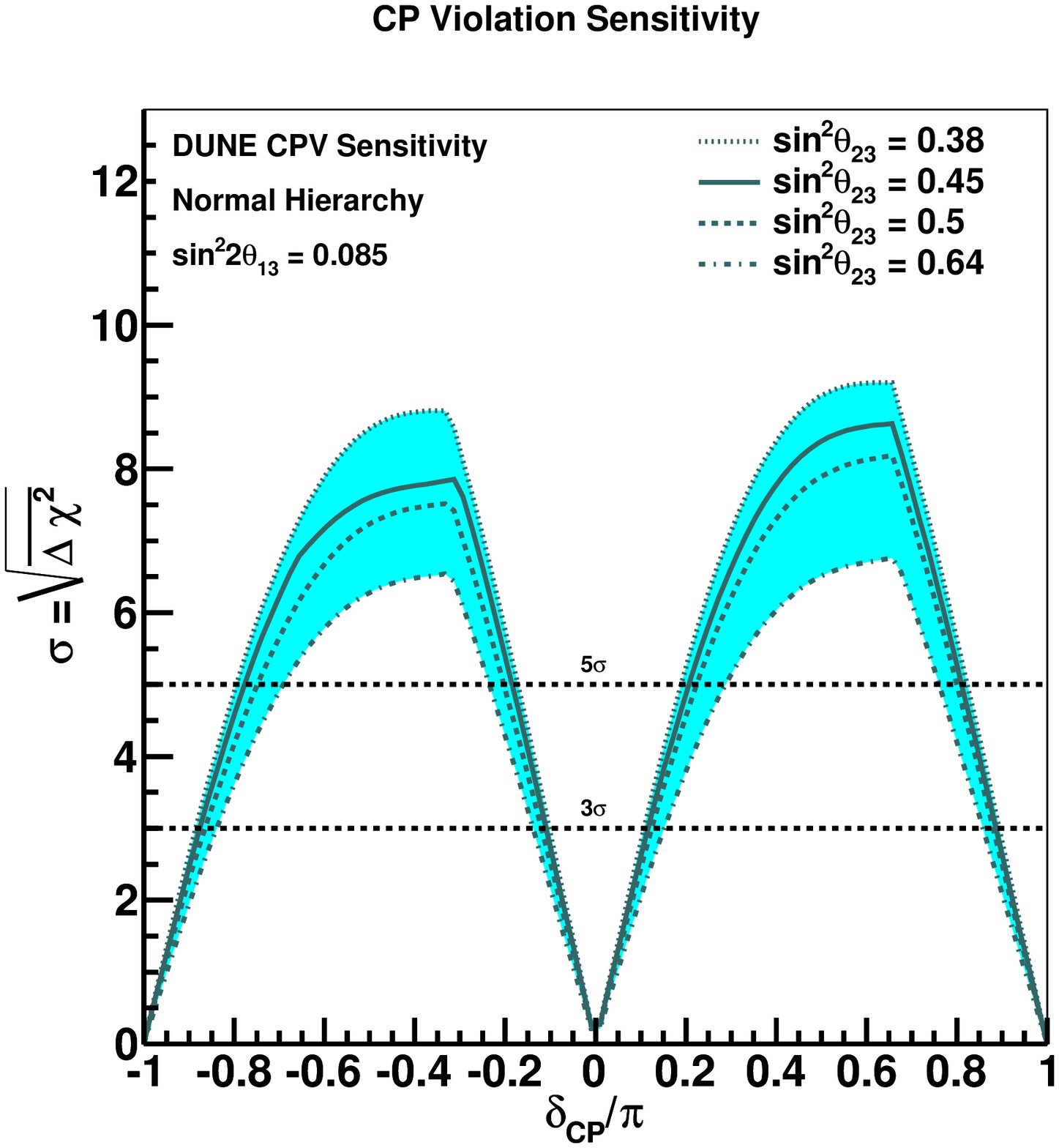}
\end{cdrfigure}

\begin{cdrfigure}[Variation in CP sensitivity due to $\theta_{13}$]{cpv_theta13}{The variation in the CP sensitivity due to different values of $\theta_{13}$ within the allowed range.  In this figure, the nominal value of $\sin^22\theta_{13} = 0.085$ provides a significance of at least 3$\sigma$ for 75\% of \deltacp values. (See Figure~\ref{fig:cpv_exposure} for the possible range of exposures to achieve this level of significance.)}
 \includegraphics[width=0.7\textwidth]{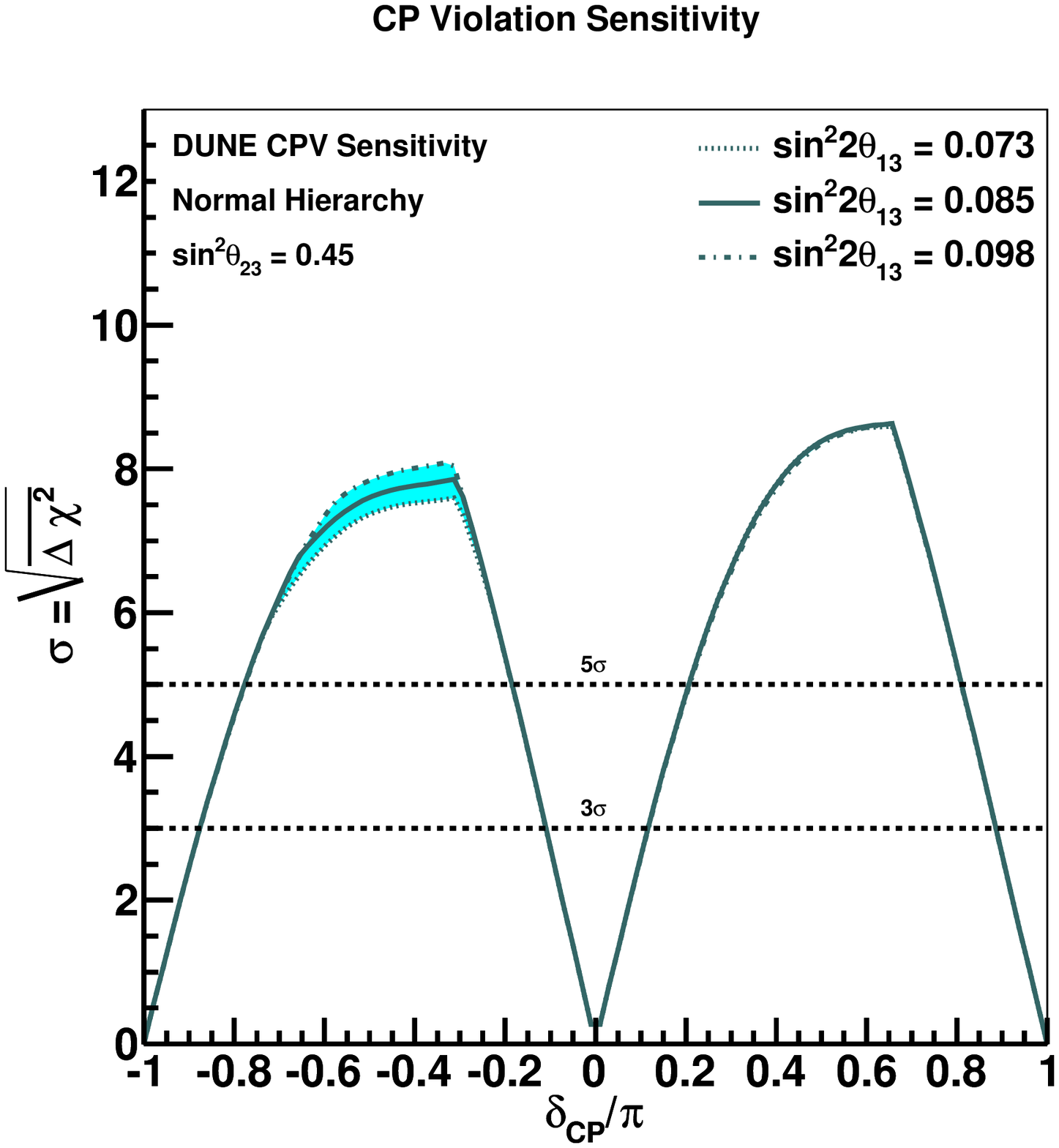}
\end{cdrfigure}

\begin{cdrfigure}[Variation in CP sensitivity due to $\Delta m^{2}_{31}$]{cpv_deltamsq}{The variation in the CP sensitivity due to different values of \dm{31} within the allowed range.  In this figure, the nominal value of \dm{31} = $2.46\times 10^{-3}$~eV$^2$ provides a significance of at least 3$\sigma$ for 75\% of \deltacp values.  (See Figure~\ref{fig:cpv_exposure} for the possible range of exposures to achieve this level of significance.)}
 \includegraphics[width=0.7\textwidth]{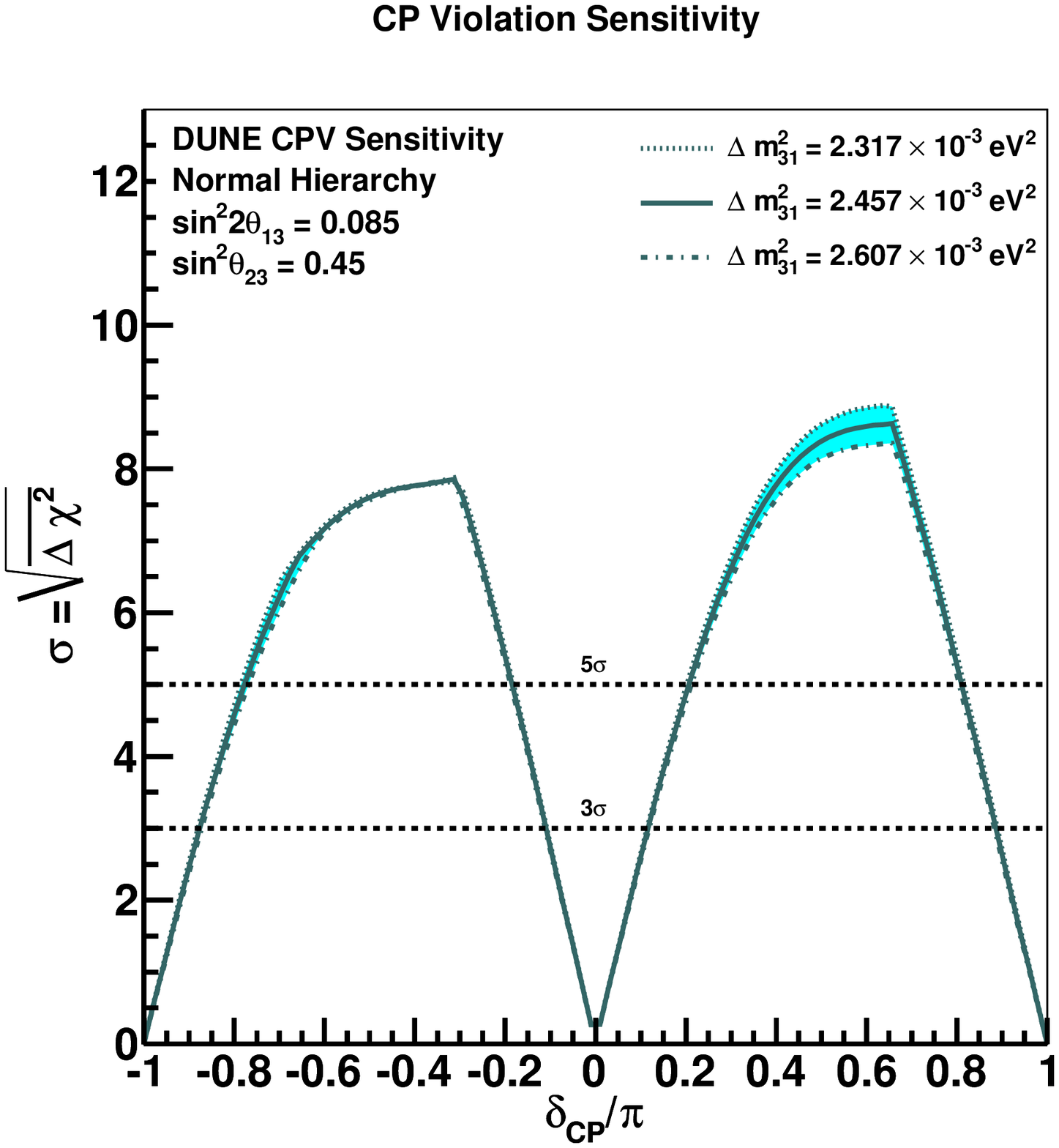}
\end{cdrfigure}

\section{Precision Oscillation Parameter Measurements}

In addition to the discovery potential for neutrino mass hierarchy and CP-violation, 
DUNE will improve the precision on key parameters that govern neutrino oscillations, including:
\begin{itemize}
 \item $\sin^2\theta_{23}$ and the octant of $\theta_{23}$
 \item \deltacp
 \item $\sin^22\theta_{13}$
 \item \dm{31}
\end{itemize}

Higher-precision measurements of the known oscillation parameters
improves sensitivity to physics beyond the three-flavor oscillation model,
particularly when compared to independent measurements by other
experiments, including reactor measurements of $\theta_{13}$ and
measurements with atmospheric neutrinos.

The most precise measurement of $\sin^2\theta_{23}$ to date comes from
T2K, $\sin^2\theta_{23} = 0.514^{+0.055}_{-0.056}$ (normal hierarchy)
and $\sin^2\theta_{23} = 0.511~\pm~0.055$ (inverted
hierarchy)~\cite{Abe:2015awa}.  This corresponds to a value of
$\theta_{23}$ near 45\mbox{$^{\circ}$}, but leaves an ambiguity as to
whether the value of $\theta_{23}$ is in the lower octant (less than
45\mbox{$^{\circ}$}), the upper octant (greater than
45\mbox{$^{\circ}$}), or exactly 45\mbox{$^{\circ}$}.  The value of
$\sin^2 \theta_{23}$ from the global fit reported
by~\cite{Gonzalez-Garcia:2014bfa} is $\sin ^2 \theta_{23} = 0.452
^{+0.052} _{-0.028} (1 \sigma)$ for normal hierarchy (NH), but the
distribution of the $\chi^2$ from the global fit has another local
minimum -- particularly if the MH is inverted -- at $\sin^2
\theta_{23} = 0.579 ^{+0.025} _{-0.037} (1 \sigma)$. A \emph{maximal}
mixing value of $\sin^2 \theta_{23} =0.5$ is therefore still allowed
by the data and the octant is still largely undetermined.  A value of
$\theta_{23}$ exactly equal to 45\mbox{$^{\circ}$} would indicate that
$\nu_{\mu}$ and $\nu_{\tau}$ have equal contributions from $\nu_3$,
which could be evidence for a previously unknown symmetry.  It is
therefore important experimentally to determine the value of $\sin ^2
\theta_{23}$ with sufficient precision to determine the octant of
$\theta_{23}$.  The measurement of $\nu_\mu \rightarrow \nu_\mu$
oscillations is sensitive to $\sin ^2 2 \theta_{23}$, whereas the
measurement of $\nu_\mu \rightarrow \nu_e$ oscillations is sensitive
to $\sin^2 \theta_{23}$.  A combination of both $\nu_e$ appearance and
$\nu_\mu$ disappearance measurements can probe both maximal mixing and
the $\theta_{23}$ octant.  The $\Delta\chi^2$ metric is defined as:
\begin{eqnarray}
\Delta\chi^2_{octant} & = & |\chi^2_{\theta_{23}^{test}>45^\circ} - \chi^2_{\theta_{23}^{test}<45^\circ}|, \\ \nonumber
\end{eqnarray}
where the value of $\theta_{23}$ in the \emph{wrong} octant is constrained 
only to have a value within the \emph{wrong} octant (i.e., it is not required
to have the same value of $\sin^22\theta_{23}$ as the true value).
Figure~\ref{fig:octant} shows the sensitivity to determining the octant as a function of $\theta_{23}$.  Figure~\ref{fig:res_th23} shows the resolution of $\sin^2\theta_{23}$ as a function of exposure, assuming the true value is $\sin^2\theta_{23} = 0.45$ from the current global fit.

\begin{cdrfigure}[Octant sensitivity]{octant}{The significance with which DUNE can resolve the $\theta_{23}$ octant as a function of the true value of $\theta_{23}$. The green shaded band around the curve represents the range in sensitivity due to potential variations in the beam design and in the true value of \deltacp. The yellow shaded regions indicate the current 1$\sigma$ and 3$\sigma$ bounds on the value of $\theta_{23}$ from a global fit.  The same exposure that gives a 3$\sigma$ measurement of CP violation for 75\% of the values of \deltacp is assumed.  See Figure~\ref{fig:cpv_exposure} for the possible range of exposure to achieve this significance.}
 \includegraphics[width=0.7\textwidth]{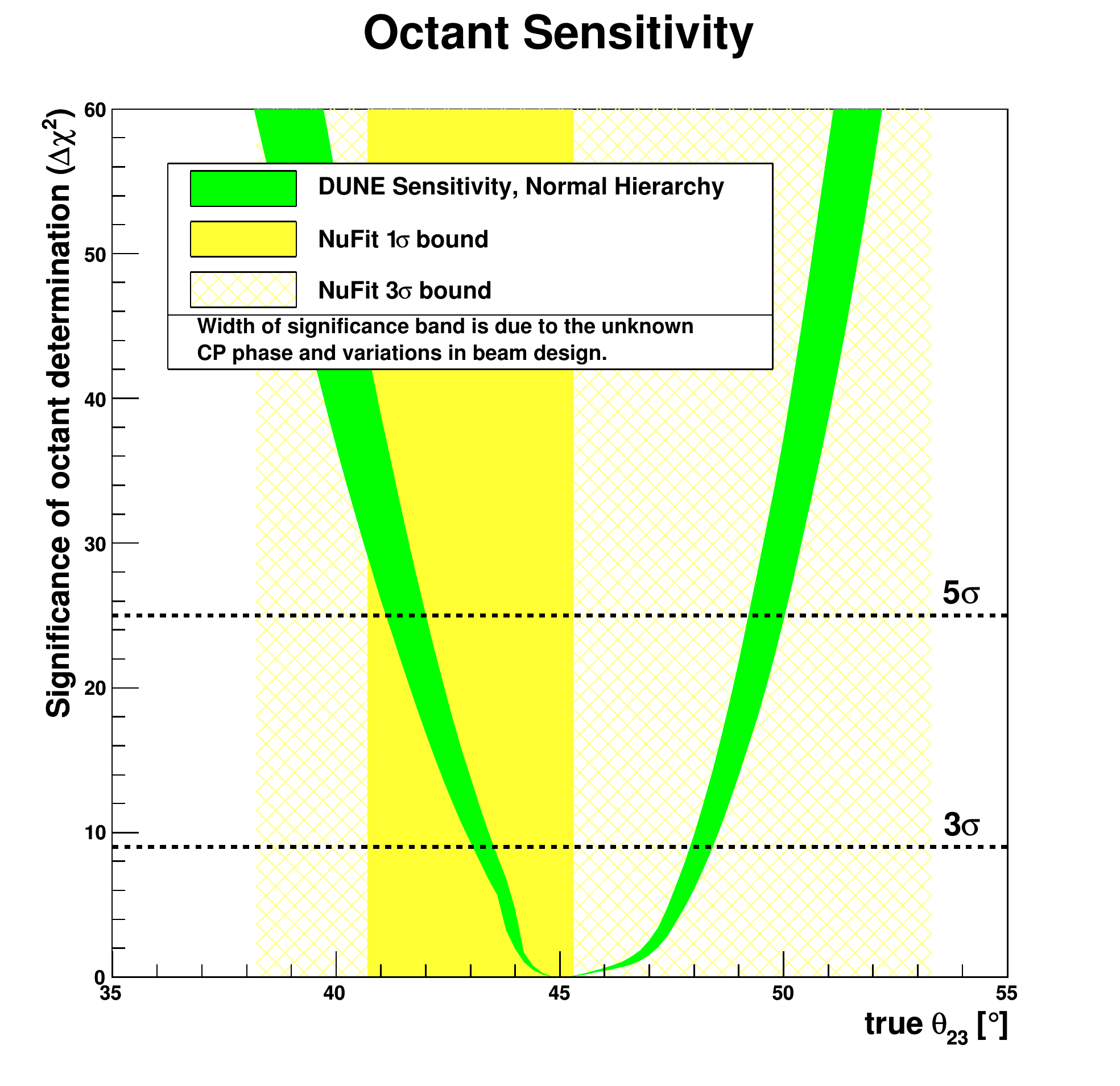}
\end{cdrfigure}

\begin{cdrfigure}[Resolution of $\sin^2\theta_{23}$ as a function of exposure]{res_th23}{The resolution of a measurement of $\sin^2\theta_{23}$ as a function of exposure assuming normal MH and $\sin^2\theta_{23} = 0.45$ from the current global fit. The shaded region represents the range in sensitivity due to potential variations in the beam design.  }
 \includegraphics[width=0.7\textwidth]{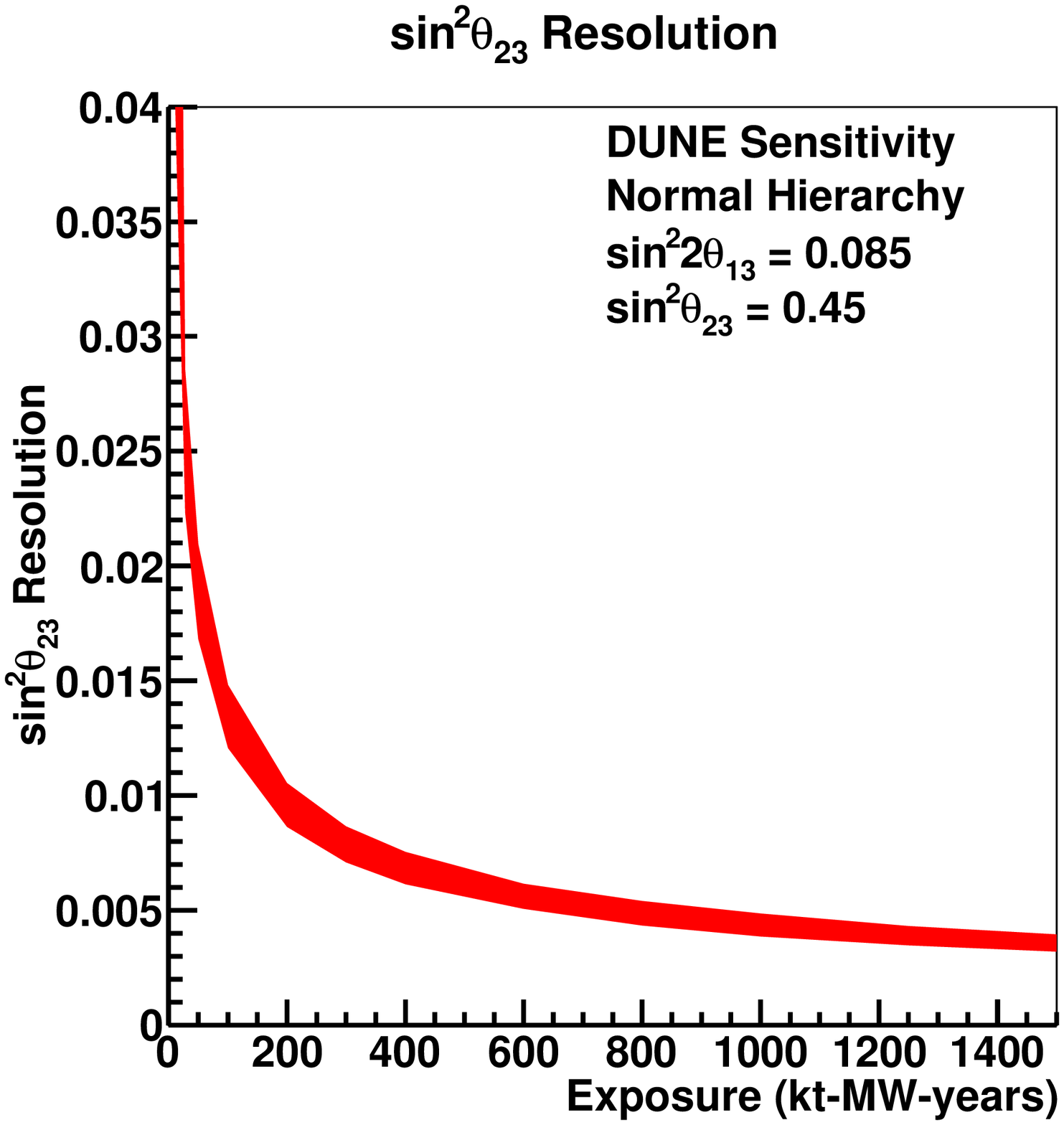}
\end{cdrfigure}

As mentioned in Section~\ref{sec:physics-lbnosc-cpv}, DUNE will seek
not only to demonstrate explicit CP violation by observing a
difference in the neutrino and antineutrino oscillation probabilities,
but also to measure the value of the parameter \deltacp.
Figure~\ref{fig:res_cp} shows the resolution of \deltacp as a function
of exposure for a CP-conserving value (\deltacp = 0) and the value
that gives the maximum CP violation for normal MH (\deltacp =
90\mbox{$^{\circ}$}).  Minimum exposures of approximately
\SI{450}~\ktMWyr{} and \SI{290}~\ktMWyr{} are required to measure
\deltacp with a resolution of 10\mbox{$^{\circ}$} for the CDR
reference beam design and the optimized beam design, respectively, for
a true value $\mdeltacp = 0$.

\begin{cdrfigure}[Resolution of \deltacp as a function of exposure]{res_cp}{The resolution of a measurement of \deltacp as a function of exposure assuming normal MH.  The resolution is shown for a CP-conserving value (\deltacp = 0) and the value that gives the maximum CP violation for normal MH (\deltacp = 90\mbox{$^{\circ}$}). The shaded region represents the range in sensitivity due to potential variations in the beam design.  }
 \includegraphics[width=0.7\textwidth]{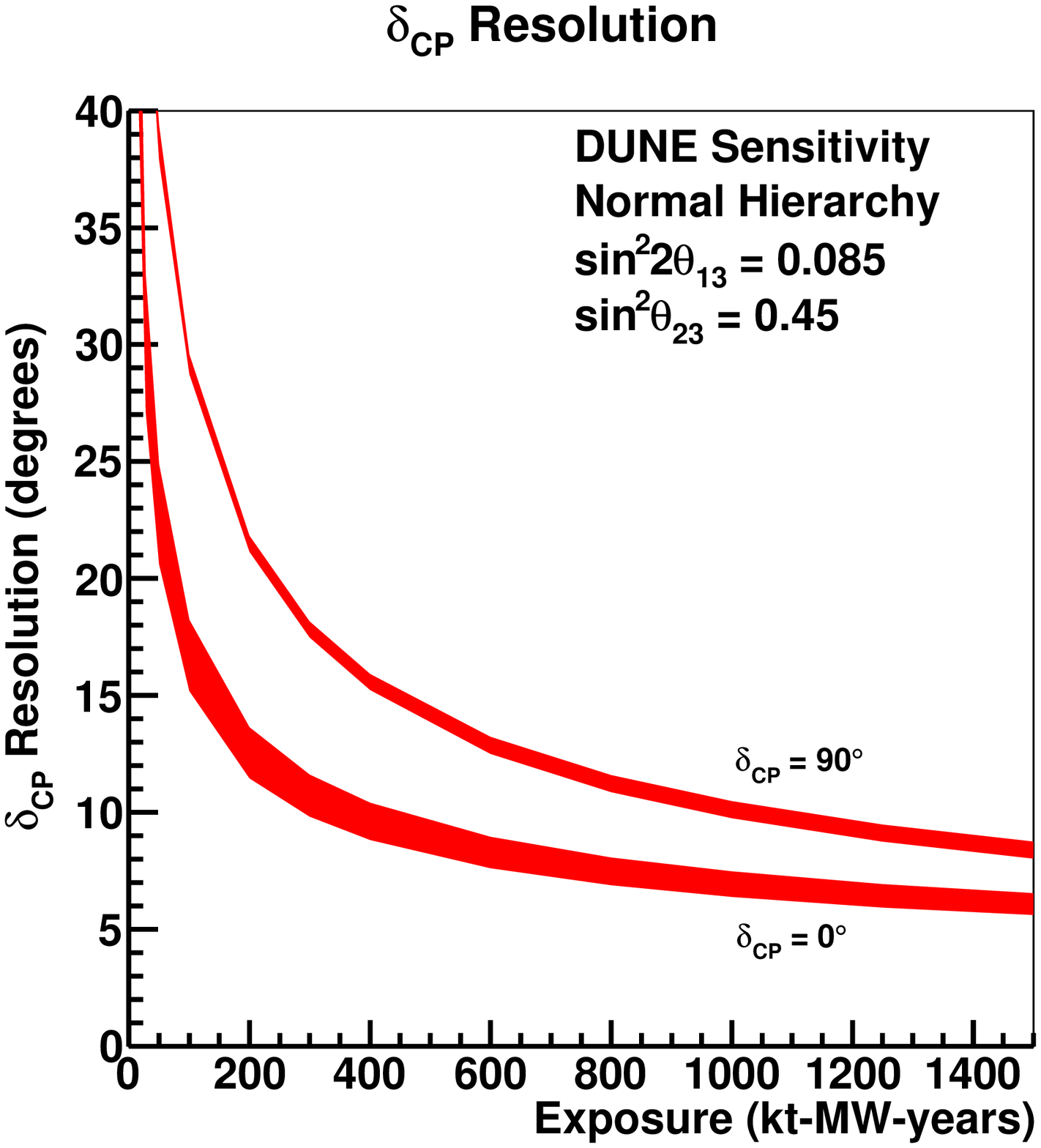}
\end{cdrfigure}

The rich oscillation structure that can be observed by DUNE and the
excellent particle identification capability of the detector
will enable precision measurement  in a single experiment of all the mixing parameters
governing $\nu_1$-$\nu_3$ and $\nu_2$-$\nu_3$ mixing. Theoretical models probing quark-lepton
universality predict specific values of the mixing angles and the
relations between them. The
mixing angle $\theta_{13}$ is
expected to be measured accurately in reactor experiments by the end
of the decade with a precision that will be limited by
systematics. 
The combined statistical and systematic uncertainty on the value of \sinstt{23} 
from the Daya Bay reactor neutrino experiment, which has
the lowest systematics, is currently $\sim6$\% (\sinstt{13} $= 0.084\pm0.005$),
with a projected uncertainty of $\sim$3\% by 2017~\cite{Zhang:2015fya}.
While the constraint on $\theta_{13}$ from the reactor experiments will be
important in the
early stages of DUNE for determining CP violation, measuring
\deltacp and determining the $\theta_{23}$ octant, 
DUNE itself will eventually be able to measure
$\theta_{13}$ independently with a similar precision to that expected from the reactor experiments. 
Whereas the reactor experiments measure $\theta_{13}$ using $\bar{\nu}_e$
disappearance, DUNE will measure it through $\nu_e$ and
$\bar{\nu}_e$ appearance, thus providing an independent constraint on
the three-flavor mixing matrix.   Figure~\ref{fig:res_th13} shows the resolution of \sinstt{13} as a function of exposure, assuming the true value is \sinstt{13}$ = 0.085$ from the current global fit.

\begin{cdrfigure}[Resolution of \sinstt{13} as a function of exposure]{res_th13}{The resolution of a measurement of \sinstt{13} as a function of exposure assuming normal MH and \sinstt{13}$ = 0.085$ from the current global fit. The shaded region represents the range in sensitivity due to potential variations in the beam design.  }
 \includegraphics[width=0.7\textwidth]{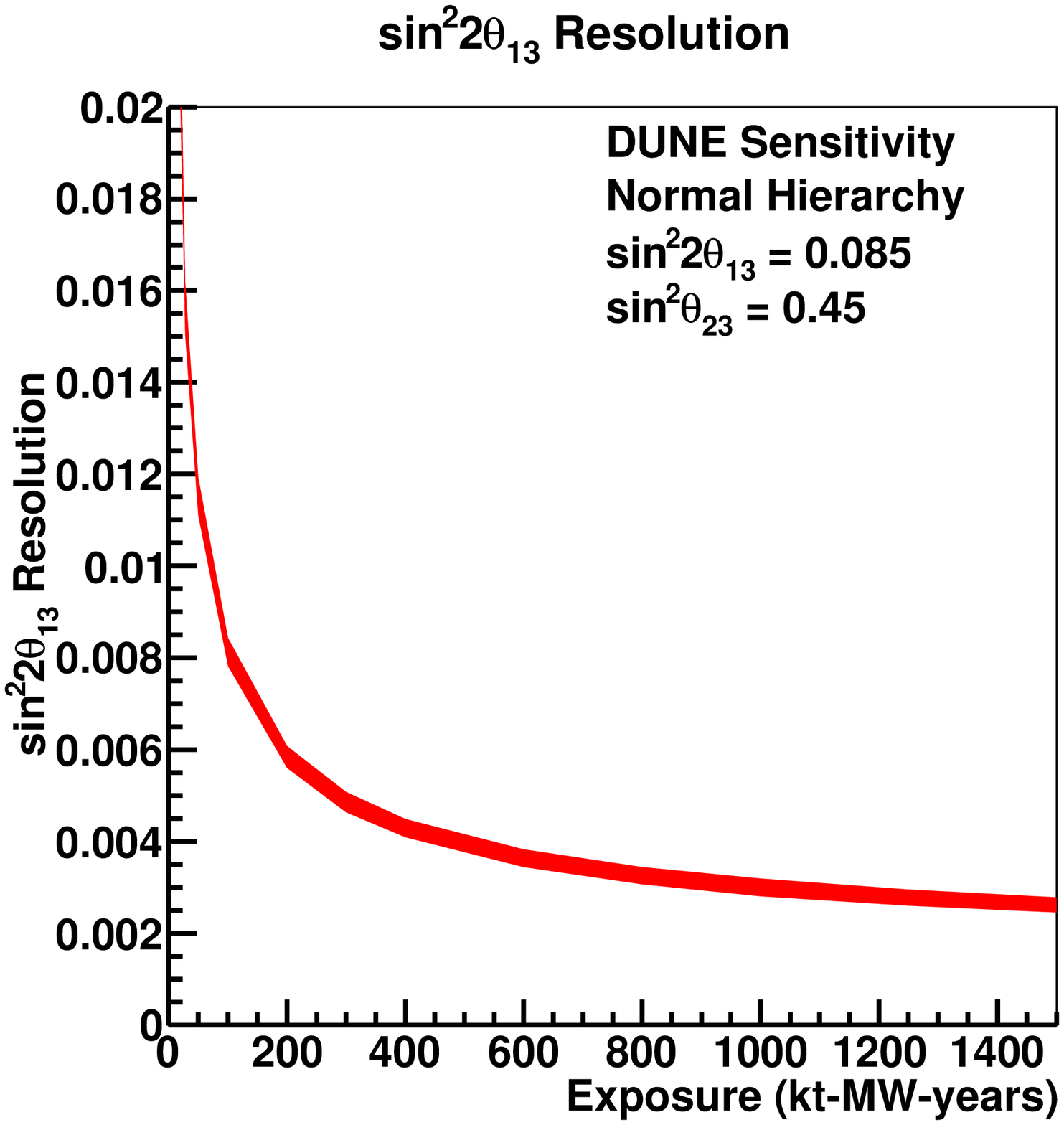}
\end{cdrfigure}

DUNE can also significantly improve the
resolution on the larger mass splitting beyond the precision of current experiments.  The current best-fit value for 
\dm{32} from MINOS is $|\Delta m^2_{32}| = (2.34\pm0.09)\times10^{-3}$~eV$^2$ (normal hierarchy) and $|\Delta m^2_{32}| = (2.37^{+0.11}_{-0.07})\times10^{-3}$~eV$^2$ (inverted hierarchy)~\cite{Sousa:2015bxa}, with comparable precision achieved by both Daya Bay and T2K. The
precision on \dm{31} will ultimately depend on tight control
of energy-scale systematics.  Figure~\ref{fig:res_dm2} shows the expected resolution of \dm{31} as a function of exposure, assuming the true value is \dm{31} = $2.457\times10^{-3}$~eV$^2$ from the current global fit.

\begin{cdrfigure}[Resolution of \dm{31} as a function of exposure]{res_dm2}{The resolution of a measurement of \dm{31} as a function of exposure assuming the true value is \dm{31} = \num{2.457e-3}~eV$^2$ from the current global fit. The shaded region represents the range in sensitivity due to potential variations in the beam design.  }
 \includegraphics[width=0.7\textwidth]{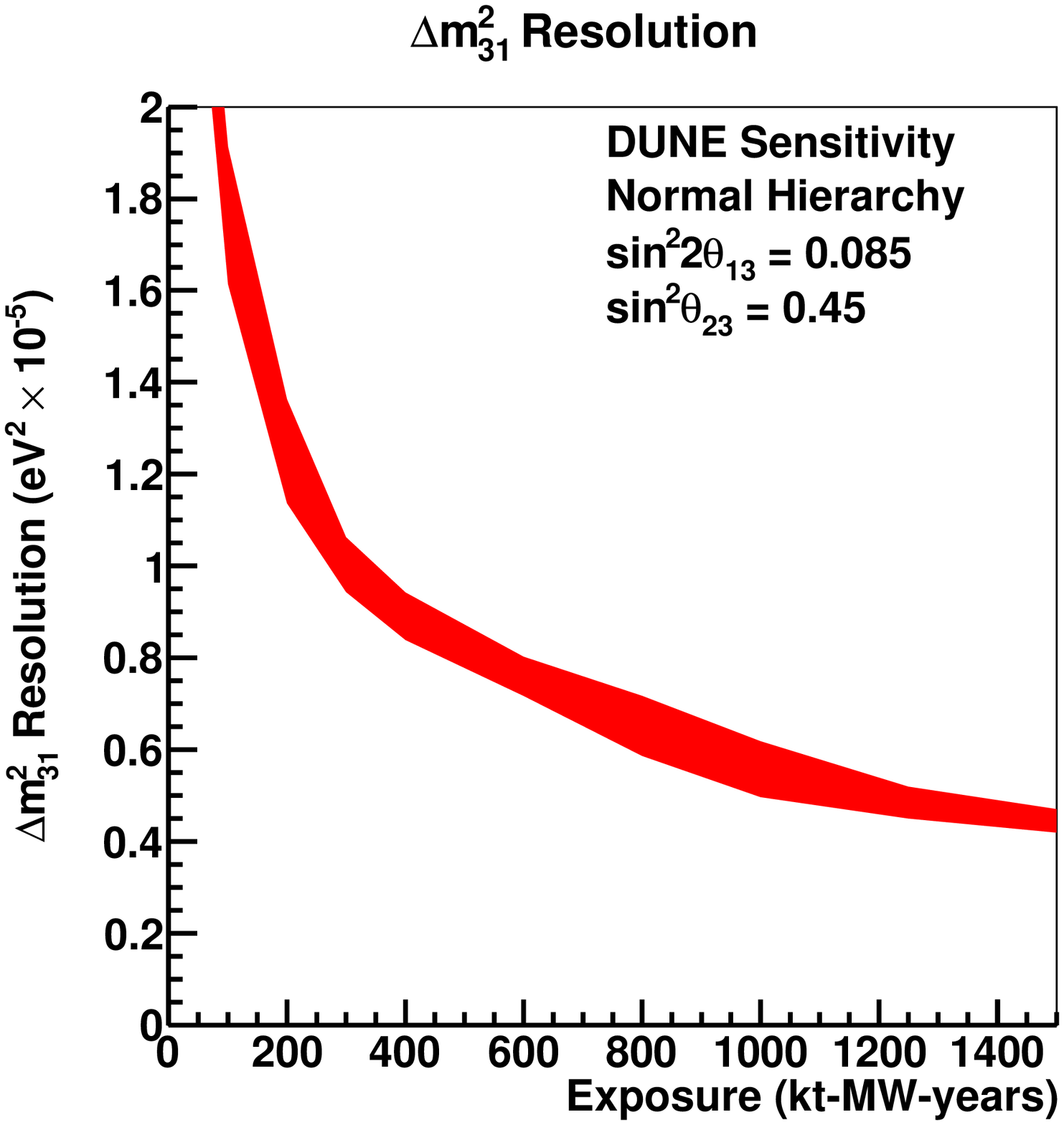}
\end{cdrfigure}

\section{Effect of Systematic Uncertainties}
\label{sec:physics-lbnosc-beamnd-req}

Sensitivity studies presented in Section~\ref{sec:physics-lbnosc-senscalc} test the ability to distinguish
the expected number of \nue appearance and \numu disappearance events given a set of oscillation parameters
from the expectations given an alternate set of parameters. For example, the CP-violation and 
MH-sensitivity
studies test the spectral differences induced by shifting \deltacp away from 0.0 and $\pi$ and by changing the
mass hierarchy. These differences are quantified with a test statistic (see Equation~\ref{eq:dx2_MH}~-~\ref{eq:dx2_CP}) 
which accounts for statistical and systematic uncertainties. 

The effect of systematic uncertainty in the models used to 
predict these spectra is included by allowing the parameters to vary within Gaussian ranges. In the fits,
these systematic nuisance parameters are profiled, i.e., the set of nuisance parameters that produces the
minimum value of the test statistic is chosen.  The central values of the oscillation
parameters and their relative uncertainties are taken from the Nu-Fit~\cite{Gonzalez-Garcia:2014bfa} global
fit to neutrino data; these values are given in Table~\ref{tab:oscpar_nufit}. Uncertainty in non-oscillation
parameters is approximated using
normalization uncertainties on each constituent interaction mode that comprise the signal and background
in each sample. The values for these normalization uncertainties are chosen based on current constraints
on underlying model parameters, the ability of previous experiments to constrain these quantities,
and the expected capability of the DUNE near detector (ND) as outlined in the Near Detector Reference Design chapter of \voldune. 
Consideration is also given to the sources of uncertainty that go into each of the effective normalization
parameters and how they
may be correlated among the different far detector (FD) analysis samples that will be fit in combination.


In the following sections, a justification is presented for the chosen values of the signal and background
normalization uncertainties and their respective correlations.
Studies that consider the effect of varying the size of the residual normalization
uncertainties on the \nue and \anue samples are also presented.
Finally the ongoing effort to characterize and evaluate the effect of individual sources
of uncertainty when propagated to oscillation parameter measurements in the DUNE experiment
is described.

\subsection{Far Detector Samples}
\label{sec:syst_just}
Uncertainties in DUNE will be constrained by external data, near detector data, and a combined
fit to the four (\nue appearance, \anue appearance, \numu disappearance, \anumu disappearance) far detector samples.
This four-sample fit is alternatively referred to as a three-flavor analysis, because the constraints depend
upon the validity of the three-flavor model of neutrino oscillation.

The \numu disappearance analysis sample is composed of \numu CC interactions with backgrounds from NC
interactions in which a charged pion is misidentified as a muon and $\nu_{\tau}$ CC interactions in which the resulting
tau decays to a muon and two neutrinos.
The unoscillated \numu rate and spectrum are expected to be well constrained by the near detector.
The uncertainty on the neutral current (NC) background comes primarily from uncertainty in pion production rates
for the coherent, resonance, and DIS channels, as well as modeling of pion topological signatures that
determine the likelihood of it being misidentified as a muon.
Uncertainties in the $\nu_{\tau}$ CC background level arise from the uncertainty in the $\nu_{\tau}$/\numu
cross section ratio, which cannot be directly constrained by ND measurements.

The \nue appearance sample is composed of \nue CC interactions resulting from \numu$\rightarrow$\nue oscillation
and background from intrinsic beam \nue interactions, NC and
\numu CC interactions in which a photon from a final-state neutral pion is
misidentified as an electron, and $\nu_{\tau}$ interactions in which the resulting $\tau$
decays to an electron and
two neutrinos. Since the
\numu disappearance signal and the \nue appearance signal are produced by the same flux,
the \nue appearance signal is constrained relative to the \numu
signal. The residual uncorrelated uncertainty on the \nue signal results from the statistical
limitations of the \numu constraint, differences in energy scale and selection efficiency between the samples,
and theoretical uncertainties on the \nue/\numu cross section ratio.
The uncertainty on the intrinsic
beam \nue background is dominated by flux uncertainties which are constrained by the near detector and the
observed \numu events.
Predictions for NC and \numu CC background rates are limited by the uncertainties on pion
production rates,
the $\pi^{0}/\pi^{\pm}$ production ratio, and
differences in selection efficiencies.
Again, the $\nu_{\tau}$ background uncertainties are related to cross section ratio uncertainties
which are treated as 100\% correlated among samples.

The far detector samples for the antineutrino beam mirror those described above for the neutrino beam samples.
Additional constraints are expected to occur in a fit to both neutrino and antineutrino beam samples;
variations in \deltacp induce opposite effects (in both shape and rate) in the \nue and \anue samples,
while most systematic uncertainties have a positively correlated effect.
In the neutrino and antineutrino samples,
NC background to the \numu and \anumu samples is treated as correlated,
as is NC and \numu CC background to the \nue and \anue samples, because the dominant source
of uncertainty is expected to be modeling of pion production.
Signal and beam \nue background normalization is treated as uncorrelated.
The normalization for \nutau CC background is treated as 100\% correlated among all samples.


Energy-scale uncertainties in these samples, which can affect the shape of the reconstructed energy spectra,
result from
inaccurate models of detector response, missing energy in the hadronic systems (primarily from neutron production),
and from final-state interactions (FSI). The dominant source of uncertainty is the hadronic energy scale,
which is the same for both \nue and \numu samples, so relative energy-scale uncertainties are limited to
differences in kinematics between \numu and \nue interactions and differences in detector response for
muons and electrons, which will be highly constrained by test beam experiments.
Systematic uncertainties stemming from the FSI model are different
between the $\nu$ and $\bar\nu$ samples, which provides enough freedom in the three-flavor fit to
potentially mimic the effect of a CP violation signal, thus degrading experimental sensitivity.
However, the effect will be the same in the \nue (\anue) and \numu
(\anumu) samples, allowing the relative \nue to \numu (\anue to \anumu) energy scales to be fixed by comparing
the energies of the appearance peak and the disappearance trough. Additional constraints on the FSI 
model will be required from ND analyses and external data.

\subsection{Anticipating Uncertainties Based on Previous Experience}

Table~\ref{tab:nuesysts} shows the uncertainties in
analyses of the \nue appearance rate achieved by MINOS~\cite{Adamson:2013ue}
and T2K~\cite{Abe:2015awa} compared to the uncertainties anticipated in a similar DUNE analysis.
The goals for normalization uncertainties represent the total expected uncertainty on
an analysis of \nue appearance rate in DUNE; the actual DUNE analysis will be based on
a three-flavor spectral fit to all four far detector samples, so that the portions of these
uncertainties that are correlated among far detector samples is expected to largely
cancel. The portions of these uncertainties that are not correlated among samples and
the effect of energy reconstruction on this analysis must be well understood.
The goals for each source of systematic uncertainty are chosen by determining which
of the existing experiments is more representative of DUNE for that source
of uncertainty and, based on that comparison, setting a reasonable goal for a next-generation
experiment. The goals are based on expected capabilities of the high-resolution
LArTPC far detector, precise measurements expected from a highly capable near detector,
and well-understood analysis techniques developed in the existing generation of experiments.
Explanations of the choices in Table~\ref{tab:nuesysts} follow.
\begin{cdrtable}[Systematic uncertainty in current experiments]{lcccl}{nuesysts}{
    Systematic uncertainties on the $\nu_e$ appearance
    signal rate prediction in MINOS and T2K and a projection of the
    anticipated uncertainties in DUNE. In each case, the quoted uncertainty is
    the effect on the $\nu_e$ appearance signal rate only. These uncertainties
    are the \emph{total} expected uncertainties on the $\nu_e$ appearance signal
    rate; this includes both those uncertainties that are correlated and those that
    are uncorrelated in the
    three-flavor fit. For reference, the uncertainties assumed in the nominal
    DUNE sensitivity calculations are also provided.}
Source of & MINOS & T2K & DUNE & Comments \\  \rowtitlestyle
Uncertainty & $\nu_e$ & $\nu_e$ & $\nu_e$ & \\  \toprowrule
Beam Flux & 0.3\% & 3.2\% & 2\% & See ``Flux Uncertainties'' in Section \ref{sec:syst_just_flux}\\
after N/F & & & & \\
extrapolation & & & & \\  \colhline
Interaction & 2.7\% & 5.3\% & $\sim 2\%$ & See ``Interaction Model Uncertainties''  \\
Model & & & & in Section \ref{sec:syst_just_sim} \\  \colhline
Energy scale  & 3.5\% & included& (2\%) & Included in 5\% $\nu_\mu$ sample normalization\\
($\nu_\mu$) & & above & &  uncertainty in DUNE 3-flavor fit. \\  \colhline
Energy  scale & 2.7\% & 2.5\% & 2\% & See ``\nue Energy-Scale Uncertainties''\\
($\nu_e$) & & includes & &  in Section\ref{sec:syst_just_fd}\\
 & & all FD & & \\
 & & effects & & \\   \colhline
Fiducial & 2.4\% & 1\% & 1\% & Larger detectors = smaller uncertainty. \\
volume & & & & \\   \colhline  \colhline
Total  & 5.7\% & 6.8\% & 3.6 \% & \\  \colhline  \colhline
Used in DUNE & & & $5\% \oplus 2\%$ & Residual \nue uncertainty: 2\% \\
Sensitivity & & & & \\
Calculations & & & & \\ 
\end{cdrtable}

\subsubsection{Flux Uncertainties}
\label{sec:syst_just_flux}
DUNE plans to take advantage of spectral analysis,
meaning that absolute and relative flux normalization is required. Since the MINOS \nue appearance analysis
is based on normalization only, in terms of the \nue appearance analysis, DUNE will be more like T2K,
which has achieved 3.2\% normalization uncertainty on its \nue sample from uncertainties in the flux.
Additionally, the inclusive neutrino charged-current cross section measurement from the MINOS
near detector reported in \cite{Adamson:2009ju} has achieved a normalization uncertainty of $\sim$2\% in the
range $3 < E_\nu < 9$ GeV and the near-to-far \numu unoscillated-spectrum extrapolation errors in MINOS
are $<$3\% without any independent constraints on hadron production or muon-flux measurements at the near
site. Therefore, as DUNE is planned to have a highly capable near detector, beamline
muon detectors, dedicated hadronization measurements, and improved simulation of beam flux based on
\minerva~\cite{Aliaga:2013uqz} measurements in the NuMI beam, a goal uncertainty of 2\% has been set on \nue signal
normalization from uncertainties in the flux determination.
As described in Chapter~\ref{ch:physics-nd} and summarized in Section~\ref{sec:syst_studies_ind},
preliminary simulations of the fine-grained tracker ND suggest this is an appropriate goal,
predicting a 2.5\% uncertainty on the absolute flux and
a 1--2\% uncertainty on the flux shape from ND analyses.

\subsubsection{Interaction Model Uncertainties}
\label{sec:syst_just_sim}
Interaction model uncertainties result from uncertainties in modeling neutrino interactions with the target
nuclei in the near and far detectors. These uncertainties include \nue and \numu cross section uncertainties,
uncertainties from modeling the structure of the target nucleus, and the impact of
hadronization model uncertainties in simulating the break up of the target nucleus in higher-energy inelastic
interactions. DUNE will employ argon nuclear targets in both the near and far detectors, allowing for a larger
cancellation of interaction model uncertainties than in T2K, in which the target nuclei in the near detector are
carbon while those in the far detector are oxygen. Additionally, the angular resolution, vertex resolution,
and particle identification capability of the DUNE near detector are expected to increase its ability to
constrain those cross section uncertainties that are common between near and far detectors, but for which
the T2K near detector could not provide significant constraint. DUNE's high-resolution near
detector is expected to enable further constraints on hadronization uncertainties, relative to MINOS, by
resolving many of the individual particles produced in the resonance and deep-inelastic scattering interactions,
which represent the majority of the DUNE data sample. Finally, significant improvements to neutrino interaction
models are anticipated as a result of the intermediate neutrino program~\cite{Adams:2015ogl},
in which measurements will be made
across a range of different nuclei and the resulting models will be tested on argon in LArTPCs.
Therefore, 2\% is taken as a goal for the effect of
interaction model uncertainties on the DUNE \nue signal normalization. It is important to note that this level of
uncertainty depends upon the ability to isolate neutrino-argon interactions in the near detector to facilitate
cancellation of near-far uncertainties; this is a requirement of the ND design.

Additionally, in considering the effect of the three-flavor analysis on the final uncertainty,
the neutrino beams in DUNE and MINOS have energy
spectra that peak around 2.5--3.0~GeV, compared to 600~MeV in T2K. 
The theoretical uncertainty on the \nue/\numu cross section ratio is
less than 1\% above neutrino energies of 1.0--1.5 GeV~\cite{Day-McFarland:2012},
a factor of about three smaller than at T2K's median energy,
so the uncertainty on the \nue normalization with respect to the \numu spectrum in DUNE will be
significantly improved compared to T2K. Uncertainty in the $\nu/\bar{\nu}$ cross section ratio is
somewhat more difficult to quantify given the existing discrepancies between data and currently
implemented models, though this is expected to improve as more complete models are introduced.
As described in Section~\ref{sec:syst_studies_ind}, preliminary studies with a
Fast MC demonstrate the potential for significant cancellation of cross section uncertainties
in the DUNE three-flavor
analysis, even when uncertainties in the \nue/\numu and $\nu/\bar\nu$ cross section ratios
are as large as 20\%.

\subsubsection{Uncertainty from \nue Energy Scale}
\label{sec:syst_just_fd}
MINOS and T2K have achieved uncertainty in the \nue signal normalization from \nue energy scale
of 2.7\% and 2.5\% respectively,
where the 2.5\% from T2K actually includes most far detector effects. DUNE's LArTPC far detector
is expected to outperform both the MINOS sampling calorimeter and the T2K water Cerenkov detector
in reconstruction of \nue interactions. Purity of the quasielastic-like event selection
should be improved relative to T2K's by the capability of the LArTPC to detect hadronic showers
that would be below threshold in SuperK, as described in~\cite{Mosel-Lalakulich-Gallmeister:2014}. For non-quasielastic-like
events, the low thresholds and high resolution of the DUNE LArTPC will significantly improve
calorimetric reconstruction over the MINOS sampling calorimeter.
Significant experience with simulation, reconstruction, and calibration
of neutrino interactions in LArTPCs is expected from the Intermediate Neutrino Program, particularly
Fermilab's SBN program~\cite{Antonello:2015lea},
which will include three LArTPCs: SBND~\cite{Admas:2013xka}, $\mu$BooNE~\cite{microboonetdr},
and ICARUS-T600~\cite{Rubbia:2011ft}. An active program of
prototypes and test-beam measurements is planned to study the reconstruction of charged and neutral particles
in LArTPCs; this suite of experiments includes the
DUNE 35-t prototype 
LArIAT~\cite{Adamson:2013/02/28tla},
CAPTAIN~\cite{Berns:2013usa}, and
the CERN neutrino platform single 
phase prototypes.
(The 35-t and CERN prototypes are discussed in the Prototyping Strategy chapter of \voldune.) 
Finally, an improved model of neutrino interactions will reduce the impact of imperfect reconstruction
of energy from neutrons and low-momentum protons on the DUNE analysis.
Therefore, a goal has been set of using the superior detector performance and the improvements
in understanding of LArTPC energy response and neutrino interactions
expected in the next five to ten years to reduce the normalization uncertainty
from the \nue energy scale to 2\%.

In considering the effect of the three-flavor analysis on the final uncertainty, hadronic energy is expected
to contribute more than half of the total energy deposit for many \nue and \numu interactions in the DUNE
far detector. Since the hadronic energy scale does not depend on neutrino flavor, the uncertainties on this
portion of the LArTPC energy response are expected to largely cancel in the DUNE three-flavor analysis, up
to kinematic differences in the \nue and \numu samples. However, uncertainty in the \nue and \numu energy
scales will also reduce the sensitivity of the fit to spectral shape.
The effect of one such uncertainty on experimental sensitivity is shown in
Section~\ref{sec:syst_var}, but the full impact of energy-scale uncertainty has not yet been explored.
The fraction of the total energy carried by neutrons will be different between the $\nu$ and $\bar{\nu}$
samples both because of the different probabilities for neutrinos and antineutrinos to interact with
protons and neutrons and because of differing
kinematics. The contribution from neutrons will also be different between the \nue and \numu samples because
these samples peak at different energies due to oscillation effects. For this reason, understanding
of both neutron production and detector response to neutrons
will be important for constraining uncertainty in
the three-sample fit. Deployment of the CAPTAIN detector in a neutron beam at LANL is planned to
determine the response of a LArTPC to neutrons. With the neutron response well understood, measurements
by CAPTAIN and other detectors in the intermediate neutrino program will be able to determine
average neutron production rates, which will allow for appropriate corrections to the energy-scale bias at
a statistical level.

\subsubsection{Total Uncertainties Assigned to the Normalization Parameters}
Based on the preceding considerations, the DUNE signal normalization uncertainty is taken to be
$5\% \oplus 2\%$ in both neutrino and antineutrino mode, where 5\% is the normalization uncertainty
on the FD \numu sample and 2\% is the effective uncorrelated
uncertainty on the FD \nue sample after fits to both near and far detector data and all external constraints.
These signal normalization parameters are treated as 100\% uncorrelated between neutrinos and antineutrinos.
The normalization uncertainties on background to these samples and their respective correlations
are given in Table~\ref{tab:bgnormsys}.
These assumptions for the non-oscillation systematic uncertainties 
are used to calculate the sensitivities presented in Section~\ref{sec:physics-lbnosc-senscalc}.
The goal for the \emph{total} uncertainty on the \nue sample in
DUNE is less than 4\%, so the $5\% \oplus 2\%$ signal normalization uncertainty
used for sensitivity calculations is appropriately conservative.
Additionally, cancellation of the correlated portion of the uncertainty is expected in the four-sample fit, so the
residual uncorrelated normalization uncertainty on the \nue sample is expected to be reduced to the 1--2\% level,
such that the 2\% residual normalization uncertainty used in the sensitivity calculations
is also well-justified. 
Variations on these assumptions are explored in Section~\ref{sec:syst_var}.

\begin{cdrtable}[Background normalization uncertainties]{lcl}{bgnormsys}{Normalization uncertainties and
correlations for background to the \nue, \anue, \numu, and \anumu data samples}
      Background & Normalization Uncertainty & Correlations \\ \toprowrule
      \multicolumn{3}{l}{For \nue/\anue appearance:} \\ \colhline
      Beam \nue & 5\% & Uncorrelated in \nue and \anue samples \\ \colhline
      NC      & 5\%  & Correlated in \nue and \anue samples \\ \colhline
      \numu CC & 5\% & Correlated to NC \\ \colhline
      $\nu_\tau$ CC & 20\% & Correlated in \nue and \anue samples \\ \toprowrule
      \multicolumn{3}{l}{For \numu/\anumu disappearance:} \\  \colhline
      NC & 5\% & Uncorrelated to \nue/\anue NC background \\ \colhline
      $\nu_\tau$ & 20\% & Correlated to \nue/\anue $\nu_\tau$ background \\
  \end{cdrtable}

\subsection{Effect of Variation in Uncertainty}
\label{sec:syst_var}
Figure \ref{fig:exp_systs} shows DUNE sensitivity to determination of
neutrino mass hierarchy and discovery of CP violation
as a function of exposure for several levels of signal normalization uncertainty.
As seen in Figure~\ref{fig:exp_systs}, for early phases of DUNE
with exposures less than \num{100}~\ktMWyr, the experiment
will be statistically limited.
The impact of systematic uncertainty on the CP-violation sensitivity for large exposure
is obvious in Figure~\ref{fig:exp_systs}; the \nue signal normalization uncertainty must
be understood at the level of $5\% \oplus 2\%$ in order to reach 5$\sigma$ sensitivity for
75\% of \deltacp values with exposures less than $\sim$\num{900}~\ktMWyr{} in the case of the
Optimized Design. Specifically, the absolute normalization of the \numu sample must be known to
$\sim$5\% and the normalization of the \nue sample,
relative to the \anue, \numu, and \anumu samples after all constraints from
external, near detector, and far detector data have been applied, must be determined 
at the few-percent level. This level of systematic uncertainty sets the capability and
design requirements for all components of the experiment, including the beam design and the
near and far detectors.
\begin{cdrfigure}[Variation in sensitivity due to systematics variations]{exp_systs}{
  Expected sensitivity of DUNE  to determination of the neutrino mass
  hierarchy (top) and discovery of CP violation, i.e., $\delta_{CP} \ne$ 0 or $\pi$,
  (bottom) as a function of exposure in~\ktMWyr, assuming 
  equal running in neutrino and antineutrino mode, for a range of values for
  the \nue and \anue signal normalization uncertainties from $5\%\oplus3\%$ to
  $5\%\oplus1\%$. The sensitivities quoted
  are the minimum sensitivity for 100\% of \deltacp values in the case of 
  mass hierarchy and 50\% (bottom left) or 75\% (bottom right) of \deltacp values 
  in the case of CP violation. The two bands on each plot represent a range of potential
  beam designs: the blue hashed band is for the CDR Reference Design and the solid green
  band is for the Optimized Design. 
  Sensitivities are for true normal hierarchy; neutrino mass hierarchy
  and $\theta_{23}$ octant are assumed to be unknown.}
\includegraphics[width=0.45\linewidth]{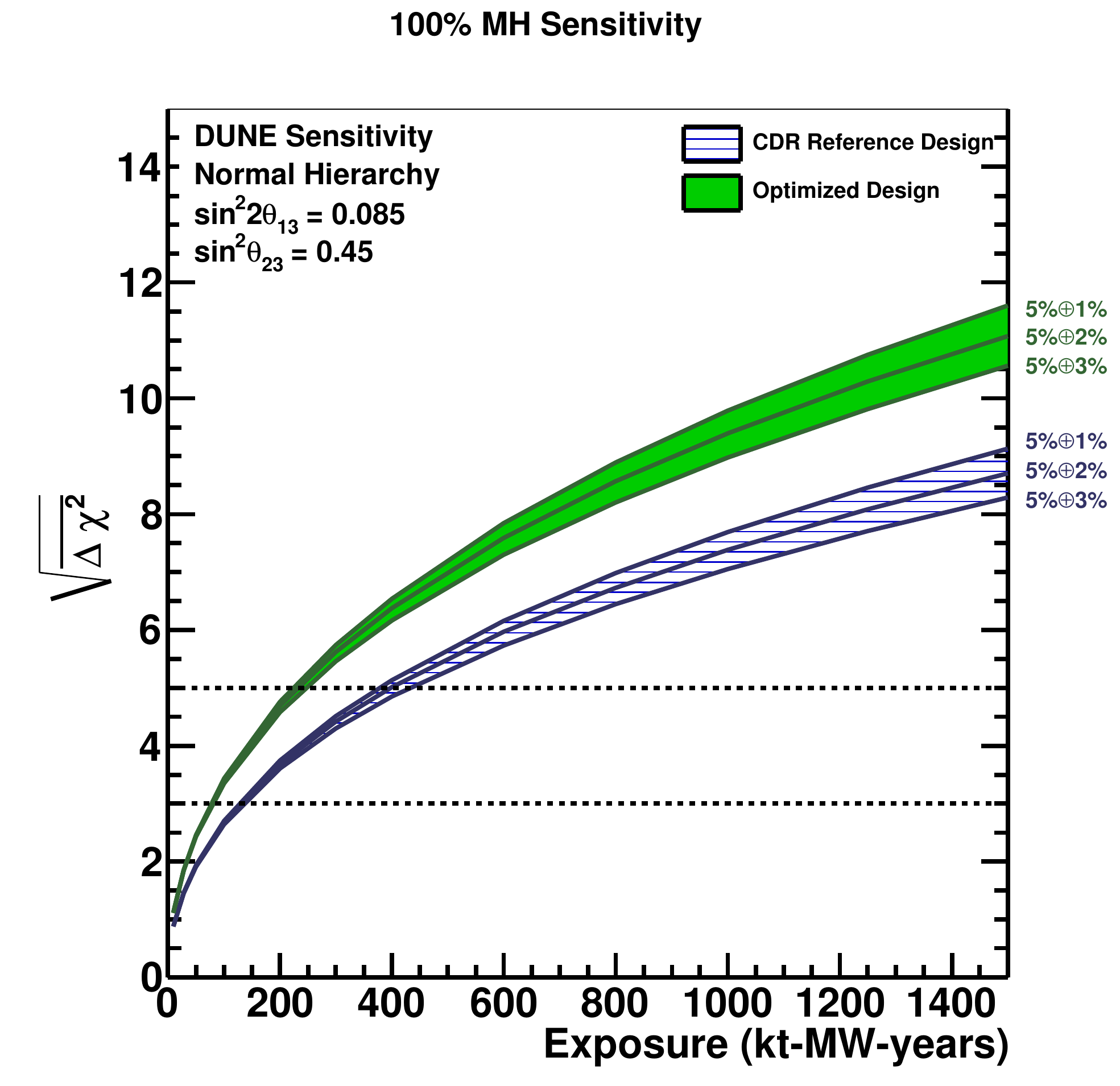} \\
\includegraphics[width=0.45\linewidth]{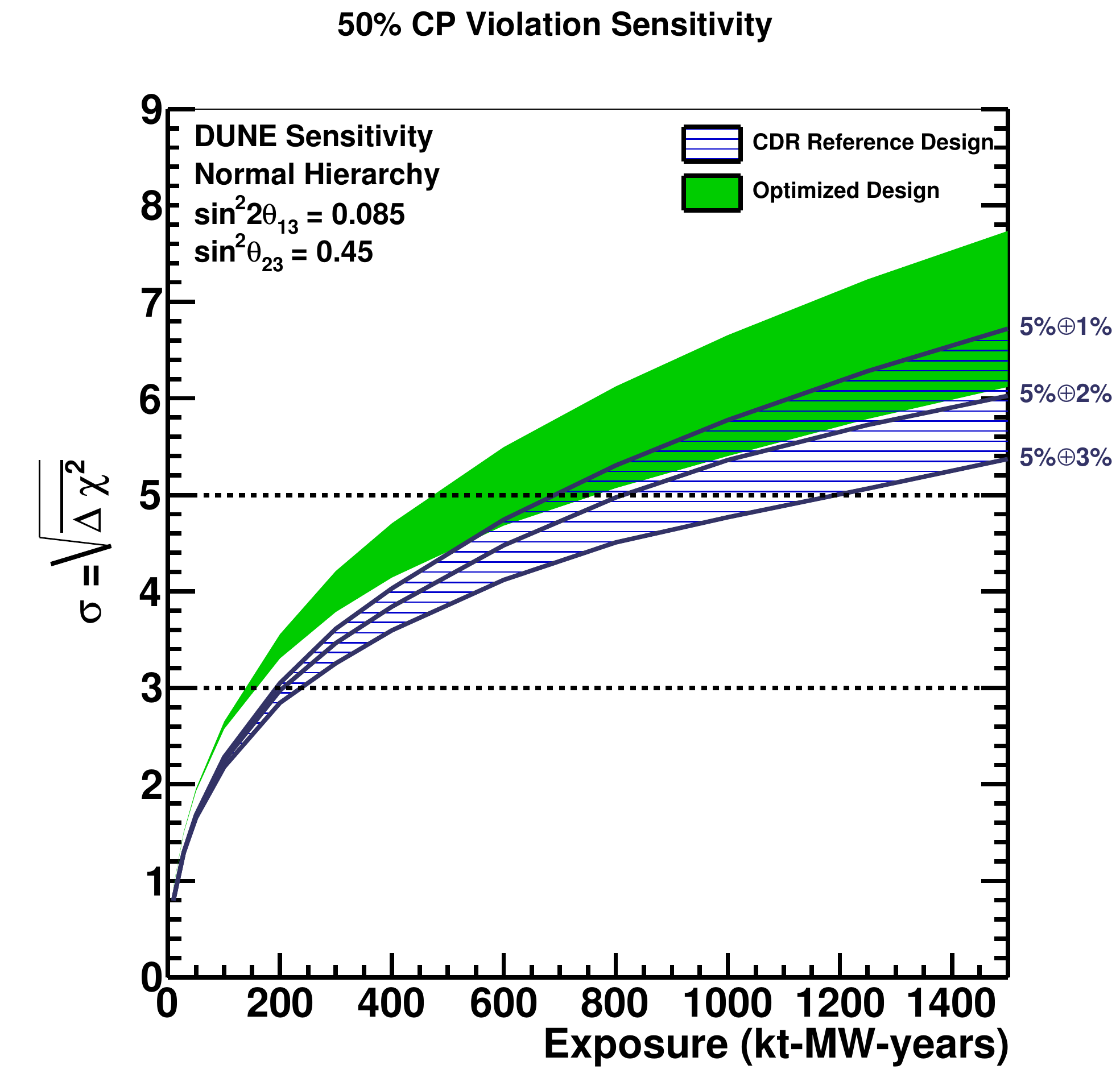}
\includegraphics[width=0.45\linewidth]{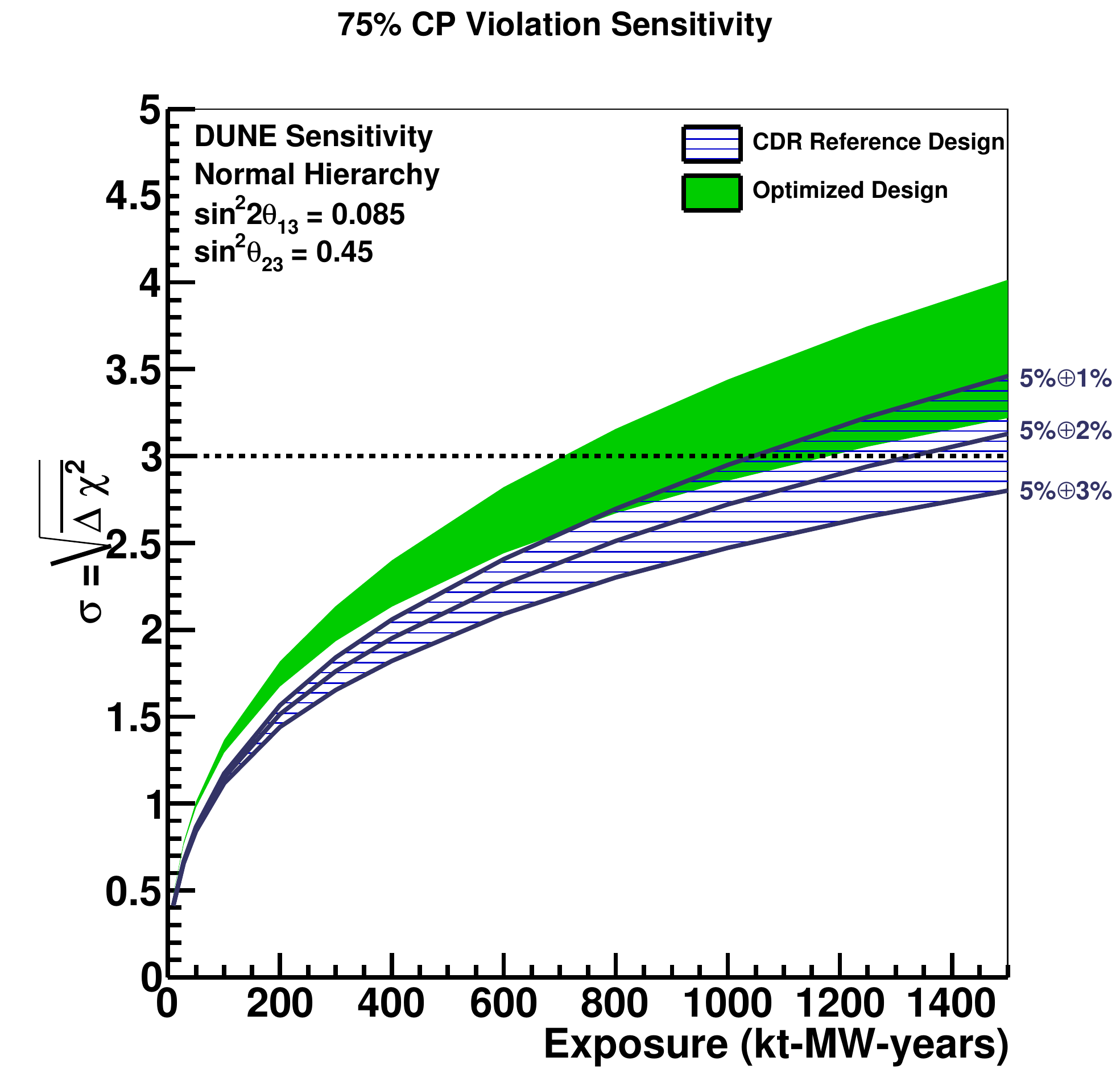}
\end{cdrfigure}

Signal and
background normalization uncertainties remain
relatively unimportant for the mass hierarchy measurement, even at large exposure, when considering
minimum sensitivity for 100\% of \deltacp values. This is because the minimum sensitivity 
occurs in the near-degenerate region where it is difficult to determine
whether one is observing \deltacp $= + \pi/2 $ in the normal hierarchy
or \deltacp $=-\pi/2$ in the inverted hierarchy. Spectral analysis will
help resolve this near-degeneracy, but is dependent on as-yet
unexplored uncertainties in the spectral shape, which are expected to be dominated
by energy-scale uncertainty. Figure~\ref{fig:escale_syst} shows the
impact on MH and CP-violation sensitivity of one possible energy-scale variation, in which
energy bins are adjusted by N[E]$\rightarrow$N[(1+a)E], while keeping the total number of
events fixed. This is only one possible type of energy-scale uncertainty; more comprehensive
study of energy-scale uncertainty is in progress and will be included in
future analyses of experimental sensitivity.
\begin{cdrfigure}[Variation in sensitivity due to an energy-scale uncertainty]{escale_syst}{
Expected sensitivity of DUNE to determination of the neutrino mass
  hierarchy (left) and discovery of CP violation, i.e. $\delta_{CP} \ne$ 0 or $\pi$,
  (right) as a function of the true value of \deltacp, assuming 
  equal running in neutrino and antineutrino mode, for a range of values assigned to the
  ``a'' parameter in the energy-scale variation described in the text. In the MH figure, the case with no
  energy-scale systematic provides a significance of at least $\sqrt{\Delta\chi^2}$ = 5 for
  all values of \deltacp. In the CPV figure, the case with no energy-scale systematic provides
  a significance of at least 3$\sigma$ for 75\% of \deltacp values.
  (See Figures~\ref{fig:mh_exposure} and \ref{fig:cpv_exposure} for the possible range of exposures
  to achieve this level of significance.)
  Sensitivities are for true normal hierarchy; neutrino mass hierarchy
  and $\theta_{23}$ octant are assumed to be unknown.}
\includegraphics[width=0.44\linewidth]{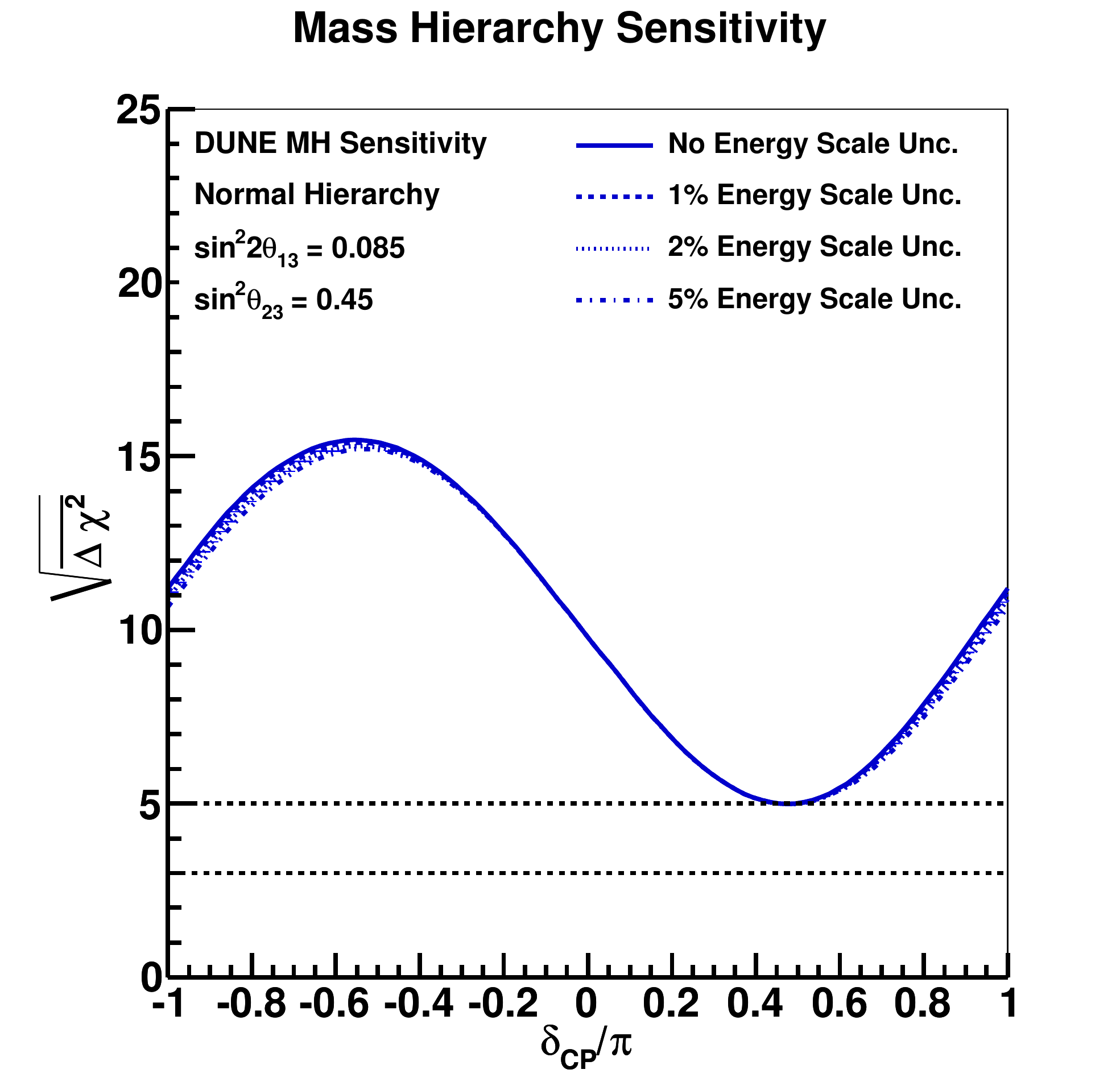}
\includegraphics[width=0.44\linewidth]{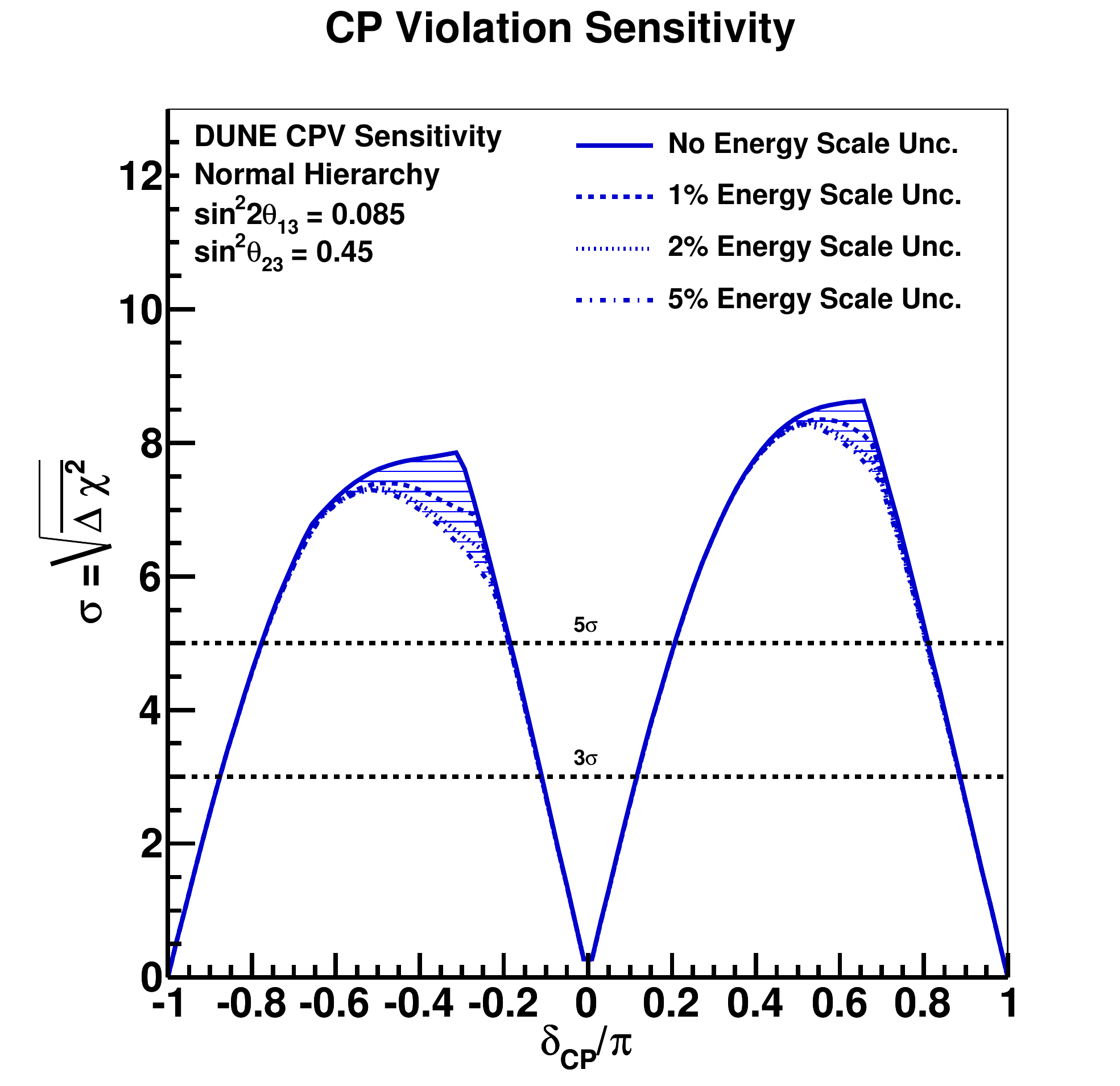}
\end{cdrfigure}
\subsection{Ongoing and Planned Studies of Systematic Uncertainty}
\label{sec:syst_studies_ind}
Detailed evaluation of systematic uncertainties for DUNE is ongoing. In many cases plans for studies
have been developed but have not yet been executed. In general, each systematic will be studied both by
propagating its uncertainty to oscillation analyses to evaluate the resultant degradation of the sensitivity
and by ensuring the considered variations give proper coverage, i.e., truly encapsulating
the lack of knowledge of the processes/effects in question. Estimates of systematic uncertainty for the 
propagation studies will be varied between the constraints available from current external knowledge
and a range of projections for ND performance. In cases where systematic uncertainty is shown to
degrade the oscillation parameter
measurement sensitivities, the required constraints will become detector performance requirements.
The details of these studies are beyond the scope of this document; however, conclusions from some
initial studies and an overview of each source of systematic uncertainty is laid out in the remainder of
this section.

Initial studies using a Near Detector Fast Monte Carlo with a parameterized detector response
predict 3\% statistical uncertainties on the absolute flux using fully 
leptonic neutrino interactions for which high-precision cross section predictions 
exist. Specifically,
the statistical uncertainty is expected to be $\sim$3\% for neutrino-electron
scattering ($E_\nu<5$~GeV) and inverse muon decay ($E_\nu>11$~GeV).
Relative normalization using the low-$\nu_0$ method is
expected to constrain the flux shape to 1--2\%; this level of
precision in the \numu flux was achieved by NOMAD\cite{Wu:2007ab,Lyubushkin:2008pe}, enabled by its 0.2\%
uncertainty in the muon energy scale. This flux shape measurement will be made for both \nue and \numu, therefore,
in combination with measurements from hadron production experiments, it can determine the distribution of
the parent mesons which will constrain the near/far flux ratio. Detailed discussion of the planned program
of ND measurements is in Chapter~\ref{ch:physics-nd} and~\cite{Mishra:2008nx,Adams:2013qkq}.
Studies using a multi-sample fit  to constrain the flux with simulated DUNE near detector
event samples show significant constraints on all flux
uncertainties; the post-fit uncertainty in most flux bins for this preliminary fit is less
than 5\%, which is the uncorrelated \numu signal normalization
uncertainty assumed by the sensitivity calculations.


The two main sources of uncertainty in the beam simulation come from variations in the beam optics,
$\mathcal{O}($1\%), and uncertainties in the hadron production models, $\mathcal{O}$(10\%).
Beam optics variations have been studied in detail
and are found to be easily constrained by the ND. Software tools that
allow re-weighting of neutrinos based on their parent hadrons have been developed by \minerva; work is progressing with
them to implement these tools in the DUNE simulation to evaluate the impact of uncertainty in
hadron production models.
In the meantime, \minerva has agreed to provide its flux covariance matrix
that details the flux rate and shape uncertainties prior to ND constraints. This will be combined with DUNE
simulations to project reasonable hadron-production uncertainties to ND and FD analyses. 
Ultimately these uncertainties will be constrained by near detector measurements and
dedicated hadron-production measurements such as those at NA61/SHINE~\cite{NA61:2014fnalbeams}
will provide additional external constraints.

Primary interaction uncertainties are specific to each model, and each of the three major
cross section components (quasi-elastic processes, resonance production, and deep inelastic scattering)
contribute roughly equally to the \nue and \anue
appearance signal. In most cases, uncertainty in modeling primary interactions comes from the
hadronic interaction part of the calculation, which includes form factors in the hadron tensor,
the nuclear initial state, and FSI. 

  \emph{Coherent scattering}: Coherent models built upon partially conserved axial current
  theory relate the neutrino scattering 
  cross section to pion-nucleon or pion-nucleus scattering data~\cite{Rein-Sehgal:1983}\cite{Berger-Sehgal:2009}. The choice and characterization
  of that data can have large effects on the calculated cross section. Alternate ``microscopic model'' 
  formulations are valid only over limited kinematic ranges and are not adequate to describe this process 
  for DUNE~\cite{Alvarez-Ruso-Geng-Hirenzaki-Vicente-Vacas:2007}\cite{Alvarez-Ruso-Geng-Vicente-Vacas:2007}. 
  Both types of model suffer from limited data constraints over a range of neutrino energy
  and nuclear targets. 
  However, the low hadron thresholds and good angular resolution of the DUNE ND should be able
  to produce world-leading measurements and provide adequate constraints for this interaction channel and 
  its relatively small contribution to the overall cross section. Data from \minerva~\cite{Higuera:2014azj},
  T2K, and
  upcoming LArTPC experiments will provide constraints for a variety of target nuclei over the
  relevant energy ranges required to constrain this sub-dominant process.

  \emph{Quasi-elastic processes (QE)}: Models for this type of interaction require that the target nucleon 
  is neither excited nor fragmented because the 4-momentum transfer to the hadronic system ($Q^{2}$) is low.
  For these low-$Q^{2}$ interactions, details of the nuclear initial state are important. However,
  current implementations of nuclear initial state models are inadequate, therefore the uncertainty in the only
  free parameter in the free-nucleon cross section model, $M_{A}^{QE}$, has been expanded to absorb the
  differences between simulations and $\nu$-nucleus scattering data. Better models of the nuclear
  initial state have been developed
  and are currently being implemented in GENIE and other generators. These new models will be compared with current 
  and future data from \minerva, T2K and upcoming LArTPC experiments, and the effect of variations in $M_{A}^{QE}$
  on FD spectra will be compared to the effect of introducing the new models.
  Eventually the set of models that best agrees with data will be adopted in the DUNE simulations and
  the uncertainties assigned to these models will reflect the level of agreement with data.

  \emph{Resonance production}:
  There are two important sources of uncertainty in this model. The first is the uncertainty on the free-nucleon
  cross section due to unconstrained form factors and their use as effective parameters to absorb nuclear
  modeling effects.
  The second is the disagreement in outgoing pion kinematics between simulations and data.
  Data from T2K, \minerva~\cite{Eberly:2014mra}\cite{Aliaga:2015wva}, upcoming LArTPC experiments, and the DUNE ND should provide good constraints for DUNE
  oscillation analyses, but model improvements will be required to help propagate these constraints
  to the FD signal and background predictions. Model improvements are needed for the principal interaction
  model, so-called ``background'' interactions where the pion is produced at the interaction vertex
  rather than through an intermediate $\Delta$ (or higher resonance), the interference between the two
  models, and the contributions to single-pion production from low-multiplicity DIS. Improved nuclear models
  are also required in order to estimate the impact of processes like pion-less delta decay and FSI. New models, which are
  available for some relevant regions of phase space, must be incorporated into generators and compared
  with data~\cite{Hernandez-Nieves-Valverde:2007}.

  \emph{Deep inelastic scattering (DIS)}: The inclusive DIS cross section on iron has been very well constrained by 
  data but individual final states have not. The primary source of uncertainty is in modeling the
  content and kinematics of the hadronic system as a function of its invariant mass. The resulting
  uncertainty on the DIS contribution to signal samples is relatively small, but it is nonetheless important
  to better constrain these models because the DIS contribution to background via pion production is significant.
  Data from \minerva and upcoming LArTPC experiments should help to constrain
  the exclusive cross sections, as well as nuclear effects on the inclusive cross section~\cite{Tice:2014pgu}.
  Current studies are focused on building parameterized re-weighting 
  functions for the hadronization model based on GENIE samples generated with 1$\sigma$ changes to each relevant 
  model parameter.

Nuclear models enter into the simulation of neutrino interactions both through modeling of initial-state interactions,
i.e., interactions between the neutrino and the initial state of the nucleons and virtual particles within the nucleus,
and modeling of final-state interactions (FSI), i.e., interactions of the particles exiting the
primary interaction vertex with the nuclear medium. 

  \emph{Nuclear initial state}: Uncertainties in initial-state interactions due to
  naive modeling of the environment of the nucleus have thus far been taken into account through inflation
  of the uncertainties on the free nucleon or quark interaction
  model. New models~\cite{Alvarez-Ruso-Hayato-Nieves:2014} are being added to generators and will soon be incorporated into the Fast MC
  to study how the impact on sensitivity of these models compares with uncertainties in the current nominal model.
  Data from upcoming LArTPC neutrino experiments will provide detailed information on nucleon production
  rates and kinematics, which will help to distinguish which of the new models best describes the data.

  \emph{Final-state interactions}: FSI can alter event reconstruction in two distinct ways. The first is a smearing
  of the total energy available to be deposited in the detector. The second is the misidentification of
  event topologies used to classify the neutrino flavor and interaction mode. Uncertainties in selection
  efficiencies and event-sample migrations
  due to intranuclear rescattering can be studied with existing DUNE tools. The predictions and
  uncertainties on GENIE's ``hA'' model~\cite{Dytman:2011zz} of intranuclear interaction
  are being tested against the detailed FSI model in the GiBUU~\cite{Buss:2011mx} event
  generator. Studies of correlations among the free model parameters and and how variations in those
  parameters propagate differently for $\nu$ and $\bar{\nu}$ are also needed. 
  Electron-argon scattering data~\cite{Benhar:2014nca}  and studies of hadron production
  in upcoming LArTPC experiments are expected to further constrain the effects of FSI in argon nuclei.

A fit to Fast MC simulation of all four far detector samples
(\nue, \anue, \numu, \anumu) significantly
constrains cross section systematic uncertainty even in the case where many
cross section parameters are allowed to vary simultaneously within their
GENIE uncertainties. As seen in the example shown in Figure
\ref{fig:MAresqesyst}, 
a fit in which both $M_A^{QE,CC}$ and 
$M_A^{RES,CC}$ are allowed to vary within their GENIE uncertainties 
($\pm$20\%), which could significantly alter the energy distribution of the 
the selected events, results in a dramatic reduction in sensitivity if one 
considers only the $\nu_e$ appearance signal without constraint from the 
$\bar{\nu}_e$ and $\nu_{\mu}$/$\bar{\nu}_{\mu}$ samples.
In contrast, for a four sample fit,
this same parameter variation results in a smaller reduction in
sensitivity to CP violation.
This result includes a 10\% uncertainty in the $\nu/\bar{\nu}$
cross section ratio and a 2.5\% uncertainty in the $\nu_e/\nu_{\mu}$
cross section ratio; uncertainties in these ratios as large as 20\% have
been considered and did not produce dramatically different results.
More details on this analysis are available in \cite{Bass:2014vta}.
Preliminary studies also demonstrate significant constraint on cross section systematics
from the near detector.
\begin{cdrfigure}[Example of cross section uncertainty cancellation in FD fit]{MAresqesyst}{
An example CP violation sensitivity calculated using inputs from the 
  FastMC in a fit to all four ($\nu_e$, $\overline\nu_e$, $\nu_{\mu}$, 
  $\overline\nu_{\mu}$) samples (red) and a fit to the $\nu_e$ appearance sample 
  only (blue), for the case of no systematic uncertainty (solid) and the case in
  which both $M_A^{QE,CC}$ and $M_A^{RES,CC}$ are allowed to vary with a
  1$\sigma$ uncertainty of 20\% (dashed). This example was taken from an earlier
  DUNE study, so the absolute sensitivity can not be compared with the DUNE 
  sensitivities presented in this document.}
\includegraphics[width=0.5\linewidth]{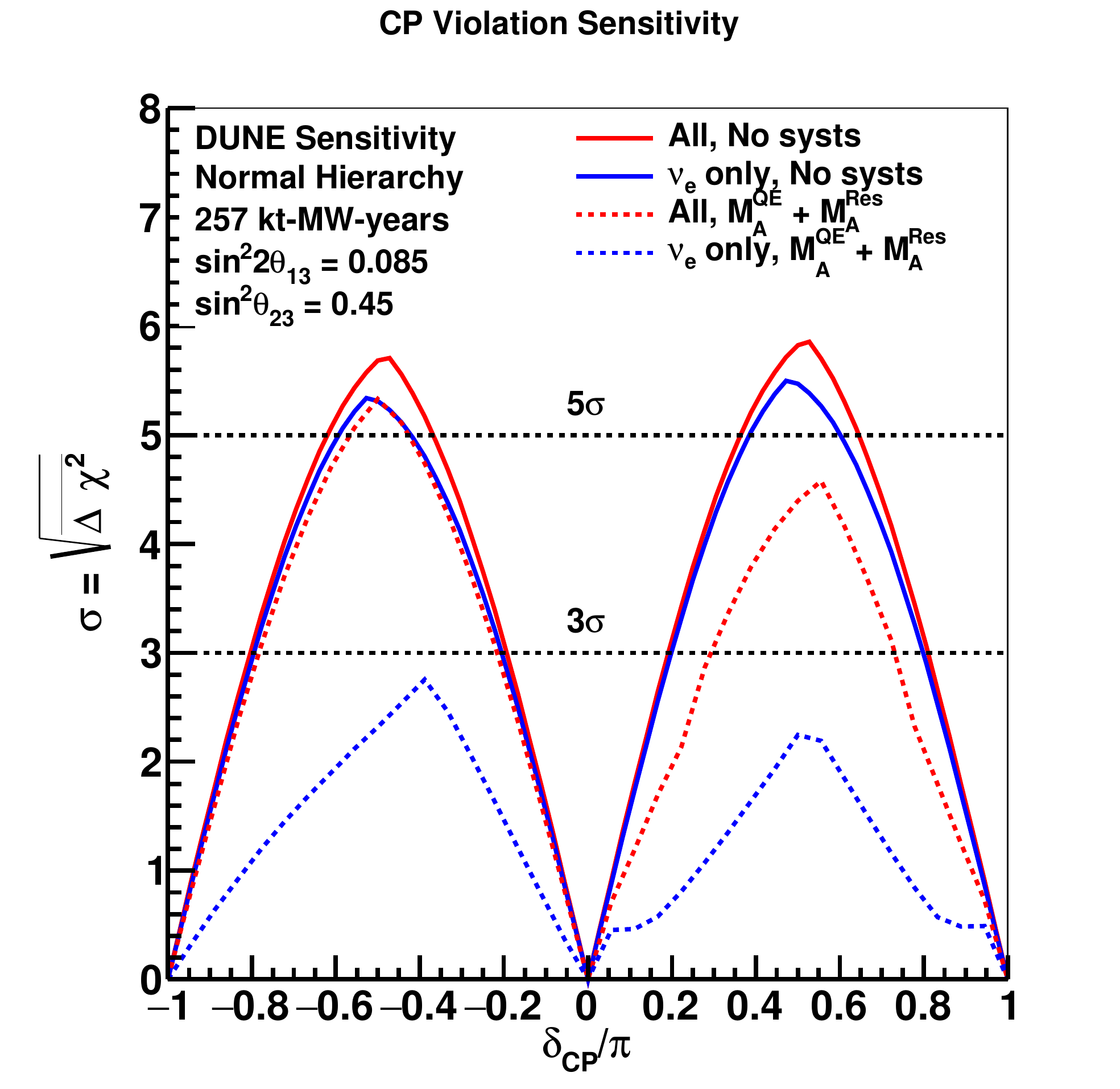}
\end{cdrfigure}

Uncertainty stemming from detector effects
are somewhat more difficult to address with existing
simulation efforts. Tools to evaluate the effect of uncertainty in single-particle resolutions,
detection and particle-identification efficiencies, and energy scale are in development within
the Fast MC framework. The results of these studies will provide performance requirements
for the DUNE detectors, but more complete understanding of the expected size of these effects
will require comparison between data and a full Monte Carlo.
The status of efforts to develop reconstruction and analysis tools for a full Monte Carlo simulation
of DUNE is described in 
the Software and Computing chapter of \voldune. At the same time,
a number of test-beam and prototype experiments, including the DUNE 35-t prototype,
LARIAT, CAPTAIN, and the CERN neutrino platform experiments, are being designed and built to reduce these
uncertainties with experimental data. The status of some of these efforts is described in the Prototyping
Strategy chapter of \voldune.

These ongoing studies to improve models of neutrino interactions in LArTPC detectors and
to evaluate the remaining uncertainties, by comparisons to data and alternate models, 
are considered high priority not only by the DUNE collaboration, but by the global neutrino community.
It is reasonable to expect that model improvements and new data will provide DUNE with improved inputs
and reduced uncertainties compared to current knowledge. Following the plan described in the preceding
paragraphs, DUNE collaborators will actively participate in the global effort to improve understanding
of neutrino interactions, will propagate what is learned
in the intermediate neutrino program to DUNE analyses, and will
evaluate the effect of remaining uncertainties on the DUNE analyses.


\section{Optimization of the LBNF Beam Designs}
\label{sec:physics-lbnosc-beam}

The LBNF neutrino facility at Fermilab utilizes a conventional
horn-focused neutrino beam produced from pion and kaon
decay-in-flight. It will aim the neutrino beam toward the DUNE far
detector located \num{1300}~km away at the Sanford Underground Research
Facility.  The design of the LBNF neutrino beamline is a critical
component for the success of DUNE. As demonstrated in earlier
sections, the optimization of the beam design can have significant
impact on the exposures needed to achieve the desired physics goals
independent of additional improvements to the accelerator complex such
as upgrades to 2.4~MW and improvements to uptime and efficiency, and
reductions in systematic uncertainties. The reference beam design is
described in detail in \vollbnf. In this section a summary of the
ongoing efforts to optimize the beam focusing system and decay pipe
geometry designs for the primary oscillation physics measurements is
presented.

\subsection{Reference Beam Design}
\label{sec:reference-design-focusing-system}
The reference beam design is based on the designs of targets and
focusing systems for NuMI. These designs are well understood and have
proven track records of reliability and performance. The LBNF
reference design includes a target similar to the one used for the
low-energy tune of the NuMI beam~\cite{Anderson:1998zza},  but with a larger
thickness to accommodate the \MWadj{1.2} primary proton beam, and focusing
horns essentially identical to those currently in operation in the
NuMI beamline. 
The target consists of 47 graphite
segments, for a total length of 95~cm including the space between
segments, corresponding to two interaction lengths. The upstream face
of the first segment is positioned 45~cm upstream of the first
focusing horn to ensure sufficient clearance of the target's
downstream end from the horn inner conductor. The separation of the
upstream faces of the two horns has been decreased to 6.6~m, compared
to the 10-m distance for the low-energy tune of the NuMI beam, to
slightly enhance the neutrino flux at lower energies. A helium-filled
decay pipe, 4~m in diameter and 204~m in length, provides the decay
volume for the secondary pions to decay to muon neutrinos.

Neutrino fluxes for the reference beam are shown in
Figure~\ref{fig:beam_req_reference_flux} for a \GeVadj{120} primary proton
beam.  Lowering the momentum of the primary proton beam increases
right-sign neutrino flux at low energies and decreases wrong-sign
contamination.  Figure~\ref{fig:beam_req_proton_energy} shows current
estimates of CP and mass hierarchy sensitivities versus proton
momentum. It is estimated that these quantities improve very slightly as
proton momentum decreases, until approximately 60~GeV, at which point the Main
Injector cycle time becomes constant with proton momentum, causing
beam power and physics sensitivities to drop sharply.  Unless
otherwise noted, results throughout this volume assume an \GeVadj{80}
proton beam, corresponding to the optimal momentum whose technical
feasibility has been thoroughly studied.

Another handle for tuning the neutrino energy
spectrum of the reference beam is modification of the distance between
the target and the first focusing horn.  The expected fluxes for three
different configurations are shown in
Figure~\ref{fig:beam_req_lemehe}. The improvement in flux for each configuration with a longer decay pipe length of 250~m is also shown.

\begin{cdrfigure}[Neutrino fluxes for the reference focusing 
  system.]{beam_req_reference_flux}{Neutrino fluxes for the reference 
    focusing system operating in neutrino mode (left) and antineutrino 
    mode (right), generated with a \GeVadj{120} 
    primary proton beam.} 
\centering 
\begin{minipage}{0.45\textwidth}
\centering 
\includegraphics[width=1.0\textwidth]{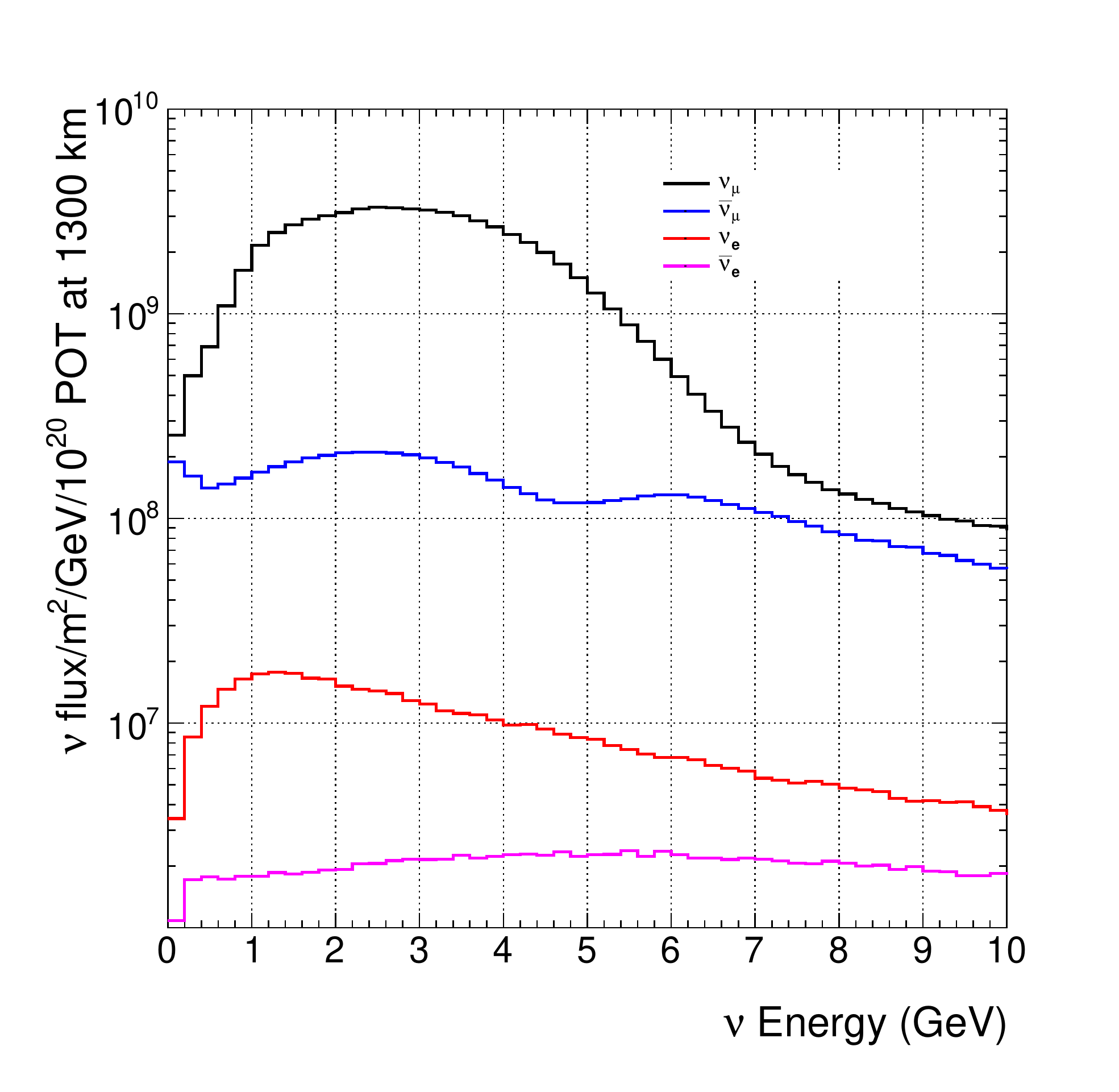}
\end{minipage}\hfill 
\begin{minipage}{0.45\textwidth}
\centering 
\includegraphics[width=1.0\textwidth]{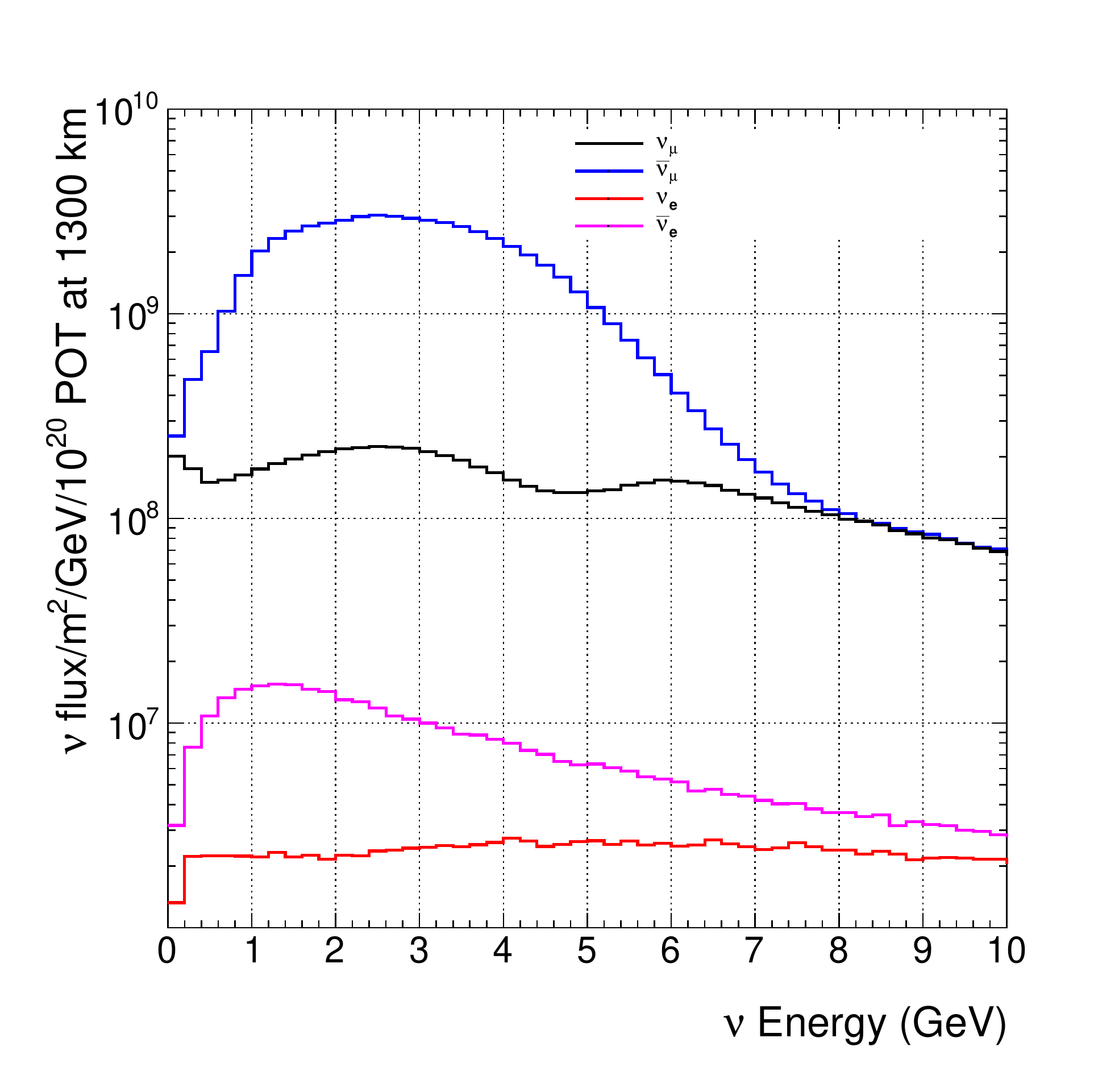}
\end{minipage}
\end{cdrfigure}

\begin{cdrfigure}[CP and mass hierarchy sensitivity versus proton 
  momentum]{beam_req_proton_energy}{Minimum mass hierarchy
    sensitivity (left) and 
    coverage of 75\% of possible values of $\delta_{CP}$ (right) as a
    function of proton momentum assuming an exposure of \num{280}~\ktyr}
\centering 
\begin{minipage}{0.45\textwidth}
\centering 
\includegraphics[width=1.0\textwidth]{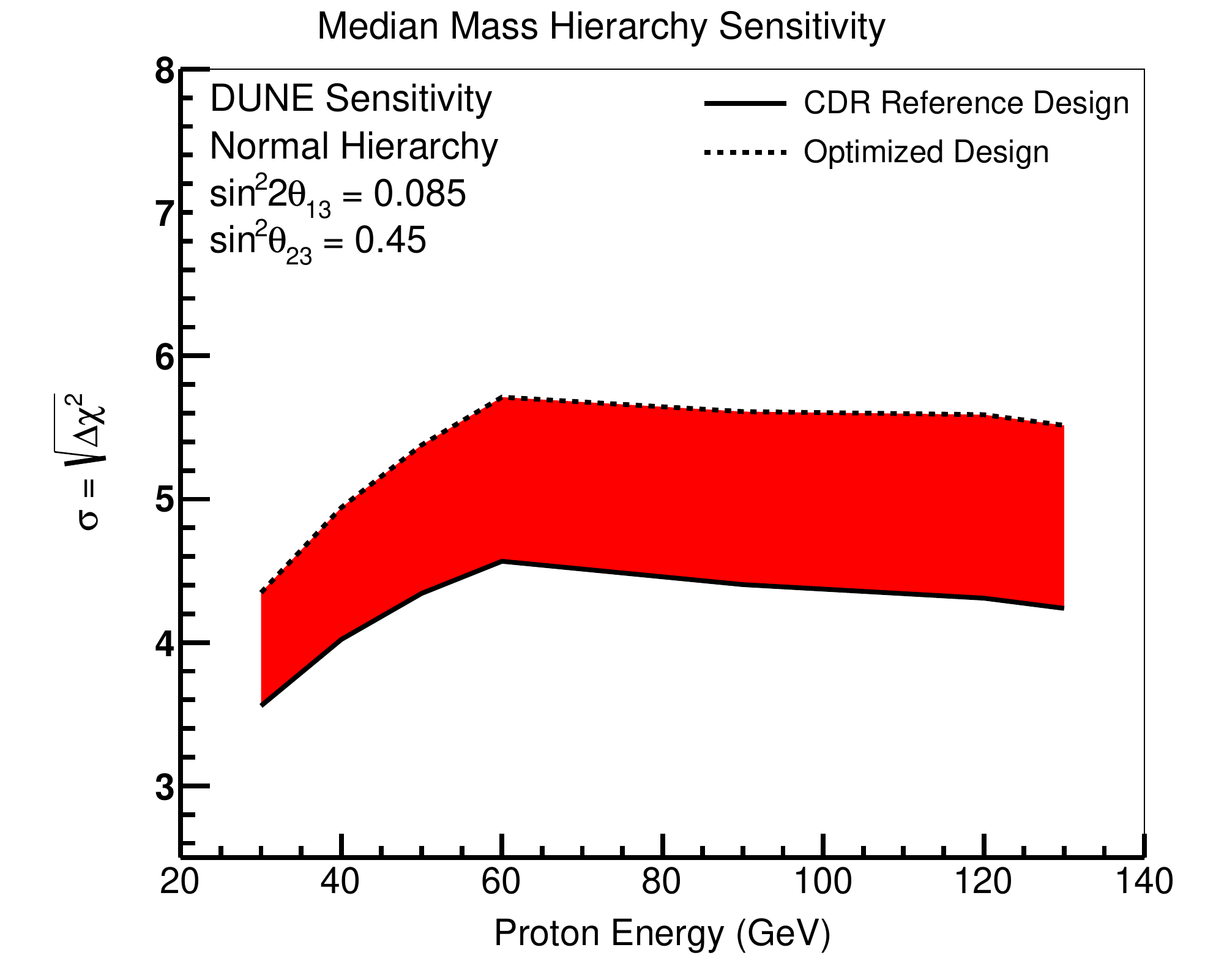}
\end{minipage}\hfill 
\begin{minipage}{0.45\textwidth}
\centering 
\includegraphics[width=1.0\textwidth]{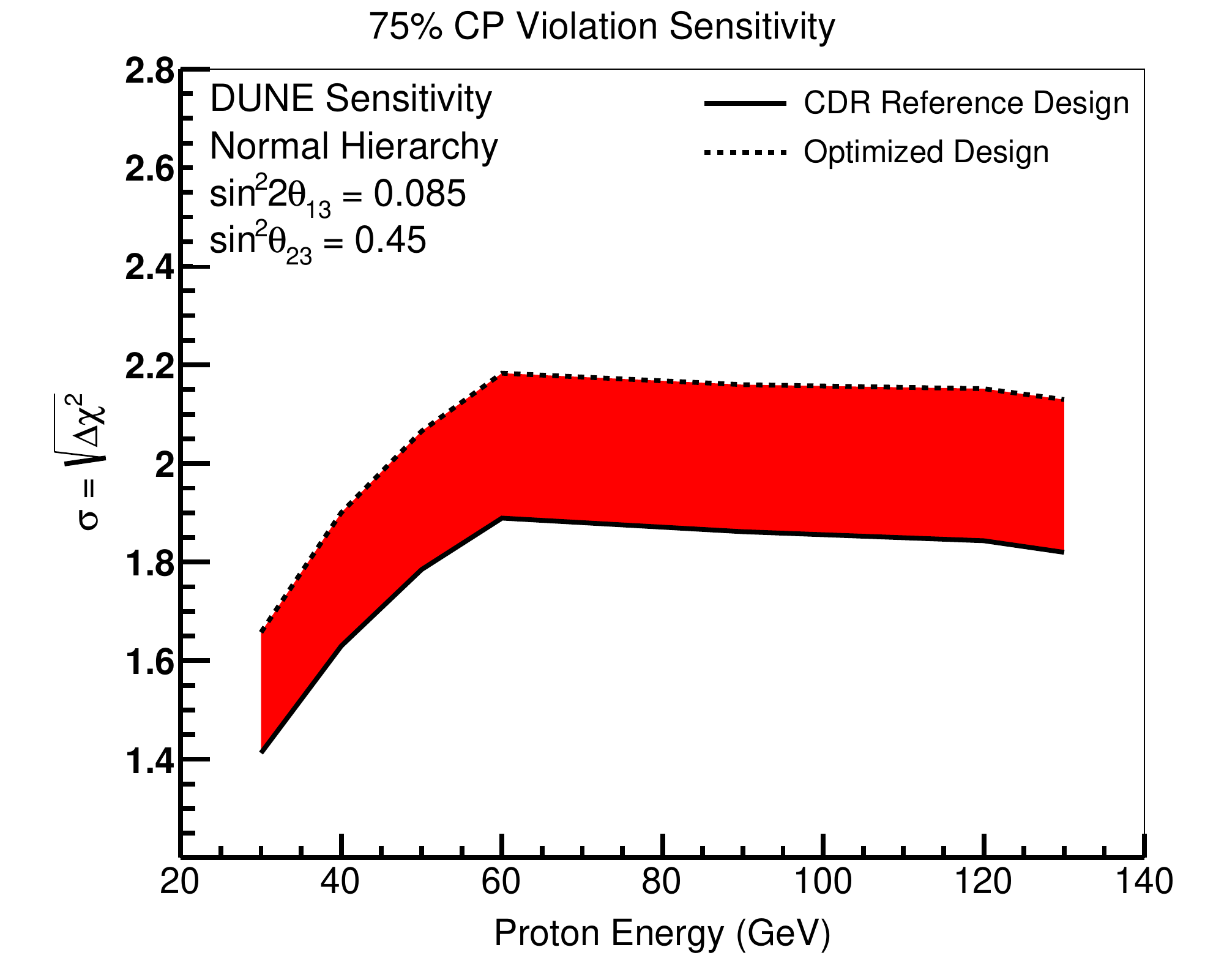}
\end{minipage}
\end{cdrfigure}

\begin{cdrfigure}[Neutrino flux comparison for various target 
  positions and decay pipe lengths] {beam_req_lemehe}{Neutrino fluxes for the 
    reference beam design with \GeVadj{120} protons, with the target starting 45 (LE), 135 (ME) 
    and 250 (LE)~cm upstream from the start of Horn~1 (left), and the 
   fractional improvement in flux for the same beam configurations 
    when the decay pipe is lengthened to 250~m (right).}
    \includegraphics[width=0.75\textwidth]{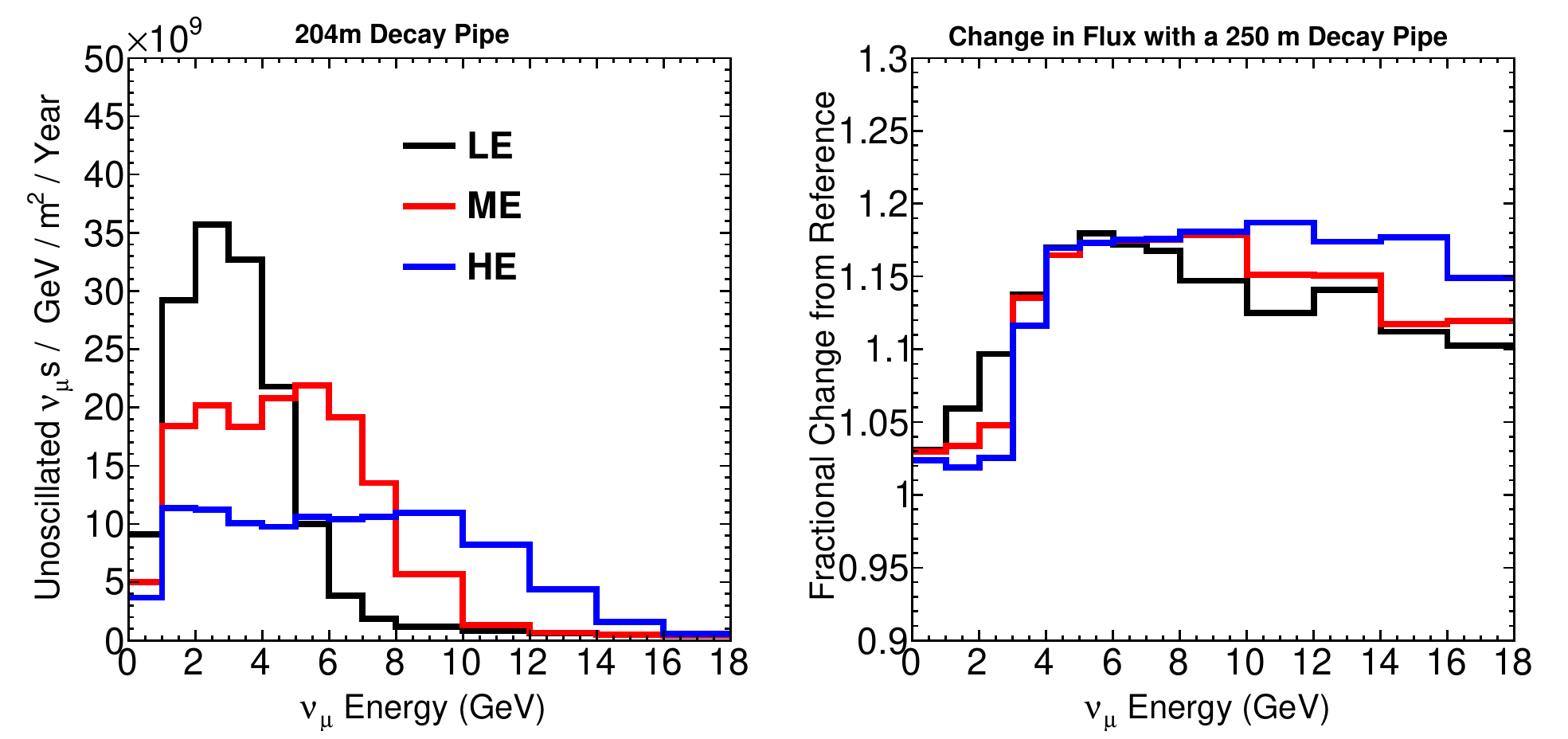}
  \end{cdrfigure}
 
\subsection{Improved Beam Options}
\label{sec:alternative-focusing-systems}

There are several potential modifications to the reference beam design
that would improve the experiment's sensitivity to the CP-violating
phase and mass hierarchy.  One option, which will be referred to as the \emph{enhanced}
reference beam, 
is based on the NuMI focusing horn design but uses a
thinner and shorter cylindrical beryllium target positioned 25~cm
upstream of the first focusing horn. We consider two decay pipe
configurations for this enhanced reference beam, one 204 m long and 6~m 
in diameter and the other 250 m long and 4~m in diameter.  The
neutrino fluxes for these options, generated with an \GeVadj{80} 
primary proton beam, are shown in Figure~\ref{fig:beam_req_focusing_comp}.
  
The option offering the largest gains in sensitivity is a redesign of
the focusing system, including target and horns, which would require a
modification of the dimensions of the present target chase.  To
identify optimal designs, a genetic algorithm has been implemented to
search for beam configurations that maximize sensitivity to CP
violation.  The procedure is inspired by a similar one developed by
the LBNO collaboration~\cite{Agarwalla:2014tca}, and considers 20
beamline parameters governing the primary proton momentum, target
dimensions, and horn shapes, positions and
current. Figure~\ref{fig:beam_req_opthorn} shows the approximate shape
and the 12 shape parameters (5 radial and 7 longitudinal dimensions)
used for the optimization of the first focusing horn, while the second
focusing horn is modeled as a NuMI-style horn, but allowed to rescale
both in radial and longitudinal dimensions. The procedure yields horn
size and shapes similar to those found by the LBNO collaboration. The
first focusing horn is $\sim$5.5~m in length and $\sim$1.3~m in
diameter. This optimized beam configuration includes a second focusing
horn that is 32\% longer and 7.8~m further downstream than that of the
reference focusing design.  This option would require an increase both
in length, by $\sim$9~m, and in width, by $\sim$60~cm, of the target
chase in the reference design. The increase in target chase length can
be compensated by a reduction in decay pipe length, so that the total
length of the combined chase and decay pipe remains the same as in the
reference design.
Figure~\ref{fig:beam_req_focusing_comp} shows a comparison of fluxes
for the reference, enhanced reference and optimized beam
configurations. This optimized beam, with a decay pipe 195 m long and
4~m in diameter, produces a muon neutrino flux that is 20\% greater
than the nominal configuration at the first oscillation maximum
(between 1.5 and 4~GeV), 53\% greater at the second oscillation
maximum (between 0.5 and 1.5~GeV), and reduces the antineutrino
contamination of the beam.  Sensitivity to $\delta_{CP}$ and the mass
hierarchy as a function of the exposure for reference and alternative
beam options are shown in Figure~\ref{fig:beam_req_sensitivities}.
The optimized beam leads to improvements in sensitivity to both mass
hierarchy and CP violation. Refinement of the optimization procedure,
including verification of the results with alternate hadron production
models, and evaluation of the feasibility of the optimized designs are
in progress.

\begin{cdrfigure}[Neutrino fluxes with alternative beam designs]
{beam_req_focusing_comp}{Neutrino mode muon 
    neutrino fluxes for several beam designs, including the 
    reference, the enhanced reference, and the optimized beam
    described in section~\ref{sec:alternative-focusing-systems} (top). The total CC interaction rate per year at \SI{1300}{\km} from the optimized focusing design with a \SIadj{195}{\m} decay pipe, and the ratio to the reference beam design (bottom).  
    All beams use \GeVadj{80} protons.}
  \includegraphics[width=0.6\textwidth]{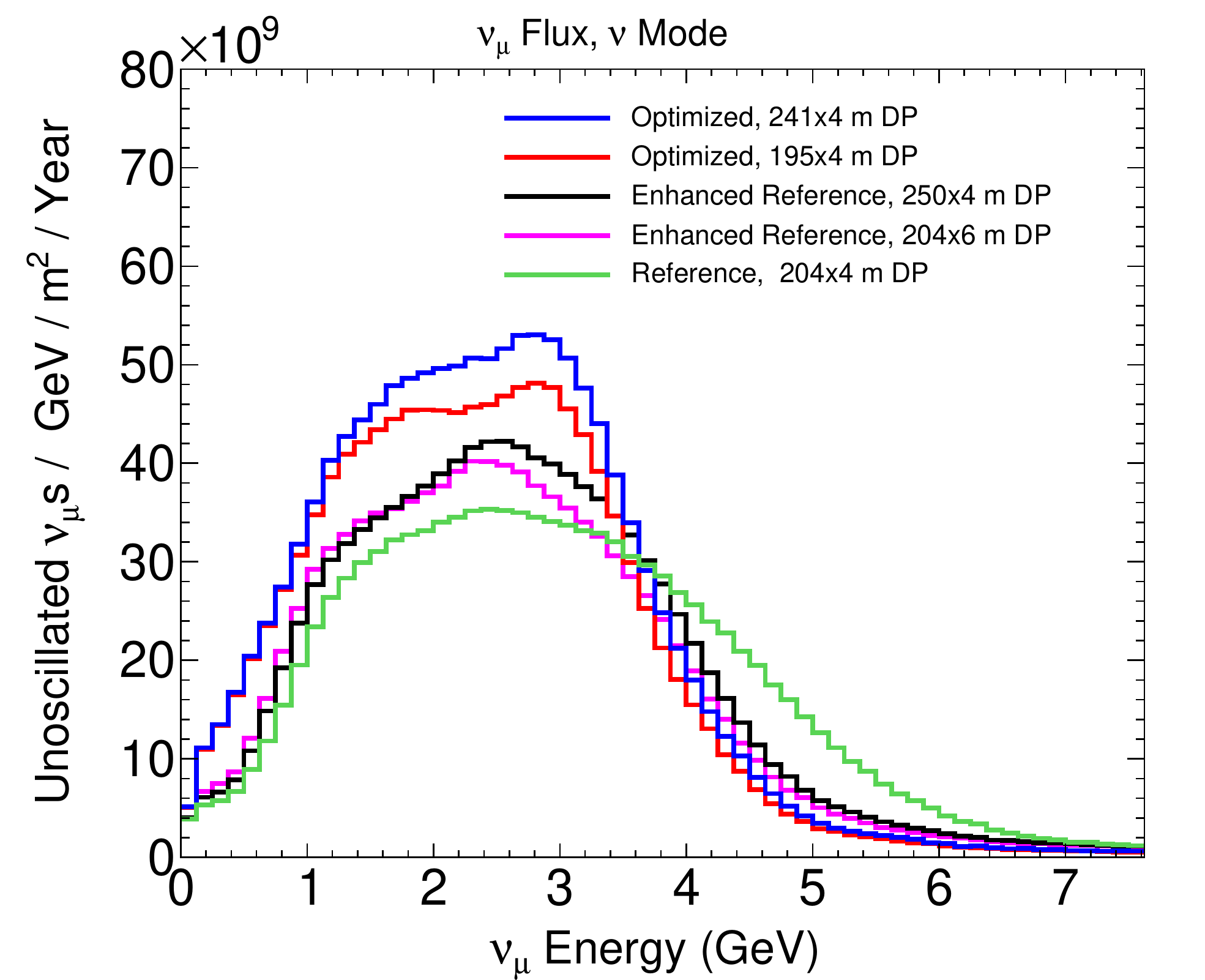}
  \includegraphics[width=0.6\textwidth]{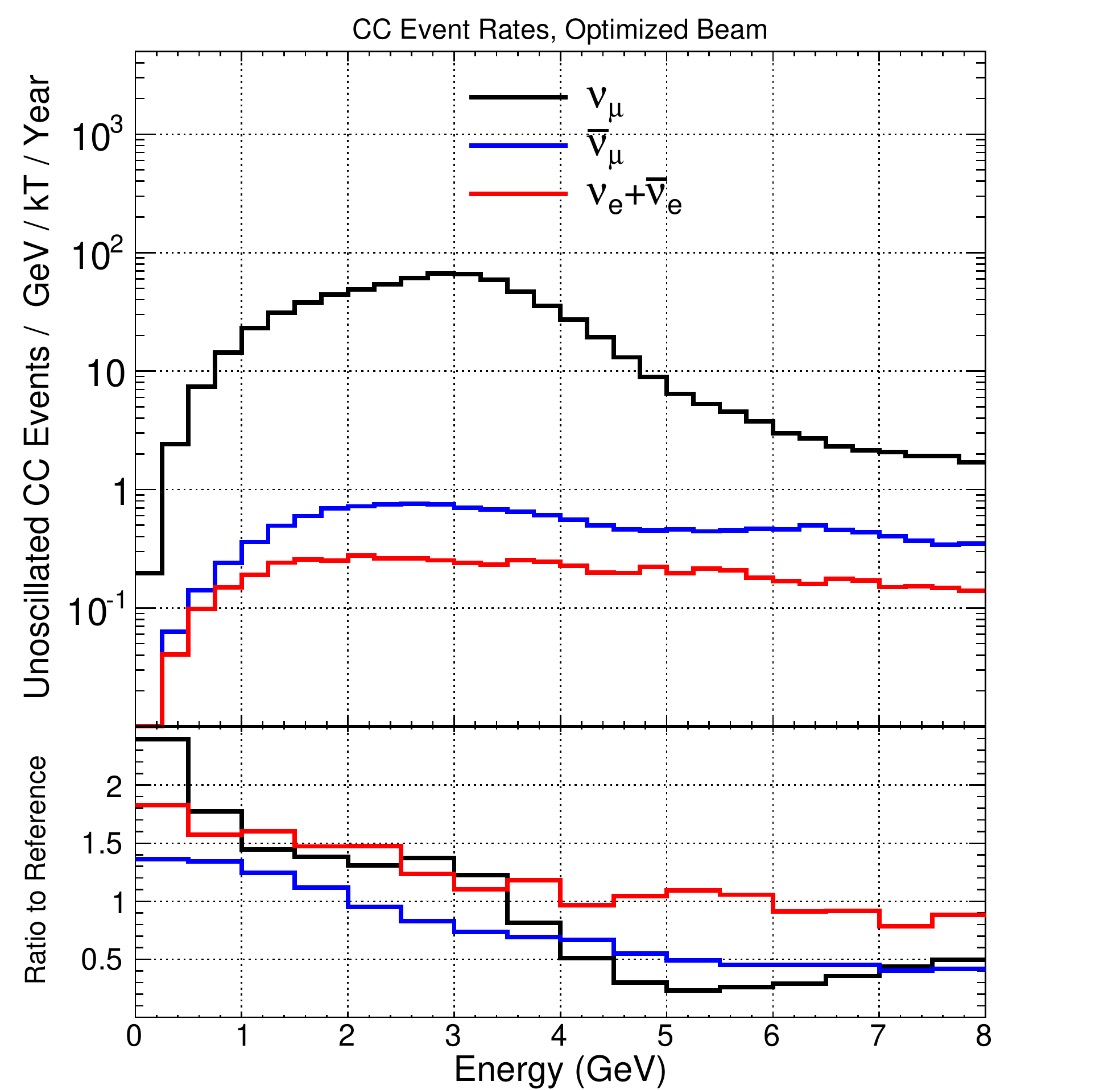}
\end{cdrfigure}

\begin{cdrfigure}[Radial view of the first horn shape considered in focusing system 
  optimization]{beam_req_opthorn}{First focusing horn design considered 
    in the alternative focusing optimization. 
}
  \includegraphics[width=0.75\textwidth]{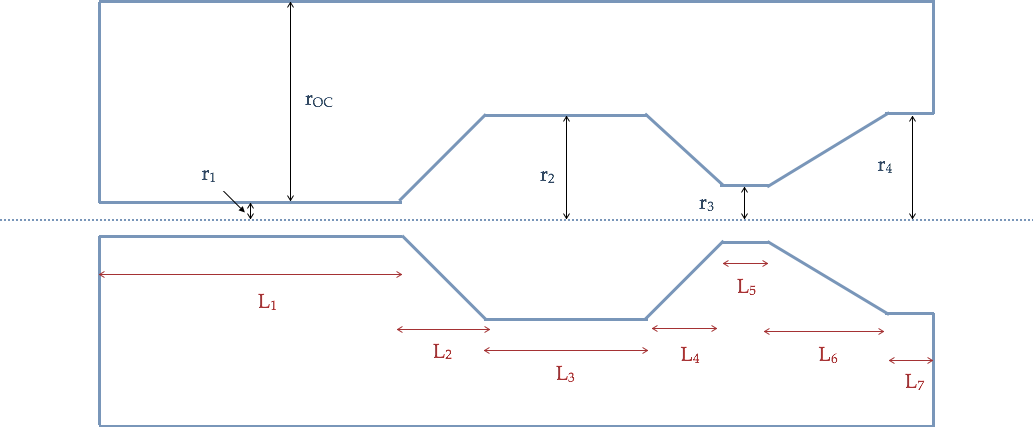}
\end{cdrfigure}

\begin{cdrfigure}[Sensitivities to the mass hierarchy and
  $\delta_{CP}$ for reference and alternative beam designs]
{beam_req_sensitivities}{Sensitivity to the mass hierarchy (left) and
  $\delta_{CP}$ (right) as a function of exposure for the reference beam and
  the beam options discussed in
  Section~\ref{sec:alternative-focusing-systems}.  All beams use \GeVadj{80} protons.}

\centering 
\begin{minipage}{0.5\textwidth}
\centering 
\includegraphics[width=1.0\textwidth]{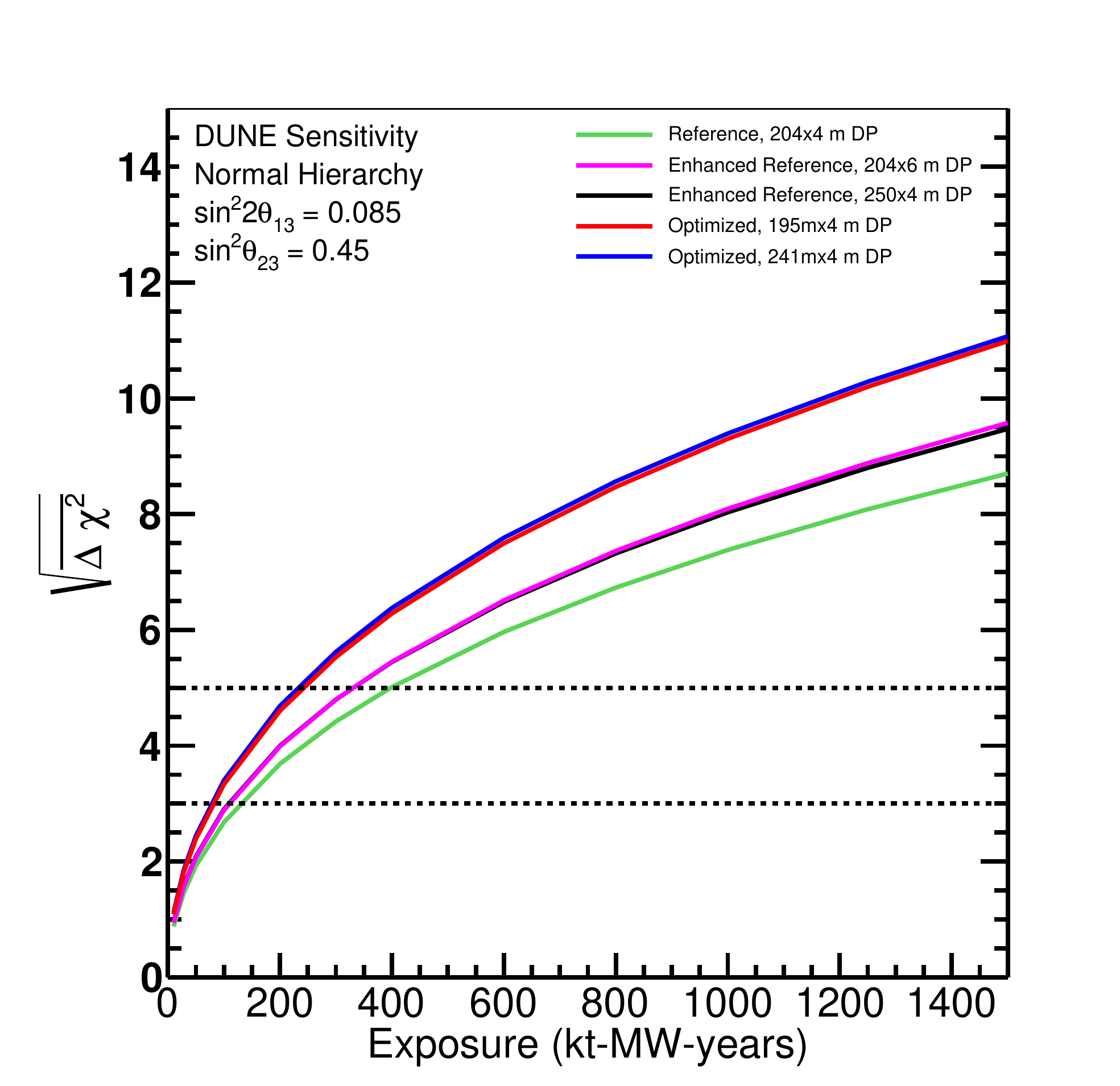}
\end{minipage}\hfill 
\begin{minipage}{0.5\textwidth}
\centering 
\includegraphics[width=1.0\textwidth]{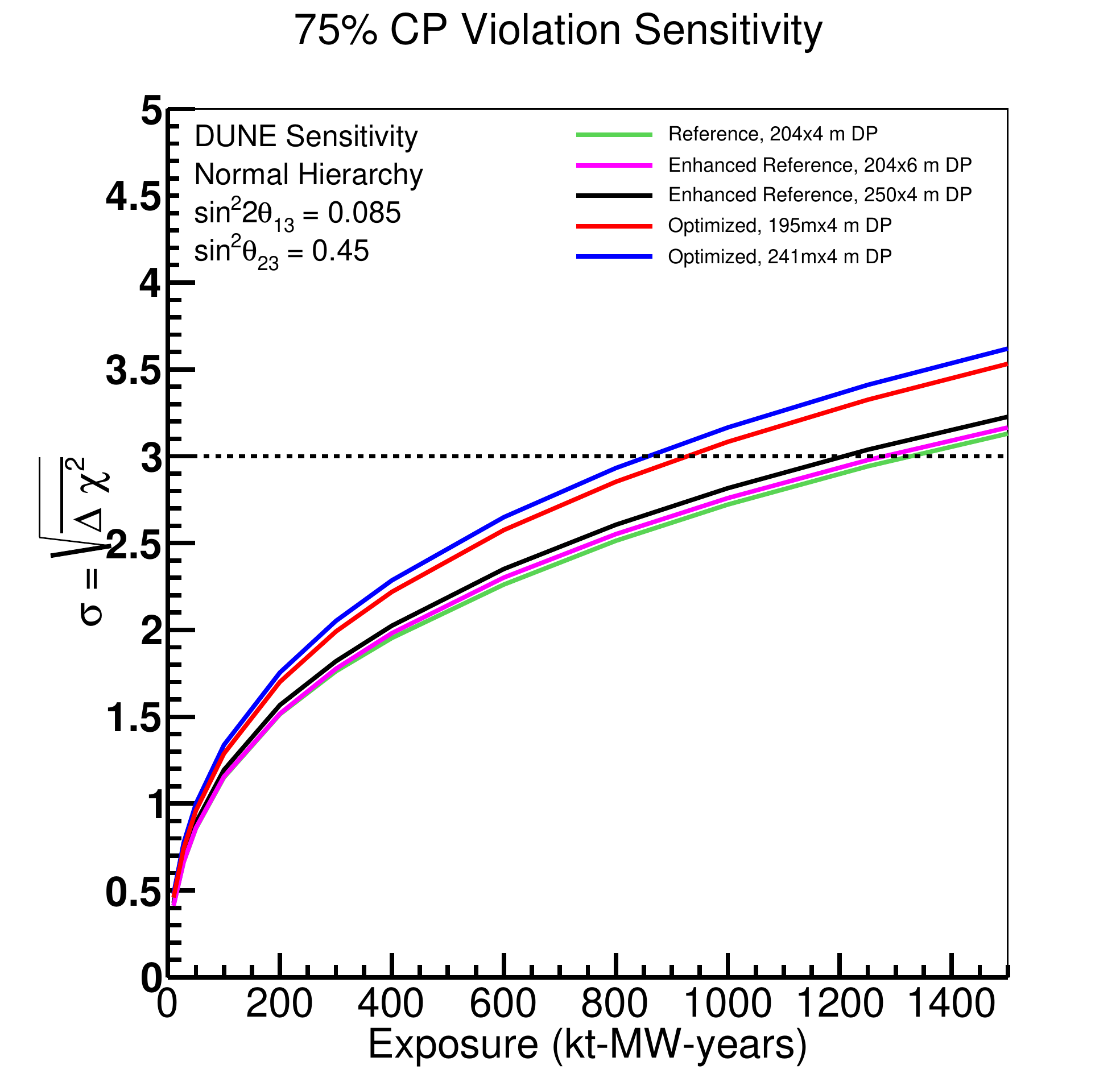}
\end{minipage} 
\end{cdrfigure}

\section{Testing the Three-Flavor Paradigm and the Standard Model}
\label{sec:physics-lbnosc-3nutests}

Due to the very small masses and large mixing of neutrinos, their oscillations over a long distance
  act as an exquisitely precise interferometer with high sensitivity to very small perturbations caused by 
  new physics phenomena, such as:
  \begin{itemize}
  \item nonstandard interactions in matter that manifest in
    long-baseline oscillations as deviations from the three-flavor mixing model
  \item new long-distance potentials arising from discrete symmetries
    that manifest as small perturbations on neutrino and antineutrino
    oscillations over a long baseline
  \item sterile neutrino states that mix with the three known active neutrino states
  \item large compactified extra dimensions from String Theory models that manifest through mixing
    between the Kaluza-Klein states and the three active neutrino
    states
   \item Lorentz and CPT violation due to an underlying Planck-scale theory that manifest through sidereal dependence on the neutrino oscillation probability
  \end{itemize}
    Full exploitation of DUNE's sensitivity to such new phenomena
  will require high-precision predictions of the unoscillated
  neutrino flux at the far detector and large exposures. Studies will be conducted to understand the
limits that DUNE could impose relative to current limits and those expected from other experiments.
  
\subsection{Search for Nonstandard Interactions}

For $\nu_{\mu,e} \rightarrow \nu_{e,\mu}$ 
oscillations that occur as the neutrinos propagate through matter,  
the coherent forward scattering of $\nu_e$'s on electrons in matter 
modifies the energy and path-length dependence of the vacuum oscillation 
probability in a way that depends on the magnitude \emph{and} sign of $\Delta m^2_{31}$. 
This is  the Mikheyev-Smirnov-Wolfenstein (MSW) effect~\cite{Mikheev:1986gs,Wolfenstein:1977ue}.
NC nonstandard interactions (NSI) may be interpreted as nonstandard
matter effects that are visible only in a far detector at a
sufficiently long baseline. 
They can be parameterized as new contributions
to the MSW matrix in the neutrino-propagation Hamiltonian~\cite{Valle:1987gv,Roulet:1991sm}:

\begin{equation}
  H = U \left( \begin{array}{ccc}
           0 &                    & \\
             & \Delta m_{21}^2/2E & \\
             &                    & \Delta m_{31}^2/2E
         \end{array} \right) U^\dag + \tilde{V}_{\rm MSW} \,,
\end{equation}
with
\begin{equation}
  \tilde{V}_{\rm MSW} = \sqrt{2} G_F N_e
\left(
  \begin{array}{ccc}
    1 + \epsilon^m_{ee}       & \epsilon^m_{e\mu}       & \epsilon^m_{e\tau}  \\
        \epsilon^{m*}_{e\mu}  & \epsilon^m_{\mu\mu}     & \epsilon^m_{\mu\tau} \\
        \epsilon^{m*}_{e\tau} & \epsilon^{m*}_{\mu\tau} & \epsilon^m_{\tau\tau}
  \end{array} 
\right)
\end{equation}

Here, $U$ is the leptonic mixing matrix, and the $\epsilon$ parameters give the
magnitude of the NSI relative to standard weak interactions.  For new physics
scales of a few hundred GeV,  a value of $|\epsilon| \leq 0.01$ is
expected~\cite{Davidson:2003ha,GonzalezGarcia:2007ib,Biggio:2009nt,Barranco:2007ej,Escrihuela:2011cf}.
DUNE's \kmadj{1300} baseline provides an advantage in the detection of NSI relative
to existing beam-based experiments with shorter baselines.
Only atmospheric-neutrino experiments have longer baselines, but the sensitivity
of these experiments to NSI is limited by systematic effects. See \cite{Adams:2013qkq}
for potential sensitivities to these parameters at a \kmadj{1300} baseline.

\subsection{Search for Long-Range Interactions}

The small scale of neutrino-mass differences implies that minute
differences in the interactions of neutrinos and antineutrinos with
currently unknown particles or forces may be detected through 
perturbations to the time evolution of the flavor eigenstates.  
The longer the experimental
baseline, the higher the sensitivity to a new long-distance potential
acting on neutrinos. For example, some of the models for such
long-range interactions (LRI) as described in~\cite{Davoudiasl:2011sz} could contain discrete symmetries that
stabilize the proton and give rise to a dark-matter candidate particle,
thus providing new
connections between neutrino, proton decay and dark matter
experiments. The longer baseline of DUNE improves the sensitivity to
LRI beyond that possible with the current generation of long-baseline
neutrino experiments. The sensitivity will be determined by the amount
of $\nu_\mu/\bar{\nu}_\mu$-CC statistics accumulated and the accuracy
with which the unoscillated and oscillated $\nu_\mu$ spectra can be
determined.

\subsection{Search for Mixing between Active and Sterile Neutrinos}

Several recent anomalous experimental results count among their possible
interpretations phenomena that do not fit in the three-flavor mixing
model~\cite{Aguilar:2001ty,AguilarArevalo:2007it,Aguilar-Arevalo:2013pmq,Mention:2011rk,Giunti:2010zu}, 
and searches for evidence of one or more sterile neutrino states are ongoing.

At DUNE, searches for active-sterile neutrino mixing can be
conducted by examining the NC event rate at the far detector and
comparing it to a precise estimate of the expected rate extrapolated
from $\nu_\mu$ flux measurements from the  
near detector and from 
beam and detector simulations. Observed deficits in the NC rate could be evidence for mixing between the
active neutrino states and unknown sterile neutrino states. The most recent such search
in a long-baseline experiment was conducted by the MINOS
experiment~\cite{Sousa:2015bxa,Adamson:2011ku}.  DUNE will provide a unique
opportunity to revisit this search over a large
range of neutrino energies and a longer baseline. The large detector mass and high beam power will
allow a high-statistics sample of NC interactions to be collected.
The high-resolution LArTPC far detector
will enable a coarse measurement of the incoming neutrino
energy in a NC interaction by using the event topology and correcting
for the missing energy of the invisible neutrino.  Both the
energy spectrum and the rate of NC interactions will be measured
with high precision at both near and far detectors.

Long-baseline experiments are sensitive to sterile neutrinos not only
through NC measurements but also via appearance searches.  In
particular, as recently shown in~\cite{Klop:2014ima}, long-baseline
appearance searches are the sole ones that are sensitive to the
additional CP-violating phases that appear in a framework with more
than three neutrinos.  In the event of a discovery of sterile neutrino
states, long-baseline experiments --- and DUNE in particular ---  will have a
unique role in completing the picture of the new enlarged framework,
because short-baseline experiments have almost no sensitivity to these
additional CP-violating phases.  It has been shown that T2K already
provides some information on one such phase~\cite{Klop:2014ima}, 
 and the sensitivity has been studied
for LBNE~\cite{Hollander:2014iha}.

\subsection{Search for Large Extra Dimensions}

Several theoretical models propose that right-handed neutrinos
propagate in large compactified extra dimensions, whereas the standard
left-handed neutrinos are confined to the four-dimensional
brane~\cite{Machado:2011wx}. Mixing between the right-handed
\emph{Kaluza-Klein} modes and the standard neutrinos would change the
mixing patterns predicted by the three-flavor model. The effects could
manifest, for example, as distortions in the disappearance spectrum of
$\nu_\mu$.  The rich oscillation structure visible in DUNE, measured
with its high-resolution detector using both beam and atmospheric
oscillations, could provide further opportunities to probe for this
type of new physics.

\subsection{Search for Lorentz and CPT Violation}

Lorentz invariance and its associated CPT symmetry are foundational
aspects of the Standard Model. However, the Standard Model is thought
to be a low-energy limit of a more fundamental theory that unifies
quantum physics with gravity at the Planck scale. As a result, an
underlying theory can induce violations of Lorentz invariance and
violations of CPT symmetry that can generate experimentally observable
signals of Planck-scale physics.  The Standard-Model Extension
(SME)~\cite{Colladay:1996iz,Colladay:1998fq,Kostelecky:2003fs} is an
effective field theory that contains the Standard Model, general
relativity, and all possible operators that break Lorentz
symmetry. (Since CPT violation implies Lorentz violation, the SME
necessarily includes operators that break CPT symmetry.)  Within the
SME framework, the probability for neutrino oscillations depends on
the direction of neutrino propagation within a Sun-centered inertial
frame~\cite{Kostelecky:2003cr,Kostelecky:2011gq}.  For a long-baseline
experiment with both the neutrino beam source and detector fixed to
the surface of the Earth, the Earth's rotation causes the direction of
neutrino propagation to change with sidereal time.  The SME theory
thus predicts a sidereal dependence of the observed beam neutrino
rate.  DUNE has the potential to perform studies that explore regimes
never previously investigated and to improve existing sensitivities
obtained in other neutrino experiments.  For example, the baseline of
1300~km offers an advantage because the sensitivity to
Lorentz and CPT violation grows linearly with the baseline.  The beam
orientation for DUNE is different from other experiments such as MINOS
or T2K, so the combinations of coefficients for Lorentz and CPT
violation appearing in the DUNE mixing probabilities are
distinct. Additionally, the wide range of energy for the beam
neutrinos and the ability to investigate both neutrino and
antineutrino channels are advantageous.

\section{Experimental Requirements}
\label{sec:physics-lbnosc-det-req}

The technical designs of LBNF and the DUNE detectors must fulfill the
scientific objectives described in Chapter~\ref{ch:physics-goals}. The
following is a summary of the high-level scientific requirements for
the neutrino oscillation physics. Details of the scientific and technical 
requirements for LBNF/DUNE deriving from the scientific objectives can be found in reference~\cite{lbnfdune-cdr-req}.

\subsection{Neutrino Beam Requirements}
\label{sec:physics-lbnosc-beam-req}

LBNF must be designed for approximately twenty years of
operation, in order to provide adequate exposure for the DUNE
experiment. During its lifetime, the facility must be able to
accommodate various target and focusing configurations to enable
tuning of the neutrino energy spectrum, and must be suitable for
upgraded targets and horns as technology improves and the primary
proton beam power increases. Such flexibility is an essential
requirement for a facility that will operate over multiple
decades. The energy range of the neutrino beam must be adaptable, in
order to address new questions in neutrino physics that may come up
during such a long period.

The DUNE experiment requires that the LBNF facility provides a
neutrino beamline and the conventional facilities to support it. The
global science requirements on the LBNF beamline are as follows:

\begin{itemize} 
\item The neutrino beam spectrum shall cover the energy
  region of the first two oscillation maxima affected by muon-neutrino
  conversion from the atmospheric parameters. For a baseline of \SI{1300}{\km}, 
  with the current knowledge of parameters, the first two nodes
  are expected to be approximately 2.4 and 0.8 GeV. The matter effects
  dominate over the CP effects above 3 GeV and the CP effect dominates
  below 1.5 GeV. Adequate number of electron neutrino events with good
  energy resolution will allow DUNE to exploit this spectral
  information to determine mass hierarchy, CP phase, and a precise
  value of $\theta_{13}$ unambiguously. The beam spectrum will also
  allow muon disappearance measurement with two nodes.

\item The beam shall be sign-selected to provide separate neutrino and
  antineutrino beams with high purity to enable measurement of CP
  violation mass hierarchy, and precision oscillation measurement.

\item The electron neutrino content in the beam shall be kept small so
  that the systematic errors on the additional background have a small
  impact on the CP phase measurement (compared to the statistical
  error).

\item The neutrino beam spectrum shall extend beyond the first maximum
  to higher energies, while maintaining a high signal-to-background
  ratio to obtain the maximum number of charged-current signal events.
  This will allow precision probes of the PMNS parameters that govern
  neutrino oscillations.

\item The beam shall be aimed at the far detector with an angular
  accuracy that allows the determination of the far detector spectrum
  using the near detector measurements.  The angular accuracy shall
  not be the dominant factor in the determination of oscillation
  parameters.
  
\item The beam shall be capable of operating with a single-turn,
  fast-extracted primary proton beam from the Main Injector with
  greater than 2~MW of power. The fast extraction enables short spills
  which are essential for good cosmic ray background rejection for
  detectors. 

\item The beamline shall be able to accept a range of Main Injector
  proton energies that is well matched to the oscillation physics
  requirements. Proton beam energies of around \SI{60}\GeV{} are
  optimal for measurement simultaneous measurements of CP violation
  and MH, while higher energy beams (the maximum possible from the
  Main Injector is \SI{150}\GeV) can probe physics beyond the 3-flavor
  mixing, and probe $\nu_\tau$ appearance with higher statistics. The
  power and protons-on-target available from the MI as a function of
  proton beam energy is summarized in Table~\ref{tab:beam_req_pot}.
 
\end{itemize}

\subsection{Far Detector Requirements}
\label{sec:physics-lbnosc-fd-req}

The DUNE Far Detector (FD) requirements relevant to long-baseline
neutrino oscillation physics include:
\begin{itemize}
\item Identification of Electron Neutrino and Antineutrino Events: The
  FD shall be capable of identifying electron neutrino and
  antineutrino charged current beam events in sufficient numbers 
  within the fiducial volume of the detector to enable precision
  measurements of the parameters that govern $\nu_\mu \rightarrow
  \nu_e$ oscillations. The neutrino flavor of the event will be
  identified by clearly identifying the primary final state charged
  electron. The total energy of the charged current event shall be
  measured.
\item Muon Neutrino and Antineutrino Events: The FD shall be capable
  of identifying muon neutrino and antineutrino charged current beam
  events in sufficient numbers 
within the fiducial volume of the
  detector and identify the primary muon particle emerging from the
  main event vertex. 
   The total energy of the charged current event
  shall be measured.
\item Multiple tracks and Electromagnetic Showers: The FD shall be
  capable of identifying events with multiple electromagnetic showers
  and non-showering particles produced within the fiducial volume of
  the detector.
\item Baseline Length: A baseline of sufficient length shall be
  established between the neutrino beam facility and a far detector
  facility so that the difference between muon to electron neutrino
  conversion for the two cases of neutrino mass ordering can be
  clearly separated from the variation due to the CP phase, leading to
  unique determination of the CP phase.
\item Cosmic Ray Shielding: The FD shall be located at a depth to
  reduce the number of in-time (within the beam spill time) cosmic ray
  background so that it does not contribute more than 1\% of the final
  beam neutrino sample.
 \item CP Phase Measurement: The total number of observed
   electron-neutrino and electron-antineutrino type events --
   including consideration of background -- shall be sufficient to
   measure the CP phase to better than 3$\sigma$ at the maximum CP
   violation.
 \item Time Accuracy: Individual event times shall be measured with
   sufficient time accuracy to allow correlation of event times
   between detectors that are geographically separated. In the case of
   long-baseline oscillations, this would include correlation between
   the DUNE near and far detectors.
\end{itemize}

\subsection{Near Detector Requirements}
\label{sec:physics-lbnosc-nd-req}

The DUNE Near Detector (ND) requirements relevant to long-baseline
neutrino oscillation physics include:
\begin{itemize}
\item FD measurements not limited by ND: ND measurements shall be of
  sufficient precision to ensure that when extrapolated to FD to
  predict the FD event spectra without oscillations, the associated
  systematic error must be significantly less than the statistical
  error over the lifetime of the experiment.
\item Muon Neutrino and Antineutrino Flux measurements: The ND shall
  measure the absolute and relative muon neutrino and antineutrino
  spectra separately.  See Sections~\ref{sec:syst_just_flux}
  \ref{sec:syst_studies_ind} for discussions of the required flux
  uncertainty and current studies.
\item Electron Neutrino and Antineutrino Flux measurements: The ND
  shall measure the electron-neutrino and antineutrino contamination
  spectra of the beam separately in order to render the CP measurement
  as precise as possible.
 \item Background Measurements: The ND shall measure rates, kinematic
   distributions and detailed topologies of physics processes that
   could mimic signal events in the FD nuclear targets. This
   measurement shall be made with sufficient resolution to allow FD
   background calculation with precision that does not limit the
   oscillation measurements.
 \item Cross section measurements I: The ND will measure CC and NC
   differential cross sections separately as a function of energy.
 \item Cross section measurements II: The ND shall characterize
   various exclusive and semi-exclusive processes such as
   quasi-elastic interactions, resonance production, deep inelastic
   scattering, and neutrino-electron and neutrino-proton elastic
   scattering.
 \item Cross section measurements III: The ND shall measure the
   neutrino nucleus cross section off various targets like Hydrogen,
   Ar, Fe, Ca, and C.
\end{itemize}

\cleardoublepage

\chapter{Nucleon Decay and Atmospheric Neutrinos}
\label{ch:physics-atmpdk}

\section{Nucleon Decay}
\label{sec:physics-atmpdk-ndk}

\subsection{Physics Motivation}
Grand Unified Theories (GUTs)  unite the three gauge interactions of particle 
physics -- strong, weak, and electromagnetic -- into one single force, and as a 
consequence, make predictions about  baryon number violation and proton lifetime that may be within reach of DUNE.  
The theoretical motivation for the study of proton decay has a long and
distinguished history~\cite{Pati:1973rp,Georgi:1974sy,Dimopoulos:1981dw} and
has been reviewed many times~\cite{Langacker:1980js,deBoer:1994dg,Nath:2006ut}.
Early GUTs provided the original motivation for proton-decay searches in
kiloton-scale detectors placed deep underground to limit backgrounds~\cite{homestake:depth}.  The
\ktadj{22.5} \superk\ experiment extended the search for proton decay by more
than an order of magnitude relative to the previous generation of experiments.
Contemporary reviews~\cite{Raby:2008pd,Senjanovic:2009kr,Li:2010dp} discuss the
strict limits already set by \superk\ and the context of the proposed next
generation of larger underground
experiments such as Hyper-Kamiokande and DUNE.

Although no evidence for proton decay has been detected, lifetime
limits from the current generation of experiments already constrain
the construction of many contemporary GUT models. 
In some cases, these limits are approaching the upper bounds of what these models will allow.
This situation points naturally toward continuing 
the search with new, highly capable underground detectors, especially those 
with improved sensitivity to specific proton decay modes favored by GUT models.
In particular, the exquisite imaging, particle identification and calorimetric 
response of the DUNE LArTPC Far Detector opens the possibility of obtaining 
evidence for nucleon decay on the basis of a single well reconstructed event.

\subsection{Proton Decay Modes} 

\tikzset{
particle/.style={draw=blue, postaction={decorate},
    decoration={markings,mark=at position .6 with {\arrow[blue]{triangle
45}}}},
noarrow/.style={draw=blue},
boson/.style={draw=blue,dashed},
}

\begin{figure}[!htb]
  \begin{center}
\begin{tikzpicture}
  \coordinate (cross);
  \coordinate[above left=of cross, label={[yshift=-6mm]$d$}](d);
  \coordinate[below left=of cross, label={[yshift=1mm]$u$}](u);  
  \coordinate[above right=of cross](tau);
  \coordinate[below right=of cross](tee);
  \coordinate[right=of tau, label=below:$\overline{\nu}_\tau$] (vtau);
  \coordinate[right=of tee, label=above:$\overline{s}$] (sbar);
  \coordinate[below=of u, label=above:$u$](u2);
  \coordinate[below=of sbar, label=above:$u$](u3);

  \draw[particle] (d) -- (cross);
  \draw[particle] (u) -- (cross);
  \draw[noarrow] (cross) -- node[label=above left:$\widetilde{\tau}$] {} (tau);
  \draw[noarrow] (cross) -- node[label=below left:$\widetilde{t}$] {} (tee);
  \draw[boson] (tau) -- node[label=right:$\widetilde{W}$] {} (tee);
  \draw[particle] (vtau) -- (tau);
  \draw[particle] (sbar) -- (tee);
  \draw[particle] (u2) -- (u3);

  \draw [decorate,decoration={brace,mirror,raise=3mm}] (d) -- (u2) node
[black,midway,xshift=-8mm] {$P$};
  \draw [decorate,decoration={brace,raise=3mm}] (sbar) -- (u3) node
[black,midway,xshift=8mm] {$K^+$};

\end{tikzpicture}%
\begin{tikzpicture}
  \coordinate (x);
  \coordinate[above left=of x, label={[yshift=-5mm]$u$}](u);
  \coordinate[below left=of x, label={[yshift=1mm]$u$}](u2);  
  \coordinate[right=of x](x2);  
  \coordinate[above right=of x2, label={[yshift=-6mm]$e^+$}](eplus);  
  \coordinate[below right=of x2, label={[yshift=1mm]$\overline{d}$}](dbar);
  \coordinate[below=of u2,label=above:$d$](d);  
  \coordinate[below=of dbar, label=above:$d$](d2);  

  \draw[particle] (u) -- (x);
  \draw[particle] (u2) -- (x);
  \draw[boson] (x) -- node[label=above:$X$] {} (x2);
  \draw[particle] (eplus) -- (x2);
  \draw[particle] (dbar) -- (x2);
  \draw[particle] (d) -- (d2);

  \draw [decorate,decoration={brace,mirror,raise=3mm}] (u) -- (d) node
[black,midway,xshift=-8mm] {$P$};
  \draw [decorate,decoration={brace,raise=3mm}] (dbar) -- (d2) node
[black,midway,xshift=8mm] {$\pi^0$};

\end{tikzpicture}
  \end{center}
\caption[Proton decay modes from SUSY and gauge-mediation models]{Feynman
diagrams for proton decay modes from
supersymmetric GUT, $p^+ \rightarrow K^+ \overline{\nu}$  (left) and
gauge-mediation GUT models, $p^+ \rightarrow e^+ \pi^0$ (right).}
\label{fig:pdk_feyn}
\end{figure}
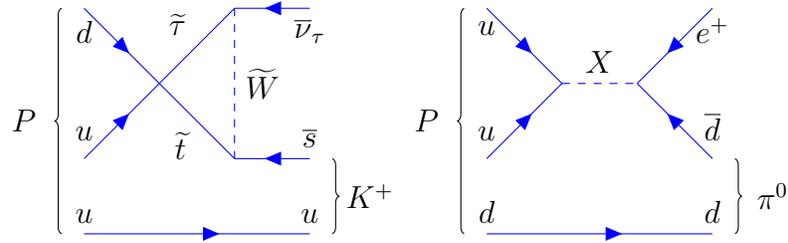
The strength of the DUNE experiment is particularly evident in its capabilities to detect
two prominent decay modes, shown in Figure~\ref{fig:pdk_feyn}. 
The decay $p \rightarrow e^+ \pi^0$ arises from gauge mediation and 
is often predicted to have the higher branching fraction.  In this mode,
the total mass of the proton is converted into the electromagnetic
shower energy of the positron and two photons from $\pi^0$ decay,
with a net momentum vector near zero.  
This channel is demonstrably the more straightforward 
experimental signature for a water Cherenkov detector. 

The second key mode is $p \rightarrow K^+ \overline{\nu}$.  This mode is
dominant in most supersymmetric GUTs, many of which also favor other modes
involving kaons in the final state.  
Among the modes with a \emph{charged} kaon in the final state, 
$p \rightarrow K^+ \overline{\nu}$ is
uniquely interesting for DUNE: since stopping kaons have a higher ionization density
than lower-mass particles, an LArTPC could identify the $K^+$ track with 
high efficiency.  In addition, many final states of $K^+$ decay would be 
fully reconstructible in an LArTPC.

Although significant attention will be focused on the above benchmark 
modes, the nucleon decay program at DUNE will be a broad effort.
Many other allowed modes of proton or bound neutron into 
antilepton plus meson that also conserve $B-L$ have been identified.  
And other modes that conserve $B+L$, or that decay only into leptons, have been 
hypothesized.  
In addition to nucleon decay, another promising way of probing baryon number violation in DUNE is through the search for the spontaneous conversion of neutrons into antineutrons in the nuclear environment. 
While these are less well motivated theoretically, opportunistic 
experimental searches cost little and could have a large payoff.

Figure~\ref{fig:PDK-limits-theory} shows a comparison of experimental limits on key decay modes to the
ranges of lifetimes predicted by an assortment of GUTs. The limits are dominated by recent results from \superk.
\begin{cdrfigure}[Current \& projected nucleon decay lifetime limits 
                  compared with GUT-predicted ranges]
                 {PDK-limits-theory}
                 {Current nucleon decay lifetime 
                  limits~\cite{Beringer:1900zz,Nishino:2012ipa} (90\% C.L.)
                  compared with ranges predicted by 
                  Grand Unified Theories. The upper section is for
                  $p \rightarrow e^+ \pi^0$, most commonly caused by 
                  gauge mediation.  The lower section is for SUSY-motivated 
                  models, which commonly predict decay modes with kaons 
                  in the final state.  Marker symbols other than stars 
                  indicate published experimental limits, 
                  as labeled by the 
                  colors on top of the figure.  The stars represent projected 
                  limits for several recently proposed future experiments,   
                  calculated based on Poisson statistics including background, 
                  assuming that detected event yields equal the expected 
                  background.}
\includegraphics[width=\textwidth]{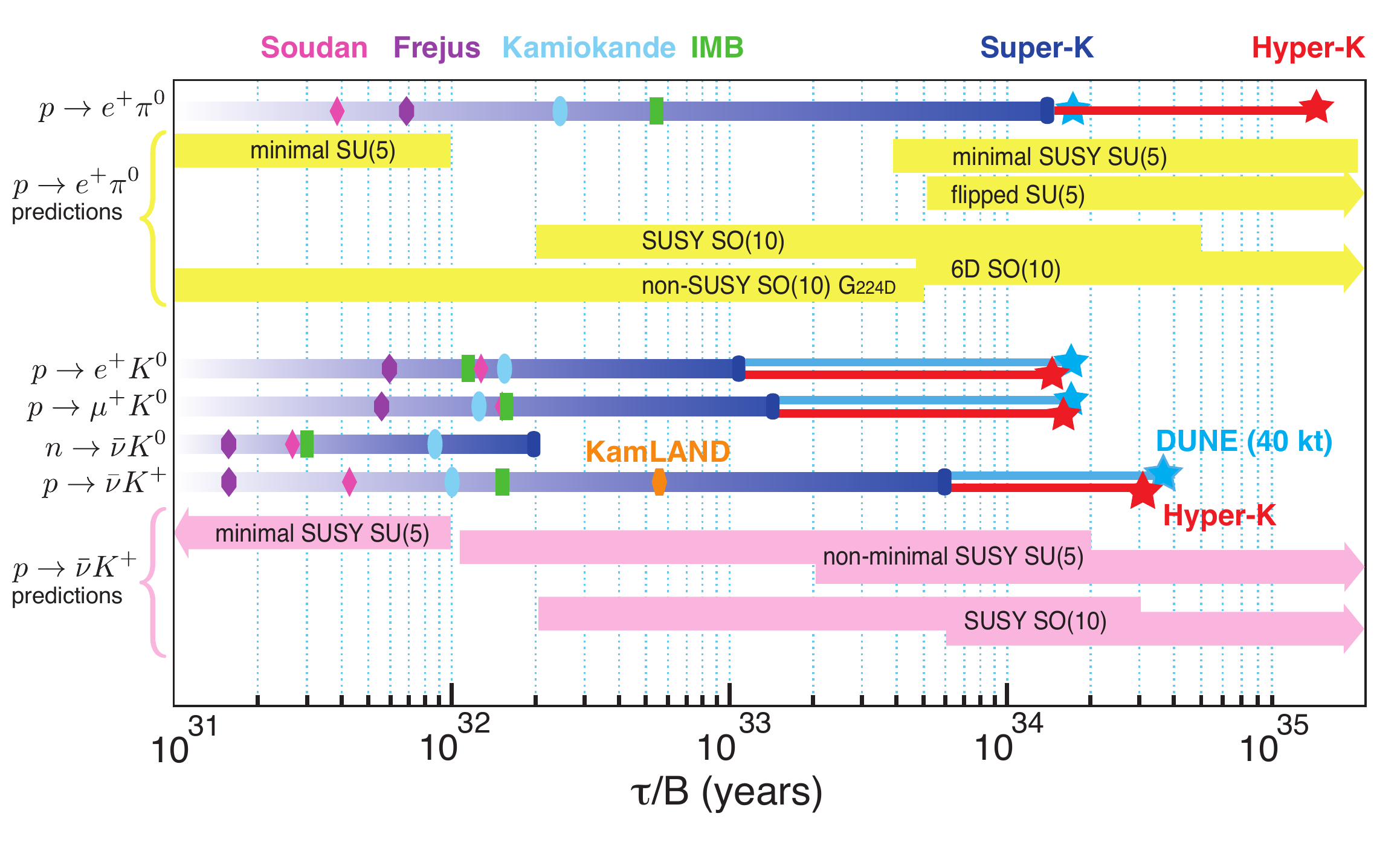}
\end{cdrfigure}
From this figure it is clear that an experiment such as DUNE with sensitivity to proton lifetimes
between $10^{33}$ and $10^{35}$ years will probe a large number of GUT models, and 
thus will present a compelling opportunity for discovery.
Even if no proton decay is detected, stringent lifetime limits will constrain 
the models: minimal SU(5) was ruled out by the early work of IMB and
Kamiokande, and minimal SUSY~SU(5) is considered to be ruled out by \superk.
In most cases, another order of magnitude in improved limits will not rule out
specific models but will constrain their allowed parameters;
this could allow identification of less favored models that 
would require fine-tuning in order to accommodate the data. 

It is also clear from Figure~\ref{fig:PDK-limits-theory} that it will not be easy for 
a LArTPC-based detector to make significant inroads on the $p \rightarrow e^+ \pi^0$ 
channel, where background-free high-efficiency searches are possible with large 
water Cherenkov detectors at a lower cost per kt.  For this reason, the 
focus of the remaining discussion is on the channels with kaons, in particular 
$p \rightarrow K^+ \overline{\nu}$.  However, it is important to note that 
the full-scale DUNE far detector would be able to provide confirming evidence 
for $p \rightarrow e^+ \pi^0$ should a signal for this channel start to develop 
in the next-generation water detector at the few-times-$10^{34}$-year level. 

\subsection{Signatures for Nucleon Decay in DUNE}

Extensive surveys~\cite{Bueno:2007um,Klinger:2015kva} of nucleon decay efficiency 
and background rates for large LArTPCs with various depth/overburden 
conditions provide the starting point for the 
assessment of DUNE's capabilities.  Table~\ref{tab:pdecay} lists selected
modes where LArTPC technology exhibits a significant performance 
advantage (per kt) over the water Cherenkov technology.
This section focuses on the capabilities 
of DUNE for the $p\to K^+\overline{\nu}$ channel, which is seen as the most 
promising from theoretical and experimental 
considerations.  Much of the discussion that follows can be 
applied to the other channels with kaons listed in 
the table.
\begin{table}[!htbp]
\caption[Efficiencies and background rates for nucleon decay modes]
        {Efficiencies and background rates (events per \SI{}{\Mtyr}) for nucleon decay 
         channels of interest for a large underground LArTPC~\cite{Bueno:2007um}, and 
         comparison with water Cherenkov detector capabilities.  
         The entries for the water Cherenkov capabilities are based 
         on experience with the \superk{} detector~\cite{kearns_isoups}.  
        }
\begin{center}
\begin{tabular}{$L^c^c^c^c} 
\toprule
\rowtitlestyle
Decay Mode   & \multicolumn{2}{^>{}c}{Water Cherenkov} & 
\multicolumn{2}{^>{}c}{Liquid Argon TPC} \\
\rowtitlestyle
   & Efficiency &   Background & Efficiency &   Background \\ \toprowrule
$p \rightarrow K^+ \overline{\nu}$       & 19\%  &  4   &  97\%   &     1  \\ \colhline
$p \rightarrow K^0 \mu^+$      & 10\%  &  8   &  47\%   &  $<2 $ \\ \colhline
$p \rightarrow K^+ \mu^- \pi^+$ &       &      &  97\%   &     1  \\ \colhline
$n \rightarrow K^+ e^- $        & 10\%  &  3   &  96\%   &  $<2$  \\ \colhline
$n \rightarrow e^+\pi^-$      & 19\%  &  2   &  44\%   &  0.8   \\
\bottomrule
\end{tabular}
\end{center}
\label{tab:pdecay}
\end{table}

The key signature for $p\to K^+\overline{\nu}$ is the presence of an
isolated charged kaon (which would also be monochromatic 
for the case of free protons, with $p=$\SI{340}{\MeV}$/c$).  
Unlike the case of $p\to e^+\pi^0$, where the maximum
detection efficiency is limited to 40--45\% because of inelastic
intranuclear scattering of the $\pi^0$, the kaon in $p\to
K^+\overline{\nu}$ emerges intact (because the kaon momentum is 
below threshold for inelastic reactions)
from the nuclear environment of the decaying proton $\sim 97\%$ of the
time.  Nuclear effects come into play in other ways, however: the kaon
momentum is smeared by the proton's Fermi motion and shifted downward
by re-scattering~\cite{Stefan:2008zi}.

In LArTPC detectors, the $K^+$ can be tracked, its momentum measured
by range, and its identity positively resolved via detailed analysis
of its energy-loss profile.  This is in sharp contrast with water 
detectors, in which the $K^+$ momentum is below Cherenkov threshold.
Additionally, all decay modes can be cleanly reconstructed 
and identified in an LArTPC, including those with neutrinos,
since the decaying proton is nearly at rest.  With this level of
detail, it is possible for a single event to provide overwhelming
evidence for the appearance of an isolated kaon of the right momentum
originating from a point within the fiducial volume.  The strength of
this signature is clear from cosmogenic-induced kaons observed by the
ICARUS Collaboration in the cosmic-ray (CR) test run of half of the T600
detector, performed at a surface installation in Pavia~\cite{Amerio:2004ze} 
and in high-energy neutrino interactions with the full T600 in the recent 
CNGS (CERN Neutrinos to Gran Sasso) run~\cite{Antonello:2012hu}.
Figure~\ref{fig:icaruskaon} shows a sample event from the CNGS run in
which the kaon is observed as a progressively heavily ionizing track 
that crosses into the active liquid argon volume, stops, and
decays to $\mu\nu$, producing a muon track that also stops and decays
such that the Michel-electron track is also visible. 
%
\begin{figure}[!htb]
\centering
\includegraphics[width=0.72\textwidth]{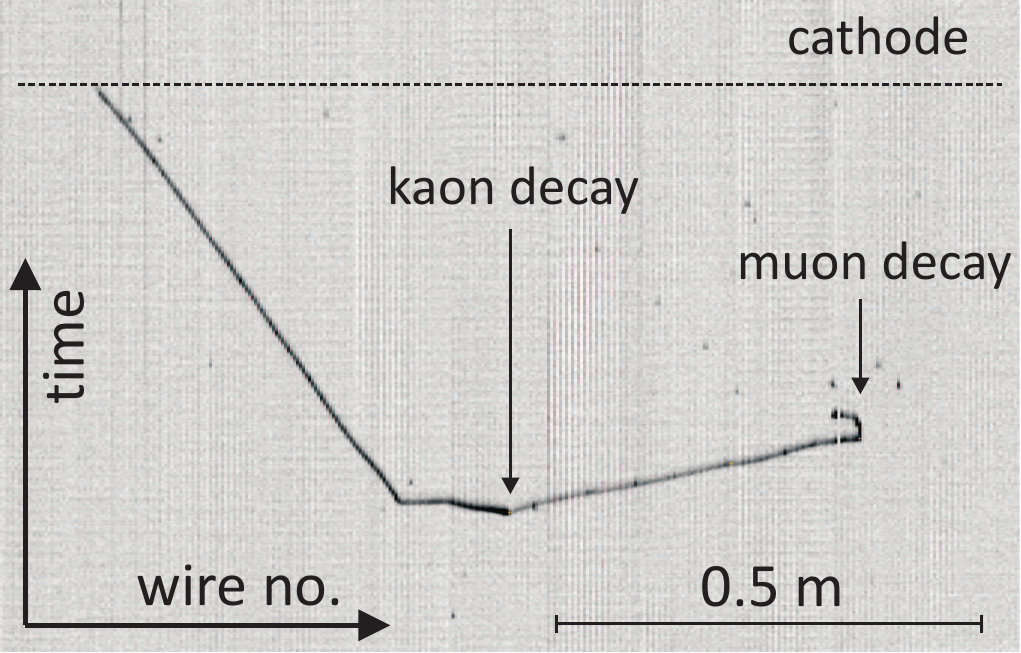}
\includegraphics[width=0.72\textwidth]{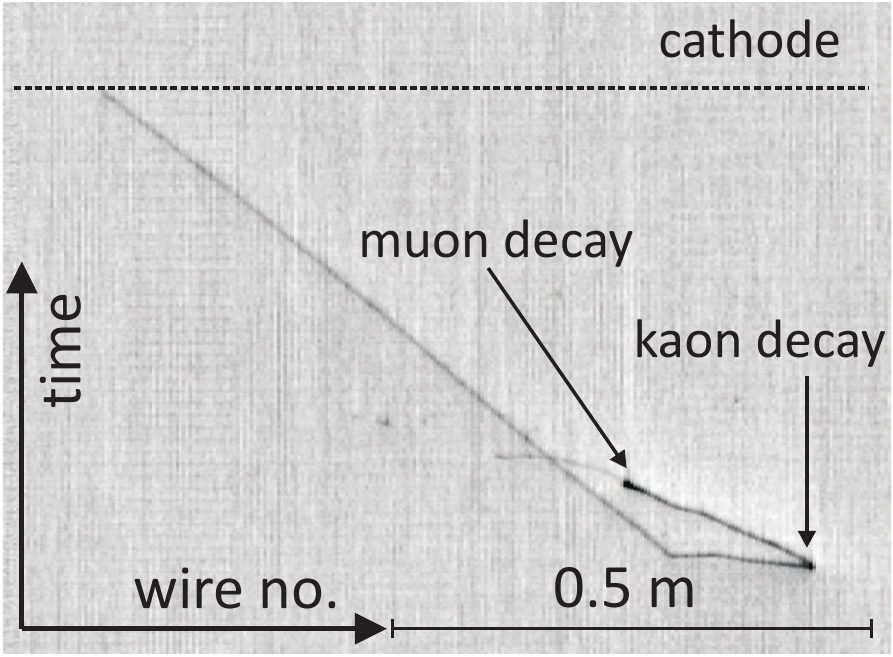}
\caption[Decaying kaon observed during the ICARUS run at CNGS]
{Event display for a decaying kaon candidate $K \rightarrow \mu \nu_\mu \ \mu \rightarrow e \nu_e \nu_\mu$ 
in the ICARUS T600 detector observed
in the CNGS data ($K$: \SI{90}{\cm}, \SI{325}{\MeV}; $\mu$ : \SI{54}{\cm}, \SI{147}{\MeV}; 
$e$ : \SI{13}{\cm}, \SI{27}{\MeV}). The top figure shows the signal on the collection plane,
  and the bottom figure shows the signal on the second induction plane~\cite{Antonello:2012hu}.}
\label{fig:icaruskaon}
\end{figure}

References~\cite{Adams:2013qkq,Klinger:2015kva,blake_doc8836} present detailed examinations 
of possible backgrounds, including those arising from cosmic ray interactions in the
detector and surrounding rock, atmospheric neutrino interactions in the
detector, and reconstruction failures. Table~\ref{tab:pdkbkds} summarizes the
results of those background studies.  All together, our estimate of total
background events in the 
$p\to K^+\overline{\nu}$ sample is less than 1 per \SI{}{\Mtyr}.


\begin{cdrtable}[Background summary for nucleon decay]{lll}{pdkbkds}
{Background sources and mitigation strategies for the $p\to K^+\overline{\nu}$ search in DUNE}
Background Source & Mitigation Strategy  \\ 
\toprowrule
Internal cosmic ray spallation      & Energy threshold \\ \colhline
External cosmogenic & & \\
$K^+$ production  & Depth, fiducialization \\ \colhline
External cosmogenic  & & \\
$K^0$ production & & \\
+internal charge-exchange  & & \\
to $K^+$ & Cuts on other secondaries    \\ \colhline
Atmospheric $\nu$ & & \\ 
$\Delta S=0$ processes & Cut on associated strange baryon \\ \colhline
Atmospheric $\nu$ & Cabibbo-suppressed, & \\ 
$\Delta S=1$ processes &lepton ID \\ \colhline
Atmospheric $\nu$ & $dE/dx$ discrimination, & \\
with $\pi$ mis-ID & 236 MeV muon track \\
\colhline
Reconstruction pathologies & $dE/dx$ profiles vs track length \\
\end{cdrtable}
%
\subsection{\boldmath Summary of Expected Sensitivity to Key Nucleon Decay Modes}

Based on the expected signal efficiency and upper limits on the
background rates, the expected limit on the proton
lifetime as a function of running time in DUNE for $p \rightarrow K^+
\overline{\nu}$ is shown in Figure~\ref{fig:kdklimit}. 
\begin{figure}[!htb]
\centering
\includegraphics[width=0.8\textwidth]{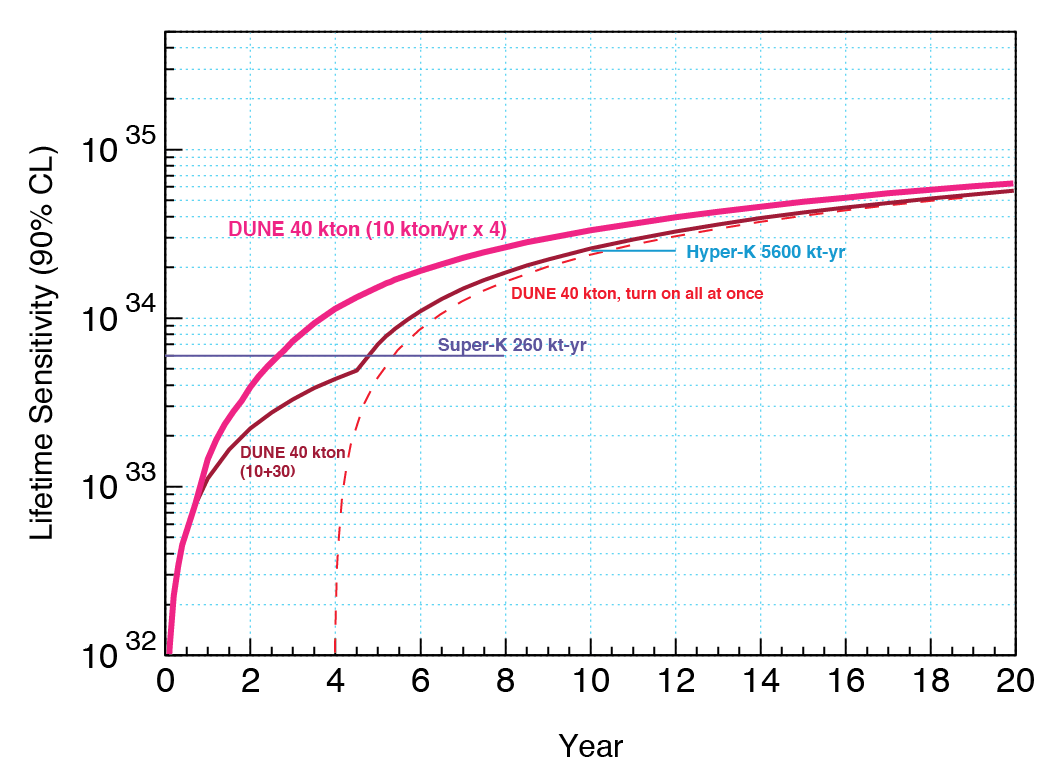}
\caption[Proton decay lifetime limit for $p \rightarrow K^+ \overline{\nu}$
  versus time]{Proton decay lifetime limit for $p
  \rightarrow K^+ \overline{\nu}$ as a function of time for
  underground LArTPCs starting with an initial 10 kt and adding another 10 kt
each year for four years, for a total of 40 kt. 
  For comparison, the current limit from SK and a projected limit from Hyper-K 
is also shown.
  The limits are at 90\% C.L., calculated for
  a Poisson process including background, assuming that the detected events
  equal the expected background.}
\label{fig:kdklimit}
\end{figure}
%


The current limits on
the $p \rightarrow \overline{\nu} K^+$ were set by \superk. This figure demonstrates that 
improving these limits significantly
beyond that experiment's sensitivity would require 
a LArTPC
detector of at least \SI{10}{kt}, installed deep underground. 
A \ktadj{40} detector will improve the current limits by an order of
magnitude after running for two decades.  Clearly a larger detector
mass would improve the limits even more in that span of time.

\section{Atmospheric Neutrinos}
\label{sec:physics-atmpdk-atmnu}

Atmospheric neutrinos 
provide a unique tool to study neutrino oscillations: the 
oscillated flux contains all flavors of neutrinos and antineutrinos, is very sensitive to 
matter effects and to both $\Delta m^2$ values, 
and covers a wide range of L/E. In principle, 
all oscillation parameters could be measured, with high complementarity to 
measurements performed with a neutrino beam. 
Atmospheric 
neutrinos are of course available all the time, 
which is particularly important before the beam becomes 
operational. 
They also provide a laboratory in which to search 
for exotic phenomena where the dependence of the flavor-transition and survival 
probabilities on energy and path length can be defined. The DUNE far detector, 
with its large mass and the overburden to protect it from backgrounds, is an 
ideal tool for these studies. The following discussion will focus on the 
measurement of the oscillation parameters in which the role of atmospheric neutrinos is 
most important. 

The sensitivity to oscillation parameters has been evaluated with a 
dedicated simulation, reconstruction and analysis chain. 
The fluxes of each neutrino species were computed at the far detector location, after 
oscillation. Interactions in the LAr medium were simulated with the GENIE event 
generator. Detection thresholds and energy resolutions based on full 
simulations were applied to the outgoing particles, to take into account 
detector effects. Events were classified as Fully Contained (FC) or 
Partially Contained (PC) by placing the vertex at a random position inside the 
detector and tracking the lepton until it reached the edge of the detector. 
Partially Contained events 
are those where a final state muon exits the detector.  The number of events expected 
for each flavor and category is summarized in Table~\ref{tab:atmos_rates}.

\begin{cdrtable}
[Atmospheric neutrino event rates]{lc}{atmos_rates}{Atmospheric neutrino event rates including oscillations in \SI{350}{\ktyr} with a LArTPC, fully or partially contained in the detector fiducial volume. }
Sample   &  Event Rate \\ \toprowrule
fully contained electron-like sample   &14,053 \\ \colhline
fully contained muon-like sample       &20,853 \\ \colhline
partially contained muon-like sample   & 6,871 \\ 
\end{cdrtable}
\begin{sloppypar}
Figure~\ref{fig:lovere} shows the expected L/E distribution for high-resolution, muon-like 
events from a \SI{350}{\ktyr} exposure. The data provide excellent resolution of the 
first two oscillation nodes, even when taking into account the expected statistical uncertainty.
In performing oscillation fits, the data in each flavor/containment category are 
binned in energy and zenith angle. 
\end{sloppypar}

\begin{cdrfigure}[Reconstructed L/E Distribution of `High-Resolution' Atmospheric Neutrinos]{lovere}
{Reconstructed L/E Distribution of `High-Resolution'
$\mu$-like atmospheric neutrino events in a \SI{350}{\ktyr} exposure with and
without oscillations (left), and the ratio of the two (right), with the
shaded band indicating the size of the statistical uncertainty.}
\includegraphics[width=0.45\linewidth]{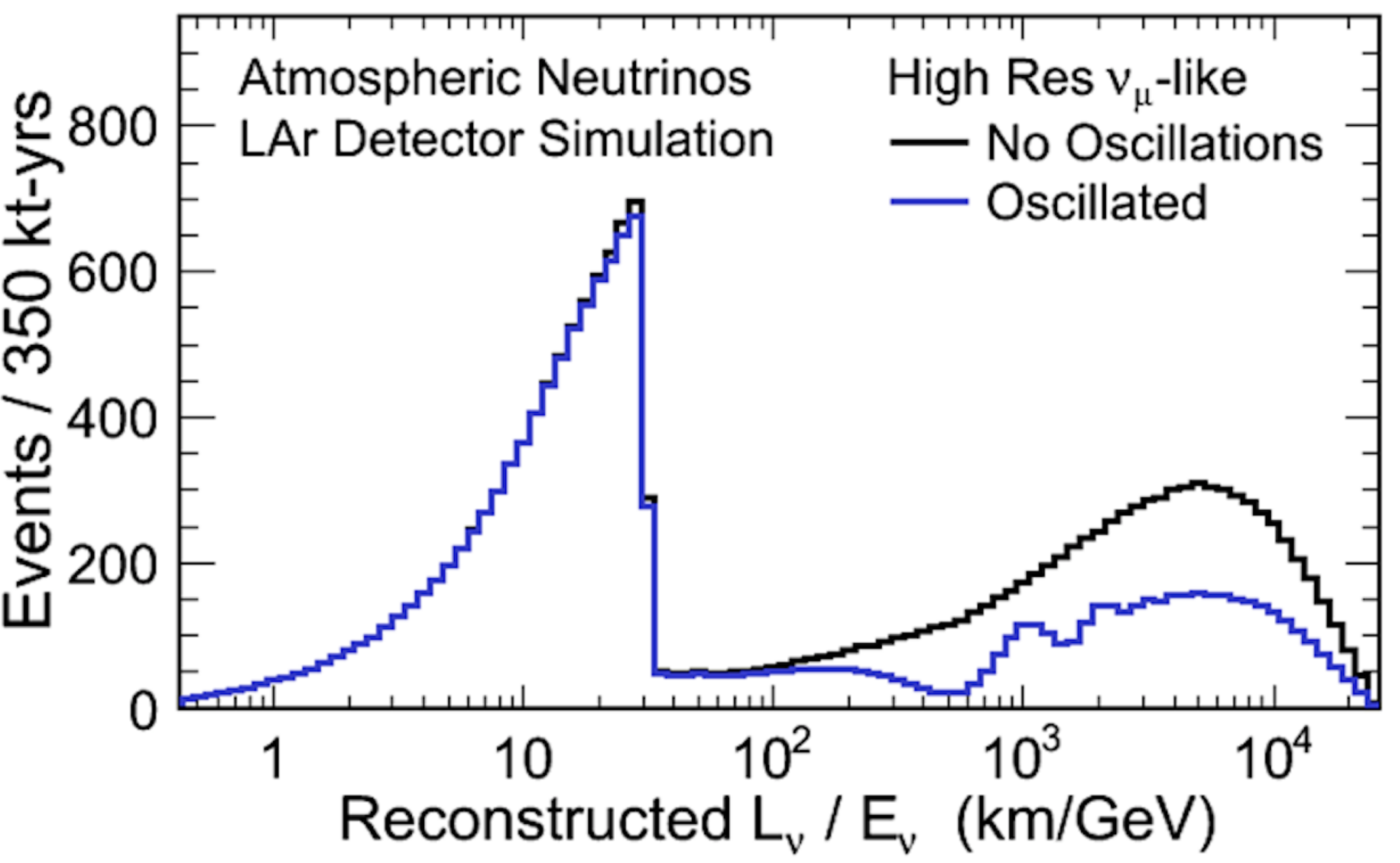}
\includegraphics[width=0.45\linewidth]{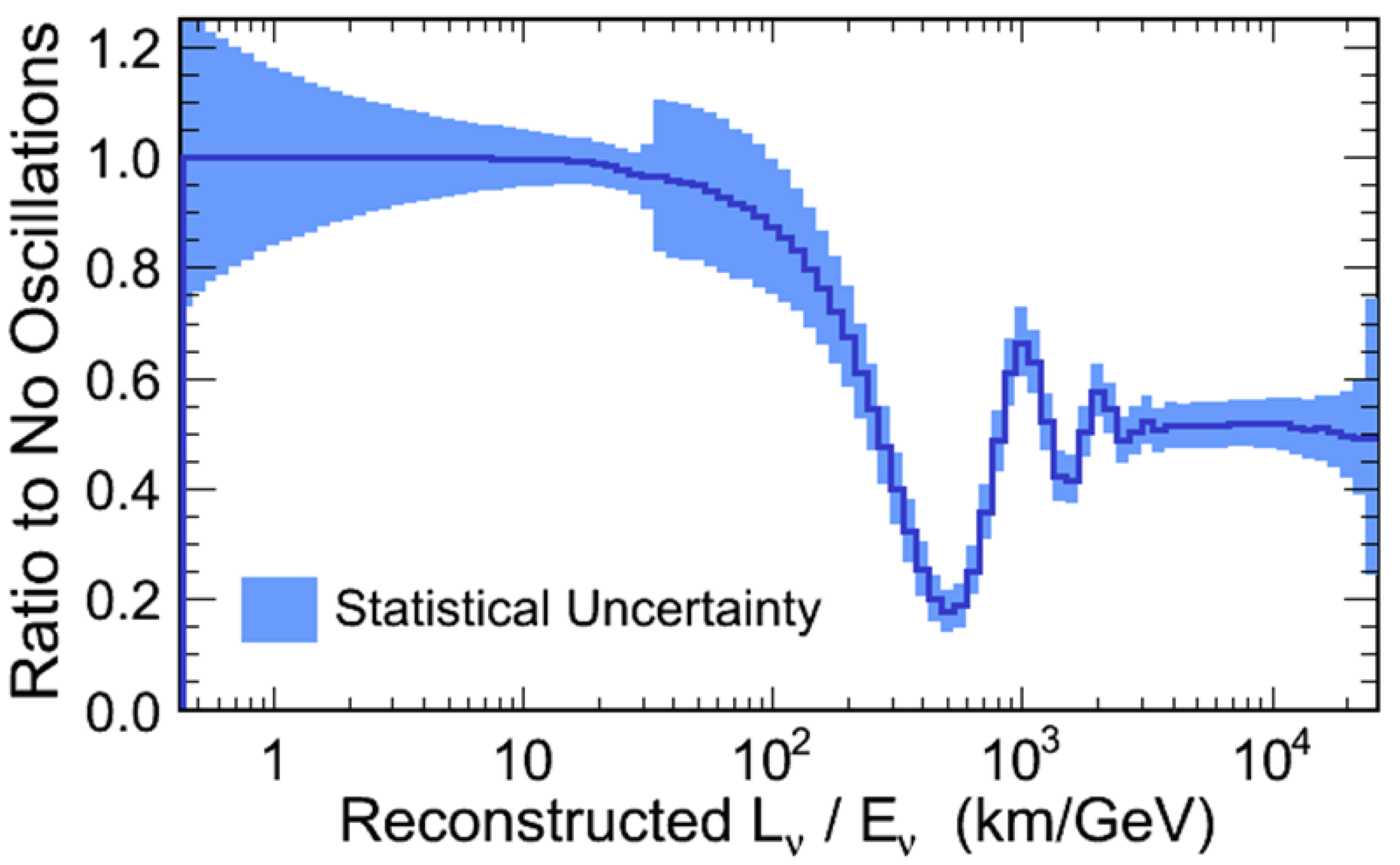}
\end{cdrfigure}

When neutrinos travel through the Earth, the MSW resonance influences 
electron neutrinos in the few-GeV energy range. More precisely, the resonance 
occurs for $\nu_e$ in the case of normal mass hierarchy (NH, $\Delta m^2_{32} > 0$), and for 
$\overline{\nu}_e$ in the case of inverted mass hierarchy (IH, $\Delta m^2_{32} < 0$). This is 
illustrated in Figure~\ref{fig:atm_e_zenith}. 

\begin{cdrfigure}[Zenith Angle vs. Energy For Atmospheric Neutrinos]{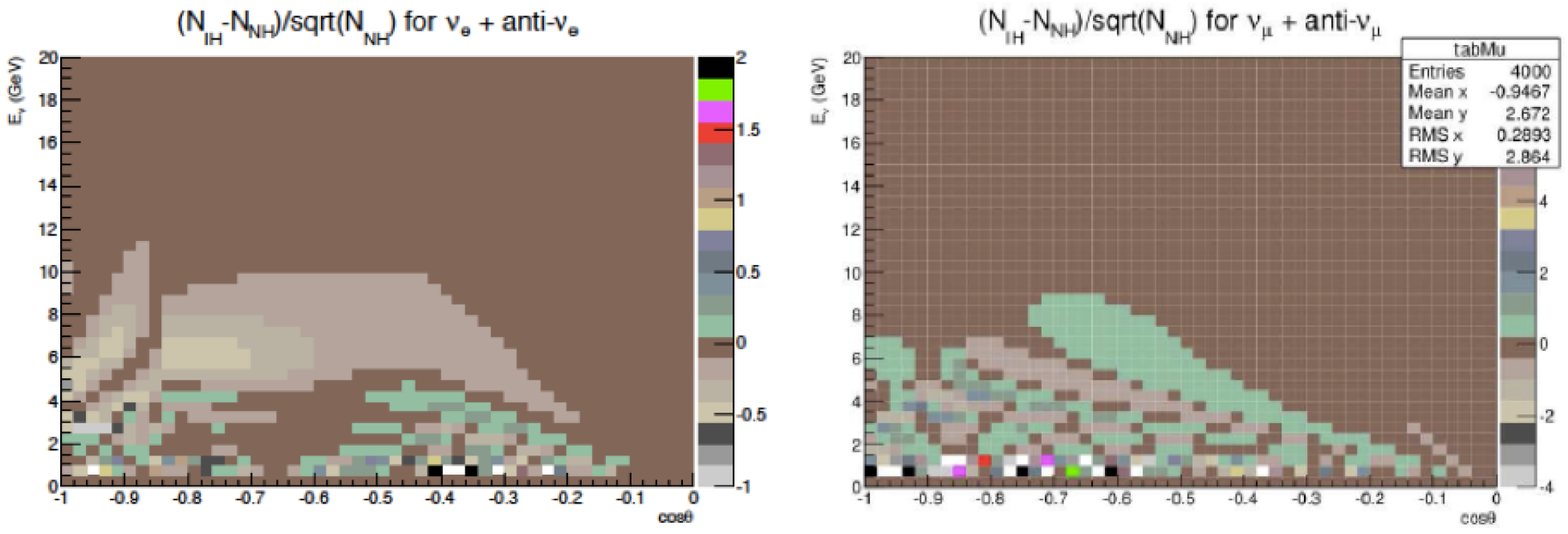}
{Statistical significance of the difference in expected event rates for NH and IH for 
electron neutrino events (left) and muon neutrino events (right), as a function of neutrino
energy and zenith angle, for a \SI{350}{\ktyr} exposure.}
\includegraphics[width=0.45\linewidth]{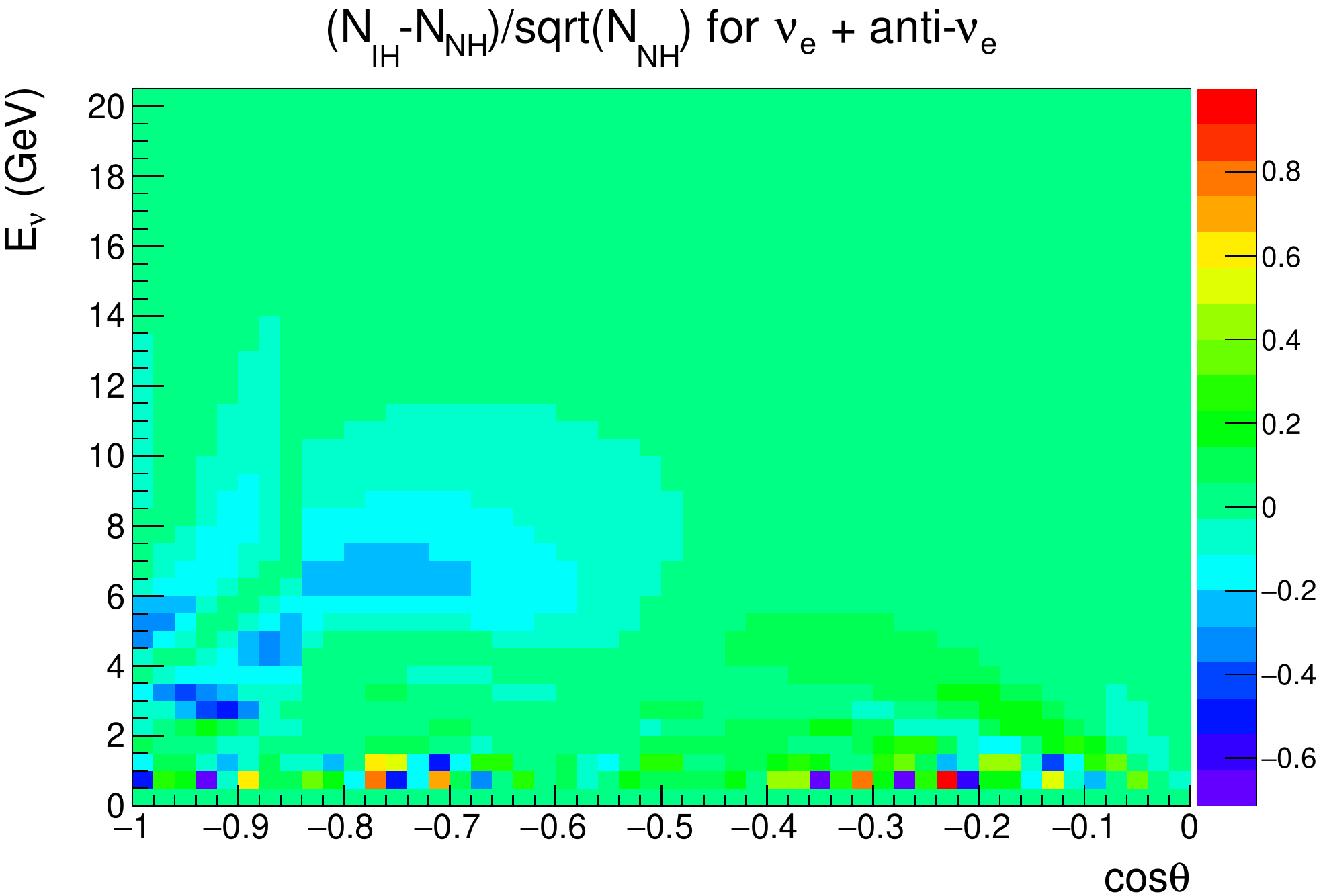}
\includegraphics[width=0.45\linewidth]{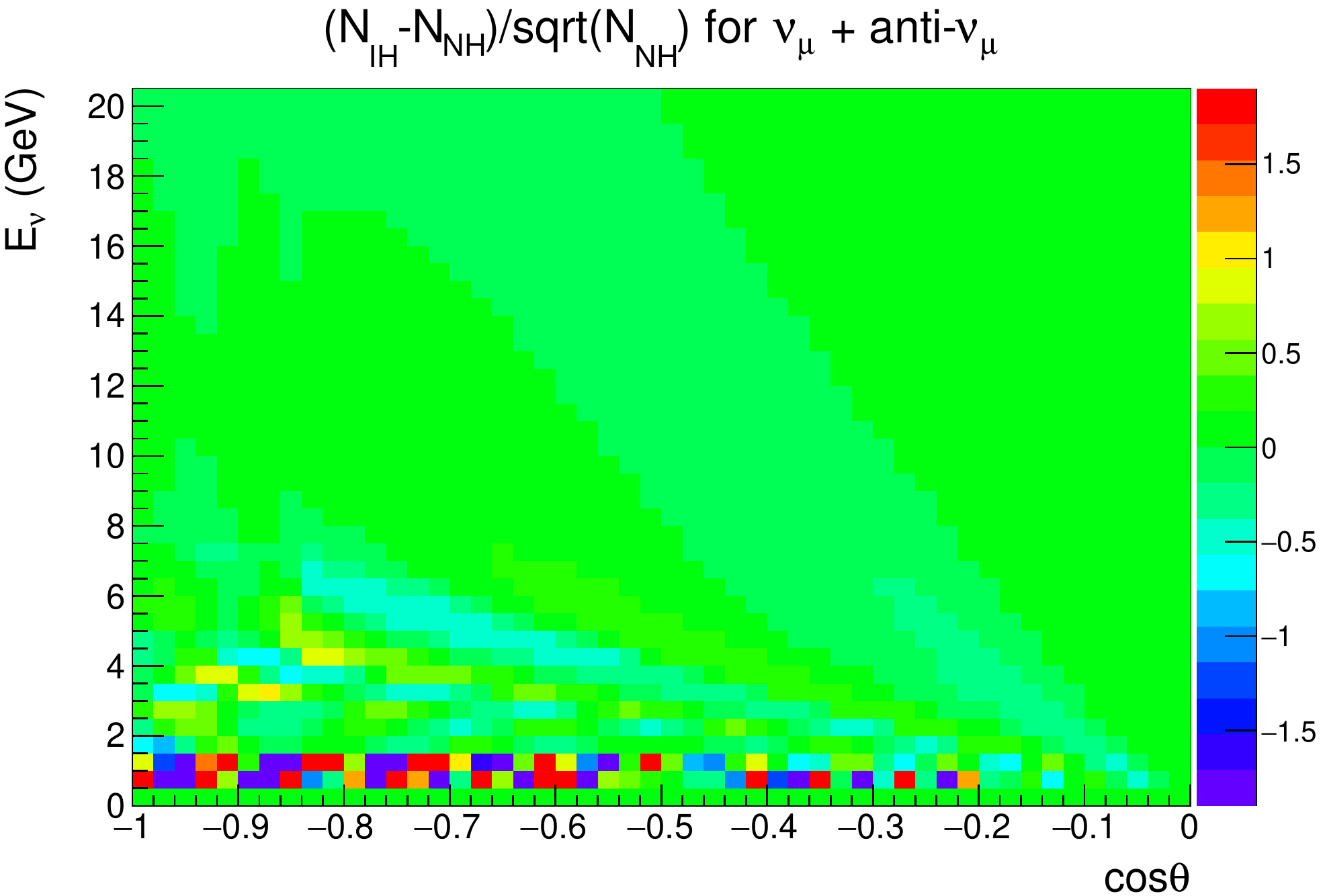}
\end{cdrfigure}

The mass hierarchy (MH) sensitivity can be greatly enhanced if neutrino and antineutrino events can be 
separated. The DUNE detector will not be magnetized; however, its high-resolution 
imaging offers possibilities for tagging features of events that provide statistical 
discrimination between neutrinos and antineutrinos. For the sensitivity calculations 
that follow, two such tags were included: a proton tag and a decay electron tag. 


Figure~\ref{fig:atm_mh} shows the MH sensitivity as a function of the fiducial exposure. 
Over this range of fiducial exposures, the sensitivity goes essentially as the square 
root of the exposure, indicating that the measurement is not systematics-limited. 
Unlike for beam measurements, the sensitivity to MH with atmospheric neutrinos is 
nearly independent of the CP-violating phase.  The sensitivity comes from both 
electron neutrino appearance as well as muon neutrino disappearance, and is strongly 
dependent on the true value of $\theta_{23}$, as shown in Figure~\ref{fig:atm_mh}.  Despite the
much smaller mass, DUNE would have comparable sensitivity to Hyper-Kamiokande regarding atmospheric 
neutrino analyses~\cite{Kearns:2013lea} due to the higher detector resolution.    

\begin{cdrfigure}[MH Sensitivity vs. Exposure for Atmospheric Neutrinos]{atm_mh}
{Sensitivity to mass hierarchy using atmospheric neutrinos as a function of fiducial 
exposure in a liquid argon detector (left), and as a function of the true value of 
$\theta_{23}$ (right).  For comparison, Hyper-K sensitivities are also shown \cite{Kearns:2013lea}.}  
\includegraphics[width=0.42\linewidth]{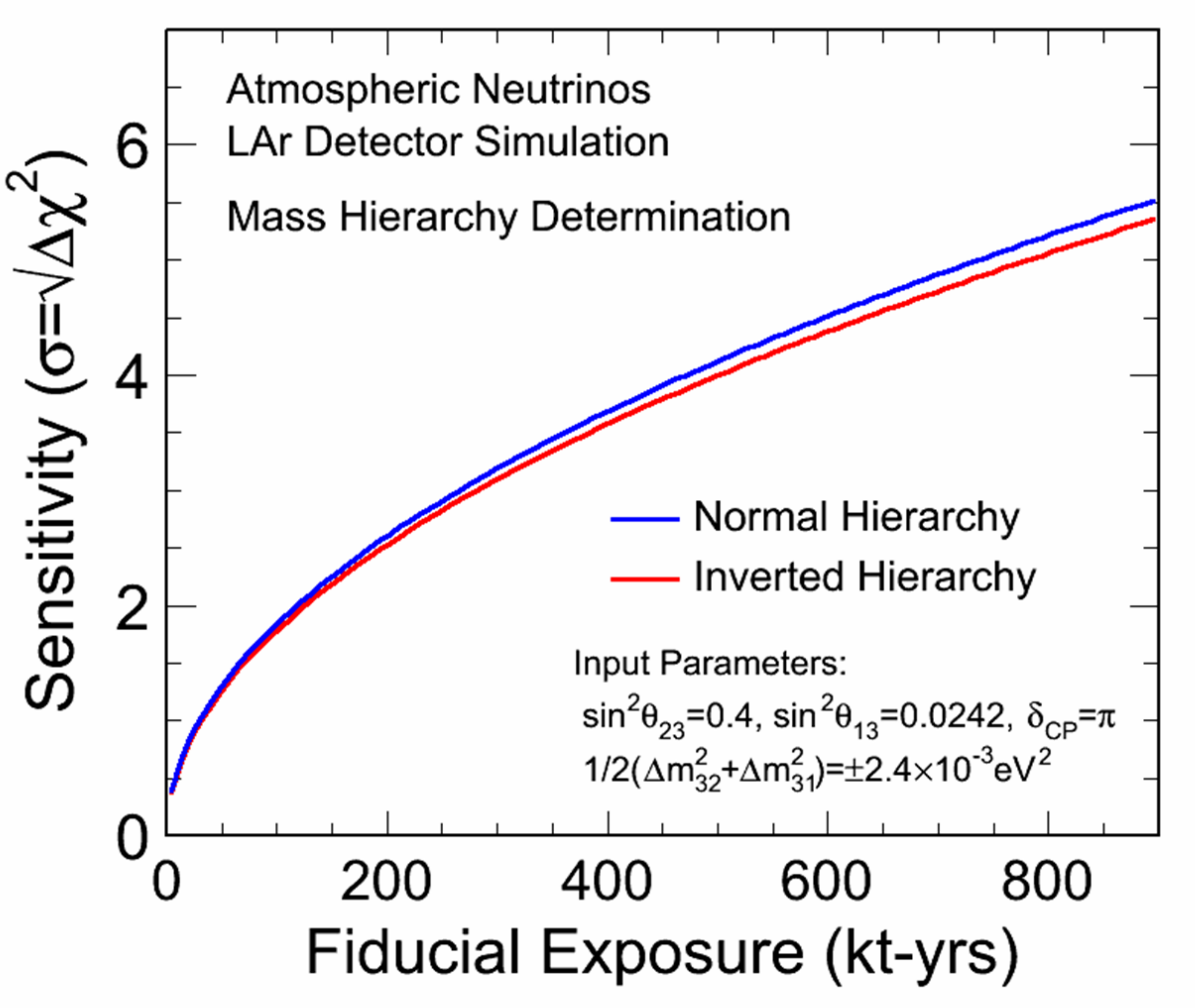}
\includegraphics[width=0.48\linewidth]{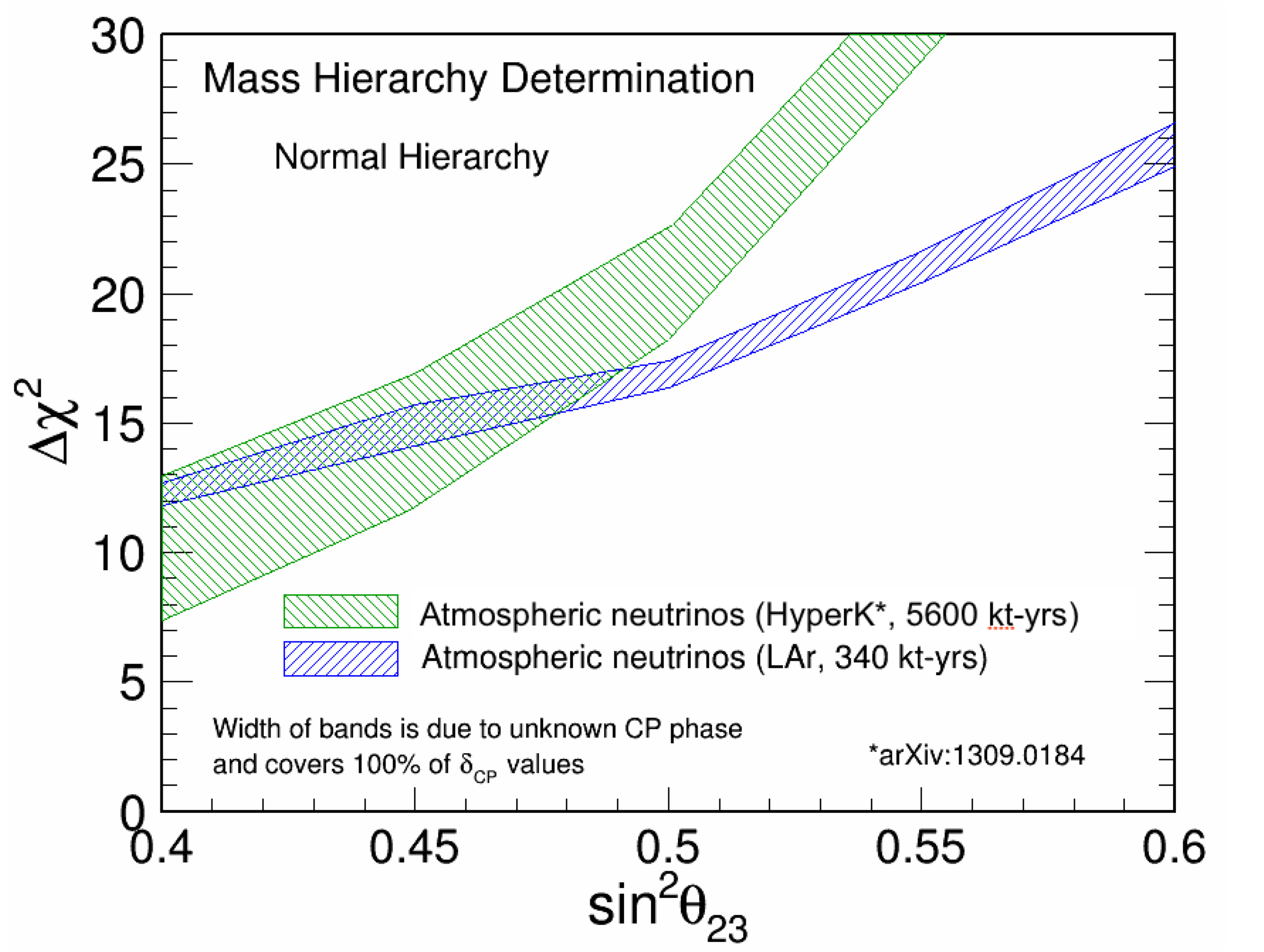}
\end{cdrfigure}

In the two-flavor approximation, neutrino oscillation probabilities depend on 
$\sin^2(2\theta)$, which is invariant when changing $\theta$ to $\pi/2-\theta$. In this case, the octant 
degeneracy remains for $\theta_{23}$ in the leading order terms of the full 
three-flavor oscillation probability, making it impossible to determine whether $\theta_{23}< \pi/4$ or 
$\theta_{23}> \pi/4$. Accessing full three-flavor oscillation with atmospheric neutrinos 
will provide a handle for solving the ambiguity.




These analyses will provide an approach complementary to that of beam neutrinos. 
For instance, they should enable resolution of 
degeneracies that can be present in beam analyses, since 
the MH sensitivity is essentially independent of $\delta_{CP}$.   Atmospheric neutrino data will be acquired 
even in the absence of the beam, and will provide a useful sample for the development of 
reconstruction software and analysis methodologies.  
Atmospheric neutrinos provide a window into a range of new physics scenarios, and 
may allow DUNE to place limits on CPT violation~\cite{Kostelecky:2003cr}, 
non-standard interactions~\cite{Chatterjee:2014gxa}, mass-varying neutrinos~\cite{Abe:2008zza}, 
sterile neutrinos~\cite{Abe:2014gda}, and
Lorentz invariance violation~\cite{Kostelecky:2011gq}.

\section{Indirect Search for WIMPs at the DUNE Far Detector}

If the true nature of DM does indeed involve a weakly interacting particle (WIMP) with a mass in the 100's of GeV range, one of the main search strategies involves looking in astrophysical data for anomalous signals from the 
annihilation (or decay) of a WIMP into SM particles, like neutrinos. Signals of DM via neutrinos can come from such distant objects as the galactic center, the center of the Sun or even the Earth~\cite{Press:1985ug,Silk:1985ax,Gaisser:1986ha,Gould:1987ir,Cirelli:2005gh}. As our solar system moves through the DM halo, WIMPs interact with nuclei 
 and become trapped in a body's gravitational well.  Over time, the WIMPs accumulate near the core of the body, enhancing the possibility of annihilation. The high-energy neutrinos ($E\sim m_{\rm WIMP}$) from these annihilations can free-stream through the astrophysical body and emerge roughly unaffected (although oscillation and matter effects can slightly alter the energy spectrum).  For the Sun, the background of neutrinos is produced at much lower energies via the nuclear fusion process. Thus, the detection of high-energy neutrinos pointing from the Sun and detected in the DUNE far detector would be clear evidence of DM annihilation.  Since the DUNE far detector has relatively large mass, of the order tens of kt, it can act as a ``neutrino telescope'' and be used to search for signals of DM annihilations coming from the Sun and/or the core of the Earth.  IceCube~\cite{Aartsen:2012kia} and Super-Kamiokande~\cite{Choi:2015ara} have searched for WIMPs with masses from a few to a few hundred GeV$/c^2$ using this method, but have not observed a signal of DM annihilation into neutrinos.  These indirect-detection experiments  are limited by atmospheric neutrino background.  Compared to these experiments, which are based on Cherenkov light detection, the DUNE LArTPC can provide much better angular resolution. This would substantially reduce the background in the direction of the expected WIMP-induced neutrino signal, and could potentially provide competitive limits in the low-WIMP-mass range.  Studies are needed to investigate the sensitivity of DUNE for indirect WIMP detection.

\section{Detector Requirements}
\label{sec:physics-atmpdk-detector-requirements}

Physics with atmospheric neutrino interactions and searches for
nucleon decays have several requirements that are not necessarily in
common with the beam-related physics program.  Detector mass and depth plays a
more critical role here; the atmospheric ``beam'' is fixed, and the
number of nucleons available to decay obviously depends on the number
of nuclei in the detector.  The DUNE Far Detector
Requirements~\cite{lbnfdune-cdr-req} specific to searches for proton
decay and measurements of the atmospheric neutrino flux are as
follows:

\begin{itemize}

\item Far Detector Depth: The far detector shall be located at
  sufficient depth to allow detection of atmospheric neutrinos and
  proton decay with negligible backgrounds from cosmogenic sources.
  Depth plays a greater role for these physics topics than for
  long-baseline neutrino oscillations, because of the lack of a beam
  gate coincidence. As discussed in the previous sections, cosmic ray
  interactions in the surrounding rock and detector dead regions
  can lead to critical backgrounds, or create difficulties for
  reconstruction algorithms. Analyses of
  backgrounds~\cite{homestake:depth, Bueno:2007um, Klinger:2015kva, Adams:2013qkq} has
  shown that a depth of 4850 ft is sufficient to reduce these
  backgrounds to negligible levels.


\item Far Detector Mass: The far-detector fiducial size multiplied by
  the duration of operation [expressed in kiloton-years] shall be
  sufficient to yield a scientifically competitive result on proton
  decay. As Figure~\ref{fig:kdklimit} shows, a \ktadj{10} detector can
  exceed the existing Super-Kamiokande limits for the $p\rightarrow
  K^+ + \bar{\nu}$ nucleon decay channel in five years, and the full \ktadj{40} 
  scope in a much shorter time than that.  For atmospheric neutrinos, a
  detector mass of 40 kt will achieve a mass hierarchy determination
  of better than 3$\sigma$ in 10 years.


\item Far Detector DAQ: The far detector DAQ must enable continuous recording 
of data, outside of any beam gate, and retain enough
  information from the front end (including photon system) 
  through the DAQ chain to enable a trigger on events of interest. The DAQ must keep any
  information that would allow the identification of putative events, 
  and its livetime fraction should be higher than the
  efficiency of all other cuts placed on the data.  The trigger system
  itself must be able to provide information that allows linking of
  tracks in different modules of the far detector.

\item Far Detector Particle ID: It is required that the separation
  of $K^+$'s and $\pi^+$'s in the Far Detector be sufficient to ensure that
  much less than one event leak into the 
  $p\to K^+\overline{\nu}$ sample.  Unlike the
  long-baseline physics, proton decay physics requires a rare-process search;
  therefore small tails from e.g., atmospheric neutrino
  interactions creating $\pi^+$s that are mis-identified could be very
  damaging.

\item Far Detector Energy Resolution: It is required that the energy
  resolution be known well enough that its uncertainty is a negligible
  contribution to the measurement of the atmospheric neutrino energy
  spectrum of all flavors and that this uncertainty have a negligible impact on
  background predictions for proton decay.

\end{itemize}

\cleardoublepage

\chapter{Supernova Neutrino Bursts and Low-energy Neutrinos}
\label{ch:physics-snblowe}

\section{Overview}
\label{sec:physics-snblowe-overview}

The DUNE experiment will be sensitive to neutrinos in the few
tens of MeV range, which create short electron tracks in liquid argon, potentially accompanied by a few
gamma rays. 
This regime is of
particular interest for detection of the burst of neutrinos from a galactic
core-collapse supernova (the primary focus of this chapter). 
The sensitivity of DUNE is primarily to \textit{electron flavor} supernova neutrinos, and this capability is unique among existing and proposed supernova neutrino detectors for the next decades.  
Neutrinos from other astrophysical sources are also potentially detectable.  
The low-energy event regime has several reconstruction, background and triggering challenges. 

The observation of neutrinos from the celebrated SN1987A core
collapse~\cite{Bionta:1987qt,Hirata:1987hu} in the Large Magellanic
Cloud outside the Milky Way 
provided qualitative validation of the basic physical picture of core-collapse and provided powerful constraints on numerous models of new physics. At the same time, the statistics were sparse 
and many questions remain.  A high-statistics observation of a
nearby supernova neutrino burst would be possible with the current
generation of detectors. Such an observation would shed light on 
the nature of the astrophysical event, as well as on the nature of
neutrinos themselves.  Sensitivity to the different flavor components
of the flux is highly desirable.

\subsection{The Stages of Core Collapse} 

As a result of nuclear burning throughout a massive star's lifetime, the 
inner region of the star forms an ``onion'' structure, with an iron core at the center surrounded by concentric shells of lighter elements (silicon, oxygen, neon, magnesium, carbon, etc.). Eventually the core collapses, causing 
a core-collapse supernova\footnote{In this chapter ``Supernova'' always refers to a ``core-collapse supernova.''}.
 
As the star ages, its iron core, at temperatures of $T\sim 10^{10}$ K and densities of $\rho \sim 10^{10}$ g/cm$^{3}$, continuously loses energy through neutrino emission caused by pair annihilation and plasmon decay. Since iron does not burn, there is no mechanism to replenish this lost energy within the core, and the core continues to contract and heat up. Meanwhile, the shells around it burn, producing iron that gravitates to the core, adding mass to it.  When the core reaches the critical mass of about $1.4 M_{\odot}$ of Fe, a stable configuration is no longer possible. At this point, as electrons are absorbed by the protons and some iron is disintegrated by thermal photons, the pressure support is suddenly removed and the core collapses essentially in free fall, reaching speeds of about a quarter of the speed of light. 
\footnote{Other collapse mechanisms are possible: an ``electron-capture'' supernova does not reach the final burning phase before highly degenerate electrons break apart nuclei and trigger a collapse.}

The collapse of the core suddenly halts after $\sim 10^{-2}$ seconds, as the density reaches nuclear (and up to supra-nuclear)  values. The core then bounces and a shock wave forms. The extreme physical conditions of this core, in particular the densities of order $10^{12}-10^{14}$ g/cm$^{3}$, create a medium that is opaque even to neutrinos; 
the temperature of this core is 
$\lesssim$ 30 MeV, which is relatively \emph{cold}. At this stage, the gravitational energy of the collapse is stored mostly in the degenerate Fermi sea of electrons ($E_{F}\sim 200$ MeV) and electron neutrinos, which are in equilibrium with each other, and the core's lepton number is trapped.

A point is reached where the trapped energy and lepton number both escape 
from the core, carried by the least interacting particles, i.e., neutrinos, according to the Standard Model.  A tremendous amount of energy, some $10^{53}$ ergs, is released in a time span of a few seconds by $10^{58}$ neutrinos and antineutrinos of all flavors, with energies of $\sim 10$~MeV. 
A small fraction of this energy is absorbed by beta reactions that form a shock wave. This shock wave blasts away the rest of the star creating a spectacular explosion, which, curiously enough, is only a tiny perturbation from the energetics point of view. 

Over 99\% of all gravitational binding energy of the $1.4 M_{\odot}$ collapsed core -- some 10\% of its rest mass -- has now been emitted as neutrinos. The resulting central object then settles to a neutron star or a black hole.

\subsection{Observable Signals from the Explosion} 

The flavor content and spectra of the neutrinos emitted from the neutrinosphere (the surface of neutrino trapping) change
throughout the phases of the core collapse, and the neutrino signal provides information on the supernova's evolution.  

The signal starts with a short, sharp
\emph{neutronization} (or \emph{break-out}) burst primarily composed of
$\nu_e$. 
This quick and intense burst is followed by an
\emph{accretion} phase lasting some hundreds of milliseconds, depending on the progenitor star mass, as matter falls onto the collapsed core and the shock is stalled at the distance of perhaps $\sim 200$ km. The gravitational binding energy of the accreting material powers the neutrino luminosity during this stage. The 
\emph{cooling} phase that follows, 
lasting $\sim$10~seconds, represents the main part of
the signal, over which the proto-neutron star sheds its trapped energy.  

Some fairly generic features of the neutrinos emitted in each stage are illustrated in Figure~\ref{fig:spectrum}, based on a 1-dimensional model of~\cite{Fischer:2009af} and reproduced from~\cite{Wurm:2011zn}.
\begin{cdrfigure}[Expected core-collapse neutrino signal]{spectrum}{Expected
  core-collapse neutrino signal from the ``Basel''
  model~\cite{Fischer:2009af}, for a
  10.8 $M_{\odot}$ progenitor.  The left plots show the very early
  signal, including neutronization burst; the middle plots show
  the accretion phase, and the right plots show the cooling
  phase. Across the top, luminosities as a function of time are shown. 
  Across the bottom, the plots show average energy as a function of time for the
  $\nu_e$, $\overline{\nu}_e$ and $\nu_{\mu,\tau}$ flavor components of the
  flux (fluxes for $\nu_\mu$, $\overline{\nu}_\mu$, $\nu_\tau$,
  and $\overline{\nu}_\tau$ should be identical).  Figure courtesy of~\cite{Wurm:2011zn}.}
\includegraphics[width=0.9\textwidth]{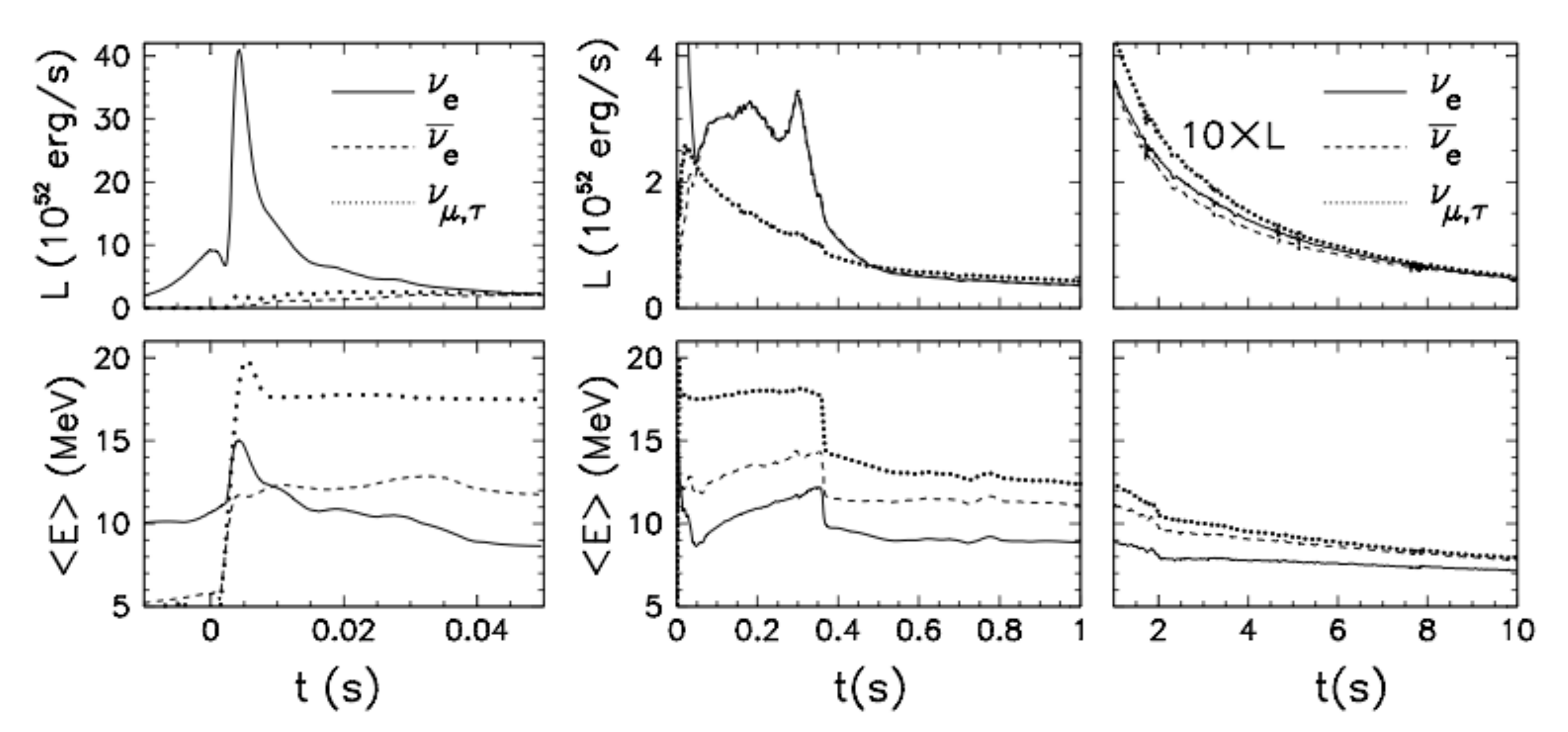}
\end{cdrfigure}

The physics of neutrino decoupling and spectrum formation is far from trivial, owing to the energy dependence of the cross sections and the roles played by both charged- and neutral-current reactions.
Detailed transport calculations using methods such as Monte Carlo or Boltzmann solvers have been employed. It has been observed that spectra coming out of such simulations can typically be parameterized at a given moment in time by the following ansatz (e.g.,~\cite{Minakata:2008nc,Tamborra:2012ac}):
\begin{equation}
        \label{eq:pinched}
        \phi(E_{\nu}) = \mathcal{N} 
        \left(\frac{E_{\nu}}{\langle E_{\nu} \rangle}\right)^{\alpha} \exp\left[-\left(\alpha + 1\right)\frac{E_{\nu}}{\langle E_{\nu} \rangle}\right] \ ,
\end{equation}
where $E_{\nu}$ is the neutrino energy, $\langle E_\nu \rangle$ is the
mean neutrino energy, $\alpha$ is a ``pinching parameter,'' and
$\mathcal{N}$ is a normalization constant.
Large $\alpha$ corresponds to a more \emph{pinched} spectrum (suppressed
high-energy tail). This parameterization is referred to as a
\emph{pinched-thermal} form. The different $\nu_e$, $\overline{\nu}_e$ and
$\nu_x, \, x = \mu, \tau$ flavors are expected to have different
average energy and $\alpha$ parameters and to evolve differently in
time. 

The initial spectra get further processed (permuted) by flavor oscillations; understanding these oscillations is very important for extracting physics from the detected signal.

\subsection{Detection Channels and Interaction Rates in Liquid Argon}

Liquid argon has a particular sensitivity to the $\nu_e$ component of a supernova neutrino burst, via charged-current (CC)
absorption of $\nu_e$ on $^{40}$Ar,
\begin{equation}
\nu_e + ^{40}{\rm Ar} \rightarrow e^- + ^{40}{\rm K^*},
\label{eq:nueabs}
\end{equation}
for which the observables are the $e^-$ plus de-excitation products from the excited $K^*$ final state, as well as a $\bar{\nu}_e$ interaction and elastic scattering on electrons.
Cross sections for the most
relevant interactions are shown in Figure~\ref{fig:xscns}.

\begin{cdrfigure}[Cross sections for supernova-relevant interactions in argon]{xscns}{Cross sections for supernova-relevant interactions in argon~\cite{GilBotella:2003sz,snowglobes}.}
\includegraphics[width=0.6\textwidth]{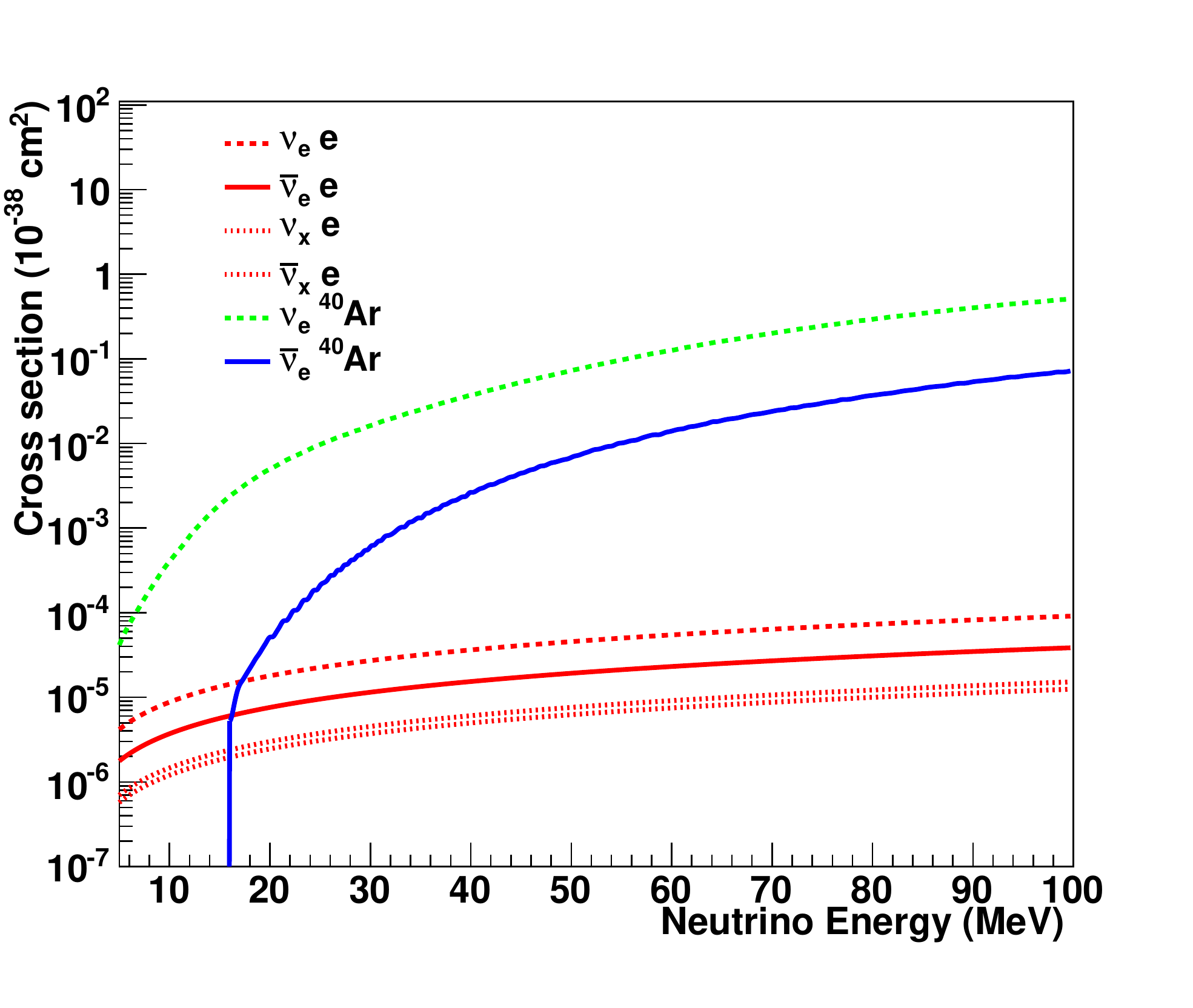}
\end{cdrfigure}


Neutral-current (NC) scattering on Ar nuclei by any type of neutrino, $\nu_x + {\rm Ar} \rightarrow \nu_x + {\rm Ar}^*$, is another process of interest for supernova detection in LAr detectors that is not yet fully studied. The signature is given by the cascade of de-excitation $\gamma$s from the final-state Ar nucleus. A dominant 9.8-MeV Ar$^*$ decay line has been recently identified as a spin-flip M1 transition~\cite{Hayes}.   At this energy the probability of $e^+e^-$ pair production is relatively high, offering a potentially interesting neutral-current tag.

The predicted event rate (NC or CC) from a supernova burst may be calculated by
folding expected neutrino differential energy spectra in with cross
sections for the relevant channels and with detector response; this is done using SNOwGLoBES~\cite{snowglobes}, which uses Icarus detector resolution~\cite{Amoruso:2003sw} and assumes a detection threshold of 5~MeV.

Table~\ref{tab:argon_events} shows rates calculated  for the dominant interactions in argon for
the ``Livermore'' model~\cite{Totani:1997vj} (no longer preferred, but included for comparison with literature), and the ``GKVM''
model~\cite{Gava:2009pj}; for the former, no oscillations are assumed; the latter assumes collective oscillation effects (see Section~\ref{sec:physics-snblowe-neutrino-physics}). There is a rather wide variation --- up to an order of magnitude --- in event rate for different models, due to different numerical treatment (e.g., neutrino transport, dimensionality), physics input (nuclear equation of state, nuclear correlation and impact on neutrino opacities, neutrino-nucleus interactions) and oscillation effects. In addition, there is intrinsic variation in the nature of the progenitor and collapse mechanism.  
 Furthermore, neutrino emission from the supernova may exhibit an emitted lepton-flavor asymmetry~\cite{Tamborra:2014aua}, which would lead to observed rates being direction-dependent.
\begin{cdrtable}[Event rates for different models in \SI{40}{\kt} of LAr for
    a core-collapse at 10~kpc]{lcc}{argon_events}{Event rates for different
    supernova models in \SI{40}{\kt} of liquid argon for a core collapse at 10~kpc, for $\nu_e$ and $\bar{\nu}_e$ charged-current channels and elastic scattering (ES) on electrons.
    Event rates will simply scale by active detector mass and inverse square of supernova distance. The ``Livermore'' model assumes no oscillations; ``GKVM'' assumes collective oscillation effects.  Oscillations (both standard and ``collective'') will potentially have a large, model-dependent effect.}
Channel & Events & Events \\
\rowtitlestyle
& ``Livermore'' model & ``GKVM'' model  \\ 
\toprowrule

$\nu_e + ^{40}{\rm Ar} \rightarrow e^- + ^{40}{\rm K^*}$ & 2720  & 3350 \\ \colhline

$\overline{\nu}_e + ^{40}{\rm Ar} \rightarrow e^+ + ^{40}{\rm Cl^*}$ & 230 & 160\\ \colhline

$\nu_x + e^- \rightarrow \nu_x + e^-$                           & 350 &  260\\ \colhline

Total &  3300 & 3770 \\ 
\end{cdrtable}

Figure~\ref{fig:garching} shows another example of an expected burst
signal, for which a calculation with detailed time-dependence of the
spectra is available~\cite{Huedepohl:2009wh} out to 9~seconds
post-bounce.  This model has relatively low luminosity but includes the standard robust
neutronization burst.   Note that the relative fraction of
neutronization-burst events is quite high.
Figure~\ref{fig:eventrates} shows the event channel breakdown for the same model.  Clearly, the $\nu_e$
flavor dominates.  Although other types of detectors, i.e., water and scintillator, have the capability to record $\nu_e$ events~\cite{Laha:2013hva,Laha:2014yua}, liquid argon offers the only prospect for observation of a large, clean supernova $\nu_e$ sample~\cite{Scholberg:2012id}.

\begin{cdrfigure}[Garching flux signal with neutronization burst]{garching}{Expected
  time-dependent signal for a specific flux model for an
  electron-capture supernova~\cite{Huedepohl:2009wh} at 10~kpc.  No oscillations are assumed. The
  top plot shows the luminosity as a function of time, the second plot
  shows average neutrino energy, and the third plot shows the $\alpha$
  (pinching) parameter.  The fourth (bottom) plot shows the total number of
  events (mostly $\nu_e$) expected in 40 kt of liquid argon, calculated using
  SNOwGLoBES.  Note the logarithmic binning in time; the plot shows
  the number of events expected in the given bin and the error bars
  are statistical. The vertical dashed line at 0.02 seconds indicates
  the time of core bounce, and the vertical lines indicate different
  eras in the supernova evolution.  The leftmost time interval
  indicates the infall period.  The next interval, from core bounce to
  50~ms, is the neutronization burst era, in which the flux is
  composed primarily of $\nu_e$.  The next period, from 50 to 200~ms,
  is the accretion period. The final era, from 0.2 to 9~seconds, is
  the proto-neutron-star cooling period.}
\includegraphics[width=0.9\textwidth]{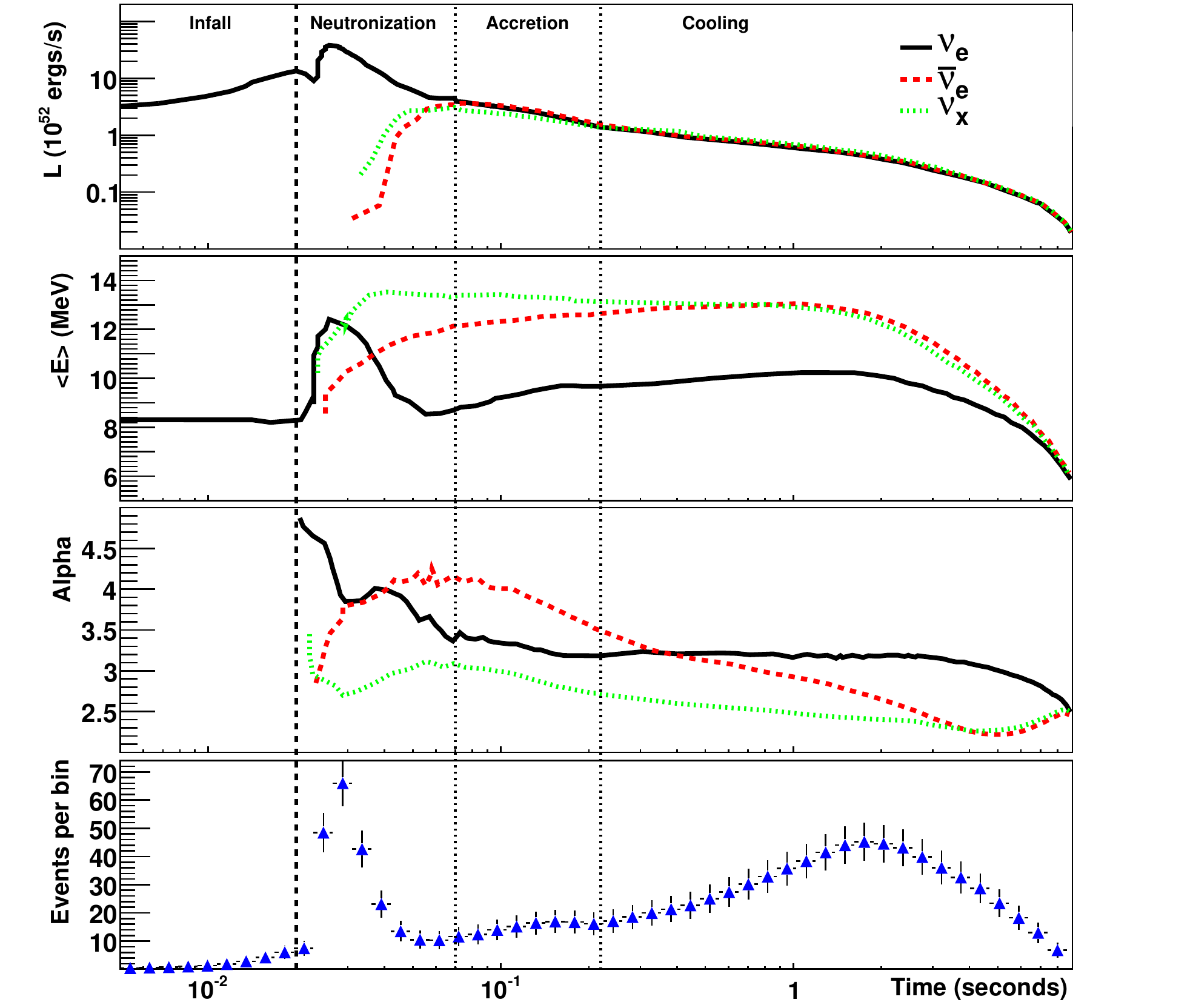}
\end{cdrfigure}

\begin{cdrfigure}[Supernova $\nu$ event rates in \SI{40}{kt} of LAr for Garching flux]{eventrates}{Left: Expected
  time-dependent signal in 40 kt of liquid argon for the electron-capture supernova~\cite{Huedepohl:2009wh} at 10~kpc, calculated using SNoWGLoBES~\cite{snowglobes}, showing breakdown of event channels.  Right: expected measured event spectrum for the same model, integrated over time.}
\includegraphics[width=2.5in]{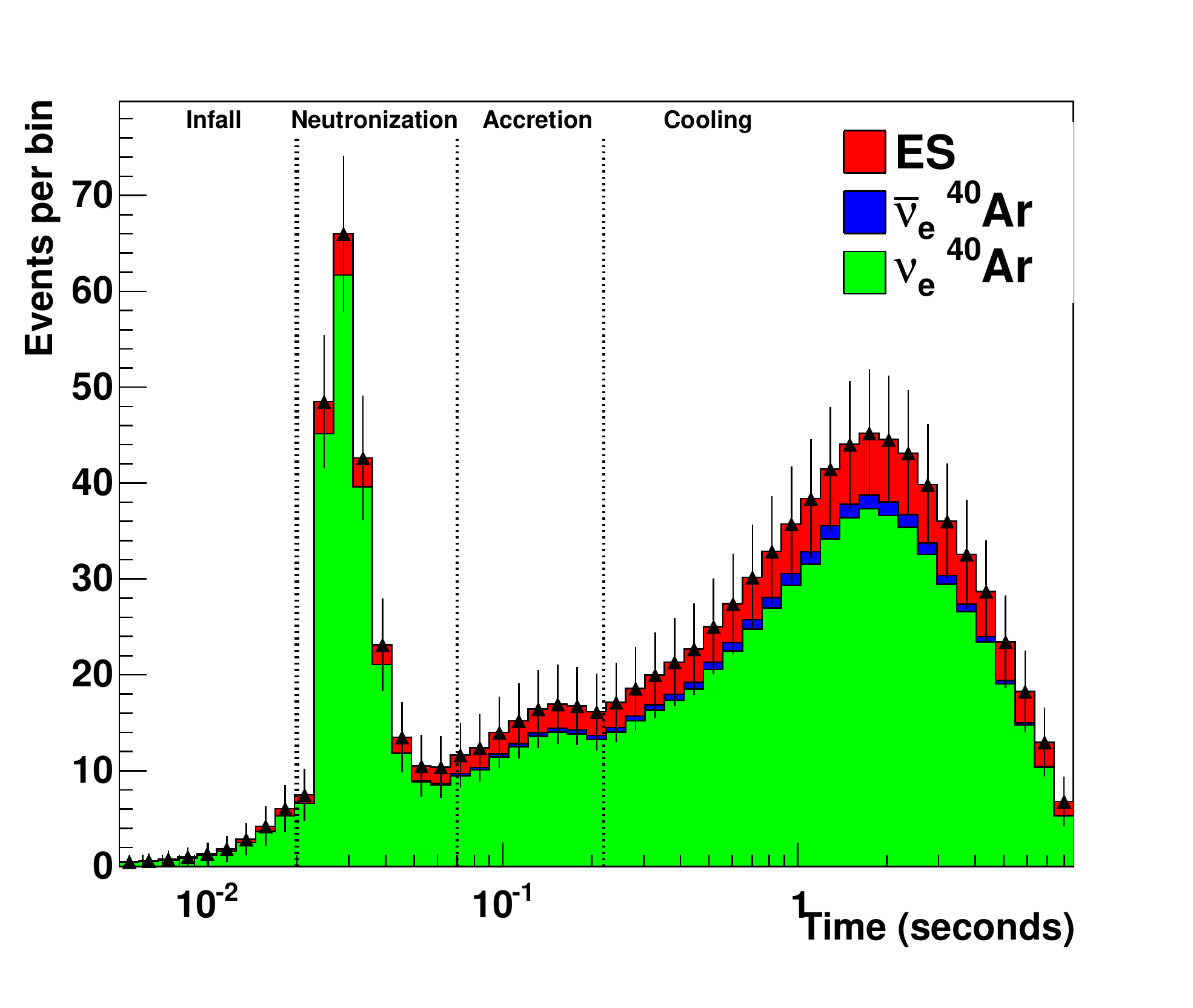}
\includegraphics[width=2.5in]{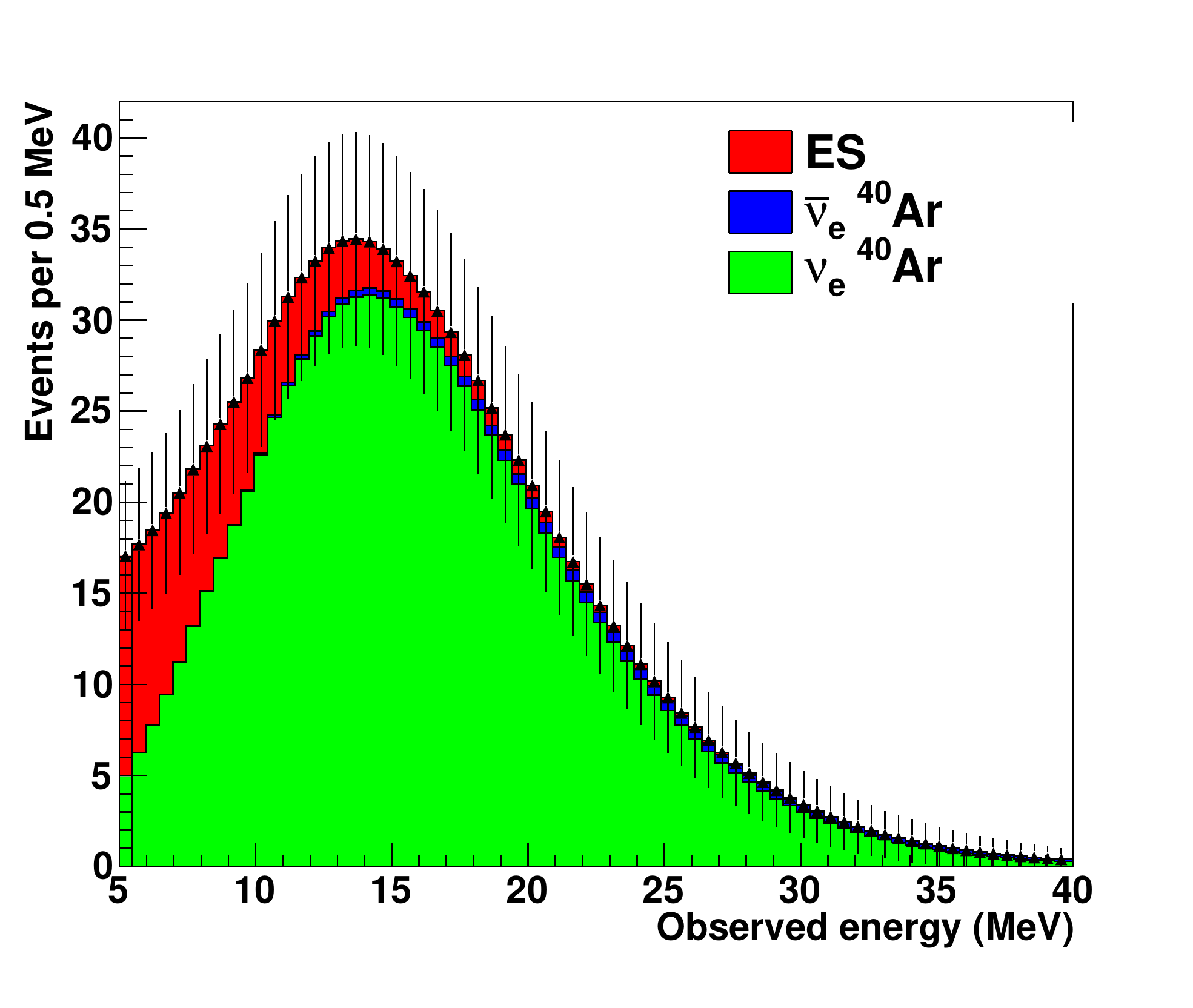}
\end{cdrfigure}

The number of signal events scales with mass and inverse square of distance as shown in Figure~\ref{fig:ratesvsdist}.  For a collapse in the Andromeda galaxy, a 40-kt detector would observe a few events.

\begin{cdrfigure}[Supernova neutrino rates vs. distance]{ratesvsdist}{Estimated numbers of supernova neutrino interactions in DUNE as a function of distance to the supernova, for different detector masses ($\nu_e$ events dominate). The red band 
represents expected events for a 40-kt detector and the green band represents expected events for a 10-kt detector. The borders of these bands (dashed lines) limit a fairly wide range of possibilities for ``Garching-parameterized'' supernova flux spectra (Equation~\ref{eq:pinched}) with luminosity $0.5\times 10^{52}$ ergs over ten seconds. The optimistic upper line of a pair gives the number of events for average $\nu_e$ energy of $\langle E_{\nu_e}\rangle =12$~MeV, and ``pinching'' parameter $\alpha=2$; the pessimistic lower line of a pair gives the number of events for $\langle E_{\nu_e}\rangle=8$~MeV and $\alpha=6$. (Note that the luminosity, average energy and pinching parameters will vary over the time frame of the burst, and these estimates assume a constant spectrum in time. Oscillations will also affect the spectra and event rates.) The solid lines represent the integrated number of events for the specific time-dependent neutrino flux model in~\cite{Huedepohl:2009wh} (see Figs.~\ref{fig:garching} and \ref{fig:eventrates}; this model has relatively cool spectra and low event rates). Core collapses are expected to occur a few times per century, at a most-likely distance of around 10 to 15 kpc.}
\includegraphics[width=5in]{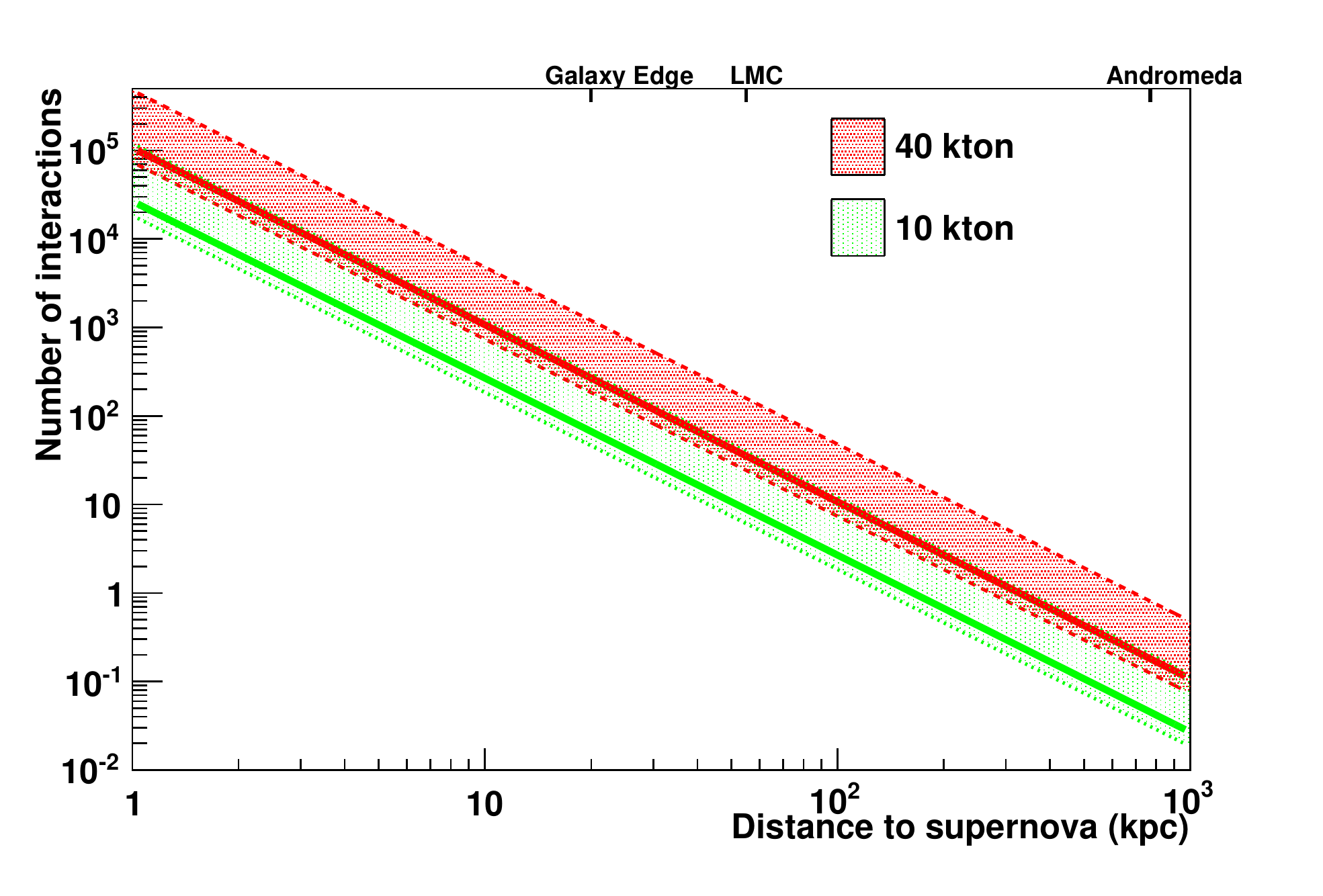}
\end{cdrfigure}

\section{Neutrino Physics and Other Particle Physics}
\label{sec:physics-snblowe-neutrino-physics}

The key property of neutrinos that leads to a dominant role in supernova dynamics is the feebleness of their interactions. It then follows that should there be unknown, even weaker interactions or properties of neutrinos, or 
new, light ($< 100$ MeV) weakly interacting particles, they could alter the energy transport process and the resulting evolution of the nascent proto-neutron star. 
A core-collapse supernova can be thought of as 
a hermetic system that can be used to search for numerous types of new physics (e.g.,~\cite{Schramm:1990pf,Raffelt:1999tx}). The list includes various Goldstone bosons (e.g., Majorons), neutrino magnetic moments, new gauge bosons (\emph{dark photons}), \emph{unparticles}, and extra-dimensional gauge bosons. %
The existing data from SN1987A already provide significant constraints on these scenarios, by confirming the basic energy balance of the explosion. At the same time, more precision is highly desirable and should be provided by the next galactic supernova. 

\begin{cdrfigure}[Simulated cooling curves from the Garching light progenitor model]{coolingcurves}{ Average $\nu_{e}$ energy from a simulated fit to the oscillated fluxes predicted by the Garching 1D model with a light (10.8 $M_{\odot}$) progenitor. DUNE's oscillation calculations included full multi-angle treatment of collective evolution, for two
different mass hierarchy assumptions. The predicted events were then smeared with SNOwGLoBES and fit with a pinched-thermal spectrum as a function of time (assuming a supernova at 10 kpc and a 34 kt LAr detector). The bands represent $1\sigma$ error bars from the fit (assuming only statistical uncertainties). The solid black line is the true
$\langle E_{\nu} \rangle$ for the unoscillated spectrum. Clearly, the rate of energy escape from the proto-neutron star can be gleaned by tracking $\nu_{e}$ spectra as a function of time.}
\includegraphics[width=0.9\textwidth]{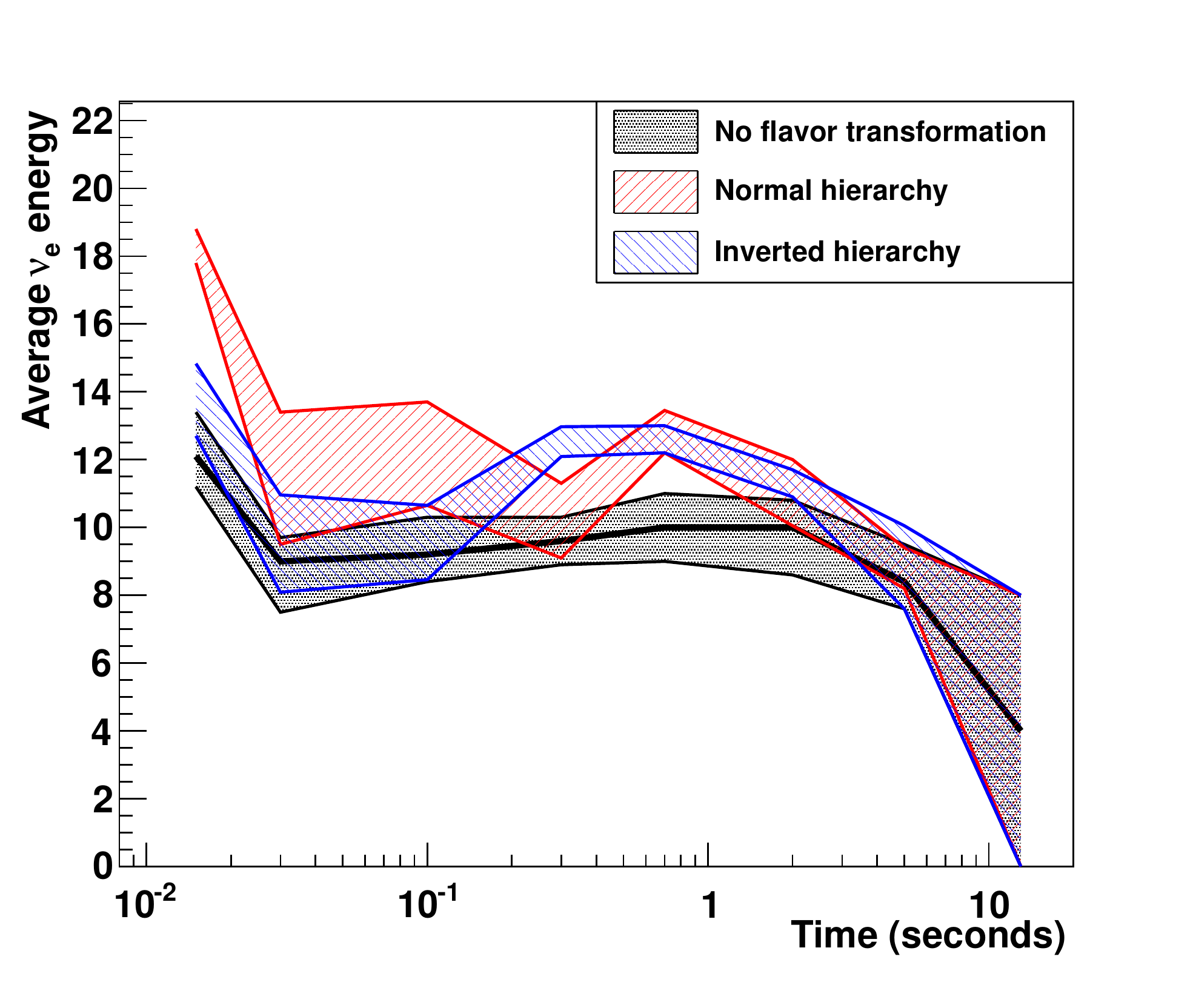}
\end{cdrfigure}

The analysis of possible supernova events will make use of two types of information. First, the total energy of the emitted neutrinos will be compared with the expected release in the gravitational collapse.  Note that measurements of all flavors, including $\nu_e$, are needed for the best estimate of the energy release.
Second, the rate of cooling of the proto-neutron state should be measured and compared with what is expected from diffusion of the standard neutrinos. This requires comparing one-second-interval time-integrated spectra successively 
as illustrated in Figure~\ref{fig:coolingcurves}. 

Because DUNE is mostly sensitive to $\nu_e$, in order to enable inference of the fluxes of $\mu$ and $\tau$ flavors complementary $\bar\nu_{e}$ measurements are needed from water Cherenkov and scintillator detectors, as is a careful analysis of the oscillation pattern (see below). Measuring the energy loss rate will require sufficient statistics at late times and, once again, an understanding of the oscillation dynamics; this is evident in Figure~\ref{fig:coolingcurves} where oscillated and unoscillated cases are shown.


The flavor oscillation physics and its signatures are a major part of the physics program. Compared to the well understood case of solar neutrinos, supernova neutrino flavor transformations are much more involved. Besides the facts that neutrinos and antineutrinos of all flavors are emitted and there are two mass splittings -- ``solar'' and ``atmospheric'' --  the physics of the transformations is significantly richer. For example, several seconds after the onset of the explosion, the flavor conversion probability is affected by the expanding shock front and the turbulent region behind it. The conversion process in such a stochastic profile is qualitatively different from the adiabatic MSW effect in the smooth, fixed-density profile of the Sun. 

Even more complexity is brought about by the coherent scattering of neutrinos off each other. This neutrino ``self-refraction'' 
 results in highly nontrivial flavor transformations close to the neutrinosphere, typically within a few hundred kilometers from the center, where the density of streaming neutrinos is very high. Since the evolving flavor composition of the neutrino flux feeds back into the oscillation Hamiltonian, the problem is \emph{nonlinear}. Furthermore, as the interactions couple neutrinos and antineutrinos of different flavors and energies, the oscillations are characterized by \emph{collective} modes.  
This leads to very rich physics that has been the subject of intense theoretical interest over the last decade. A voluminous literature exists exploring these collective phenomena,
e.g.,~\cite{Duan:2005cp,Fogli:2007bk,Raffelt:2007cb,Raffelt:2007xt,EstebanPretel:2008ni,
Duan:2009cd,Dasgupta:2009mg,Duan:2010bg,Duan:2010bf,Wu:2014kaa} the effects of which are not yet fully understood. 
A supernova burst is the only opportunity to study neutrino-neutrino interactions experimentally.

\begin{cdrfigure}[Simulated cooling curves from the Garching light progenitor model]{shockandcollective}{Both panels show solid lines that represent the simulated unoscillated $\nu_e$ (green, cooler) and $\nu_x$ (blue, hotter) fluxes. The filled curve shows the observed flux after the collective oscillations in the absence of (left) and presence of (right) the shock front. (Flux is shown in arbitrary units)}
\includegraphics[width=0.9\textwidth]{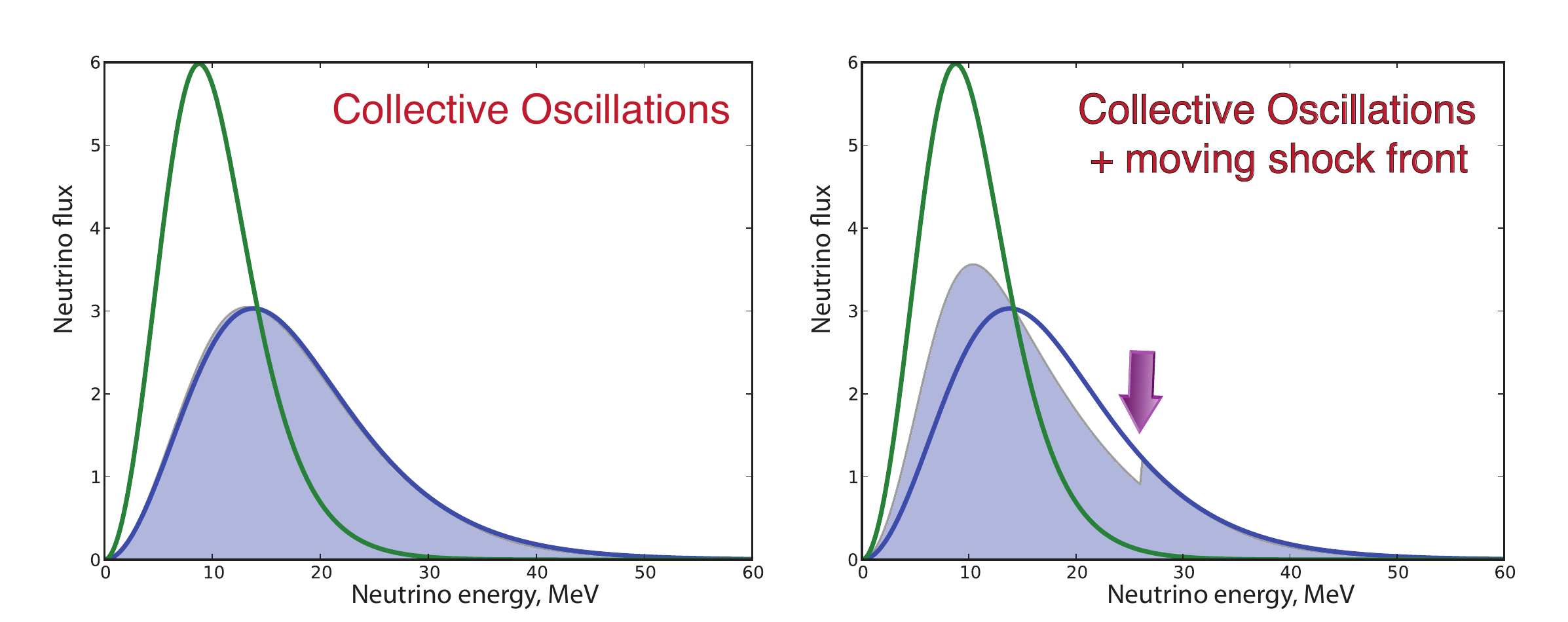}
\end{cdrfigure}

Matter effects in the Earth can also have a MH-dependent effect on the signal (e.g.,~\cite{Choubey:2010up}).

One may wonder whether all this complexity will impede the extraction of useful information from the future signal. In fact, the opposite is true: the new effects can \emph{imprint} information about the inner workings of the explosion on the signal. The oscillations can modulate the characteristics of the signal (both event rates and spectra as a function of time), as seen in Figure~\ref{fig:coolingcurves}. Moreover, the oscillations can imprint \emph{non-thermal} features on the energy spectra, potentially making it possible to disentangle the effects of flavor transformations and the physics of neutrino spectra formation. This in turn should help 
illuminate the development of the explosion during the crucial first 10 seconds.   It is important to note that the features depend on the unknown mass hierarchy, and therefore may help reveal it. 

Figure~\ref{fig:shockandcollective} illustrates the effects of collective oscillations. 
These oscillations serve to permute almost completely the original $\nu_{e}$ and $\nu_{\mu,\tau}$ spectra, so that the flux of observed electron neutrinos is noticeably hotter than the original one. Moreover, the shock front modulates the MSW conversion probability and imprints a non-thermal \emph{step} in the spectrum.  Below this step, the swap between the original $\nu_{e}$ and $\nu_{\mu,\tau}$ spectra is only partial. As the shock expands, the feature moves to higher energies, creating a ``smoking-gun'' signature that exists only in the neutrino channel.

As another example of a probe of new physics with supernova neutrinos or antineutrinos,
a class of tests of Lorentz and CPT violation involves comparing the propagation of neutrinos with other species or of neutrinos of the same flavor but different energies~\cite{Kostelecky:2003cr,Kostelecky:2003xn,Kostelecky:2011gq,Diaz:2009qk}. These amount to time-of-flight or dispersion studies.

Time-of-flight and dispersion effects lack the interferometric resolving power available to neutrino oscillations, but they provide sensitivity to Lorentz- and CPT-violating effects that cannot be detected via oscillations. The corresponding SME coefficients controlling these effects are called oscillation-free coefficients~\cite{Kostelecky:2011gq}.
Supernova neutrinos are of particular interest in this context because of the long baseline, which implies sensitivities many orders of magnitude greater than available from time-of-flight measurements in beams. Observations of the supernova SN1987A yield constraints on the difference between the speed of light and the speed of neutrinos, which translates into constraints on isotropic and anisotropic coefficients in both the minimal and nonminimal sectors of the SME. Knowledge of the spread of arrival times constrains the maximum speed difference between SN1987A antineutrinos of different energies in the approximate range 10--40 MeV, which restricts the possible antineutrino dispersion and yields further constraints on SME coefficients~\cite{Kostelecky:2011gq}.

Analyses of this type would be possible with DUNE if supernova neutrinos are observed. Key features for maximizing sensitivity would include absolute timing information to compare with photon spectral observations and relative timing information for different components of the neutrino energy spectrum. Significant improvements over existing limits are possible.
Figure~\ref{fig:snliv} displays DUNE supernova sensitivities to coefficients for Lorentz and CPT violation that leave neutrino oscillations unaffected and so cannot be measured using atmospheric or long-baseline neutrinos. The figure assumes a supernova comparable to SN1987A (optimistically at a distance of 50~kpc). Studies of supernova neutrinos using DUNE can measure many coefficients (green) at levels that improve on existing limits (grey).

\begin{cdrfigure}[DUNE SN sensitivities to oscillation-free coefficients for Lorentz and CPT violation]{snliv}{DUNE supernova sensitivities to oscillation-free coefficients for Lorentz and CPT violation. Studies of DUNE supernova neutrinos can measure many coefficients (green) at levels improving over existing limits (grey). These Lorentz- and CPT-violating effects leave oscillations unchanged and so are unobservable in atmospheric or long-baseline measurements~\cite{kostelecky}.}
\includegraphics[width=0.9\textwidth]{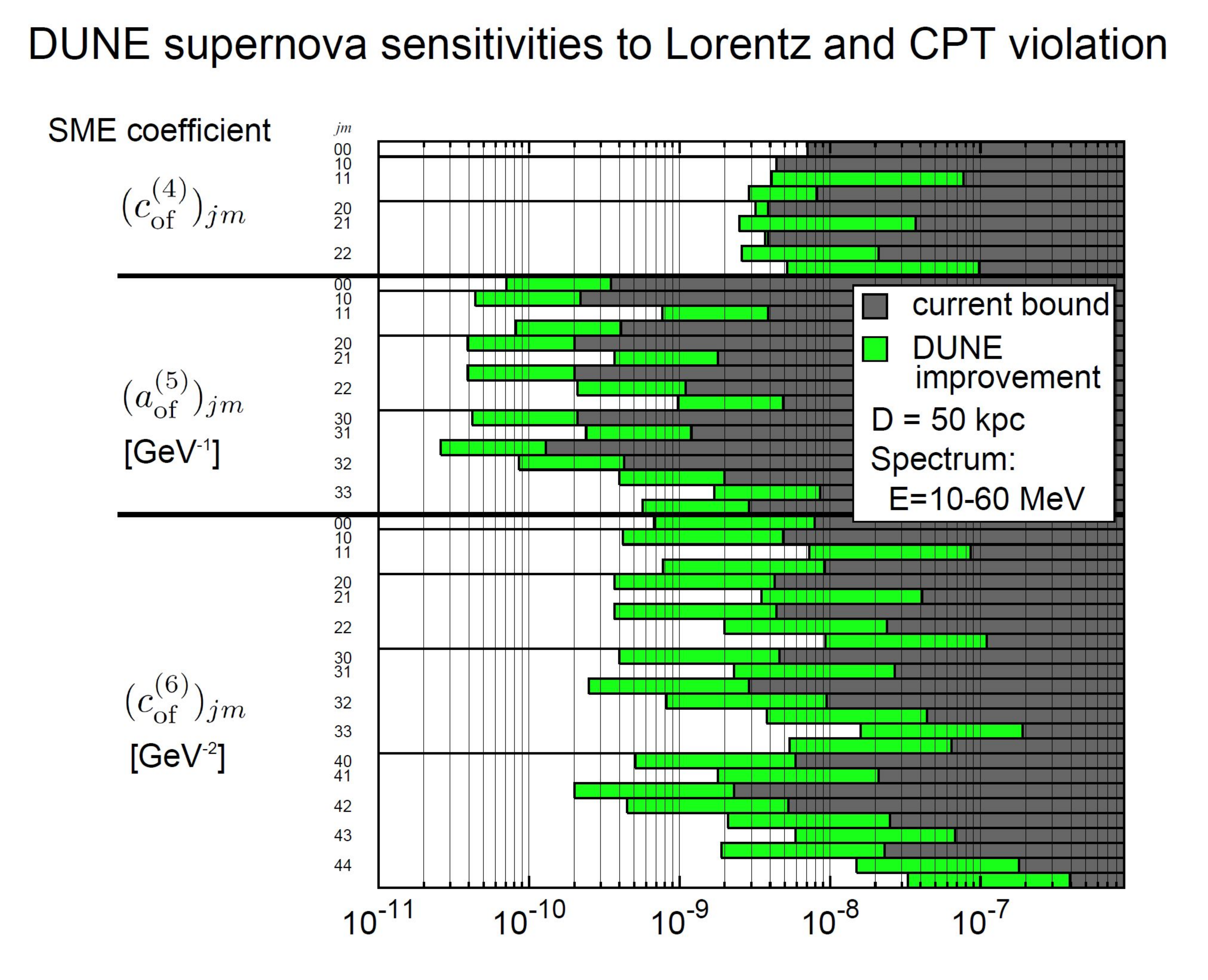}
\end{cdrfigure}

Finally, via detection of time-of-flight delayed $\nu_e$ from the  neutronization burst,  DUNE will be able to probe neutrino mass bounds of $\mathcal{O}(1)$~eV for a 10-kpc supernova~\cite{Rossi-Torres:2015rla} (although will likely not be competitive with near-future terrestrial kinematic limits). If eV-scale sterile neutrinos exist, they will likely have an impact on astrophysical and oscillation aspects of the signal (e.g.,~\cite{Keranen:2007ga,Tamborra:2011is,Esmaili:2014gya}), as well as time-of-flight observables.

\section{Astrophysics}
\label{sec:physics-snblowe-astrophysics}

A number of astrophysical phenomena associated with supernovae are expected to be observable
in the supernova neutrino signal, providing a remarkable window into the event.  In particular, the supernova explosion mechanism, which in the current paradigm involves energy deposition via neutrinos, is still not well understood, and the neutrinos themselves will bring the insight needed to confirm or refute the paradigm.

There are many other examples of astrophysical observables.
\begin{itemize}
\item The short initial ``neutronization'' burst, primarily composed of $\nu_e$, 
  represents only a small component of the total signal.  However,
  oscillation effects can manifest themselves in an observable manner
  in this burst, and flavor transformations can be modified by the
  ``halo'' of neutrinos generated in the supernova envelope by
  scattering~\cite{Cherry:2013mv}. 
  
\item The formation of a black hole would cause a sharp signal cutoff
  (e.g.,~\cite{Beacom:2000qy,Fischer:2008rh}).
\item Shock wave effects (e.g.,~\cite{Schirato:2002tg}) would cause a
  time-dependent change in flavor and spectral composition as the
  shock wave propagates.
\item The standing accretion shock instability
  (SASI)~\cite{Hanke:2011jf,Hanke:2013ena}, a ``sloshing'' mode
  predicted by three-dimensional neutrino-hydrodynamics simulations of
  supernova cores, would give an oscillatory flavor-dependent
  modulation of the flux.
\item Turbulence effects~\cite{Friedland:2006ta,Lund:2013uta} would
  also cause flavor-dependent spectral modification as a function of
  time.
\end{itemize}

Observation of a supernova neutrino burst in coincidence with gravitational waves (which would also be prompt, and could indeed provide a time reference for a time-of-flight analysis) would be especially interesting~\cite{Arnaud:2003zr,Ott:2012jq, Mueller:2012sv, Nishizawa:2014zna}.

The supernova neutrino burst is prompt with respect to the
electromagnetic signal and therefore can be exploited to provide an
early warning to astronomers~\cite{Antonioli:2004zb,Scholberg:2008fa}.  
Additionally, a liquid argon signal~\cite{Bueno:2003ei} is expected to
provide some pointing information, primarily from elastic scattering
on electrons.
We note that not every core collapse will produce an observable supernova, and observation of a neutrino burst in the absence of light would be very interesting. 

Even non-observation of a burst, or non-observation of
a $\nu_e$ component of a burst in the presence of supernovae (or other
astrophysical events) observed in electromagnetic or gravitational
wave channels, would still provide valuable information about the
nature of the sources.  Further, a long-timescale, sensitive search
yielding no bursts will also provide limits on the rate of
core-collapse supernovae.

We note that the better one can understand the astrophysical nature of core-collapse supernovae, the easier it will be to extract information about particle physics.  DUNE's capability to characterize the $\nu_e$ component of the signal is unique and critical.

\section{Additional Astrophysical Neutrinos}
\label{sec:physics-snblowe-other}

\subsection{Solar Neutrinos}

Intriguing questions in solar neutrino physics remain,
even after data
from the Super-K and SNO~\cite{Fukuda:2001nj,Ahmad:2001an}
experiments explained the long-standing mystery of missing solar
neutrinos~\cite{Cleveland:1998nv} as due to flavor
transformations. 
Some unknowns, such as the fraction of energy production via the CNO
cycle in the Sun, flux variation due to helio-seismological modes that
reach the solar core, or long-term stability of the solar core
temperature, are astrophysical in nature. Others directly impact
particle physics. Can the MSW model explain the amount of flavor
transformation as a function of energy, or are non-standard neutrino
interactions required?  Do solar neutrinos and reactor antineutrinos
oscillate with the same parameters? 

Detection of solar and other low-energy neutrinos is challenging in
a LArTPC because of relatively high intrinsic detection energy thresholds for
the charged-current interaction on argon ($>$\SI{5}{\MeV}). 
Compared with other technologies, a LArTPC offers a large
cross section and unique potential signatures from de-excitation
photons. Aggressive R\&D efforts in low-energy triggering and
control of background from radioactive elements may make detection
of solar neutrinos in DUNE possible.

Signatures of solar neutrinos in DUNE
are elastic scattering on electrons as well as CC absorption of $\nu_e$ on $^{40}$Ar (equation~\ref{eq:nueabs}), which has a 4.5-MeV energy threshold and a large cross section compared to elastic scattering on electrons.  Furthermore, the CC absorption differential cross section (the interaction products track neutrino energy closely) potentially enables precise solar-neutrino spectral measurements.
The solar neutrino event rate in a
\ktadj{40} LArTPC, assuming a roughly \MeVadj{4.5} neutrino energy
threshold and 31\% $\nu_e$ survival, is 122 per day.

The solar neutrino physics potential of a large LArTPC depends
on the ability to pick up a low-energy electron, light collection of the photon-triggering system,
and, critically, on background suppression. 
The decay of the naturally occurring $^{39}$Ar
produces $\beta$'s with a \keVadj{567} endpoint and an expected rate
of \SI{10}{\MHz} per \SI{10}{\kt} of liquid argon. This limits the
fundamental reach of DUNE to neutrino interactions with visible
energies above \SI{1}{\MeV}. 
Cosmic-muon and fast-neutrino  interactions with the $^{40}$Ar nucleus (which are rather complex
compared to interactions on $^{16}$O or $^{12}$C) are likely to generate many long-lived spallation products which could limit the
detection threshold for low-energy neutrinos.
$^{40}$Cl, a beta emitter with an
endpoint of \SI{7.48}{\MeV}, is a dominant source of background at
energies above \SI{5}{\MeV}, and is expected to be produced with a rate on the order of 10 per kiloton of LAr per day at 4850 ft.

The ICARUS collaboration has reported a \MeVadj{10}
threshold~\cite{Guglielmi:2012}. Assuming the detector itself
has low enough radioactivity levels, this threshold level would enable
a large enough detector to measure the electron flavor component of
the solar $^8$B neutrino flux with high statistical accuracy. It could 
thereby further test the MSW flavor transformation curve with higher statistical precision and
potentially better energy resolution. 
In addition to these solar
matter effects, solar 
neutrinos also probe terrestrial matter effects
with the variation of the $\nu_e$ flavor observed with solar zenith
angle while the Sun is below the horizon --- the day/night effect (reported recently in ~\cite{Renshaw:2013dzu}). 
The comparison of solar and reactor disappearance tests CPT invariance as well as other new physics.

\subsection{Diffuse Supernova Background Neutrinos}

Galactic supernovae are relatively rare, occurring somewhere between
once and four times a century. In the Universe
at large, however, thousands of neutrino-producing explosions occur
every hour.  The resulting neutrinos --- in fact most of the neutrinos
emitted by all the supernovae since the onset of stellar formation ---
suffuse the Universe.  Known as the \emph{diffuse supernova neutrino background
  (DSNB)}, their energies are in the few-to-\MeVadj{30} range.  The DSNB
has not yet been observed, but an observation would greatly enhance
our understanding of supernova-neutrino emission and the overall
core-collapse rate~\cite{Beacom:2010kk}.

A liquid argon detector such as DUNE's far detector is sensitive to
the $\nu_e$ component of the diffuse relic supernova neutrino flux,
whereas water Cherenkov and scintillator detectors are sensitive to
the antineutrino component.  However, backgrounds in liquid argon are as
yet unknown, and a huge exposure ($>$\SI{500}{\ktyr}s)
would likely be required for observation.  
With tight control of
backgrounds, 
DUNE --- in the long term --- could play a unique and
 complementary role in the physics of relic neutrinos.

Background is a serious issue for DSNB detection.
The solar {\em hep} neutrinos, which have an                
endpoint at \SI{18.8}{\MeV}, will determine the lower bound of the DSNB
search window ($\sim$ \SI{16}{\MeV}).  The upper bound is determined
by the atmospheric ${\nu}_{e}$ flux and
is around \SI{40}{MeV}.
Although the LArTPC provides a unique sensitivity to the
electron-neutrino component of the DSNB flux, early studies indicate
that due to this lower bound of $\sim$ \SI{16}{\MeV}, DUNE would need a huge
mass of liquid argon --- of order \SI{100}{\kt} --- to get more than 4$\sigma$
evidence for the diffuse supernova flux in five
years~\cite{Cocco:2004ac}.
The expected number of relic
supernova neutrinos, $N_{\rm DSNB}$, that could be observed in a
\SIadj{40}{\kt} LArTPC detector in ten years~\cite{Cocco:2004ac}
assuming normal hierarchy is:
\begin{equation}
N_{\rm DSNB} = 46 \pm 10  \ \ \ 16 \, {\rm MeV} \leq E_e \leq 40 \, {\rm MeV}
\label{eqn:srnrate}
\end{equation}
where $E_e$ is the energy of the electron from the CC interaction as
shown in Equation~\ref{eq:nueabs}. 

 The main challenge for detection of such
a low rate of relic neutrinos in a LArTPC is understanding how much of
the large spallation background from cosmic-ray interactions with the
heavy argon nucleus 
leaks into the search window.   Some studies have been done~\cite{Barker:2012nb} but more work is needed.

\subsection{Other Low-Energy Neutrino Sources}

We note some other potential sources of signals in the tens-of-MeV range that may be observable in DUNE.  These include neutrinos from accretion disks~\cite{Caballero:2011dw} and black-hole/neutron star mergers~\cite{Caballero:2009ww}.  These will create spectra not unlike those from core-collapse events, and with potentially large fluxes.  However they are expected to be considerably rarer than core-collapse supernovae within an observable distance range.  There may also be signatures of dark-matter WIMP annihilations in the low-energy signal range~\cite{Rott:2012qb, Bernal:2012qh}.

\section{Detector Requirements}
\label{sec:physics-snblowe-detector-requirements}

For supernova burst physics, the detector must be able to detect and reconstruct 
events in the range 5--100~MeV.  As for proton decay and atmospheric neutrinos, no beam trigger will be available; therefore there must be special triggering and DAQ requirements that take into account the short, intense nature of the burst, and the need for prompt propagation of information in a worldwide context.
The DUNE Far Detector
Requirements~\cite{lbnfdune-cdr-req} specific to supernova burst neutrinos are as follows:

\begin{itemize}

\item Far Detector Depth: The signal to background ratio shall be sufficiently large to identify the  burst ($<$100 seconds)  from a core-collapse supernova within 20~kpc (within the Milky Way). This will require a detector located at sufficient depth for cosmic-ray-related background, including spallation-induced events, to be sufficiently low.  Furthermore, backgrounds from radioactivity or other sources must also be sufficiently low.  Preliminary studies~\cite{gehman} indicate that backgrounds at 4850 ft (including both cosmogenics and intrinsic radioactivity) will be sufficiently low, although more work is needed. 

\item Far Detector Triggering and DAQ:  The far detector shall be capable of collecting information for a supernova burst within the Milky Way.  Events are expected within a time window of approximately 10 seconds, but possibly over an interval as long as several tens of seconds; a large fraction of the events are expected within approximately the first second of the burst.
The data acquisition buffers shall be sufficiently large and the data acquisition system sufficiently robust to allow full capture of neutrino event information for a supernova as close as 0.1 kpc.
At 10~kpc, one expects thousands of events within approximately 10 seconds, but a supernova at a distance of less than 1~kpc would result in $10^5-10^7$  events over 10 seconds.    

The far detector shall have high uptime ($>$90\%) with little event-by-event deadtime to allow the capture of low-probability astrophysical events that could occur at any time with no external trigger. 
Supernova events are expected to occur a few times per century within the Milky Way galaxy. For any 10-year period, the probability of a supernova could be 20 to 30\%.  Capturing such an event at the same time as many of the other detectors around the Earth is very important.  

The DUNE detector systems shall be configured to provide  information to other observatories on possible astrophysical events (such as a galactic supernova) in a short enough time to allow global coordination.   
To obtain maximum scientific value out of a singular astronomical event, it is very important to inform all other observatories (including optical ones) immediately, so that they can begin observation of the evolution of the event. 

\item Far Detector Event Reconstruction:   
The far detector shall be capable of collecting low energy ($<$\SI{100}{\MeV})  charged-current electron neutrino interactions on $^{40}$Ar nuclei that arrive in a short period of time. The final-state electron (or positron) shall be detected and its energy measured.   An energy threshold of 5 MeV or better is highly desirable; most supernova burst events are expected to have energy depositions in the range 5--50~MeV.
Energy and event time resolution must be sufficient to resolve interesting physics features of the burst.  Preliminary studies suggest that resolution measured by Icarus for low-energy events~\cite{Amoruso:2003sw} should be adequate, and that approximately millisecond event time resolution should be sufficient to resolve features such as the neutronization burst and the preceding short notch due to neutrino trapping in the $\nu_e$ spectrum (see the luminosity curve of Figure~\ref{fig:garching}), given adequate statistics.   

Detection of gamma ray photons from the final-state excited nucleus could lead to additional electronics requirements.  

\end{itemize}

The other low-energy physics described in Section~\ref{sec:physics-snblowe-other} typically requires event reconstruction capabilities similar to supernova-burst physics; however, background requirements are much more stringent for these (especially for DSNB).  Realistic background conditions in the few-tens-of-MeV range are not currently  very well understood.  
These physics topics do not drive detector requirements, although it may still be possible for DUNE to address them if backgrounds can be kept sufficiently well under control.

\cleardoublepage

\chapter{Near Detector Physics}
\label{ch:physics-nd}

\section{Introduction and Motivation}
\label{sec:physics-nd-introduction}

The LBNF neutrino beam used to study neutrino oscillations in DUNE is
an extended source at the near site, therefore every single spectrum
induced by the neutrino charged (CC) and neutral (NC) current
interactions --- $\nu_\mu$-CC, $\bar \nu_\mu$-CC, $\nu_e$-CC, $\bar
\nu_e$-CC, and the NC --- is different when measured at the far
detector versus the near detector.  In order to achieve the systematic
precision for the signal and background events in the far detector,
which ideally should always be lower than the corresponding
statistical error, the near detector measurements --- including
neutrino fluxes, cross sections, topology of interactions and smearing
effects --- must be unfolded and extrapolated to the far detector
location.  The
charge, ID, and the momentum vector resolution of particles produced
in the neutrino interactions are 
key to constraining the systematic uncertainties in the predictions at
far detector.

To this end, it is useful to recall that for the LBNF low-energy reference beam
(80-GeV protons, 1.07 MW, $1.47 \times 10^{21} $ POT/year), the 
event rates expected at the 40-kt far detector per year are 
2900 (1000) for the $\nu_\mu (\bar \nu_\mu)$
disappearance channel and 230 (45) for the $\nu_e(\bar \nu_e)$
appearance channel (for $\delta_{CP}=0$, normal hierarchy and assumed best fit values of
the mass-squared differences and mixings). For comparison, the raw
event rates per ton of near detector target mass (without detector
effects) for various neutrino interactions in the near detector at
\SI{574}{\m} from the proton beam target are summarized in
Table~\ref{tab:rates}. The rates are indicated for a ton of target
mass of Ar (Carbon) per $10^{20}$ protons-on-target.  The mass of Ar
in the near detector targets is required to have sufficient mass to
provide $\times$ 10 the statistics of the far detector. Although the
Ar-target design is preliminary, the Ar mass is expected to be
approximately 100kg.

\begin{cdrtable}
[Interaction rates, $\nu$ mode, per ton
for \SI{1e20}{\POT}, \SI{574}{\meter}, \SI{120}{\GeV}]
{lrr}{rates}
{Estimated interaction rates on Ar (Carbon) in the neutrino (second column) and antineutrino (third column) beams per ton of detector 
  for \SI{1e20}{\POT} at \SI{574}{\meter} assuming neutrino
  cross-section predictions from GENIE~\cite{GENIE} and a \GeVadj{120}
  proton beam using the optimized design (Section~\ref{sec:alternative-focusing-systems}).  Processes are defined at the initial neutrino
  interaction vertex and thus do not include final-state effects. These estimates do not
  include detector efficiencies or acceptance~\cite{DOCDB740,DOCDB783}. 
}
Production mode & $\nu_\mu$ Events  & $\overline\nu_\mu$ Events \\
\rowtitlestyle
                & on Ar (Carbon) & on Ar (Carbon)\\
\toprowrule
CC QE ($\nu_\mu n \rightarrow \mu^- p$)                                       & 30,000 (28,000) & 13,000 (15,000) \\ \colhline  
NC elastic ($\nu_\mu N \rightarrow \nu_\mu N$)                                & 11,000 (11,000) & 6,700 (68,00) \\ \colhline  
CC resonant ($\nu_\mu p \rightarrow \mu^- p \pi^+$)                           & 21,000 (24,000) &      0 (0) \\ \colhline  
CC resonant ($\nu_\mu n \rightarrow \mu^- n \pi^+\,(p\pi^0)$)                 & 23,000 (21,000) &      0 (0) \\ \colhline   
CC resonant ($\bar\nu_\mu p \rightarrow \mu^+ p \pi^-\,(n\pi^0)$)             &      0 (0)      &  83,00 (7,800) \\ \colhline
CC resonant ($\bar\nu_\mu n \rightarrow \mu^+ n \pi^-$)                       &      0 (0)      & 12,000 (8,100) \\ \colhline
NC resonant ($\nu_\mu p \rightarrow \nu_\mu p \pi^0\,(n\pi^+)$)               & 7,000 (9,200) &      0 (0) \\ \colhline
NC resonant ($\nu_\mu n \rightarrow \nu_\mu n\pi^+\,(p\pi^0)$)                & 9,000 (11,000) &      0 (0) \\ \colhline
NC resonant ($\bar\nu_\mu p \rightarrow \bar\nu_\mu p\pi^-\,(n\pi^0)$)        &  0 (0)          & 3,900 (4,300)     \\ \colhline
NC resonant ($\bar\nu_\mu n \rightarrow \bar\nu_\mu n \pi^-$)                 &  0 (0)          & 4,700 (4,300)     \\ \colhline
CC DIS ($\nu_\mu N \rightarrow \mu^- X$ or 
$\overline{\nu}_\mu N \rightarrow \mu^+ X$)                                   & 95,000 (92,000) & 24,000 (25,000) \\ \colhline
NC DIS ($\nu_\mu N \rightarrow \nu_\mu X$ or 
$\overline{\nu}_\mu N \rightarrow \overline{\nu}_\mu X$)                      & 31,000 (31,000)   &  10,000 (10,000) \\ \colhline
CC coherent $\pi^+$ ($\nu_\mu A \rightarrow \mu^- A \pi^+$)                   & 930 (1,500)     &      0 (0) \\ \colhline
CC coherent $\pi^-$ ($\overline{\nu}_\mu A \rightarrow \mu^+ A \pi^-$)        &     0 (0)         &  800 (1,300) \\ \colhline
NC coherent $\pi^0$ ($\nu_\mu A \rightarrow \nu_\mu A \pi^0$ or 
$\overline{\nu}_\mu A \rightarrow \overline{\nu}_\mu A \pi^0$)                & 520 (840)       &  450 (720) \\ \colhline
NC elastic electron ($\nu_\mu e^- \rightarrow \nu_\mu e^-$  
or  $\overline{\nu}_\mu e^- \rightarrow \overline{\nu_\mu} e^-$)              & 16 (18)           & 11 (12) \\ \colhline
Inverse Muon Decay ($\nu_\mu e \rightarrow \mu^- \nu_e$)                      & 9.5 (11)           & 0 (0) \\ \colhline
\toprowrule
Total CC                                                         & 170,000 (170,000) & 59,000 (61,000) \\ 
Total NC+CC                                                      & 230,000 (230,000) & 84,000 (87,000) \\
\end{cdrtable}

Importantly, given the scale and ambition of LBNF/DUNE, the near
detector must offer a physics potential that is as rich as those
offered by collider detectors.  One of the main advantages of a
high-resolution near detector built according to the reference design
(detailed in Volume 4, Chapter 7 of this CDR) is that it will offer a
rich panoply of physics 
spanning an estimated 100 topics and resulting in over 200
publications and theses during a ten-year operation.


\section{Physics Goals of the Near Detector}
\label{sec:physics-nd-goals}

The physics goals of the DUNE near detector fall under three categories: 

\begin{itemize}
\item constraining the systematic uncertainties in  oscillation studies
\item  offering a generational advance in the precision measurements of neutrino interactions, e.g., 
cross sections, exclusive processes, electroweak and isospin physics, structure of nucleons and nuclei 
\item conducting searches for new physics covering unexplored regions, 
including heavy (sterile) neutrinos, large $\Delta m^2$ neutrino oscillations, light Dark Matter 
candidates, etc. 
\end{itemize}

These three broad goals possess significant synergy. The physics
requirements for the near detector are driven by the oscillation
physics. However, the unprecedented neutrino fluxes available at LBNF
and the challenging constraints required by the long-baseline program,
especially those related to the CP measurement, also make the near
detector imminently suitable for short-baseline precision physics. And
conversely, conducting precision measurements of neutrino interactions
will actually 
result in a reduction of systematic uncertainties on signal and
background predictions in the far detector~\cite{HIRESMNU, DPR,
  Adams:2013qkq}.

\section{The Role of the Near Detector in Oscillation Physics} 
\label{sec-nd-oscl} 

As illustrated in Chapter~\ref{ch:physics-lbnosc}, studies on the
impact of different levels of systematic uncertainties on the
oscillation analysis indicate that uncertainties exceeding 1\% for
signal and 5\% for backgrounds may result in substantial degradation
of the sensitivity to CP violation and mass hierarchy.
The near detector physics measurements discussed in this section are
needed in order to match this level of systematic uncertainty.

The near detector will need to determine the relative abundance and
energy spectrum of all \textit{four} species of neutrinos in the LBNE
beam.
This requires measurement of $\nu_\mu$, $\bar \nu_\mu$, $\nu_e$, and
$\bar \nu_e$ via their CC-interactions, which in turn demands precise
measurement of $\mu^-$, $\mu^+$, $e^-$, and $e^+$ in the near
detector. Specifically, to measure both the small $\nu_e$ and $\bar
\nu_e$ contamination in the beam with high precision, the detector
would need to be able to distinguish $e^+$ from $e^-$.  This last
requirement is motivated by
\begin{enumerate}
\item the need to measure and identify $\nu - e$ NC elastic scattering
  (and calibrate the corresponding backgrounds) for the absolute flux
  measurements
\item the redundancy in determining the momentum distributions of the
  neutrinos' parent mesons, in particular, the $K^0$ mesons using $\bar
  \nu_e$-events, which are essential ingredients for predicting the
  far detector/near detector flux ratio as a function of energy
\item the measurement of the $\pi^0$ yield in CC and NC interactions 
from converted photons
\item the different composition in terms of QE, single-pion resonance,
  multi-pion resonance, and deep inelastic scattering (DIS) of CC and
  NC events originated by each of the four species in the far detector
  and near detector, respectively
\end{enumerate}

Quantifying asymmetries between neutrinos and antineutrinos, such as
energy scales and interaction topologies, which are relevant for the
measurement of the CP-violating phase, is another job of the near
detector.
Since the reference near neutrino detector, the fine-grained tracker
(FGT), is not identical to the far detector, it is not possible in
long-baseline analyses to ``cancel'' the event reconstruction errors
in a near-to-far ratio.  The extent to which such a cancellation will
limit the ultimate precision of the experiment has yet to be fully
explored.  Because of the low average density of the FGT (0.1
g/cm$^3$), however, DUNE will be able to measure the missing
transverse momentum ($p_{T}$) vector in the CC processes, in addition
to accurately measuring the lepton and hadron energies.  This
redundant missing-$p_{T}$ vector measurement provides a most important
constraint on the neutrino and antineutrino energy
scales. Measurements of exclusive topologies like quasi-elastic,
resonance and coherent meson production offer additional constraints
on the neutrino energy scale.

In the disappearance studies, the absolute $\nu_\mu$- and $\bar
\nu_\mu$ flux should be determined to $\simeq 3\%$ precision in $0.5
\leq E_\nu \leq 8$~GeV so as to eliminate uncertainties in the
neutrino and antineutrino cross sections affecting the oscillation
measurements. For precision measurements of electroweak and QCD
physics, a similar precision is required at higher energies.

 
For precision $\nu_\mu$- and $\bar \nu_\mu$-disappearance channels,
the far detector/near detector ratio of the number of neutrinos
($\nu_\mu$ and $\bar \nu_\mu$) at a given $E_\nu$ bin in $0.5 \leq
E_\nu \leq 8$ GeV range should be known to $\simeq 1-2\%$
precision. The capability to precisely measure the muon momenta and
the low-hadronic energy of the near detector will enable this
precision.


NC processes constitute one of the largest backgrounds to all
appearance and disappearance oscillation channels. It is therefore
important for the near detector to make a precise measurement of the
NC cross section relative to CC as a function of the hadronic energy,
$E_{Had}$.

A precise measurement of $\pi^0$ and photon yields by the near
detector in \textit{both} $\nu$-induced NC and CC interactions is
essential; this is the most important background to the $\nu_e$- and
$\bar \nu_e$ appearance at low energies.


The $\pi^{\pm}$ content in CC and NC hadronic jets is the most important background to 
the $\nu_\mu$- and $\bar \nu_\mu$ disappearance coming from the hadronic   $\pi^{\pm} \rightarrow \mu^{\pm}$; 
it can also be a background to the appearance channel at lower energies. 
By separately measuring the momenta of $\pi^+$ and $\pi^-$ produced in each of 
the CC- and NC-induced hadronic shower, the ND 
will in situ measure the $\pi^{\pm}$ content in CC and NC events, and hence the corresponding backgrounds 
to the disappearance signal.


Precise near detector measurements leading to characterization of 
 various exclusive (semi-exclusive) 
channels such as Quasi-Elastic (QE), resonance (Res), 
coherent-mesons and Deep-Inelastic-Scattering (DIS)
will yield in situ constraints on the nuclear effects from both
initial and final interaction (FSI).


The near detector will quantify the neutrino-argon cross section by
measuring interactions off Ar, Ca, C, H, etc. targets. The goal is to
provide a consistent model, as opposed to an empirical
parameterization, for the nuclear effects.

Finally, the near detector will constrain NC and CC backgrounds to the
$\tau-$ appearance in the far detector.  
One of the
unique capabilities of the liquid argon (LAr) far-detector is 
identification of the tau appearance with high fidelity. The ND measurements
would considerably reduce the error in the production of the
tau-lepton (large $x_{BJ}$ data) and would constrain the backgrounds
that mimic the tau signal.

The above requirements suggest a high-resolution, magnetized near
detector for identifying and measuring $e^+$, $e^-$, $\mu^-$, $\mu^+$,
$\pi^0$, $\pi^{+,-}$ and protons with high efficiency.

\section{Precision Measurements at the Near Detector} 
\label{sec-nd-sbp} 

Over a five-year run in neutrino mode, 
the intense neutrino source at LBNF will provide ${\cal {O}}$(100)
million neutrino interactions in a 7-t near detector; and about $0.4$
times as many in antineutrino mode. 
The high-resolution, fine-grained near detector described in \voldune
would offer not only the requisite systematic precision for
oscillation studies, but also a generational advance in the precision
measurements and unique searches that a neutrino beam can provide.
This section outlines the salient physics reach of this detector;
further details can be found in~\cite{DPR}
and~\cite{Adams:2013qkq}. Discussed first are precision measurements
that would support and impact the oscillation physics program,
followed by examples of other non-oscillation-related physics
measurements that would extend our knowledge of important aspects of
particle physics.

\subsection{Precision Measurements Related to Oscillation Physics}


Using $\nu$-electron NC scattering, the absolute neutrino flux can be
determined to $\leq 3\%$ precision in the range $0.5 \leq E_\nu \leq
10$ GeV. Additionally, the $\nu$-electron CC scattering leading to
inverse muon decay would determine the absolute flux to $\simeq 3\%$
precision in $E_\nu \geq 20$ GeV region~\cite{ABS-FLUX}.  The DUNE
near detector's ability to determine the background (primarily from
$\nu_{\mu}$ quasi-elastic scattering) to inverse muon decay
\textit{without} relying on $\bar \nu_\mu$ measurements or ad hoc
extrapolations, such as made in
CCFR~\cite{CCFR-IMD-Mishra-89},~\cite{CCFR-IMD-Mishra-90} and
CHARM~\cite{CHARM-IMD-95}, allows such precisions, which are dominated
by statistics.  Importantly, the ability to extract the quasi-elastic
$\bar \nu_\mu$-H interaction, via subtraction of hydrocarbon and pure
carbon targets, would allow an extraction of the $\bar \nu_\mu$
absolute flux to a few percent precision. Furthermore, novel
techniques such as the use of coherent-$\rho$ meson production in the
near detector combined with photo-production could provide a
constraint on the absolute flux to $\simeq 5\%$ in the intermediate $5
\leq E_\nu \leq 20$ GeV region.  We note that this near detector will
be the first 
to in-situ-constrain the absolute flux to a level approaching
$\sim$2.5\% precision.

The most promising method of determining the shape of $\nu_\mu$ and
$\bar \nu_\mu$ flux is by measuring the low-hadronic energy CC, the
Low-$\nu_0$ method~\cite{MISHRA-Nu0}. The method, when combined with
the empirical parameterization of the $\pi^{\pm}$, $K^{\pm}$, and such
hadroproduction data as would be available in the coming decade,
permits a bin-to-bin precision of 1--2\% on the flux spanning $1 \leq
E_\nu \leq 50$ GeV.  Recent model
calculations by Bodek et al. ~\cite{Bodek:2012uu} confirm these error
estimates.  Specifically, for the $\nu_\mu$ and $\bar \nu_\mu$
disappearance the Low-$\nu_0$ method would predict the far
detector/near detector($E_\nu$) to 1--2\% precision.

By precisely measuring the $\nu_\mu$-, $\bar \nu_\mu$-, $\nu_e$-,
$\bar \nu_e$-CC, the near detector will decompose the $\pi^+$, $K^+$,
$\pi^-$, $K^-$, $\mu^+$, $\mu^-$, and $K^0_L$ contents of the beam,
thus allowing a precise far detector/near detector($E_\nu$) prediction
to a few percent.  This ability lends a unique power to not only
measure the cross sections of all four neutrino species, but also to
allow sensitive searches for new physics, e.g., violation of
universality and large-$\Delta$m$^2$ oscillations, as
discussed 
below.

\noindent
For the $\delta_{CP}$ in particular, the near detector measurements
would constrain the $\nu_e$/$\nu_\mu$ to $<1\%$ and $\bar \nu_e$/$\bar
\nu_\mu$ to $\simeq 1\%$ precision, thereby vastly reducing the
associated error.

The high resolution measurements of $E_\mu$ and $E_e$ and those of
charged hadrons and the reconstruction of about one million $K^0_S$
and several million $\pi^0$ will provide a tight constraint on the
(anti)neutrino energy scale.  However, nuclear physics, including
initial and final state interactions, affects the $E_\nu$-scale and
can affect $\nu$ differently from $\bar\nu$ interactions thus
producing a spurious contribution to any measured CP-violating
observable.  The unique experimental handle on these seemingly
intractable effects comes from a precise measurement of the missing
transverse momentum {\it vector},{ \bf{ $P^m_T$}} afforded by the
high-resolution near detector.

The ability to determine {\bf{$P^m_T$}} affords an event-by-event 
identification of NC events. This is particularly crucial in order to decompose the background contributions to 
the $\nu_e$ or $\bar \nu_e$ appearance and the disappearance measurements.

The yields and momentum vector measurements of 
$\pi^0$, $\pi^+$, $\pi^-$, $K^+$, $K^-$, proton, $K^0_S$ and $\Lambda$
particles in CC and NC, as a function of visible energy, will provide
an ``event generator'' $measurement$ for the far detector and
constrain the ``hadronization'' error to $\leq 2.5\%$ associated with
the far detector prediction.


In any long-baseline neutrino oscillation program, including
LBNF/DUNE, the quasi-elastic (QE) interactions are special. First, the
QE cross section is substantial, especially at the second oscillation
maximum.  Second, because of the simple topology --- a muon and a
proton --- QE provides, to first order, a close approximation to
$E_\nu$.  Precise momentum measurements of this two-track topology
impose direct constraints on nuclear effects associated with both
initial and final state interactions.

Resonance is the second most dominant interaction mode, besides deep
inelastic scattering (DIS), in the LBNF/DUNE oscillation range, $0.5
\leq E_\nu \leq 10$ GeV.
By measuring the complete topology of resonance, a high-resolution
near detector will offer an unprecedented precision on the resonance
cross section, and will provide in situ constraints on the nuclear
effects.

Precision measurements of structure functions and differential cross
sections would directly affect the oscillation measurements by
providing accurate simulations of neutrino interactions. They would
also offer measure of that background processes that are dependent
upon the angular distribution of the outgoing lepton,
i.e. $x_{BJ}$-distribution, in the far detector.  Furthermore, the
differential cross section measurements at the ND would constitute one
of the most important ingredients to the `QCD-fitting' enterprise:
extraction of the parton distribution functions (PDF), benefiting not
only neutrino physics but also hadron-collider analyses.  Under the
rubric of nucleon structure and QCD, the topics include:
\begin{enumerate}
\item Measurement of form factors and structure functions
\item QCD analysis,  tests of perturbative QCD and quantitating the non-perturbative 
QCD effects
\item d/u Parton distribution functions at large $x$, which is the limiting error in the 
$\nu_\tau$-CC measurements/searches at the far detector
\item Sum rules and the strong coupling constant; and v) Quark-hadron duality
\end{enumerate}

An integral part of the near detector physics program is a set of
detailed measurements of (anti)neutrino interactions in argon and in a
variety of nuclear targets including calcium, carbon, hydrogen (via
subtraction), and steel.  The goals are
twofold, (1) obtain a model-independent direct measurement of nuclear
effects in Ar using the FGT's ability to isolate $\nu (\bar \nu)$
interactions off free hydrogen via subtraction of hydrocarbon and
carbon targets; and (2) measure the neutrino-nuclear interactions so as
to allow an accurate modeling of initial and final state effects. The
studies would include the nuclear modification of form factors and
structure functions, effects in coherent and incoherent regimes,
nuclear dependence of exclusive and semi-exclusive processes, and
nuclear effects including short-range correlations, pion-exchange
currents, pion absorption, shadowing, initial-state interactions and
final-state interactions.

\subsection{Other Precision Measurements}

Neutrinos and antineutrinos are the most effective probes for
investigating electroweak physics.  Interest in a precise
determination of the weak mixing angle ($\sin^2 \theta_W$) at DUNE
energies via neutrino scattering is twofold: (1) it provides a direct
measurement of neutrino couplings to the $Z$ boson, and (2) it probes
a different scale of momentum transfer than LEP did by virtue of not
being at the $Z$ boson mass peak.
The weak mixing angle can be extracted experimentally from several
independent NC physics processes: (1) deep inelastic scattering off
quarks inside nucleons: $\nu N \to \nu X$; (2) elastic scattering off
electrons: $\nu e^- \to \nu e^-$; (3) elastic scattering off protons:
$\nu p \to \nu p$; iv) coherent $\rho^0$ meson production.  Note that
these processes involve substantially different scales of momentum
transfer, providing a tool to test the running of $\sin^2 \theta_W$
within a single experiment.

The most sensitive channel for $\sin^2 \theta_W$ in the DUNE near
detector is expected to be the $\nu N$ DIS through a precision
measurement of the NC/CC cross-section
ratio. This measurement will be dominated by
systematic uncertainties, which can be accurately constrained by
dedicated in situ measurements using the large CC samples and
employing corresponding improvements in theory that will have evolved
over the course of the experiment. Using the existing knowledge of
structure functions and cross-sections we expect a relative precision
of about $0.35\%$ on $\sin^2 \theta_W$, with the default low-energy
LBNF beam. An increase of the fluxes with a beam upgrade and/or a one
year run with a high energy tuning of the neutrino spectrum would
allow a substantial reduction of uncertainties down to about $0.2\%$.
This level of precision is comparable to colliders (LEP) and offers a
shot at discovery.
 
The various independent channels measured in the DUNE near detector
can be combined though global electroweak fits, further optimizing the
sensitivity to electroweak parameters. The level of precision
achievable as well as the richness of the physics measurements put the
DUNE near detector electroweak program on par with the gold standard
electroweak measurements at LEP.

One of the most compelling physics topics accessible to the DUNE-near
detector in the LBNF-beam is the isospin physics using neutrino and
antineutrino interactions. Given the statistics and a commensurate
resolution of near detector, for the first time we have a chance to
test the Adler sum-rule to a few percent level, and perhaps claim a
discovery.  Precision test of sum-rules is a rich ground for finding
something new, refuting the prevalent wisdom.  An added motivation is
the possibility of isospin asymmetry in nucleons .

To accomplish this, neutrino and anti-neutrino scattering off hydrogen
is needed.  Whereas the Adler sum rule is the prize, the $\bar
\nu_\mu$-H and $\nu_\mu$-H scattering will provide (a) the absolute
flux normalization via low-Q$^2$ $\bar \nu_\mu$-QE interactions, and
(b) will be crucial to achieve a model-independent measurement of
nuclear effects in the neutrino-nuclear interactions.

The question of whether strange quarks contribute substantially to the
vector and axial-vector currents of the nucleon remains unresolved. A
large observed value of the strange-quark contribution to the nucleon
spin (axial current), $\Delta s$, would enhance our understanding of
the proton structure.

The strange \emph{axial vector} form factors are poorly
determined. The most direct measurement of $\Delta s$, which does not
rely on the difficult measurements of the $g_1$ structure function at
very small values of the Bjorken variable $x$, can be obtained from
(anti)neutrino NC elastic scattering off protons.  %

The low-density magnetized tracker in DUNE near detector can provide a
good proton reconstruction efficiency as well as high resolution on
both the proton angle and energy, down to $Q^2\sim0.07$~GeV$^2$.  This
capability will reduce the uncertainties in the extrapolation of the
form factors to the limit $Q^2 \to 0$. About $2.0 (1.2) \times 10^6$
$\nu p (\overline{\nu} p)$ events are expected after the selection
cuts in the low-density tracker, yielding a statistical precision on
the order of 0.1\%.

\section{New Physics Searches} 
\label{sec-nd-np} 

A search for heavy neutrinos is intriguing.  The most economic way to
handle the problems of neutrino masses, dark matter and baryon
asymmetry of the universe in a unified way may be to add to the SM
three Majorana singlet fermions with masses roughly on the order of
the masses of known quarks and leptons.
The lightest of the three new leptons is expected to have a mass from
1~keV to 50~keV and play the role of the Dark Matter particle (for
details and additional references, see~\cite{DPR} and
~\cite{Adams:2013qkq}).

The most effective mechanism of sterile neutrino production is through
weak two body and three body decays of heavy mesons and baryons. In
the search for heavy neutrinos, the strength of the proposed
high-resolution near detector, compared to earlier experiments lies in
reconstructing the exclusive decay modes, including electronic,
hadronic and muonic channels. Furthermore, the detector provides a
means to constrain and measure the backgrounds using control
samples. Preliminary investigations of these issues are ongoing and
suggest that the FGT will have an order of magnitude higher
sensitivity in exclusive channels than previous experiments did.


The near detector could potentially search for large $\Delta$m$^2$
oscillations. As has become evident over the past decade or more,
there may be evidence from several distinct experiments that points
towards the existence of sterile neutrinos with mass in the range
1~eV$^2$ (for details and additional references, see~\cite{DPR}
and~\cite{Adams:2013qkq}).  A short-baseline neutrino program has been
initiated at Fermilab and elsewhere to clear the questions raised by
these varying pieces of evidence.

Since the DUNE near detector is located at a baseline of several
hundred meters and uses the LE beam, it has values of $L/E \sim 1$,
which render it sensitive to these oscillations --- if they exist. Due
to the differences between neutrinos and antineutrinos, four
possibilities have to be considered in the analysis: $\nu_\mu$
disappearance, $\bar \nu_\mu$ disappearance, $\nu_e$ appearance and
$\bar \nu_e$ appearance.  It must be noted that the search for the
high $\Delta$m$^2$ oscillations must be performed simultaneously with
the in situ determination of the fluxes.  To this end, it is necessary
to obtain an independent prediction of the $\nu_e$ and $\bar \nu_e$
fluxes starting from the measured $\nu_\mu$ and $\bar \nu_\mu$ CC
distributions, since the $\nu_e$ and $\bar \nu_e$ CC distributions
could be distorted by the appearance signal.  An iterative procedure
has been developed to handle this, details of which
can be found in~\cite{DPR} and~\cite{Adams:2013qkq}.\\

Recently, a great deal of interest has been generated in searching for
DM at low-energy, fixed-target experiments.  High-flux neutrino beam
experiments, as DUNE is planned to be, have been shown to provide
coverage of the DM and DM-mediator parameter space that can be
covered by neither direct detection nor collider experiments. Upon
striking the target, the proton beam can produce dark photons either
directly through $pp(pn)\rightarrow \bf {V}$ or indirectly through the
production of a $\pi^{0}$ or a $\eta$ meson which then promptly decays
into a SM photon and a dark photon. For the case of $m_{V}\geq
2m_{DM}$, the dark photons will quickly decay into a pair of DM
particles.  These relativistic DM particles from the beam will travel
along with the neutrinos to the DUNE near detector.  The DM particles
can then be detected through neutral-current-like interactions either
with electrons or nucleons in the detector.

Since the signature of DM events looks just like those of the
neutrinos, the neutrino beam provides the major source of background
for the DM signal. Several ways have been proposed to suppress
neutrino backgrounds by using the unique characteristics of the DM
beam. Since DM, due to much higher mass, will travel much more slowly
than the neutrinos, the timing in the near detector becomes a
discriminator.  In addition, since the electrons struck by DM will be
much more forward in direction, the angles of these electrons may be
used to reduce backgrounds, taking advantage of the fine angular
resolution of the DUNE near detector.  Finally, a special run can be
devised to turn off the focusing horn to significantly reduce the
charged particle flux that will produce neutrinos. Further studies are
required to determine appropriate hardware-parameter choices that
could benefit these searches, including granularity, absorbers, timing
resolution, DAQ-speed, etc. Studies are also required to determine if
DUNE will effectively cover the important region in parameter space
between the MiniBooNE exclusion and the direct detection region of the
most popular candidates.
 
DUNE will be the first long-baseline experiment possessing a large
statistics of high-energy $\nu_\mu$, $\bar \nu_\mu$, and $\nu_e$ +
$\bar \nu_e$ CC and NC events measured with high precision in the
liquid argon far detector. An obvious venue for discovery is to search
for distortion at energies greater than 10~GeV, not envisioned by the
PMNS mixing.

DUNE will also be the first long-baseline experiment with the capability
to reconstruct $\nu_\tau$-appearance with high statistics. The paucity
of 
measured $\nu_\tau$-CC motivates searching for new
physics.  The role of near detector --- where no $\tau$ is expected
--- will be to accurately ``calibrate'' background topologies in the
NC and CC interactions, most notably at large $x_{bj}$.

\section{Summary}
\label{sec:physics-nd-summary}

The DUNE near detector, as embodied in the FGT reference design, will
offer a rich physics portfolio that will not only support and buttress
the oscillation program at the far detector, but also extend our
knowledge of fundamental interactions and the structure of nucleons
and nuclei, possibly leading to the discovery of new phenomena. It
will do so by providing tracking of charged particles at levels of a
precision unattained by previous experiments.  In the above, we have
tried to summarize these capabilities and provide a glimpse of its
impact on the overall DUNE physics program.


\cleardoublepage

\chapter{Summary of Physics}
\label{ch:physics-summary}

The primary science goals of DUNE are drivers for the advancement of
particle physics. The questions being addressed are of wide-ranging
consequence: the origin of flavor and the generation structure of the
fermions (i.e., the existence of three families of quark and lepton
flavors), the physical mechanism that provides the CP violation needed
to generate the Baryon Asymmetry of the Universe, and the high energy
physics that would lead to the instability of matter.  Achieving these
goals requires a dedicated, ambitious and long-term program. 

Observation of $\mathcal{O}(1000)$ $\nu_\mu \rightarrow \nu_e$ events
in the DUNE LArTPCs can be achieved with moderate exposures of around
300~\ktMWyr, depending on the beamline design. When coupled with a
highly capable near detector and sophisticated analysis techniques to
control systematics to a few percent, this level of statistics will enable
discovery ($5\sigma$) of CP violation if it is near maximal, and an
unambiguous highly precise measurement of the mass hierarchy for all
possible values of \deltacp. With an optimized beam design, a
precision of $10^\circ$ on $\delta_{\rm CP} = 0$ is also achievable
with this level of exposure.  Exposures of 850 to 1320~\ktMWyr{} 
(depending on the beam design) would be needed to reach $3\sigma$
sensitivity to CP violation for 75\% of all values of \deltacp{} --- the
goal set by the P5 advisory panel. The example staging plan for DUNE,
detailed in Chapter 2 of \volintro, would enable DUNE to meet the P5
goal in less than 14 years with an optimized beam design. It is
important to note that exposures in the range of 100--160~\ktMWyr{} are
sufficient to find evidence ($3 \sigma$) for CP violation if it is
near maximal. These exposures are consistent with the minimal
requirement of achieving 120~\ktMWyr{} set by P5. No experiment can
provide 100\% coverage of \deltacp values, since CP violation
effects vanish as $\mdeltacp\to 0$ or $\pi$. Higher exposures --- with
more detector mass or higher proton beam power --- will enable high
precision probes of the three-flavor model of neutrino mixing, improving
sensitivities to new effects including the presence of sterile
neutrinos and non-standard interactions.

The DUNE far detector will significantly extend lifetime sensitivity
for specific nucleon decay modes by virtue of its high detection
efficiency relative to water Cherenkov detectors and its low
background rates.  As an example, DUNE has enhanced capability for
detecting the $p\to K^+\overline{\nu}$ channel, where lifetime
predictions from supersymmetric models extend beyond, but remain close
to, the current (preliminary) Super-Kamiokande limit of $\tau/B >
\SI{5.9e33}{year}$ (90\% CL). Supersymmetric GUT models in which
the $p\to K^+\overline{\nu}$ channel mode is dominant also favor
other modes involving kaons in the final state, thus enabling a rich 
program of searches for nucleon decay in the DUNE LArTPC detectors.

In a core-collapse supernova, over 99\% of all gravitational binding
energy of the $1.4 M_{\odot}$ collapsed core --- some 10\% of its rest
mass --- is emitted in neutrinos.  The neutrinos are emitted in a burst
of a few tens of seconds duration, with about half in the first
second. Energies are in the range of a few tens of MeV, and the
luminosity is divided roughly equally between the three known neutrino
flavors.  Compared to existing water Cerenkov detectors, liquid argon
has a unique sensitivity to the electron-neutrino ($\nu_e$) component
of the flux, via the absorption interaction on $^{40}$Ar. The $\nu_e$
component of the flux dominates the very early stages of the
core-collapse, including the ``neutronization'' burst. The observation
of the neutrino signal from a core-collapse supernova in the DUNE
LArTPCs will thus provide unique and unprecedented information on the
mechanics of supernovas, in addition to enabling the search for new
physics. The sensitivity of the DUNE LArTPCs to low energy $\nu_e$
will also enable unique measurements with other astrophysical
neutrinos, such as solar neutrinos.

A highly capable near neutrino detector is required to provide
precision measurements of neutrino interactions, which in the medium
to long term are essential for controlling the systematic
uncertainties in the long-baseline oscillation physics
program. Furthermore, since the near detector data will feature very
large samples of events that are amenable to precision reconstruction
and analysis, they can be exploited for sensitive studies of
electroweak physics and nucleon structure, as well as for searches for
new physics in unexplored regions (heavy sterile neutrinos,
high-$\Delta m^2$ oscillations, light Dark Matter particles, and so
on).

The DUNE experiment is a world-leading international physics
experiment, bringing together the world's neutrino community as well
as leading experts in nucleon decay and particle astrophysics, to
explore key questions at the forefront of particle physics and
astrophysics. The highly capable beam and detectors will enable a
large suite of new physics measurements with potential groundbreaking
discoveries.

\cleardoublepage


\cleardoublepage
\renewcommand{\bibname}{References}
\bibliographystyle{ieeetr}
\bibliography{common/citedb}

\end{document}